\documentclass[final,authoryear,3p,times,twocolumn]{elsarticle}

\usepackage{graphicx}

\usepackage{amssymb}

\def\ltsima{$\; \buildrel < \over \sim \;$}
\def\simlt{\lower.5ex\hbox{\ltsima}}
\def\gtsima{$\; \buildrel > \over \sim \;$}
\def\simgt{\lower.5ex\hbox{\gtsima}}

\journal{New Astronomy Reviews}

\newcommand{\msun}{$M_{\odot}$}

\begin{document}

\begin{frontmatter}



\title{What Drives the Growth of Black Holes?}


\author[dur]{D.~M.~Alexander}
\ead{d.m.alexander@durham.ac.uk}
\author[dur,dart]{R.~C.~Hickox}
\ead{ryan.c.hickox@dartmouth.edu}

\address[dur]{Department of Physics, Durham University, South Road, Durham DH1 3LE, UK}
\address[dart]{Department of Physics and Astronomy, Dartmouth College, 6127 Wilder Laboratory, Hanover, NH 03755, USA}


\begin{abstract}

  Massive black holes (BHs) are at once exotic and yet ubiquitous,
  residing in the centers of massive galaxies in the local
  Universe. Recent years have seen remarkable advances in our
  understanding of how these BHs form and grow over cosmic time,
  during which they are revealed as active galactic nuclei
  (AGN). However, despite decades of research, we still lack a
  coherent picture of the physical drivers of BH growth, the
  connection between the growth of BHs and their host galaxies, the
  role of large-scale environment on the fueling of BHs, and the
  impact of BH-driven outflows on the growth of galaxies. In this
  paper we review our progress in addressing these key issues,
  motivated by the science presented at the ``What Drives the Growth
  of Black Holes?''  workshop held at Durham on 26$^{\rm
    th}$--29$^{\rm th}$ July 2010, and discuss how these questions may
  be tackled with current and future facilities.

\end{abstract}


\end{frontmatter}


\section{Introduction}
\label{}

One of the most astounding astronomical discoveries of the last
$\approx$~10--20~yrs is the finding that massive galaxies in the local
Universe host a central massive black hole (BH; $M_{\rm
  BH}\approx10^5$--$10^{10}$~$M_{\odot}$) with a mass proportional to
that of the galaxy spheroid \citep[e.g.,][]{korm95, mago98, ferr00,
  gebh00, trem02, marc03, gult09msigma}.\footnote{The evidence for
  this BH--spheroid mass relationship comes from tight connection
  between $M_{\rm BH}$ and the velocity dispersion, luminosity, and
  mass of the galaxy spheroid for $\approx$~50--100 galaxies in the
  local Universe. These constraints imply $M_{\rm
    BH}\approx$~(0.001--0.002)~$M_{\rm sph}$.} The tightness of this
BH--spheroid mass relationship suggests a symbiotic connection between
the formation and growth of BHs and galaxy spheroids. Identifying the
physical drivers behind the BH--spheroid mass relationship is one of
the major challenges in extragalactic astrophysics and cosmology.
 
BHs primarily grow through mass accretion during which the central
source is revealed as an active galactic nucleus (AGN;
e.g.,\ \citealt{sal64, lyn69, shak73, solt82, rees84agn}). A huge
amount of energy is liberated during these mass-accretion events
($\epsilon\approx$~0.05--0.42 of the rest mass is converted into
energy, depending on the spin of the BH; e.g.,\ \citealt{kerr63,
  shap83}), yielding large luminosities from comparatively modest
mass-accretion rates:

\begin{equation}
{\dot{m}_{\rm BH}} = 0.15 \Bigl({{0.1}\over{\epsilon}}\Bigr)
\Bigl({L_{\rm bol}\over{10^{45} {\rm erg {s^{-1}}}}}\Bigr) M_{\odot}
yr{^{-1}},
\end{equation}

\noindent where $\epsilon$ is the mass--energy efficiency conversion
(typically estimated to be $\epsilon\approx$~0.1;
e.g.,\ \citealt{marc04smbh, merl04smbh}) and $L_{\rm bol}$ is the
bolometric luminosity of the AGN. For example, the accretion of just
$\approx$~1~$M_{\odot}$~yr$^{-1}$ (equivalent to the mass of the Moon
each second) is sufficient for an AGN to outshine the entire host
galaxy. A broad range of AGN activity is found even in the local
Universe, with mass accretion rates ranging from
$\approx10^{-5}$--1~$M_{\odot}$~yr$^{-1}$. In Fig.~1 we illustrate
some of the diversity in the AGN population within just
$\approx$~300~Mpc; we note here that Sgr~A* is not strictly an AGN but
has been included to demonstrate that even BHs in relatively quiescent
galaxies are growing to some extent.

\begin{figure*}[!t]
\centering
\includegraphics[width=150mm]{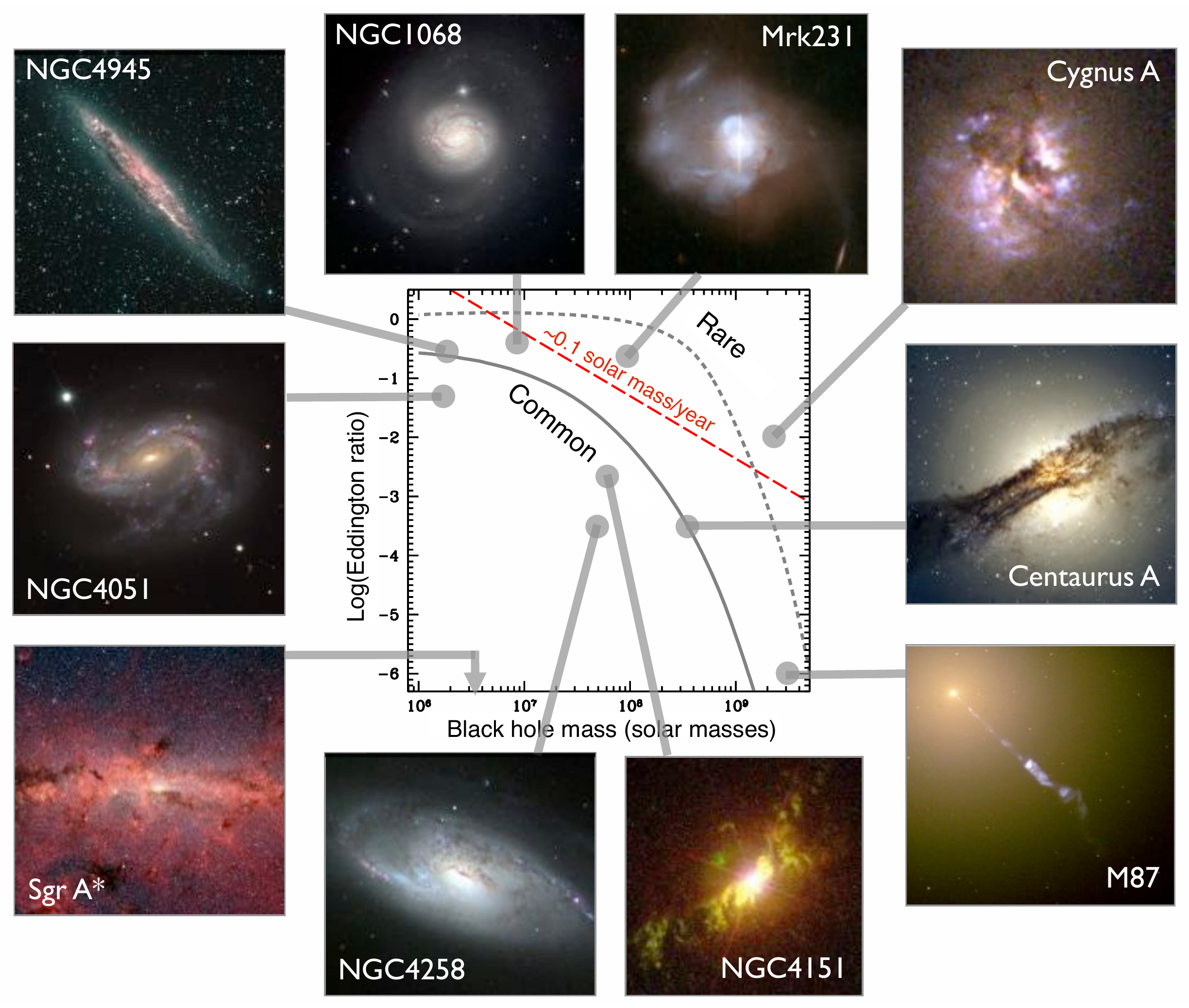}
\caption{The Eddington-ratio versus black-hole mass plane illustrating
  part of the range of mass-accretion phenomena seen in the local
  Universe ($d<300$~Mpc). The plotted systems include radio-loud AGNs
  (Centaurus~A, Cygnus~A, and M~87), an ultra-luminous IR galaxy
  (Mrk~231), as well as typical low-to-moderate luminous AGNs
  (NGC~1068, NGC~4051, NGC~4151, NGC4258, and NGC~4945). All of the
  systems plotted here have reliable black-hole masses that are
  measured independently of the host-galaxy properties. The Eddington
  ratio of Sgr~A\*, the accreting black-hole at the centre of the
  Galaxy, is $\approx10^{-8}$--$10^{-9}$ and cannot be directly
  plotted on this figure. The curves indicate the space density of BHs
  growing at a given Eddington ratio, based on the synthetic AGN model
  of \cite{merl08agnsynth}: $\Phi=10^{-5}$~Mpc$^{-3}$ (solid curve)
  and $\Phi=10^{-7}$~Mpc$^{-3}$ (dotted curve). The dashed line
  indicates a mass-accretion rate of 1~$M_{\odot}$~yr$^{-1}$, which
  corresponds to a bolometric luminosity of $L_{\rm
    bol}\approx7\times10^{44}$~erg~s$^{-1}$, which is equivalent to
  $L_{\rm 2-10 keV}\approx3\times10^{43}$~erg~s$^{-1}$. Systems lying
  below the solid curve are common mass-accretion events in the local
  Universe while systems lying above the dashed curve are rare events
  in the local Universe. Black hole mass and luminosity data are taken
  from the literature \citep{gult09msigma, kawa07bhmass, goul10census,
    baga03sgra, yama09ngc4258, brai04mrk231, youn02cyga, dima03m87}.
  Image credits: {\em NGC 4945}--2P2 Team, WFI, MPG/ESO 2.2-m
  Telescope, La Silla, ESO; {\em NGC 1068}--Francois and Shelley
  Pelletier/Adam Block/NOAO/AURA/NSF; {\em Mrk 231}--NASA, ESA, the
  Hubble Heritage (STScI/AURA)-ESA/Hubble Collaboration, and A. Evans
  (University of Virginia, Charlottesville/NRAO/Stony Brook
  University); {\em Cygnus A}--NASA, ESA, Neal Jackson (Jodrell Bank),
  Robert Fosbury (ESO); {\em Centaurus A}--ESO; {\em M87}--NASA and
  The Hubble Heritage Team (STScI/AURA); {\em NGC 4151}--John
  Hutchings (Dominion Astrophysical Observatory), Bruce Woodgate
  (GSFC/NASA), Mary Beth Kaiser (Johns Hopkins University), Steven
  Kraemer (Catholic University of America), the STIS Team., and NASA;
  {\em Sgr A*}--Susan Stolovy (SSC/Caltech) et al., JPL-Caltech, NASA;
  {\em NGC 4051}--George Seitz/Adam Block/NOAO/AURA/NSF.}
\end{figure*}

A direct connection between the growth of the BH and galaxy spheroid
might be expected since both processes are predominantly driven by a
cold-gas supply, which is provided by the host galaxy or the
larger-scale extragalactic environment. However, the nine orders of
difference in physical size scale between the BH and the galaxy
spheroid would appear to preclude a direct {\it causal}
connection. Many processes have been proposed which could forge a
direct connection between the growth of the BH and galaxy spheroid,
including galaxy major mergers, star-formation winds, and AGN-driven
outflows. There is clear observational evidence for many of these
processes occurring in individual systems. However, it is often far
from clear how universal they are and what impact they have on the
overall BH--spheroid growth. There are also many different
observational and theoretical studies providing apparently
contradictory results on the physical drivers of BH growth, with
theorists often disagreeing with observers (and observers/theorists
disagreeing with other observers/theorists).  A cause of these
discrepancies may be that (apparently) contradicting studies are
exploring systems with different ranges in BH mass, Eddington ratio,
redshift, or environment.

To address these potential conflicts we organised the ``What Drives
the Growth of Black Holes?'' workshop at Durham on 26$^{\rm
  th}$--29$^{\rm th}$ July 2010. This workshop explored the processes
that drive accretion onto BHs, from the most luminous distant quasars
to more quiescent local systems. A key aim of the workshop was to
clarify the ranges of parameter space that are probed by different
studies to better help understand how various key physical processes
vary with these parameters. The workshop was organised into four main
sessions that addressed the following key issues:

\begin{itemize}

\item How does the gas accrete onto black holes, from kilo-parsec to
  sub-parsec scales?

\item What are the links between black-hole growth and their host
  galaxies and large-scale environments?

\item What fuels the rapid growth of the most massive (and also the
  first) black holes?

\item What is the detailed nature of AGN feedback and its effects on
  black-hole fuelling and star formation?

\end{itemize}

One hundred and twenty participants attended across the 3.5 days
duration of the workshop, the vast majority of which presented
scientific results: 53 gave oral presentations [18 invited; 35
  contributed] and 49 gave poster presentations.\footnote{The
  presentations can be found at the workshop web page:
  http://astro.dur.ac.uk/growthofblackholes/index.php} In this paper
we review our progress in addressing these key issues, motivated by
the science presented at the workshop. We also provide background
material on the challenges faced in addressing these issues to
motivate discussion. However, given the large breadth of science
covered by this review, we cannot provide a complete assessment of
each paper in the literature for every component of every issue (see
\citealt{bran05, veil05, remi06bhb, done07disk, mcna07araa, ho08llagn,
  shank09, volo10bhform, bran10} for some recent indepth reviews on
individual components within individual issues).  Our aim is therefore
to provide a solid grounding in the current scientific picture and
hence a starting point for more detailed future investigations.

\section{How does the gas accrete onto black holes, from kilo-parsec to sub-parsec scales?}

The growth of the BH relies on the accretion of cool gas originally on
scales orders of magnitude larger than the BH accretion disc, either
from the host galaxy or the extragalactic environment (e.g.,\ a
companion galaxy; the dark-matter halo; cluster gas). The goal is to
determine what physical processes deliver the gas from
$\approx$~10~kpc host-galaxy scales down to the BH accretion disc at
$r<0.1$~pc -- an epic journey by any stretch of the imagination!
Substantial barriers against the gas reaching the central regions are
angular momentum and the gas collapsing and forming stars rather than
accreting onto the BH. In this section we explore the mechanisms that
can deliver gas into the nuclear region, including the competition for
gas between star formation and AGN activity, the connection between
accreting gas/dust and obscuration, and the process of mass accretion
onto the BH.

\subsection{Driving the gas into the vicinity of the black hole}

The vast difference in size scale between the host galaxy and the BH
means that the gas has to be driven down to $\approx$~10~pc before it
will come under the gravitational influence of the BH. The formidable
force that needs to be overcome to deliver the gas from the host
galaxy into the vicinity of the BH is angular momentum: the gas has to
lose $\approx$~99.9\% of its angular momentum to go from a stable
orbit at $r=$~10~kpc down to $r=$~10~pc
\citep[e.g.,][]{joge06agn}. The challenges are to map the gas inflow
from the host galaxy into the vicinity of the BH and to identify the
triggering mechanisms of AGN activity.

Large-scale gravitational torques, such as those produced by galaxy
bars, galaxy interactions, and galaxy major mergers have the potential
to remove significant amounts of angular momentum and drive the gas
into the central regions of galaxies \citep[e.g.,][]{shlo89bars,
  shlo90fuel, barn92merge, barn96merge, miho96merge, bour02accr,
  garc05agnmol}; see Fig.~2. However, the gravitational torques
exerted on the gas by these large-scale processes will have a limited
effect on sub-kpc scales, and smaller-scale processes (e.g.,\ nested
bars and nuclear spirals; \citealt{engl04bars, maci04spiral}) are
predicted to drive the gas down to $\approx$~10--100~pc.

The physics of gas inflow can be explored in some detail using
multi-scale smoothed-particle hydrodynamical simulations. For example,
adopting this approach \citet{hopk10bhgas} traced the dynamics of gas
inflow over a broad range of scales (from $\approx$~10~kpc down to
$<$~1~pc), taking into account the effects of stars, star formation,
stellar feedback, and the gravitational influence of the BH. Their
simulations revealed a much broader range of gas morphologies than
typically expected, including spirals, rings, clumps, bars within
bars, and nuclear spirals. All of these processes were effective at
driving the gas down to $\approx$~10--100~pc but the gas inflow was
found to be neither smooth nor continuous. Any individual process was
also found to be comparatively short lived, complicating attempts of
relating morphological features to the presence of AGN activity,
particularly on $>100$~pc scales.

\begin{figure}[t]
\centering \includegraphics[width=75mm]{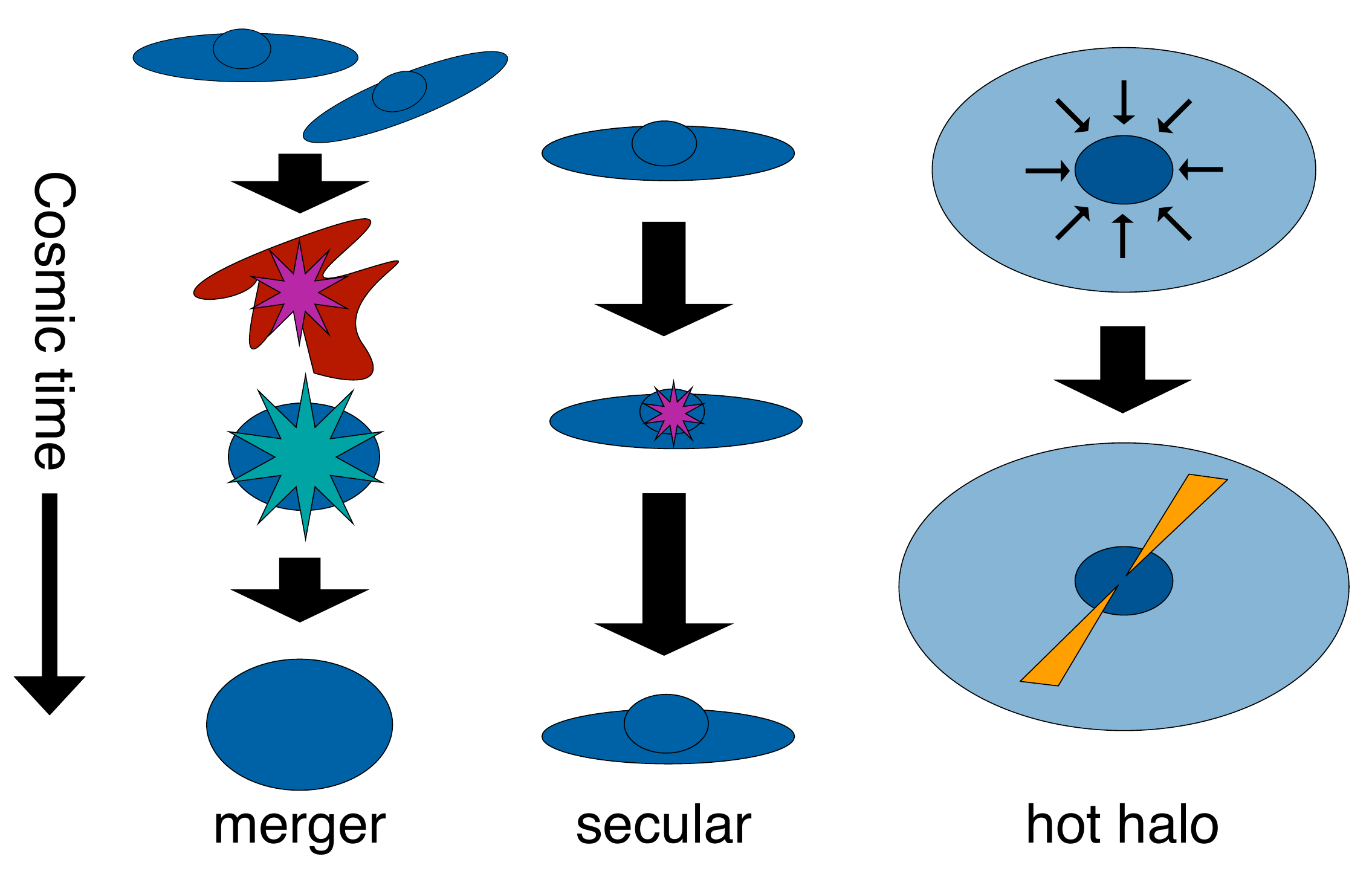}
\caption{Schematic diagrams to illustrate the large-scale processes
  that are thought to be responsible for triggering AGN activity:
  major mergers of gas-rich galaxies, secular evolution (which
  includes both internal secular evolution and external secular
  evolution, the latter of which is driven by galaxy interactions),
  and hot halo accretion, which is presumed to be the dominate BH
  growth mode for low-excitation radio-loud AGNs.}
\end{figure}

Indeed, there is at best a marginal relationship between the presence
of AGN activity and galaxy bars in the local Universe, with AGNs and
galaxies not hosting AGN activity equally likely to host kpc-scale
bars \citep[e.g.,][]{mulc97bar, knap00agn, mart03dust, simo07bhdust};
however, we note that a relationship is found for narrow-line Seyfert
1 galaxies (see \S3.2.3). On smaller $\approx$~100~pc scales,
high-resolution imaging of nearby systems has also failed to reveal
any strong differences between the circumnuclear morphologies of AGNs
and late-type galaxies not hosting AGN activity, with nuclear bars,
spirals, discs, and circumnuclear dust equally likely to be found in
all late-type systems \citep[e.g.,][]{lain02bars, pogg02nucl,
  mart03dust, hunt04seyfhst}. By contrast, a difference in the
circumnuclear regions between AGNs hosted in early type galaxies and
inactive early type galaxies is found, with the AGNs only found in
dusty environments \citep{simo07bhdust}, suggesting that dusty
structures are a necessary but not sufficient condition for BH growth.
Although imaging studies have not revealed a morphological signature
on 100-pc scales that {\it uniquely} reveals the trigger of AGN
activity, triggers of AGN activity and the identification of
significant gas inflow may be more discernable using spectroscopic
data.

Integral Field Unit (IFU) observations provide spatially resolved
spectroscopy and are a key tool to determine the dynamics of gas and
stars in nearby systems. IFU observations of nearby AGNs have revealed
clear differences between the gas and stellar kinematics on $<$~kpc
scales, with the gas velocity fields showing significant departures
from the regular circular orbits traced by the stars
\citep[e.g.,][]{duma07ifu, stok09ifu, rodr11ifu}. The gas kinematics
uncover a broad range of features not revealed by the photometric
data, including warps, counter-rotating discs, and motions out of the
galactic plane, likely due to gas inflow and outflow. However, these
features are also seen in galaxies {\it not} hosting AGN activity,
indicating that the gas inflow that feeds the BH must occur on yet
smaller scales.

IFU observations on $\approx$~10~pc scales of several nearby AGNs are
starting to reveal nuclear gas inflows, when contributions from the
stellar potential are carefully modelled and removed from the ionised
gas kinematics \citep[e.g.,][]{fath06ifu, duma07ifu, stor07ifu,
  riff08ifu, davi09ifu, schn11ifu}. The strongest evidence for a
nuclear gas inflow to date is from NGC~1097, where the gas is seen
streaming along spiral structures towards the nucleus on
$\approx3.5$~pc scales, with inflow velocities of up to
$\approx$~60~km~s$^{-1}$ (e.g.,\ \citealt{fath06ifu, davi09ifu}; see
also \citealt{prie05ngc1097}). However, a significant restriction in
the identification of nuclear-scale gas inflows is that the gas
kinematics of luminous AGNs are often dominated by gas-outflow
signatures (e.g.,\ \citealt{stor10ngc4151}; Hicks et~al. in prep), and
therefore the best constraints to date are limited to low-luminosity
AGNs where the gas-inflow rates are comparatively low.  This
observational challenge can be overcome, in at least some cases, using
spectropolarimetry, which can help distinguish between structures with
different geometries (such as gas inflow from gas outflow) even in
spatially unresolved regions \citep[e.g.,][]{youn00scat,
  smit04seyfpol}. For example, using spectropolarimetry Young
et~al. (in prep) identified small-scale gas inflows in the Seyfert 1
galaxies NGC~4151 and Mrk~509, with inward radial velocities of
$\approx$~900~km~s$^{-1}$ and mass inflow rates of
$\approx$~0.2--0.7~$M_{\odot}$~yr$^{-1}$.

IFU observations have revealed the signatures of gas inflow down to
$\approx$~10~pc scales in several nearby AGNs. However, paradoxically,
the gas inflow rates estimated from the ionised gas are often
significantly lower than the BH accretion rate
\citep[e.g.,][]{fath06ifu, stor07ifu, riff08ifu, davi09ifu,
  schn11ifu}. These discrepancies may indicate that the ionised gas
provides a poor tracer of the inflowing gas rate, which is expected to
be dominated by cold molecular gas \citep[e.g.][]{davi09ifu,
  hick09agnmol, mull09ngc1068}; the ionised gas may be the outer skin
of the molecular gas inflow \citep[e.g.,][]{riff08ifu}. Indeed, the
mass inflow rates estimated using molecular gas tracers in several
nearby AGNs are large enough to power the observed AGN activity
\citep[e.g.,][]{davi09ifu, mull09ngc1068}. The molecular gas
luminosities of AGNs on $<$~100~pc scales are also potentially larger
than those found in galaxies not hosting AGN activity, suggesting a
greater reservoir of cold circumnuclear gas in at least some AGNs
(e.g.,\ \citealt{mull09ngc1068}; Hicks et~al., in prep).

The current concensus view is therefore one where gas inflow from kpc
scales down to the central $\approx$~10--100~pc region occurs in all
gas-rich galaxies. The gas is driven into the central region through a
series of gravitational instabilities, which helps to overcome the
angular momentum of the gas. The clearest triggers of AGN activity in
the local Universe are on the smallest scales, which is perhaps not
surprising given that the gas inflow time from the host galaxy down to
the pc-scale of the BH accretion disc is likely to be
$\simgt10^8$~years, comparable to or longer than the duration of the
AGN phase \citep[e.g.,][]{mart03dust, hopk10bhgas}.

\subsection{The competition between star formation and black-hole accretion}

On scales of order $\approx$~1--10~pc the inflowing gas is expected to
hit a bottleneck. The axisymmetric structures that have driven the gas
from $\approx$~10~kpc down to $\approx$~10~pc are likely to have a
negligible effect on the gas in these inner regions
\citep[e.g.,][]{hopk10bhgas}. It is also on these scales that the
gravitational influence from the BH will overcome that of the host
galaxy; this radius is determined from the BH mass ($M_{\rm BH}$) and
the spheroid velocity dispersion ($\sigma_{\rm sph}$) as

\begin{equation}
r < {{G M_{\rm BH}}\over{\sigma_{\rm sph}^2}}.
\end{equation}

The density of the gas on these scales can be high and if it becomes
unstable then it will collapse and form stars \citep[e.g.,][]{toom64,
  kenn89sflaw, thom05rad}, potentially robbing the BH of the fuel that
it needs to grow. For example, an isothermal gas disc is expected to
become unstable when

\begin{equation}
\Sigma_{\rm g} > {{c_{\rm s} {\kappa}}\over{\pi G}},
\end{equation}

\noindent which is equivalent to $Q<1$ in terms of the Toomre
parameter \citep[e.g.,][]{toom64, thom05rad, mcke07sf}. In this
equation $c_{\rm s}$ is the effective sound speed in the gas, $\kappa$
is the epicyclic frequency, and $\Sigma_{\rm g}$ is the gas surface
density.

These size scales correspond to that expected for (and in some cases
observed; e.g.,\ \citealt{tris07torus, tris09seyf, raba09ngc1068,
  kish11torus}), the dusty molecular torus that is thought to obscure
the central regions of the AGN in inclined systems (e.g.,\ as
postulated in the unified AGN model; \citealt{anto93}). That these
structures should be on similar scales is unlikely to be a coincidence
-- the dense gaseous and dusty structure is potentially both the outer
regions of the BH fuel supply and a stellar nursery. Indeed, the
change from the galaxy potential to the BH potential can produce warps
in the gas that could obscure the accretion disc
\citep[e.g.,][]{hopk10bhgas, lawr10disk}; see also \citet{breg09warp}
for warping of the accretion disc due to dense cusps around the
BH. However, even at these small scales, the gas still has to lose
$\approx$~99\% of its angular momentum before it reaches the sub-pc
scale of the accretion disc.

Numerous studies have shown that young stars are often found in the
central regions of nearby AGNs \citep[e.g.,][]{stor01seyf,
  stor05ngc1097, cidf04llagn}, suggesting a connection (causal or non
causal) between AGN activity and star formation. Using spatially
resolved spectroscopy of the central regions ($\approx$~10--100~pc) of
nine nearby Seyfert galaxies, \citet{davi07agnsf} provided compelling
evidence for a {\it causal} connection between AGN and star-formation
activity in at least some systems (see also
\citealt{gonz01agnsf}). \citet{davi07agnsf} showed that the peak AGN
activity occurs $\approx$~50--200~Myrs after the onset of star
formation (see also \citealt{scha09agn, wild10agnsf}, and
\S3.2.3). The time delay between the star formation and AGN activity
suggests that the star-forming region may provide the fuel for the BH,
via winds from massive stars and supernovae
(e.g.,\ \citealt{voll08torus}; however, see \citealt{hopk11delay}).

These studies imply a complex interplay between AGN activity, star
formation, and stellar winds in the vicinity of the BH. A wide range
of analytical models, numerical simulations, and hydrodynamical
simulations indeed predict that stellar winds and supernovae will
enhance the BH mass accretion by injecting turbulence into the gas
disc \citep[e.g.,][]{wada02agnsf, scha09torus, hobb11bhfeed}. However,
the magnitude of the effect depends on the assumed prescriptions of
the stellar-related outflows and the gas accretion timescales. For
example, if the gas-accretion timescale is long then the majority of
the gas will collapse and form stars rather than be directly accreted
onto the BH \citep[e.g.,][]{king07agnfuel, naya09feed}. In this
scenario the gas accreted onto the BH will be predominantly from
stellar-related outflows (e.g.,\ SNe; AGB stars) and is therefore
triggered after the initial starburst activity. Assuming that the
stellar-mass loss is dominated by AGB stars, the BH accretion--star
formation ratio would be $\approx10^{-2}$--$10^{-3}$
\citep[e.g.,][]{jung01loss}. By contrast, on the basis of their
multi-scale hydrodynamical simulation, \citet{hopk10bhgas} argue that
a precessing eccentric disc will form within $\approx10$~pc of the BH
in gas-rich systems, which can efficiently deliver gas to the
accretion disc. The rate of BH accretion and star formation within the
central 10~pc in the \citet{hopk10bhgas} model is approximately equal,
and therefore $\approx$~2--3 orders of magnitude larger than that
predicted from stellar-mass loss.

While there is disagreement over how the gas is delivered to the
accretion disc from $\approx$~10~pc, it is clear that the gas inflow
is complex and cannot be easily prescribed as a simple equation
\citep[e.g.,][]{hopk10bhgas, powe11bhsim}. A popular approach in the
treatment of BH accretion in galaxy formation simulations is the
Bondi-Hoyle method \citep{bond44, bond52}, which assumes spherical
accretion onto the BH and therefore doesn't consider angular-momentum
limitations \citep[e.g.,][]{dima05qso, spri05merge, boot09bhsim}. To
address this simplification, \citet{powe11bhsim} have produced a
physically self-consistent BH accretion model suitable for galaxy
formation simulations that accounts for the angular momentum of the
gas. The difference between their model and Bondi-Hoyle accretion is
quite striking -- with the Bondi-Hoyle method, the BH is accreting
almost continuously in gas-rich systems at high mass accretion rates
while, by contrast, a sizeable fraction of the gas in the
\citet{powe11bhsim} model collapses to form stars and the mass
accretion rate onto the BH is comparatively negligible.

The effect of stellar-mass loss on the BH accretion can be explored in
even greater detail using small-scale hydrodynamical simulations of a
nuclear starburst in the vicinity of the BH. The current suite of
simulations \citep[e.g.,][]{wada09agnmol, scha09torus, scha10ngc1068}
generically predict that stellar and supernova ejecta inject
turbulence into the accretion disc, ``puffing'' it up and forming an
optically and geometrically thick molecular torus -- a single
supernova can effect the whole torus and enhance the mass-accretion
rate \citep{wada09agnmol}. Radiative heating and pressure can slow
down or stop the accretion process while radiative drag can enhance
accretion. \citet{kawa08disk} show that the fractional amount accreted
decreases with an increase in the fuel supply since star formation in
the disc itself can dominate over the growth of the BH; indeed, the
presence of nuclear stellar rings close to the BH in nearby quiescent
galaxies provide indirect evidence in support of this hypothesis
\citep[e.g.,][]{naya05sfcen, bend05m31, paum06disk, scho07nucl,
  hopk10bhgas}.

The hydrodynamical simulations also predict that the gas can cause
significant time-variable obscuration in the vicinity of the BH
(e.g.,\ \citealt{wada02agnsf, wada09agnmol}; see
\citealt{hick09agnmol} for observational constraints). The predicted
column densities and time-variable obscuration are in good agreement
with those measured in nearby obscured AGNs ($N_{\rm
  H}>10^{22}$--$10^{25}$~cm$^{-2}$ with absorbing column changes on
$<1$~year timescales; e.g.,\ \citealt{risa99, risa02agnnh, guai05,
  capp06seyfx}). Taking the properties derived for the Seyfert 2
galaxy NGC~1068 from \citet{davi07agnsf}, \citet{scha10ngc1068} used a
hydrodynamical simulation to make detailed predictions of the
properties of the obscuring torus (see Table~1 of
\citealt{scha10ngc1068} for the initial parameters). They found that
the torus can be built up through the ejection of energy from stellar
winds. Their model reproduced the dense obscuring structure observed
from interferometric mid-IR observations on $\approx$~0.5--1~pc and
compares well with the extent and mass of the rotating disc inferred
from ${\rm H_2O}$ maser observations \citep[e.g.,][]{gree97ngc1068,
  raba09ngc1068}.

Connections between the accretion onto the BH and obscuration in AGNs
can be explored using X-ray data and mid-IR observations, which will
probe the central source and the obscuring region, respectively.
Using high-spatial mid-IR imaging (typically probing $<100$~pc
scales), \citet{hors08irx, hors09seyfir} and \citet{gand09seyfir}
found a tight correlation between the unresolved mid-IR core and the
absorption-corrected X-ray luminosities of nearby AGNs, even in the
presence of extreme Compton-thick ($N_{\rm H}>10^{24}$~cm$^{-2}$)
absorption (see \citealt{lutz04irx} for similar constraints from
mid-IR spectroscopy). These results indicate that the obscuring
structure is directly heated by the AGN and furthermore they suggest
that the covering factor of the obscuring material is broadly constant
(at a given AGN luminosity; see \citealt[e.g.,][]{lawr91unified,
  simp05agn} for evidence of a luminosity-dependent covering
factor). These results are inconsistent with the simplest dusty torus
models, which assume a smooth distribution of obscuring dust
\citep[e.g.,][]{pier92torus, gran94torus, efst95disk} and therefore
predict that the mid-IR emission will be strongly absorbed in systems
when the torus is inclined to the line of sight. However, these
results are consistent with clumpy torus models
\citep[e.g.,][]{nenk02agndust, elit06torus, nenk08torus, scha08torus,
  honi06clumpy, honi10clumpy}, which predict no inclination-angle
dependence on the strength of the mid-IR emission (see
\citealt{mull11agnsed} for full IR SED constraints that also support
this view); dust in the narrow-line region can also lead to isotropic
dust emission \citep[e.g.,][]{efst95ngc1068, schw08qsodust}.

The results presented in this section show that it is challenging to
accurately predict the fate of the gas on \hbox{$\approx$~1--10~pc}
scales. Generically, the current models predict that the gas inflow on
these scales will lead to a complex interplay between AGN activity,
star formation, and stellar winds. However, different models predict
different rates of mass accretion and star formation. The current view
of the AGN dusty torus is that of a structure more dynamic than that
originally postulated in the unified AGN model
\citep[e.g.,][]{anto93}. The AGN ``torus'' is potentially both the
outer gas reservoir for the BH accretion disc and a nuclear stellar
nursery, built from the gas inflow and stellar-related winds and
outflows.

\subsection{Mass accretion onto the black hole}

The power house behind AGN activity is mass accretion onto the BH on
$\ll1$~pc scales. The basic theory of the accretion disc (the
so-called $\alpha$ disc) has been around for almost 40~years
\citep[e.g.,][]{shak73}. In the case of optically thick accretion,
viscosity (the $\alpha$ parameter, which is a major cause of
uncertainty) causes the gas to lose angular momentum and fall towards
the BH, transporting the angular momentum outwards. The optically
thick accretion disc is geometrically thin and the energy spectrum is
comprised of multi-temperature components, which reach their peak
temperature at the centre \citep[e.g.,][]{prin81disk, rees84agn}. The
local effective temperature approximates to
(e.g.,\ \citealt{good03disc})

\begin{equation}
T_{\rm eff}\approx 6.2\times10^5 \Bigl({{\lambda}\over{\epsilon_{0.1}
    M_{8}}}\Bigr)^{1/4} \Bigl({{r}\over{R_{\rm s}}}\Bigr)^{-3/4} K,
\end{equation}

\noindent where $\lambda$ is the Eddington ratio ($L_{\rm bol}/L_{\rm
  Edd}$; defined below), $\epsilon_{0.1}$ is the radiative efficiency
in units of 0.1, $M_{8}$ is $M_{\rm BH}$ in units of
$10^8$~$M_{\odot}$, $r$ is the radius, and $R_{\rm s}$ is the
Schwarzschild radius (i.e.,\ the event horizon), defined for a
spherical non-rotating BH as

\begin{equation}
R_{\rm s}= {{2 G M_{\rm BH}}\over{c^2}}.
\end{equation}

\noindent The Eddington ratio ($\lambda$~=~$L_{\rm bol}/L_{\rm Edd}$)
indicates the relative growth rate of the BH based on the AGN
bolometric luminosity ($L_{\rm bol}$) and the Eddington luminosity
($L_{\rm Edd}$), defined as

\begin{equation}
L_{\rm Edd} = {{4 {\pi} G {M_{\rm BH}} {m_{\rm p}} c}\over{{\sigma_{\rm T}}}},
\end{equation}

\noindent where $m_{\rm p}$ is the mass of the proton and $\sigma_{\rm
  T}$ is the cross section of the electron. The Eddington luminosity
is achieved when the outward radiation pressure equals the inwards
gravitational force -- if the Eddington luminosity is exceeded then
(in the absence of an additional inward pressure) the gas is expelled;
see Fig.~1 for the Eddington ratios of some nearby AGNs.

An alternative solution to the accretion-disc equations is the
optically thin accretion disc, a state achieved at low mass accretion
rates and generically referred to as Radiatively Inefficient Accretion
Flows (RIAFs; e.g,\ \citealt{nara94adaf, blan99fate}). The optically
thin accretion disc is unable to cool efficiently (the cooling time
exceeds the accretion time) and the energy is lost through
non-radiative processes such as convective transport of energy and
angular momentum to large radii or an outflowing wind. Magnetic fields
are likely to dominate viscosity and the transport of angular momentum
in accretion discs, as suggested by numerical simulations of
Magneto-Rotational Instability (MRI; e.g.,\ \citealt{balb91merge,
  balb98, balb03disk}).

Depending on the accretion disc viscosity and the mass accretion rate,
the accretion disc is predicted to change between an optically thick
and an optically thin state -- the transition from optically thin to
optically thick is likely to correspond to an Eddington ratio of
$\lambda\approx10^{-3}$--$10^{-2}$ \citep[e.g.,][]{esin97nova,
  gall03radiox, macc03bhb}. However, despite the accretion disc being
the ultimate source of power in AGNs, we lack the observational data
to robustly test these models. There are hints that the emission from
AGNs deviates from the simplest accretion-disc models. For example,
micro-lensing of the central parsec region of the accretion disc in a
number of quasars \citep[e.g.,][]{pool07qsolens, floy09qsolens} have
shown that the continuum emission profile drops more quickly that than
expected from the basic $\alpha$ accretion-disc model but is broadly
consistent with that predicted from MRI accretion-disc models
\citep[e.g.,][]{agol00disk}.

Fortunately, significantly better tests of the accretion-disc models
are available for accreting stellar-mass BHs, so called X-ray binaries
(XRBs), for several key reasons: (1) the data is typically of a
significantly higher signal-to-noise ratio than that found for AGNs
since many XRBs lie in the Galaxy, (2) the accretion-disc state of
XRBs can change on timescales of less than a day in response to
changes in the mass accretion rate, allowing for different
accretion-disc components to be identified and analysed; by comparison
the accretion-disc state of AGNs will vary on timescales of centuries
(the timescale is a function of the BH mass;
\citealt{mira98microqso}),\footnote{The X-ray variability of AGN seen
  on intra-day timescales is related to the mass accretion but is
  unrelated to the accretion-disc state \citep[e.g.,][]{ulri97,
    vaug03xvar, uttl05xvar}.} and (3) the accretion disc of XRBs peaks
in the X-ray band ($\approx$~1~keV; $\approx10^7$~K; see Eqn.~4),
where it can be studied in great detail using the current generation
of X-ray telescopes. By comparison the accretion disc of AGNs peaks at
far-UV wavelengths ($\approx$~10~eV; $\approx10^5$~K; see Eqn.~4),
where the emission is absorbed by the interstellar medium in the
Galaxy. Despite these observational challenges in comparing AGNs and
XRBs, there is good evidence that the accretion process is essentially
the same for both small and large BHs and we can therefore use studies
of XRBs to better understand mass accretion in AGNs
\citep[e.g.,][]{merl03bhplane, falc04bhplane, mcha06bhbagn}.

\begin{figure}[t]
\centering \includegraphics[width=75mm]{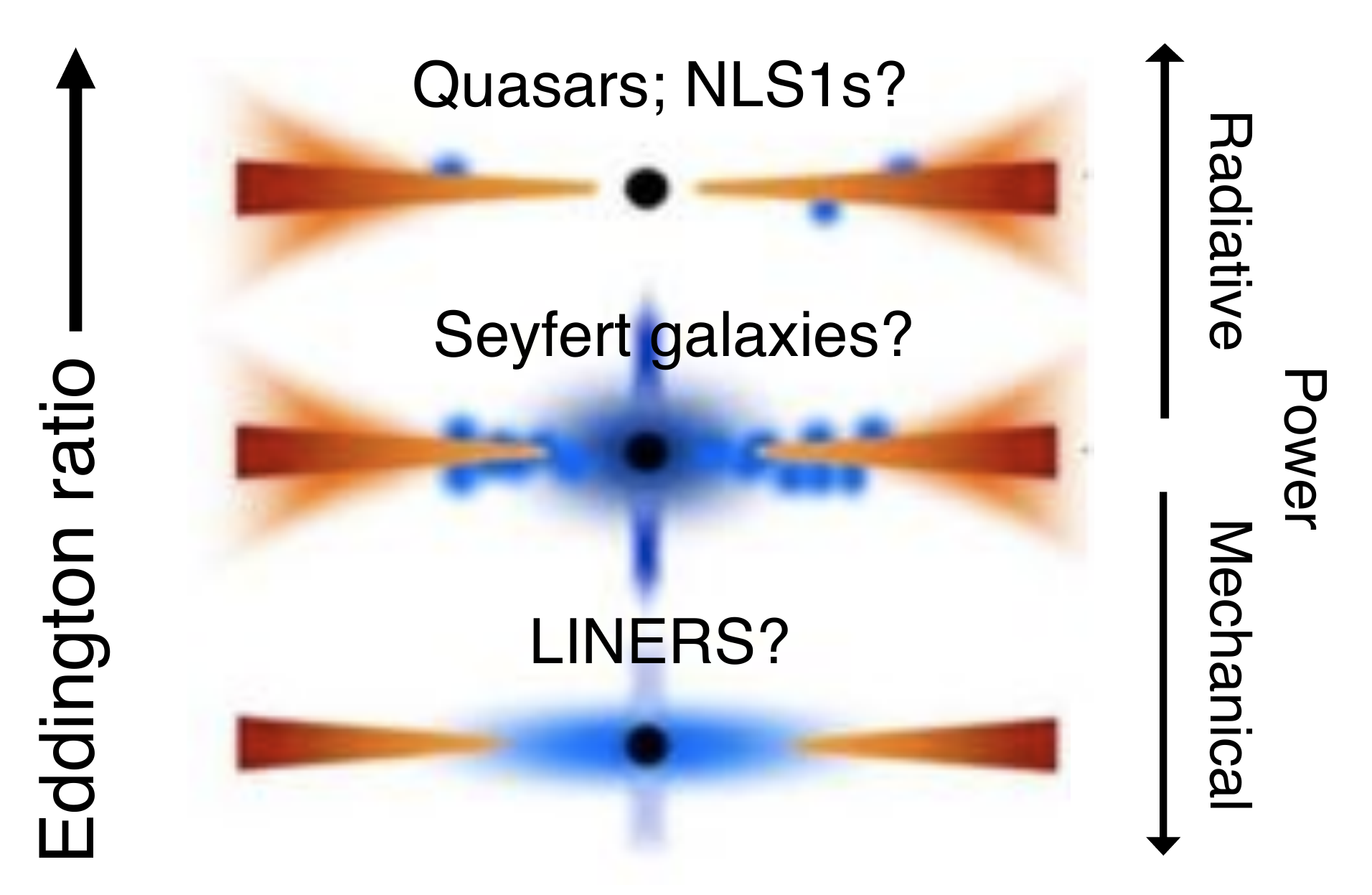}
\caption{Schematic diagrams to illustrate the expected range of BH
  accretion modes as a function of Eddington ratio. The orange-red
  regions refer to the optically thick accretion disc and the blue
  regions refers to the X-ray ``corona''. Adapted from the figure
  produced for X-ray binaries by Done, Gierlinski, \& Kubota (2007)
  and revised to show how the various optical spectroscopic classes of
  AGNs might correspond to the different BH accretion modes and the
  dominate release of energy (radiative or mechanical;
  i.e.,\ predominantly due to jets).}
\end{figure}

\subsubsection{Accretion onto stellar-mass black holes}

The X-ray spectra of XRBs are typically classified into ``X-ray
states'', which qualitatively describe their behaviour and broadly
relate to the properties of the accretion disc (e.g.,\ see Table~2 of
\citealt{remi06bhb}) -- a low-hard state, where the majority of the
X-ray emission is produced at $>10$~keV and a high-soft state, where
the emission is strong at $<10$~keV but weak at $>10$~keV. An
individual XRB can be seen to traverse from one X-ray state to another
on timescales of less than a day in response to changes to the BH
accretion rate \citep[e.g.,][]{nowa95bhb, done07disk}. At high
Eddington ratios ($\lambda\simgt10^{-2}$) the X-ray emission is often
dominated by a disc-like spectrum, assumed to be due to an optically
thick accretion disc, with a weaker power-law continuum component. At
low Eddington ratios ($\lambda\simlt10^{-2}$) the hard X-ray spectrum
is dominated by the power-law continuum and has a weak disc-like
spectrum; however, the conversion from the high-soft to the low-hard
state is more complex than that briefly implied here and should not be
taken as definitive (e.g.,\ see \citealt{remi06bhb} and
\citealt{done07disk} for a more in-depth discussion). In the high-soft
state the accretion disc is optically thick, and therefore viscous,
and emits a thermal spectrum, while in the low-hard state the
accretion disc is optically thin and radiatively inefficient and the
X-ray emission is dominated by the inverse Compton scattering of
photons from the outer regions of the accretion disc (but can also
have significant contributions from synchrotron emission and
synchrotron self-Compton emission).  Further X-ray states have also
been identified, such an intermediate very-high state (see Table~2 of
\citealt{remi06bhb}), which likely correspond to differing
contributions from the accretion disc, inverse Compton scattering, and
an associated jet.

The radio emission from XRBs is found to be linked to the X-ray state
\citep[e.g.,][]{fend04jet}. The radio emission is bright in the
low-hard state and increases in luminosity in response to an increase
in the Eddington ratio up until the low-hard--high-soft state
transition at $\lambda\approx10^{-2}$. The radio emission dramatically
declines when the XRB traverses to the high-soft state
\citep[e.g.,][]{gall03radiox, fend04jet}. The radio emission is
consistent with that expected from a synchrotron emitting jet and the
change in the radio power is possibly due to the scale height of the
accretion disc (which is geometrically thick in the optically thin
state) and provides the seed electrons to drive the
jet.\footnote{Strong radio-emitting XRBs have traditionally been
  called micro-quasars \citep[e.g.,][]{mira92jet, mira94superlum,
    mira98microqso} but it now appears likely that they are just the
  radio-bright subset of the XRB population.}  The jets in XRBs
generate large amounts of mechanical energy which can have a
significant effect on the surrounding interstellar medium.

The most luminous XRBs have X-ray luminosities $>10^{39}$~erg~s$^{-1}$
and are called ultra-luminous X-ray sources (ULXs). ULXs likely
comprise a heterogenous population (see \citealt{king01ulx,
  bege02super, robe07ulx} and \citealt{glad09ulx} for a broader
discussion of the nature of ULXs), of which a sizeable fraction are
stellar-mass BHs accreting at super-Eddington rates. ULXs therefore
provide direct insight into the properties of accretion discs at
extreme mass accretion rates. The X-ray spectra of many ULXs can be
fitted with an accretion-disc spectrum and an optically thick
``corona'' \citep[e.g.,][]{stob06ulx}. However, the identification of
the soft X-ray emission from the accretion disc in the presence of the
large amounts of absorption inferred by the optically thick corona
appears to be contradictory. A more physically plausible explanation
is that the optically thick corona is an outflowing wind, launched in
the vicinity of the accretion disc but more extended than the
accretion disc and therefore not directly obscured. Hydrodynamical
simulations of BHs accreting at super-Eddington rates indeed predict
the presence of a radiatively driven high-column density wind
\citep{ohsu05super, ohsu07agnfeed}, providing indirect support for
this model. The potential impact of these winds can be seen from the
production of ``bubbles'' or emission-line ``nebulae'' over
$\approx$~100~pc scales \citep[e.g.,][]{robe03ulx,
  paku10microqso}. These emission-line nebulae may be analogous to
those seen in AGNs (i.e.,\ the narrow-line region;
e.g.,\ \citealt{wils88seyf, schm96nlr}).

\subsubsection{Accretion onto massive black holes}

The accretion-disc in AGNs peak at far-UV wavelengths, where we lack
direct observational constraints due to absorption by interstellar gas
in the Galaxy. However, the properties of the AGN accretion disc can
be indirectly inferred from the strength of the UV and optical
emission lines. The UV continuum is efficient at photoionising the gas
since the majority of the atomic transitions occur at energies
corresponding to UV wavelengths. On the basis of this, the
low-ionisation emission lines produced by Low-Ionised Nuclear
Emission-line Regions (LINERs; e.g.,\ \citealt{heck80liner,
  bald81bpt}) would correspond to the low-hard power-law dominated
state while the high-ionisation emission lines produced by Seyfert
galaxies and quasars would correspond to the high-soft accretion-disc
dominated state \citep[e.g.,][]{kewl06agn, sobo11agnstate}; see
Fig.~3.

A major source of uncertainty in AGN studies is the mass-accretion
rate, which requires an accurate estimate of the AGN bolometric
luminosity. Since the majority of the accretion-disc emission in AGNs
is produced at far-UV wavelengths, the AGN bolometric luminosity is
typically determined using the continuum emission in a given wave band
($L_{\lambda}$; typically at X-ray or optical wavelengths) with a
wavelength dependent bolometric correction factor ($BC_{\lambda}$)
applied to calculate the total radiative output
\citep[e.g.,][]{elvi94, marc04smbh, hopk07qlf}; i.e.,\

\begin{equation}
L_{\rm bol} = L_{\lambda} BC_{\lambda}.
\end{equation}

The majority of studies assume an optically thick accretion disc when
determining BC. However, as shown for XRBs, an optically thick
accretion disc is only likely to be valid at comparatively high
Eddington ratios ($\lambda\simgt10^{-2}$--$10^{-3}$)
and therefore the mass-accretion rates estimated in studies of
low-Eddington ratio AGNs will be wrong. Indeed by fitting the
X-ray--optical SEDs of nearby AGNs, interpolating at UV wavelengths,
\citet{vasu07bolc, vasu09sed} have shown that the AGN bolometric
correction may be a strong function of the Eddington ratio: the
X-ray--bolometric conversion factors change from $\approx$~15--60 for
Eddington ratios of $\lambda\approx10^{-3}$--1. The
dominant factor in variations in the SEDs and emission-line properties
of quasars may also be due to the Eddington ratio
\citep[e.g.,][]{sule00eigen, kura09redagn}.

Even greater insight into the bolometric output of AGNs can be made by
carefully fitting the X-ray and optical spectra with physically
motivated models. Recent work has focused on fitting the X-ray spectra
of AGNs with an accretion disc, power law, and Compton component while
taking account of the various stellar and AGN components in the
optical spectra to search for connections between the underlying
physical processes \citep[e.g.,][]{jin09super, sobo11agnstate}. Hutton
et~al. (in prep) have taken this approach for AGNs identified in the
SDSS \citep[e.g.,][]{york00sdss, abaz09sdss} with excellent-quality
X-ray data obtained from the {\it XMM-Newton} serendipity survey
\citep[e.g.,][]{wats09twoxmm}. A wide diversity of X-ray--optical SEDs
are found, which would otherwise not be apparent from the X-ray data
alone. There is some evidence for correlations of the strength of the
X-ray emitting components and the width of the H$\beta$ emission line.
However, since AGNs vary significantly in the X-ray band, contemporous
X-ray and UV-optical data are required to account for intrinsic
variations \citep[e.g.,][]{vasu09sed}.

\subsubsection{Sgr~A*: the closest accreting massive black hole}

The closest massive BH to us is Sgr~A*, which lies at a distance of
$\approx$~8.3~kpc in the centre of the Galaxy
\citep[e.g.,][]{reid93sgra, meli01sgra, gill09sgra}. The BH mass of
Sgr~A* has been measured to unprecedented accuracy using
$\approx$~16~yrs of monitoring of the orbits of stars in the vicinity
of Sgr~A*: $M_{\rm BH}=(4.3\pm0.2)\times10^6$~$M_{\odot}$
(\citealt{gill09sgra}; see also \citealt{ghez08sgra}). Sgr~A* produces
emission over a broad waveband (radio--X-ray;
e.g.,\ \citealt{marr08sgra, dodd11sgra}. However, with an estimated
Eddington ratio of $\lambda\approx10^{-8}$--$10^{10}$
\citep[e.g.,][]{meli01sgra, ecka06sgra}, Sgr~A* is $\approx$~5--8
orders of magnitude below the optically thick--optically thin
accretion state and is only identifiable as ``active'' due to its
proximity. For example, none of the AGNs and LINERs investigated in
the nearby galaxy sample of \citep{ho97agnhost} have Eddington ratios
of $\lambda<10^{-8}$ (see Fig.~9 of \citealt{ho08llagn}),
indicating that Sgr~A* is in a quiescent state. Sgr~A* therefore
provides unique insight into low mass-accretion events onto a massive
BH.

The broad-band emission from Sgr~A* appears to have at least two
components \citep[e.g.,][]{dodd11sgra} -- a continuously emitting
``quiescent'' state and a flaring state. The flares are often
quasi-periodic and can vary on short timescales
\citep[e.g,.][]{baga01sgra, genz03sgra, ghez04flare, marr08sgra,
  dodd11sgra}; the radio--near-IR spectral energy distribution of the
flares observed in Sgr~A* are consistent with synchrotron emission
from transiently heated electrons \citep[e.g.,][]{mark01flare,
  gill06sgra, trip07sgra, ecka09sgra, dodd10flare, dodd11sgra}. The
origin of the flaring emission is a topic of hot debate and may be due
to an orbiting hot spot, magnetic heating/reconnection, stochastic
acceleration processes, or accretion-rate enhancements
\citep[e.g.,][]{mark01flare, liu02sgra, yuan03sgra, dodd11sgra}. The
modest BH accretion rates inferred by the emission from Sgr~A* are
consistent with that expected from the mass loss of stars in the
Galactic centre \citep[e.g.,][]{cuad06sgrawind, cuad08sgrawind}.

On the basis of the current mass accretion rate onto Sgr~A*, it would
take $\approx$~5--6 orders of magnitude longer than the Hubble time to
grow the BH to its current mass, indicating that Sgr~A* must have been
orders of magnitude more active in the past. The identification of
reflected X-ray emission off molecular clouds in the vicinity of
Sgr~A* provides a potential ``X-ray echo'' of past activity from
Sgr~A*. These data suggest that Sgr~A* may have been up-to $\approx$~3
orders of magnitude brighter $\approx$~100 years ago
\citep[e.g.,][]{muno07refl, pont10refl}. However, these implied mass
accretion rates are still extremely modest and Sgr~A* must have been
much more luminous in the past \citep{naya07sgra}; see
\citet{mici11mbh} for potential BH growth histories for Sgr~A*.

\section{What are the links between black-hole growth and their host galaxies and large-scale environments?}

In the previous section we discussed the mechanics and processes of
gas inflow from the host galaxy down to the central BH, across $>$~5
orders of magnitude in size scale. The BH--spheroid mass relationship
implies that this, otherwise seemingly unlikely journey of gas inflow,
has occured in all spheroid-hosting galaxies at some point over the
last $\approx$~13~Gyrs. The tightness of the BH--spheroid mass
relationship further suggests a connection between AGN activity and
star formation over kpc scales. In this section we explore the
triggering mechanisms and sites of AGN activity and investigate the
connection between AGN activity, star formation, and large-scale
environment. We also investigate the growth of BHs across cosmic time
and explore how the properties and triggering of distant AGN activity
differs from that found in the local Universe.

\subsection{The identification of AGN activity}

To accurately explore when, where, and how the growth of BHs has been
triggered requires the identification of AGN activity across all
environments and redshifts. The presence of dust and gas in the
vicinity of the BH and star-forming regions along the line of sight
means that penetrating observations are required to unambiguously
reveal the signatures of AGN activity in all systems. The emission
from star formation and starlight in the host galaxy can also dilute
and mask the emission from the AGN. These complications make the
construction of a complete census of AGN activity a significant
challenge.

The most extensive studies of AGN activity in the local Universe have
utilised optical spectroscopy \citep[e.g.,][]{huch92agn, maio95,
  ho97agnhost, ho97agnbroad, ho97agn, kauf03host, heck04bh,
  ho08llagn}. Since the AGN narrow-line region extends beyond the
obscured central region ($\approx$~10--1000~pc versus
$\approx$~1--10~pc), optical spectroscopy is able to identify even
heavily obscured AGNs, so long as the AGN narrow emission lines are
not obscured by dust in the host galaxy. Sensitive radio, mid-IR
spectroscopy, and X-ray observations (particularly at $>10$~keV;
e.g.,\ \citealt{sazo07agn, wint09batagn, burl11batagn}) further extend
our understanding and census of nearby AGN activity, providing the
identification of AGNs even in the presence of significant host-galaxy
dust and star formation \citep[e.g.,][]{ho01agnx, ulve02agn,
  filh06radio, saty08irs,goul09irs, zhang09census, wint10batagn,
  nard10ulirg}.  From a combination of these multi-wavelength
identification techniques, we are now close to a complete census of
AGN activity in the local Universe down to a given sensitivity limit
(i.e.,\ lower-luminosity AGNs can still be unidentified). However,
even intrinsically luminous AGNs can remain unidentified if the
obscuration towards the BH is extremely high and the central region is
heavily obscured by dust in the host galaxy (see the conflicting
evidence for Arp~220; \citealt{clem02arp220, spoo04arp220,
  iwas05arp220, down07arp220}). It is important to bear in mind these
potential limitations.

Optical spectroscopy is also effective at identifying distant obscured
AGNs \citep[e.g.,][]{stei02agn, poll06, alex08compthick, june11agn,
  yan11aegis}. However, since many of the key AGN emission-line
diagnostics move into the near-IR band at $z\simgt$~0.4, optical
spectroscopy alone can be quite limited in identifying large numbers
of AGNs (but see \citealt{june11agn, yan11aegis, trou11optx} for
techniques that extend the utility of optical spectroscopy for AGN
identification out to $z\approx$~1).\footnote{The optical signatures
  of distant AGNs are also more easily diluted from host-galaxy
  emission than nearby AGNs due to the larger angular-size distance
  for distant systems and the overall increase in star-formation
  activity at higher redshift \citep[e.g.,][]{noes07, dadd07sf,
    pann09sf, rodi10sf, elba11ms}.}  Currently the most efficient and
effective identification of distant AGNs is made with X-ray
observations, which can select AGNs almost irrespective of the
presence of obscuration \citep[e.g.,][]{baue04, tozz06, gill07cxb}. X-ray
observations are effective at identifying AGN activity because the
X-ray emission from star formation is typically weak; the positive K
correction at X-ray energies for distant obscured AGNs also means that
$<10$~keV observations are particularly effective at identifying
heavily obscured AGNs at high redshift \citep[e.g.,][]{tozz06,
  alex11stack, coma11xmmdeep, feru11kalpha}. Sensitive IR and radio
observations can further extend the census of distant AGN activity by
identifying AGNs where the absorbing column densities are so high that
not even X-ray photons can escape \citep[e.g.,][]{lacy04, ster05,
  donl05radio, donl08spitz, alon06, dadd07comp, hick07abs, fior08agn,
  fior09obsc, baue10obscagn, luo11obsc}. However, since star
formation can also produce luminous IR and radio emission, it is often
challenging to disentangle the AGN emission from that of the host
galaxy \citep[e.g.,][]{poll07agnsed, mull11agnsed}.

Differences in the various approaches in AGN identification can also
make it challenging to reconcile results from different studies. It is
therefore important to always consider how obscuration or host-galaxy
emission can effect the completeness of any AGN selection technique.
In the following sub sections we explore the processes of AGN activity
using observations at X-ray, optical, IR, and radio wavelengths. We
investigate the ubiquity of AGN activity, the AGN triggering
mechanisms, the connection between AGN activity and star formation,
the evolution of AGN activity with redshift, and the role of
environment in the triggering of AGN activity.

\subsection{AGN activity in the local Universe}

\subsubsection{Where are the massive black holes growing?} 

AGN activity is common in the local Universe, with $\approx$~5--10\%
of nearby galaxies found to clearly host optical AGN activity
(i.e.,\ identified as Seyfert galaxies from their optical
emission-line properties; e.g., \citealt{veil87line, kewl01opt,
  kewl06agn, maio95, ho97agnhost, hao05agn}. The optical AGN fraction
can be significantly higher if galaxies with Low-Ionisation Nuclear
Emission Regions (LINERs; \citealt{heck80liner}) are also taken into
account \citep[e.g.,][]{ho97agnhost}. However, LINERs comprise a
heterogenous mix of composite AGN--star-forming galaxies, optically
obscured AGNs, low-luminosity AGNs, and early type galaxies without
AGN activity \citep[e.g.,][]{bine94photoion,
  ho01agnx,ho03line,erac02liner, saty04liner, kewl06agn, shie07line,
  gonz09linerx,goul09irs,sarz10sauron,cape11liner}, leaving some
ambiguity over the true AGN fraction in the LINER population.  Mid-IR
spectroscopy can provide a more complete census of nearby AGN activity
than that obtained from optical spectroscopy due to the identification
of high-excitation emission lines (e.g.,\ [Ne~V]$\lambda$14.3~$\mu$m)
in systems where the narrow-line region is obscured by dust in the
host galaxy \citep[e.g.,][]{saty08irs, goul09irs, toma10irs,
  petr11irs}. However, at present, the number statistics obtained from
high-resolution mid-IR spectroscopy are poor compared to those
achieved from optical spectroscopy (samples of tens rather than
thousands of objects).

\begin{figure}[t]
\centering
\includegraphics[width=75mm]{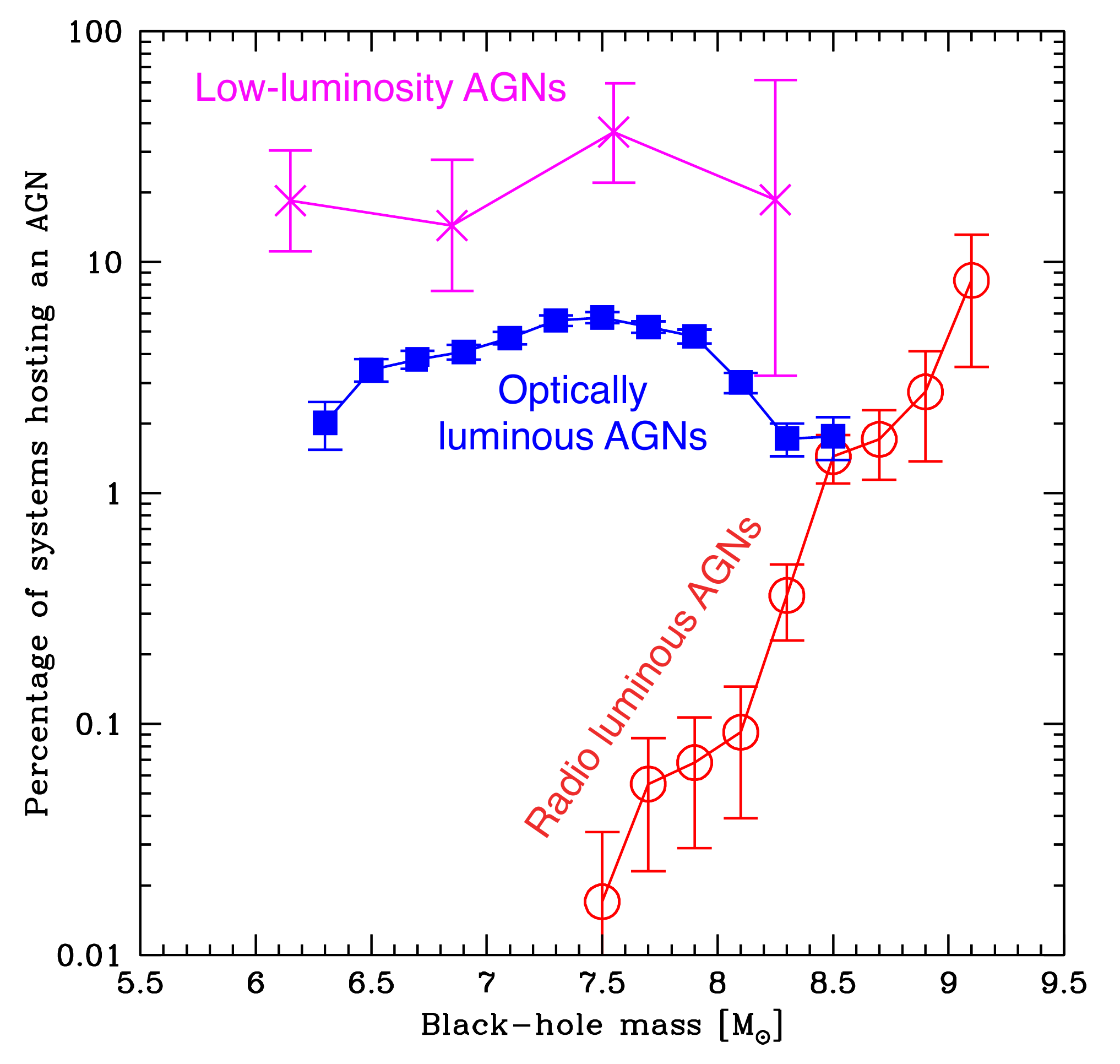}
\caption{AGN fraction as a function of BH mass for different AGN
  selection approaches: optically identified luminous AGNs (filled
  squares; Best et~al. 2005), lower-luminosity AGNs (crosses; Goulding
  et~al. 2010), and radio-luminous AGNs (open circles; Best
  et~al. 2005).}
\end{figure}

The fraction of galaxies hosting AGN activity provides a measure of
the duty cycle of BH growth. The fraction of local galaxies hosting
AGN activity as a function of BH mass is shown in Fig.~4.  The
fraction of galaxies hosting optically luminous AGN activity ($L_{\rm
  [O {\sc III}]}>3\times10^7$~$L_{\odot}$, equivalent to a 2--10~keV
luminosity of $L_{\rm X}\simgt10^{43}$~erg~s$^{-1}$, assuming the
conversion of \citealt{mulc94torus}) is approximately flat at
$\approx$~5\% over $M_{\rm
  BH}\approx$~(0.3--10)~$\times10^7$~$M_{\odot}$ and drops
substantially to lower and higher BH masses
\citep[e.g.,][]{best05}. The AGN fraction rises towards lower AGN
luminosities, as expected since there will always be more
low-accretion rate systems than high-accretion rate systems. For
example, the fraction of galaxies hosting AGN activity $\approx$~3
orders of magnitude fainter than the optically luminous AGNs is
$\approx$~20--30\% over $M_{\rm
  BH}\approx$~(0.1--10)~$\times10^7$~$M_{\odot}$
\citep[e.g.,][]{goul10census}.  Note that Sgr~A*, the active BH at the
centre of the Galaxy is still $\approx$~4 orders of magnitude fainter
than these sensitivity limits! By contrast, the fraction of galaxies
hosting luminous radio AGN activity ($L_{\rm 1.4
  GHz}>10^{24}$~W~Hz$^{-1}$) is strongly dependent on the BH mass,
rising from $\simlt0.1$~\% at $M_{\rm BH}<10^8$~$M_{\odot}$ (where
optical and mid-IR identified AGNs are common) to $\approx10$\% at
$M_{\rm BH}>10^9$~$M_{\odot}$ \citep[e.g.,][]{best05}. Luminous radio
AGNs represent the minority of the AGN population at all redshifts
(radio-loud AGNs account for $\simlt$~10\% of the total AGN energy
budget and cosmological BH growth) but they are likely to have played
a crucial role in the evolution of the most massive galaxies and BHs
(e.g., \citealt{bowe06gal, crot06, merl08agnsynth}; see
\S5.1).\footnote{Radio-loud AGNs are typically identified on the basis
  of luminous radio emission (e.g.,\ $L_{\rm 1.4
    GHz}\simgt10^{24}$~W~Hz$^{-1}$) which is not due to star-formation
  activity. By comparison, the term ``radio-quiet AGNs'' refers to the
  majority of the X-ray, optical, and IR selected AGN populations,
  which are not luminous at radio wavelengths.}

These suite of results show that mass accretion onto the BH occurs
relatively frequently in nearby galaxies. However, the AGN fraction
does not reveal how quickly the BHs are growing, which requires a
measurement of the mass accretion rate and Eddington ratio (see Eqn.~1
\& 6 and Fig.~1). Using a broad range of methods to estimate BH masses
and adopting the [O~{\sc iii}]$\lambda5007$ or [O~{\sc
    iv}]$\lambda25.9$~$\mu$m luminosity as a proxy for the
mass-accretion rate, the majority of AGNs in the local Universe have
Eddington ratios of $\lambda\approx10^{-6}$--$10^{-3}$, with a small
tail towards higher Eddington ratios (see Fig.~9 of
\citealt{ho08llagn, goul10census}; see also Fig.~1). Since the growth
time (i.e.,\ the time to double in mass) of a BH with an Eddington
ratio of $\lambda\approx10^{-3}$ is comparable to the age of the
Universe (see Eqn.~10 in \S4.3), these range of Eddington ratios are
modest and suggest that the majority of the massive BHs in the local
Universe must have grown more rapidly in the past. However, these
results can also be a bit misleading since there is a small but very
significant tail of AGNs with much higher Eddington ratios. Indeed,
$\approx$~50\% of the mass accretion onto BHs in the local Universe is
found to occurred at high Eddington ratios of $\lambda>0.1$, despite
these systems only comprising $\approx$~0.2\% of the optically
identified AGN population \citep{heck04bh}! These results therefore
suggest that, while the vast majority of the massive BHs in the local
Universe are growing slowly (potentially in the optically thin
accretion mode), at least half of the mass accretion in the local
Universe occurs is the optically thick accretion-disc mode (see \S2.3
and Fig.~3).

AGN activity is transient and therefore the measured Eddington ratio
of an individual object only provides a snapshot of its growth
path. We clearly cannot watch an individual BH grow over Myrs of time
to determine its time-average growth but we can measure the
volume-average Eddington ratio of a group of systems, which will
provide an {\it average} (and therefore typical) BH growth rate. On
the basis of optical spectroscopy from the SDSS, \citet{heck04bh}
showed that the volume-average Eddington ratio for optically
identified AGNs in the local Universe is inversely proportional to the
BH mass; see Fig.~1 for a graphical representation of this result. The
volume-average BH growth time for the lowest-mass BHs ($M_{\rm
  BH}\approx10^7$~$M_{\odot}$) is comparable to the age of the
Universe, implying that their BHs have been growing steadily over
cosmic time; \citet{goul10census} extended this work to lower BH
masses and showed that the most rapidly growing BHs in the local
Universe reside in late-type galaxies with low-mass BHs ($M_{\rm
  BH}\approx10^6$--$10^7$~$M_{\odot}$), many of which lack the clear
optical signatures of AGN activity. By comparison, more massive BHs
with $M_{\rm BH}\simgt10^8$~$M_{\odot}$ are typically growing
$\approx$~1--3 orders of magnitude more slowly than the lower-mass BHs
and must have undergone their dominant growth phases at higher
redshifts. \citet{kauf09modes} extended these results by providing
evidence for distinct regimes of BH fuelling in the local
Universe. They showed that optically identified AGNs with significant
star formation in the central kpc have a lognormal distribution of
Eddington ratios (with the peak at an Eddington ratio of
$\lambda\approx10^{-2}$) while more quiescent AGNs were found to have a
power-law distribution of Eddington ratios, where the BH accretion
rate depends on the stellar spheroid mass and the ages of the stars in
the bulge. The active and quiescent systems are broadly split on the
basis of BH mass: the active systems have $M_{\rm
  BH}\simlt10^8$~$M_{\odot}$ while the quiescent systems have $M_{\rm
  BH}\simgt10^8$~$M_{\odot}$. On the basis of these results,
\citet{kauf09modes} argued that the BHs in the active systems are
driven by a plentiful supply of cold gas while the BHs in the more
massive quiescent systems are driven by stellar-mass loss in evolved
stars, having presumably consumed their cold-gas supplies at higher
redshift.

\subsubsection{Is black hole and galaxy growth connected?}

The BH--spheroid mass relationship in the local Universe suggests that
AGN activity should be found in spheroid-hosting galaxies. Indeed, the
majority of optically identified AGNs reside in comparatively early
type galaxies (E--Sb galaxy morphologies; e.g.,\ \citealt{ho08llagn,
  gado09bhbulge}). However, a substantial fraction of the AGN
population is also hosted in late-type galaxies (Sc--Sm). This was a
somewhat unexpected result since late-type galaxies host pseudo bulges
rather than classical spheroids \citep[e.g.,][]{korm04pseudo}, and are
therefore not expected to typically host massive BHs (since the
BH--spheroid mass relationship is defined for galaxies with classical
spheroids). Focused investigations have indeed shown that there are no
significant correlations between BH mass and pseudo bulge luminosity
\citep[e.g.,][]{gree08lowmass, jian11lowmass, korm11pseudo}.  The
identification of growing BHs in late-type galaxies therefore suggests
a different growth path to systems that host classical spheroids; for
example, it is possible that these late-type systems have never
undergone a galaxy major merger, and a non-negligible fraction of the
BH mass may be from the massive BH seed formation
(e.g.,\ \citealt{korm04pseudo, volo09lowmass}; see \S4.3 for the
discussion of the formation of massive BHs). AGNs hosted in late-type
galaxies are often not revealed at optical wavelengths and mid-IR
spectroscopy or hard X-ray observations are required to unambiguously
identify them \citep[e.g.,][]{ho08llagn, saty08irs, goul09irs,
  koss11bathost}: the presence of dust and star formation in the host
galaxy often dilute or extinguish the optical AGN signatures.

The host-galaxy properties of radio-loud AGNs are connected to their
optical spectral properties. Radio-loud AGNs with low-excitation
optical spectra (i.e.,\ LINER classifications) are mostly found in
massive red early type galaxies, while radio-loud AGNs with
high-excitation optical spectra (i.e.,\ Seyfert and quasar
classifications) are typically hosted in less massive gas-rich
galaxies \citep[e.g.,][]{will01radiohost, chia05liner, evans2006,
  hard06radio,hard07radio, bald08radiohost, kauf08radio, smol09radio,
  smol09radioevol,herb10radio,smol11radiomol}. To first order the
high-excitation radio-loud AGNs are high-accretion rate systems and
are believed to evole strongly with redshift, a behaviour that would be
similar to radio-quiet Seyfert galaxies
and quasars; see \S3.3.1. By constrast, low-excitation radio-loud AGNs
are low-accretion rate systems, evolve slowly with redshift, and are
the dominant radio-galaxy population in the local Universe
\citep[e.g.,][]{sadl07radio, smol09radioevol}. The low-excitation
systems are the dominant radio-loud AGN population at modest radio
luminosities ($L_{\rm 1.4 GHz}\simlt3\times10^{25}$~W~Hz$^{-1}$) and
also dominate the global energetic output from radio galaxies
\citep[e.g.,][]{hard07radio}; these systems are typically classified
as FR~I radio galaxies \citep{fana74} and are important for the
``radio-mode'' AGN feedback (e.g.,\ \citealt{merl08agnsynth}; see
\S5.1). The high-excitation systems are typically classified as FR~II
radio galaxies and typically have the highest radio luminosities but are
comparatively rare at all redshifts.

The tightness of the BH--spheroid mass relationship suggests that
galaxies hosting classical spheroids grew in concert with their BHs
(i.e.,\ for every $\approx$~1000 units of star formation there is
$\approx$~1--2 units of BH accretion). As shown in \S2.2, a direct
link between star formation and AGN activity on $\approx$~10~pc scales
may be expected given that both processes are driven by the
availability of cold gas on nuclear scales. However, a connection
between AGN activity and star formation on the kpc scales of the
galaxy spheroid is not expected {\it a priori} since the vast
difference in size scale would preclude a direct causal link. Indeed,
although the host galaxies of many AGNs are undergoing star-formation
activity, there is large scatter in the observed AGN:star formation
ratio
\citep[e.g.,][]{netz07qsosf,wild07agnsf,baum10agnsf,diam11agnsf_aph}. The
scatter decreases when the star-formation rate is measured over
$<$~1~kpc scales \citep[e.g.,][]{wild07agnsf,diam11agnsf_aph}, and is
predicted to decrease yet further at smaller spatial scales, where a
causal connection is expected between the gas that drives star
formation and the gas that fuels the BH
\citep[e.g.,][]{hopk10bhgas}. Despite this, using optical spectroscopy
from the SDSS, \citet{heck04bh} showed that the {\it volume-average}
galaxy--BH growth rate is broadly consistent with that expected from
the BH--spheroid mass relations (a factor of $\approx$~1000) across
kpc scales (i.e.,\ the typical size scales corresponding to the SDSS
fibres), for a wide range of BH masses (see Fig.~5 of
\citealt{heck04bh}).

The fraction of galaxies hosting AGN activity is correlated to the IR
luminosity (8--1000~$\mu$m) or star-formation rate, indicating a
connection between the duty cycle of BH growth and the growth of the
galaxy. On the basis of optical spectroscopy alone, the AGN fraction
rises from $\approx$~5\% for galaxies with $L_{\rm
  IR}<10^{11}$~$L_{\odot}$ to $\approx$~15\% for Luminous IR Galaxies
(LIRGs; $L_{\rm IR}\approx10^{11}$--$10^{12}$~$L_{\odot}$) and
$\approx$~25\% for Ultraluminous IR Galaxies (ULIRGs; $L_{\rm
  IR}>10^{12}$~$L_{\odot}$; e.g.,\ \citealt{kim98ulirg, veil99ulirg,
  tran01ulirg}). However, optical spectroscopy does not identify all
of the AGNs and these AGN fractions increase to $\approx$~50--80\% in
the LIRG and ULIRG populations when including mid-IR spectroscopy and
X-ray data for the identification of AGNs
\citep[e.g.,][]{alex08bhmass,
  lehm10sfx,nard10ulirg,alon11lirg,nard11comp}. These results show
that the BH grows almost continously during periods of intense star
formation.

\subsubsection{How is AGN activity triggered?} 

Central to our understanding of the growth of BHs is the determination
of the large-scale physical mechanisms that trigger the gas inflow
towards the BH -- either external (galaxy mergers or interactions) or
internal (gas instabilties, galaxy bars, etc; see \S2.1 and Fig.~2)
processes. Internal processes and galaxy interactions are often
referred to as ``secular evolution'' and can be further divided into
internal secular evolution and external secular evolution,
respectively (see Fig.~1 of \citealt{korm04pseudo}). An accurate
determination of the fraction of AGN activity driven by these
different processes is non trivial since it depends on the depth of
the data, the spatial resolution of the data, the assumptions used on
how the observed data relates to the triggering mechanism, and how the
AGN activity is identified. Furthermore, given the gas-inflow times,
there can be a significant delay between the initial triggering event
and the onset of AGN activity ($\approx$~50--500~Myrs;
e.g.,\ \citealt{davi07agnsf, scha09agn, wild10agnsf}); see
\S2.1--2.2. These complications mean that it can be difficult
reconciling the results from one study with the results form another
study, let alone accurately measuring the true fraction of AGN
activity that is triggered by external or internal processes.

For example, even in the local Universe there are large differences
between published studies. Using morphological classifications of SDSS
galaxies from the Galaxy Zoo project \citep{lint11zoo},
\citet{darg10mergeanal} estimated that only $\approx$~1\% of AGNs
reside in systems clearly undergoing a major merger. By comparison, on
the basis of a hard X-ray selected AGN sample (from the {\it
  Swift}-BAT survey; \citealt{tuel08bat,tuel10bat}),
\citet{koss10batagn} estimated the rate of major mergers to be
$\approx$~20\%; the fraction rises to $\approx$~50\% if galaxy
interactions are also included. Part of the discrepancy in the
estimated rate of major mergers between these studies may be due to
the AGN identification method, since the {\it Swift}-BAT AGN survey
finds AGNs that lack unambiguous optical AGN signatures
\citep[e.g.,][]{wint09batagn} and may probe higher intrinsic
luminosities than the optical samples \citep{koss11bathost}. Still, a
factor $\approx$~20 difference seems difficult to reconcile from the
AGN selection method alone. In another example, \citet{liu11agnpair}
looked at the frequency of AGN pairs within $\approx$~5--100~kpc in
the SDSS and found that $\approx$~4\% of optically identified AGNs are
found in pairs. However, only $\approx$~30\% of these AGN pairs
clearly showed the morphological features of a major merger, even
though the host galaxies appear to be in the process of a galaxy
merger or interaction. The depth of the imaging data used to identify
the morphological signatures of major mergers can clearly have a large
effect on the results. Using deep imaging data, \citet{scha10merge}
identified the faint morphological signatures of galaxy major mergers
in $>50$\% of blue early type galaxies (these weak features often
referred to as ``shells''; \citealt{mali83shell, hern92shell,
  turn99shell}), which were not apparent in shallower data.

Spectroscopy can provide an alternative route to estimating the
fraction of systems in major mergers by identifying systems with
evidence for close-pair AGN activity. The identification of double
peaked emission lines (broad or narrow) with significant velocity
offset can indicate a binary system or a BH gravitational recoil
\citep[e.g.,][]{volo07recoil, komo08recoil, colp09binary_aph,
  dott09binary, civa10recoil, robi10recoil, smith10binary,
  rosa10dualagn, barr11dualagn}; however, see \citet{dott10agnpair}
and \citet{shen11doubleagn} for alternative interpretations of the
presence of double-peaked emission lines. On the basis of the
incidence of double-peaked emission lines, \citet{come09binary}
estimated a $\approx$~30\% merger fraction, similar to the upper end
of merger fractions estimated from imaging data.

These suite of studies highlight the difficulties in accurately
quantifying the triggering mechanism of AGN activity. Major mergers
clearly occur and trigger AGN activity but there are significant
uncertainties in the interpretation of the data; indeed, systems with
classical spheroids may have undergone at least one major-merger event
at some time during the past \citep[e.g.,][]{bour05merge,
  hopk10merge}. Conversely, AGN activity is also clearly often
triggered by galaxy interactions and internal processes
\citep[e.g.,][]{malk98agn, kuo08tidal,elli11agninter}. An example of
an AGN population where the majority of the BH growth may be {\it
  entirely} driven by secular processes are narrow-line Seyfert 1
galaxies (NLS1s), a subset of the Seyfert 1 population with
comparatively narrow broad emission lines (permitted broad-line widths
of $\approx$~500--3000~km~s$^{-1}$; e.g.,\ \citealt{oste85nls1}). The
majority of NLS1s reside in galaxies with pseudo bulges, in contrast
to the classical spheroids hosted by typical Seyfert galaxies
\citep{math11nls1_aph, orba11nls1}, and are therefore unlikely to have
      {\it ever} undergone a major merger
      \citep[e.g.,][]{korm04pseudo}. Pseudo bulges have more angular
      momentum than classical spheroids, which inhibits the gas
      accretion on small scales. It is therefore significant that
      NLS1s are more likely to reside in barred spiral galaxies than
      typical Seyfert galaxies \citep{cren03nls1} -- without the
      galaxy bars, the host-galaxy gas is unlikely to be efficiently
      driven into the central kpc region of NLS1s (contrast with the
      lack of a strong correlation in the overall population; see
      \S2.1).

\subsection{The distant growth of massive black holes}

\subsubsection{Evolution of the AGN population}

Extensive high-quality observations have provided valuable insight
into the processes of BH growth in the local Universe. However, the
vast majority of the mass accretion onto BHs occurred at higher
redshift ($\approx$~95--99\% of the integrated BH growth has occurred
at $z>0.1$; e.g.,\ \citealt{marc04smbh, shan04agnbh, shan09agnbh,
  hopk07qlf, merl08agnsynth, aird10xlf}), and we must therefore look
to the distant Universe to understand when, where, and how today's
massive BHs grew. Optical quasar surveys established more than four
decades ago that luminous AGNs were orders of magnitude more common at
$z\simgt$~1--2 than $z\approx$~0 \citep[e.g.,][]{schm68, schm83qso,
  hart90qso, boyl00qlf, rich06qlf}. However, the optical photometric
selection method used to identify quasars is only effective in
selecting unobscured AGNs that are bright enough to outshine the host
galaxy. Optical quasar surveys therefore provide limited constraints
on the properties and evolution of the majority of the AGN population;
i.e.,\ either obscured AGNs or lower-luminosity AGNs.

Currently the most efficient and near-complete selection for distant
AGNs is made using X-ray observations. X-ray observations provide an
almost obscuration-independent selection of AGNs and, since star
formation activity is comparatively weak at X-ray energies, can
identify even low-luminosity systems. For example, the deepest X-ray
surveys (the {\it Chandra} Deep Fields; CDFs; \citealt{bran01b,giac02,
  alex03, luo08cdfs, xue11cdfs}) can detect low-luminosity AGNs
similar to NGC~4051 (see Fig.~1 and also Fig.~6 of \citealt{bran05})
out to $z\approx$~1--2, even in the presence of large amounts of
absorption (up-to $N_{\rm H}\approx10^{23}$--$10^{24}$~cm$^{-2}$;
e.g.,\ \citealt{tozz06, raim10nh, alex11stack, coma11xmmdeep,
  feru11kalpha}); AGNs $\approx$~1--2 orders of magnitude brighter can
be identified to $z>6$, provided a sufficient number of objects exist
in the comparatively small survey volumes. Often the significant
challenge in the identification of distant X-ray selected AGNs is an
accurate measurement of source redshifts (with spectroscopic or
photometric data; e.g.,\ \citealt{barg03, szok04, zhen04, card10xhost,
  luo10cdfs, salv11xcosmos}), since the optical/near-IR counterparts
for many of the AGNs are very faint
\citep[e.g.,][]{alex01xfaint,main05xfaint, rovi10xfaint}.
Furthermore, even the deepest X-ray surveys miss the most heavily
obscured luminous AGNs where even hard X-ray photons are absorbed;
selection techniques using optical spectroscopy, IR, and radio data
are starting to identify large numbers of these systems
\citep[e.g.,][]{donl05radio, dadd07comp, alex08compthick, fior08agn,
  hick09corr, yan11aegis, june11agn, luo11obsc}. However, despite
these challenges, from a combination of X-ray observations across a
broad range of the flux--solid angle plane (e.g.,\ see Fig.~1 of
\citealt{bran05}) we are starting to piece together a more complete
picture of the evolution of AGN activity across cosmic time.

The evolution in the space density of high-luminosity AGNs broadly
tracks that found from optical quasar surveys: the space density of
luminous AGNs with $L_{\rm 2-10 keV}>10^{44}$--$10^{45}$~erg~s$^{-1}$
peaks at $z_{\rm peak}=$~1--2 and drops by $\approx$~2--3 orders of
magnitude to $z\approx$~0 and $\approx$~0--1 order of magnitude to
$z\approx$~5 \citep[e.g.,][]{fior03xevol, ueda03, shan04agnbh,
  shan09agnbh, barg05, hasi05, hopk07qlf, silv08xlf, brus09highz,
  aird10xlf}. The constraints are currently very uncertain for
$z\simgt$~5 AGNs \citep[e.g.,][]{fan01z6qso, barg03highz,
  will10z6qso}; see \S4.3. By contrast, lower luminosity AGNs evolve
more slowly and their space density peaks at lower redshifts than
high-luminosity AGNs: for example, the space density of
moderate-luminosity AGNs with $L_{\rm 2-10
  keV}\approx10^{43}$--$10^{44}$~erg~s$^{-1}$ peaks at
$z\approx$~0.5--1.0 and drops by $\approx$~1--2 orders of magnitude to
$z\approx$~0 and $z\approx$~5, respectively. This differential
redshift evolution of the AGN population is commonly modelled as
luminosity dependent density evolution. The same behaviour is also
found for optically selected and radio selected AGNs
\citep[e.g.,][]{rich05, bong07qlf, hopk07qlf, rigb08radio,
  rigb11radio, croo09twodfqz, smol09radioevol}.\footnote{Analogous
  results are also found for the star-forming galaxy population, where
  the space density of high-luminosuty star-forming galaxies peaks at
  higher redshifts than lower-luminosity systems
  \citep[e.g.,][]{lefl05irlf,pere05sfevol}. This behaviour is
  typically referred to as galaxy ``downsizing'' since there is clear
  evidence for a decrease in the mass of luminous star-forming
  galaxies with decreasing redshift
  \citep[e.g.,][]{cowi96sfev,ball06,bell05sfevol,june05sfevol,bund06sfevol}.}
Obscured AGNs are found to trace the same evolution as unobscured AGNs
of the same luminosity, although there is tentative evidence for an
increase in the obscured AGN fraction with redshift
(e.g.,\ \citealt{lafa05obsc,ball06,trei06evol,hasi08agn}; however, see also
\citealt{akyl06} and \citealt{dwel06obsc}), which would be expected if
the nuclear regions of distant AGNs are more gas rich than
lower-redshift AGNs.

The integrated growth of BHs is dominated by systems around the knee
of the AGN luminosity function \citep[e.g.,][]{hopk07qlf}, which for
X-ray detected AGNs is $L_{\rm 2-10 keV}\approx10^{44}$~erg~s$^{-1}$:
$\approx$~75\% of the growth of BHs has occurred in luminous systems
with $L_{\rm 2-10 keV}\approx10^{43}$--$10^{45}$~erg~s$^{-1}$
(i.e.,\ straddling the traditional Seyfert galaxy/quasar threshold;
see footnote 1 of \citealt{alex08compthick}). Approximately 30--50\%
of the integrated growth of BHs has occurred at comparatively low
redshifts of $z<1$, $\approx$~35-45\% has occurred at $z\approx$~1--2,
and $\approx$~15--25\% at $z>2$ \citep[e.g.,][]{marc04smbh,
  shan04agnbh, shan09agnbh, hopk06apjs, hopk07qlf, silv08xlf,
  aird10xlf}.  Optically selected quasars account for
$\approx$~35--50\% of the integrated BH growth and radio-loud AGNs
account for $\simlt$~10\% of the integrated BH growth. However, while
comprising the minority of the overall AGN populations, both of these
sub populations are important for the growth of BHs and galaxies:
optically selected quasars appear to represent a rapid growth phase of
massive BHs (see \S4.2) and radio-selected AGNs appear to have played
a significant role in the formation and evolution of galaxies (see
\S5.1).

The origin of the differential redshift evolution of the AGN
population is presumably due to a decrease in the overall cold-gas
fuel supply in the nuclear regions of galaxies, at least for the
lower-redshift drop off in space density. The global drop off in space
density at $z>2$ may be limited by the maximum BH accretion rate for
comparatively low-mass BHs (i.e.,\ the AGN luminosity cannot be higher
than the BH Eddington limit; e.g.,\ \citealt{merl08agnsynth}; see
Eqn.~6). However, it isn't immediately clear whether the decrease in
the fuel supply is across all systems or only for systems in specific
environments or BH mass ranges. Perhaps the strongest discriminator
between these different scenarios comes from studies of nearby AGNs,
where the long growth times of the most massive BHs at $z\approx$~0.1
($M_{\rm BH}>10^8$~$M_{\odot}$) implies that they must have grown more
rapidly at higher redshifts (e.g.,\ \citealt{heck04bh}); see
\S3.2.1. Direct measurements of the Eddington ratios of the BHs in
$z<2$ quasars (BH masses estimated using the virial estimator; see
\S4.1) provide further support for this view by showing a decrease in
both the characteristic active BH mass and Eddington ratio with
decreasing redshift \citep[e.g.,][]{mclu04mbh, gree07mfunc,
  netz07agnevol}. Lastly, on the basis of indirect BH mass estimates
(based on the host-galaxy luminosities or velocity dispersions and the
local BH--spheroid mass relationships; e.g.,\ \citealt{mago98, gebh00,
  ferr00, trem02, marc04smbh}), the BHs of $z\approx$~0.3--1.5 X-ray
AGNs appear to be massive and are growing more rapidly than similarly
massive BH in the local Universe ($M_{\rm
  BH}\approx10^7$--$10^9$~$M_{\odot}$ and typical Eddington ratios of
$\lambda\approx10^{-3}$--$10^{-1}$; e.g.,\ \citealt{babi07edd,
  ball07edd, rovi07xhost, alon08agn, hick09corr, raim10nh,
  simm11edd}).

The current observational data therefore suggest ``downsizing'' in the
active BH mass with decreasing redshift, in general agreement with
that found for the galaxy population (see footnote 7). However, the
current observations provide comparatively limited constraints in
isolation since accurate determinations of the BH mass and Eddington
ratio are only available for a small fraction of systems (i.e.,\ AGNs
with broad emission lines; see \S4.1), and the constraints are very
limited for $z>1.5$ AGNs, where the most massive BHs are predicted to
have been most active. To gain more detailed insight we can appeal to
models and simulations, which provide solutions to the growth of the
BH population from a broad suite of observational and physical
constraints (e.g.,\ BH mass density and distribution, evolving AGN
luminosity functions, X-ray background spectrum;
e.g.,\ \citealt{marc04smbh, merl04smbh, shan04agnbh, shan09agnbh,
  hopk06apjs, hopk08frame1, malb07smbh, dima08bhfeed, some08bhev,
  fani11agn, fani12agn_aph}). Despite a broad range of different
approaches, from simple analytical/synthetic models to more detailed
semi-analytical models and N-body/hydrodynamical simulations some
clear trends have emerged from these studies: (1) the average
Eddington ratio for BHs decreases with redshift, from rapidly growing
systems at $z\simgt$~2--3 to comparatively slow growing systems at
$z<1$, and (2) the characteristic ``active'' BH mass decreases with
redshift. However, there are significant differences between the
studies in terms of the build-up of the BH mass function with
redshift. For example, in the semi-analytical model of
\citet{fani12agn_aph}, the lower-mass end of the BH mass function
($M_{\odot}\simlt10^8$~$M_{\odot}$) is essentially in place by
$z\approx$~2 while the high-mass end is predominantly built by BH
mergers of lower-mass BHs and modest mass accretion rates through to
the present day. By comparison, in the AGN synthesis models of
\citet{marc04smbh} and \citet{merl08agnsynth} the lower-mass end of
the BH mass function is only in place by $z\simlt$~0.1--0.6. Of course
there are large differences in the complexities, scope, and
assumptions made between these different models but they serve to
illustrate some of the uncertainties that remain in modelling the
evolution of the BH mass function.

Radio-loud AGNs are the minority AGN population at all redshifts and
the majority of the growth of BHs has occurred in radio-quiet AGNs
\citep[e.g.,][]{merl08agnsynth, catt09bhgal}. However, a large
fraction of the radio emission associated with AGN activity is
produced by jets and lobes, which are powered by synchroton emission
and shocks and therefore large amounts of kinetic/mechanical
energy. If this kinetic energy is able to couple effectively to the
gas in the host galaxy or larger-scale environment then it can prevent
the gas from cooling and forming stars; see \S5.1. The evolution of
the radio luminosity density with redshift therefore provides direct
constraints on the volume-average heating rate from AGN activity
\citep[e.g.,][]{crot06, hein07klf, lehm07xevol, merl08agnsynth,
  catt09bhgal, smol09radioevol, lafa10radio}. The conversion from
radio luminosity to mechanical power is uncertain and relies on
converting the synchrotron jet luminosity into a kinetic energy
\citep[e.g.,][]{dunn04bubbles, best06feed, raff06feedback, hein07klf,
  birz08jet, cava10jet}. However, on the basis of the current
conversion factors, the overall heating rate from AGN activity is
predicted to be broadly flat over $z\approx$~0--4, with low excitation
AGNs dominating the overall heating, particularly at $z<1$ when
radio-loud high-excitation AGNs are rare
\citep[e.g.,][]{merl08agnsynth, catt09bhgal}. There is evidence for a
sharp drop in the heating rate at $z<0.5$, which would indicate a
weakening role of AGN activity towards the present day
\citep[e.g.,][]{kord08klf, lafa10radio}. However, there are also
significant differences between studies in the estimated evolution of
the AGN heating rate, which are due in part to how the current radio
luminosity functions are extrapolated down to low luminosities.

\subsubsection{The host galaxies of distant AGNs}

The strong evolution in the AGN population with redshift illustrates
that conditions in the distant Universe were different to that seen
locally. How do the host galaxies of distant AGNs compare to the host
galaxies of AGNs in the local Universe? Accurate measurements of the
host-galaxy properties of distant AGNs can be determined from
high-spatial resolution imaging, where the AGN component can be
modelled as a point source, or from photometry for obscured and
low-luminosity systems when the host-galaxy emission dominates at
rest-frame optical--near-IR wavelengths.

The concensus view is that the majority of distant AGNs are hosted in
massive galaxies. The host galaxies of X-ray AGNs with $L_{\rm
  X}\simgt10^{42}$~erg~s$^{-1}$ at $z<3$ are
$M_{*}\approx$~(0.3--3)~$\times10^{11}$~$M_{\odot}$
\citep[e.g.,][]{akiy05xhost, babi07edd, ball07edd, alon08agn,
  bund08quench, brus09xhost, hick09corr, xue10xhost}. Lower-mass
systems ($M_{*}\approx$~(0.3--2)~$\times10^{10}$~$M_{\odot}$) are also
identified but appear to comprise the minority population of X-ray
AGNs \citep[e.g.,][]{shi08xhost, xue10xhost}; however, selection and
sensitivity effects means that it is often challenging to identify
distant AGNs in low-mass galaxies (i.e.,\ for a given luminosity
threshold, a low-mass BH needs to be growing at a higher Eddington
ratio to be identified than a high-mass BH; see Fig.~1). Radio-loud
AGNs are typically hosted in the most massive elliptical galaxies with
$M_{*}\approx10^{11}$--$10^{12}$~$M_{\odot}$: systems with
low-excitation optical spectra appear to be hosted in higher-mass
galaxies than high-excitation radio-loud AGNs, as expected if the
low-excitation systems reside in the most massive dark-matter halos
and underwent their major growth phases at higher redshifts
\citep[e.g.,][]{tass08radiomode, herb10radio, floy10radio}; see
\S3.4. By comparison, optically selected quasars are predominantly
hosted in less-massive elliptical galaxies, with a non-negligible
fraction found in galaxies with a significant disc component
\citep[e.g.,][]{dunl03qsohost, floy04qsohost, koti07qsohost,
  benn08qsohost, schr08qsohost, tass08radiomode, koti09qsohost,
  veil09qsomorph, floy10radio}.

Various analyses show that the host-galaxy properties of distant AGNs
are also similar to those of distant massive galaxies. In terms of the
Colour-Magnitude Diagram (CMB; e.g.,\ \citealt{stra01galcol}), distant
X-ray AGNs predominantly lie in the ``green valley'' between the ``red
sequence'' and ``blue cloud'' \citep[e.g.,][]{nand07host, rovi07xhost,
  silv08host, hick09corr, xue10xhost, card10xhost}, consistent with
that of inactive galaxies over a similar mass range; however, we note
that some AGN host galaxies will be red due to dust obscuration
\citep[e.g.,][]{card10xhost, rovi11xhost}. The host galaxy
morphologies of X-ray detected AGNs are also comparable to those of
similarly massive inactive galaxies \citep[e.g.,][]{sanc04agnhost,
  grog05xhost,pier07morph, gabo09xmorph, geor09xmorph,
  koce11xmorph_aph, scha11xmorph}, with a mix of bulge-dominated and
disc-dominated systems. By comparison, the host galaxies of distant
radio-loud AGNs typically lie in the ``red sequence'' of the CMD
\citep[e.g.,][]{tass08radiomode, hick09corr}, as expected for massive
elliptical galaxies. The morphologies and structural properties of
distant radio-loud AGN are also typically consistent with those of
distant elliptical galaxies \citep[e.g.,][]{dunl03qsohost,
  mclu04radiohost, floy10radio, herb11radio}.

The similarity in the host-galaxy properties of distant AGNs and
coeval inactive galaxies suggests that the AGN fraction provides an
estimate of the BH-growth duty cycle. The fraction of massive galaxies
hosting X-ray AGN activity out to $z\approx$~2--3 is
$\approx$~10--20\% for systems with $L_{\rm
  X}\simgt10^{42}$~erg~s$^{-1}$ \citep[e.g.,][]{bund08quench,
  xue10xhost}. The AGN fraction drops to $\approx$~1--5\% for AGNs
with $L_{\rm X}\simgt10^{43}$~erg~s$^{-1}$, with some evidence for an
increase with redshift \citep[e.g.,][]{xue10xhost, geor11xagn},
comparable to that found in the local Universe but at higher BH and
stellar masses (see Fig.~4). These results indicate that BH growth is
recurrent and has a long duty cycle. The fraction of galaxies hosting
radio-loud AGN activity out to $z\approx$~1.3 is also comparable to or
slightly higher than that found for nearby galaxies, indicating that
the duty cycle of radio-loud AGN activity has been relatively constant
over at least the past $\approx$~9~Gyrs
\citep[e.g.,][]{tass08radiomode, smol09radioevol}.

How is the distant AGN activity triggered? Contrary to initial
expectations, only a comparatively small fraction of the X-ray AGNs
out to $z\approx$~3 clearly reside in galaxy major mergers
($\approx$~15--20\%; e.g.,\ \citealt{sanc04agnhost, pier07morph,
  geor09xmorph, cist11agn, koce11xmorph_aph}, comparable to that found
for X-ray selected AGNs in the local Universe
(e.g.,\ \citealt{koss10batagn}; see \S3.2.3). However, the host
galaxies of up-to $\approx$~45\% of the X-ray AGNs show some evidence
for disturbed morphologies, suggesting that a considerable fraction of
the distant AGN activity could be driven by external processes
(i.e.,\ either galaxy major mergers or interactions;
e.g.,\ \citealt{koce11xmorph_aph,silv11agninter_aph}). As noted in
\S3.2.3, there are potential complications in the interpretation of
the host-galaxy morphologies, making it difficult to determine the
true fraction of distant AGN activity driven by external and internal
processes. It is therefore important to note that, irregardless of
these complications, there are no clear differences in the
morphological properties of AGNs and comparably massive inactive
galaxies, suggesting that the mechanisms that trigger distant BH
growth also trigger distant galaxy growth \citep[e.g.,][]{cist11agn,
  koce11xmorph_aph, scha11xmorph}.

The fraction of optically selected quasars with evidence for
morphological distortions varies from study to study but is always
large ($\approx$~30--100\%) and, importantly, is found to be higher
than comparably massive distant galaxies
\citep[e.g.,][]{bahc97qsohost, cana01qso, dunl03qsohost,
  guyo06qsohost, benn08qsohost, urru08qsohost}. The overall fraction
of radio-loud AGNs with evidence for disturbed morphologies is also
higher than that found for distant massive galaxies
\citep[e.g.,][]{deko96radiohost, dunl03qsohost, ramo11radiohost,
  ramo11trigger}. However, \citet{ramo11radiohost, ramo11trigger}
showed that the disturbed morphology fraction of radio-loud AGNs
depends on the optical spectral properties: $>90$\% of the
high-excitation systems have disturbed host-galaxy morphologies
(i.e.,\ similar to the optically selected quasars), as compared to
only $\approx$~30\% of the low-excitation systems. Overall these
results therefore suggest that typical X-ray AGNs and low-excitation
radio-loud AGNs evolve along with the coeval massive galaxy
population, while quasars (either radio quiet or radio loud) are often
triggered by external processes and evolve more rapidly (see \S4.2 for
further discussion of quasars), as predicted by some models
\citep[e.g.,][]{hopk08frame1,hopk09fueling}.

\subsubsection{The connection with star formation}

There are a number of pieces of evidence that connects the global
evolution of star formation with that seen for AGN activity: (1) the
differential redshift evolution of the AGN population is also found
for the star-forming galaxy population (see footnote 7), (2) the
redshift distribution of the most strongly star-forming galaxies
traces that seen in the optical quasar population
\citep[e.g.,][]{chap05smg, ward11less}, and (3) the overall shapes of
the volume-average mass accretion and star-formation histories are
broadly similar \citep[e.g.,][]{boyl98qsosf,merl04bhsf,silv08xlf,
  aird10xlf}. Such a connection is not unexpected given the tightness
of the BH--spheroid mass relationship in the local Universe and the
association between AGN activity and star formation in nearby
galaxies; see \S3.2.2. However, given the different conditions between
the local and distant Universe, how does the connection between
distant AGN activity and star formation compare to that found in the
local Universe?

The volume-average mass accretion and star-formation histories are
similar up to $z\approx$~1.0--1.5, when the mass accretion history is
scaled up by a factor $\approx$~5000
\citep[e.g.,][]{silv08xlf,aird10xlf}. However, there are significant
differences at $z>2$, where the slope of the mass accretion history is
much steeper than that found the star-formation history; for example,
there is potentially $\approx$~1--2 orders of magnitude more star
formation per unit volume than that found for the mass accretion at
$z\approx$~6. At least part of the variation with redshift could be
due to incompleteness in the AGN selection (e.g.,\ there may be an
increasing fraction of X-ray undetected heavily obscured AGNs with
redshift; see \S3.3.1) but it also possible that the connection
between AGN activity and star formation was different at high
redshift.

There is a general correlation between star-formation rate (SFR) and
mass accretion rate for individual distant AGNs
\citep[e.g.,][]{schw06qsosf,lutz08qsosf,shi09qsosf,silv09xcosmos_env,
  tric09agnsf,raff11agnsf}. However, the correlation is broad,
particularly for AGNs detected in X-ray surveys where the AGN
selection is comparatively complete (i.e.,\ up-to five orders of
magnitude variation for individual X-ray AGNs; e.g.,\ see Fig.~14 of
\citealt{raff11agnsf}). Globally the star-formation rates of AGNs of
all classes are found to increase with redshift, in general agreement
with the broad trend found for the overall galaxy population
\citep[e.g.,][]{arch01agnsubmm, serj09qsosf,
  mull10agnsf,mull11agnsf,lutz10agnsf, shao10agnsf,xue10xhost,
  main11obsqso,seym11radiosf}. For example, the average star-formation
rates of moderate-luminosity AGNs (i.e.,\ systems with Seyfert galaxy
luminosities of $L_{\rm 2-10 keV}>10^{43}$~erg~s$^{-1}$) are
$\approx$~40 times higher at $z\approx$~2--3 than found for similarly
luminous AGNs at $<0.1$ \citep[e.g.,][]{mull11agnsf}. This rate of
increase is in good agreement with that found for comparably massive
galaxies over the same redshift range \citep[e.g.,][]{dadd07sf,
  pann09sf, rodi10sf, elba11ms}, suggesting that the presence of an
AGN does not have a significant influence on the total amount of star
formation in the host galaxy \citep[e.g.,][]{shao10agnsf, xue10xhost,
  mull11agnsf}; however, we note that the situtation may be different
for rapidly growing BHs in starburst galaxies, which may follow a
different evolutionary path (see Fig.~6 in \S4.2).

Taken at face value these results therefore indicate a {\it
  significant} increase in the amount of star-formation activity for a
fixed mass-accretion rate with redshift (i.e.,\ there was $\approx$~40
times more star formation per unit of AGN activity at $z\approx$~2--3
than found at $z<0.1$; e.g.,\ \citealt{mull11agnsf}), which appears
inconsistent with the concordant BH--galaxy growth ratio implied by
the BH--spheroid mass relationship. These results can be reconciled
with the relative growth implied by the BH--spheroid mass relationship
if, for example, the AGN duty cycle increases signficantly with
redshift (see below for current constraints) or if the majority of the
star formation in the distant Universe occurs in galaxy discs rather
than galaxy spheroids.

The AGN fraction is found to be a strong function of star-formation
rate in distant galaxies. For example, the fraction of
moderate-luminosity AGNs with $L_{\rm X}>10^{43}$~erg~s$^{-1}$
increases from $\approx$~3--10\% for
SFRs~$\approx$~30--200~$M_{\odot}$~yr$^{-1}$ (i.e.,\ equivalent to
LIRGs) to $\approx$~10--40\% for
SFRs~$\approx$~100--500~$M_{\odot}$~yr$^{-1}$ (i.e.,\ equivalent to
ULIRGs; e.g.,\ \citealt{alex05,lair10submmx, syme10spitzer,xue10xhost,
  geor11submmx, raff11agnsf}). The AGN fraction increases further if
lower-luminosity AGNs are considered (see Fig.~12 of
\citealt{raff11agnsf}). The high AGN fraction at the highest SFRs
indicates almost continous BH growth during vigorous star-formation
phases, as would be expected during rapid growth phases. However, the
overall AGN fractions are broadly consistent with those found for
intense star-forming galaxies in the local Universe (e.g.,\ the
fraction of nearby ULIRGs hosting AGNs with $L_{\rm
  X}>10^{43}$~erg~s$^{-1}$ is $\approx$~40\%;
\citealt{alex08compthick}), suggesting that the duty cycle of BH
growth in star-forming galaxies has remained relatively constant over
the broad redshift range of $z\approx$~0--3. This result is found
despite the large increase in SFR for AGNs over $z\approx$~0--3 and
the global increase in the cold-gas mass fraction in massive galaxies
with redshift \citep[e.g.,][]{dadd10fgas, tacc10fgas, geac11fgas}.

\subsection{The role of environment}

The BH-growth results discusssed so far are predominantly for AGNs
identified in field environments (i.e.,\ the typical regions of the
Universe). How does the growth of BHs differ as a function of
large-scale structure and environment? Theories of large-scale
structure formation predict that galaxy growth is accelerated in
regions of high density \citep[e.g.,][]{kauf96ell, delu06ell}. The
relationship between stellar age and local environment provides
support for this hypothesis, showing that the most evolved and massive
spheroids reside in the highest density regions (clusters) at the
present day \citep[e.g.,][]{bald04color,smit09galage}. Large-scale
environment is therefore also likely to have a significant effect on
the triggering of AGN activity.

In the nearby Universe, radio-quiet and radio-loud AGNs preferentially
reside in different environments. On the basis of the two-point
correlation function, which measures the clustering strength of
selected populations, the dark matter halos of nearby radio-quiet and
radio-loud AGNs are inferred to be $\approx10^{12}$~$M_{\odot}$ and
$\approx2\times10^{13}$~$M_{\odot}$ \citep[e.g.,][]{mand09agnclust,
  dono10clust}. The clustering stength of the optical AGNs is also
consistent with that found for comparably massive galaxies. By
comparison, radio-loud AGNs are preferentially found in galaxy groups
and poor-to-moderate richness galaxy clusters \citep{best04radio}. For
example, measuring environment as the local density of galaxies in a
radius of 2~Mpc, the fraction of galaxies hosting radio-quiet AGN
activity is $\approx$~2 times higher in low-density regions than in
high-density regions \citep{kauf04}, while the local galaxy density of
radio-loud AGNs is $\approx$~2 higher than that found for radio-quiet
AGNs \citep{kauf08radio}.

Broadly similar results are found for distant AGNs. The clustering of
radio-quiet AGNs is weaker than radio-loud AGNs out to
$z\approx$~1--2, with implied dark-matter halo masses of
$\approx10^{12}$--$10^{13}$~$M_{\odot}$ and
$\approx$~(0.3--1.0)~$\times10^{14}$~$M_{\odot}$, respectively
(e.g.,\ \citealt{li06agnclust, daan08clust, coil09xclust,
  gill09xclust, hick09corr, hick11qsoclust, mand09agnclust,
  krum10xclust,fine11radio}; however, also see
\citealt{brad11agnenv}), with some evidence that the clustering
strength is dependent on the adopted AGN luminosity thresholds
\citep[e.g.,][]{krum10xclust}; see Fig.~5. X-ray AGNs out to
$z\approx$~1 are found to reside in a broad range of environments
(from field galaxy environments to galaxy groups), consistent with
those found for galaxies of the same mass, colour, and star-formation
rate \citep[e.g.,][]{geor08agn, silv09agnenv, tass11xenv}. The X-ray
AGN fraction for galaxies found in the field and galaxy groups
environments are also indistinguishable \citep[e.g.,][]{wass05xenv,
  silv09agnenv}. By comparison, distant radio-loud AGNs are found to
typically reside in galaxy groups and clusters, where the radio source
may be triggered by weak accretion of the hot group/cluster gas
\citep[e.g.,][]{tass08radiomode, bard10radio, smol11radiohod}; see
Fig.~2.

\begin{figure}[t]
\centering
\includegraphics[width=75mm]{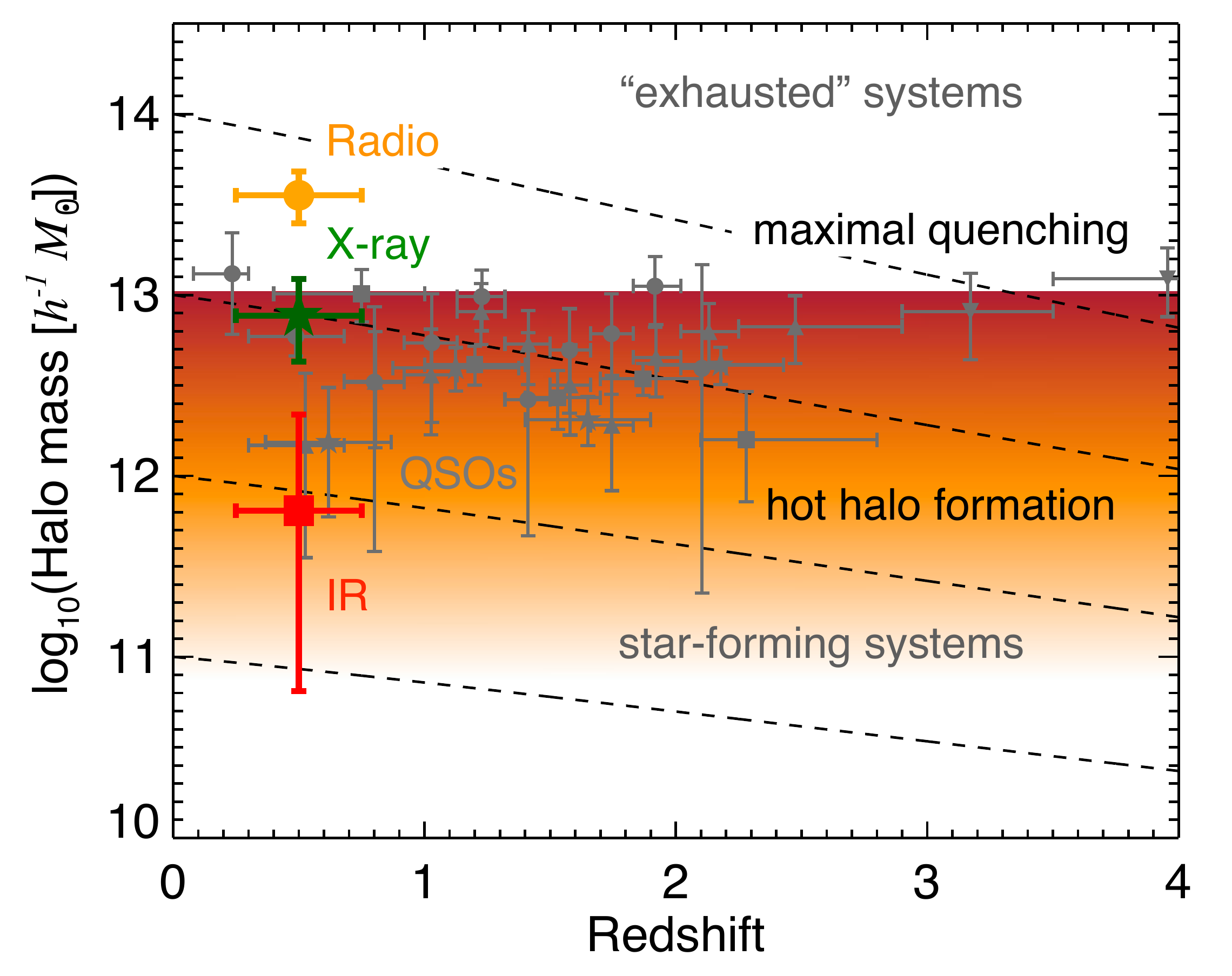}
\caption{Illustration of the evolution of dark matter halo mass versus
  redshift for AGN populations. Lines show the evolution of halo mass
  with redshift for individual halos, based on the median growth rates
  from cosmological simulations \citep{fakh10halorate}. Highlighted is
  the region of maximum ``quenching'', in which halos transition from
  having large reservoirs of cold gas to being dominated by virialized
  hot atmospheres \citep[e.g.,][]{crot09qsohalo}. The gray points show
  halo masses of optically-bright quasars derived from autocorrelation
  measurements (\citealt{croo05}, upright triangles;
  \citealt{myer06clust}, squares; \citealt{shen07clust}, inverted
  triangles; \citealt{daan08clust}, stars; \citealt{ross09qsoclust},
  circles). The colored points show the halo masses for radio, X-ray,
  and infrared-selected AGN at $z\sim 0.5$ \citep{hick09corr}. This
  figure illustrates that massive rapidly growing BHs (optical
  quasars) are found in the most massive dark matter halos that have
  not yet reached the ``exhausted'' hot halo regime, while rapidly
  growing lower-mass BHs (as traced by mid-IR samples) are found in
  somewhat lower-mass halos typical of star-forming galaxies. All halo
  masses are calculated following \citet{hick11qsoclust} assuming a
  cosmology with $(\Omega_{\rm m}, \Omega_{\Lambda}, \sigma_8) = (0.3,
  0.7, 0.84)$.}
\end{figure}

Stark differences in the incidence of luminous AGN activity is found
between distant galaxy clusters and distant field-galaxy regions. The
moderate-luminosity AGN fraction ($L_{\rm 2-10
  keV}>10^{43}$~erg~s$^{-1}$) in cluster galaxies rises by an order of
magnitude over the redshift range $0.05<z<1.3$ from $\approx$~0.1--1\%
\citep[e.g.,][]{mart06clustagn, mart09clustagn, east07clustagn}. These
luminous AGN fractions are approximately an order of magnitude {\it
  lower} than those found in field-galaxy regions over the same
redshift. This decline in AGN activity with decreasing redshift
broadly tracks that found for the star-forming population in galaxy
clusters \citep[e.g.,][]{sain08clustsf, finn10clustsf,
  atle11clustagn}, indicating a general shut-down of activity in the
densest regions of the Universe over the last $\approx$~9~Gyrs of
cosmic time.

The results found for AGN activity in galaxy clusters suggest that
growth of galaxies and BHs in overdense regions must have been
significantly more rapid at higher redshift. Indeed, we find direct
support for this view from deep X-ray observations of the densest
regions in the distant Universe (protoclusters). For example, in the
$z\approx$~3.1 SSA~22 protocluster, \citet{lehm09proto} found a factor
of $\approx$~6 increase in the fraction of galaxies hosting AGN
activity compared to the field-galaxy environment at $z\approx$~3.1,
showing the opposite trend to that found for $z<1.3$ galaxy clusters
(see also \citealt{digb10proto}). Furthermore, there is tentative
evidence that the largest incidence of AGN activity is in the densest
regions of the protocluster \citep{lehm09catalog}, suggesting that
even the local environment has a large effect on the triggering of AGN
activity, in contrast to AGNs identified in $z<1$ overdense regions
\citep[e.g.,][]{gilm07agnsuperclust,koce09xsuperclust}. The SSA~22
protocluster is predicted to become a large galaxy cluster by
$z\approx$~0, with a similar mass to that of Coma
($\approx10^{15}$~$M_{\odot}$; \citealt{stei98ssa}). The stark
difference in activity between this $z\approx$~3 protocluster and
$z<1.3$ galaxy clusters of similar mass therefore suggests a dramatic
change in fuelling mode, from a large abundance of cold gas available
at high redshift to the predominantly hot gas when the galaxy cluster
virialises.

These studies show that environment has a major role in the growth of
BHs. The optical and X-ray AGN activity is likely driven by the
availability of a cold-gas supply. At high redshift this is most
readily available in the densest environments but by lower redshifts
the majority of the cold-gas in dense regions has been heated and
cannot be accreted efficiently -- at low redshifts AGN activity is
more prevalent in typical regions of the Universe. This shut down of
activity is most dramatically seen in galaxy clusters and it may be
related to the dark-matter halo. For example, the lack of significant
AGN activity with dark-matter haloes $\gg10^{13}$~$M_{\odot}$ may be
due to the halo virialising and shock heating the gas and suppressing
vigorous mass accretion \citep[e.g.,][]{catt06crit, catt08shutdown,
  deke06shock, kere09shock}. Indeed, luminous AGNs and quasars appear
to be universally hosted by dark-matter of
$\approx10^{12}$--$10^{13}$~$M_{\odot}$; see Fig.~5.  The dark-matter
halo may therefore control the cold-gas accretion and may strongly
influence the overall evolution of AGN activity with redshift.

\section{What fuels the rapid growth of the most massive (and also the first) black holes?}

In the previous section we discussed the host galaxies and evolution
of AGNs, which provide the overall picture for the joint cosmological
growth of galaxies and their BHs. However, the work from the majority
of these studies focused on typical AGNs with luminosities near or
below the break in the AGN luminosity function. While these objects
represent the majority of AGN in the Universe and may play an
important role in galaxy growth, they are unlikely to be responsible
for the {\it bulk} of the accretion onto the most massive
BHs. Synthesis models of the BH population show that for systems with
$M_{\rm BH}> 10^8$ \msun, the major growth phase occurs in very
powerful, high-Eddington ratio quasars at high redshift (see
\S3.3.1). The large-scale process that is likely to triggers this
rapid BH growth phase is galaxy major mergers, which is also expected
to initiate luminous star-formation activity. In this section we
further discuss recent progress on the demographics of powerful
quasars and the physical mechanisms that fuel their rapid growth. We
also explore the formation and evolution of the initial BHs at high
redshift that comprise the ``seeds'' which eventually grow into the
BHs that power the quasars.

\subsection{Masses and evolution of rapidly growing black holes}

The foundations in understanding the rapid growth of massive BHs come
from measurements of demographics (BH masses, Eddington ratios, and
space densities). In recent years the most powerful observation tool
has been been spectroscopic studies of optically-selected
quasars. Their high luminosity and characteristic colors mean that
quasar luminosity functions can be traced out to high redshifts
\citep[ $z\lesssim7$; ][]{fan06, mort11z7qso}, with the peak in the
AGN space density of $z\approx$~1--3 \citep[e.g.,][]{rich05,
  croo09twodfqz}; see \S3.3.1. Compared to X-ray or IR observations,
studies of optical quasars are strongly biased against selecting
obscured sources, but they have the key advantage of enabling detailed
studies of the mass accretion \citep[e.g.,][]{floy09qsolens,
  down10qsodisk} and accurate BH masses (using virial mass estimators;
e.g.,\ \citealt{vest02bhmass, pete04bhmass, vest06, koll06,
  shen08bhmass, merl10bhmass}).

The virial BH mass estimators are based on the simple principle that
the broad-line region (BLR) is in virial equilibrium, such that
\begin{equation}
M_{\rm BH} = {{f v^2 R_{\rm BLR}}\over{G}},
\end{equation}
where $f$ is a dimensionless quantity, of order unity, that depends on
the geometry and kinematics of the BLR.\footnotemark The velocity $v$
can be measured from a single spectrum from the width of the broad
lines, but the radius $R_{\rm BLR}$ is more challenging to
estimate. All virial estimates are based on results from reverberation
mapping of local AGN, in which the time lag between the light curves
of the continuum luminosity and the broad lines allows an estimate of
$R_{\rm BLR}$ \citep[e.g.,][]{kasp05, bent09revmap}. These studies
find a correlation of roughly $R_{\rm BLR} \propto\ L_{\rm
  nuclear}^{0.5}$, implying that $M_{\rm BH} \propto v^2 L^{0.5}$,
with a normalization that needs to be calibrated against other BH mass
estimators and which depends on the kinematics and structure of the
BLR.\footnotetext{For a spherical BLR geometry, $f=3/4$ if the line
  width is expressed in terms of the full-width half maximum (FWHM),
  as is commonly adopted in virial BH mass estimates
  \citep[e.g.,][]{vest02bhmass, koll06, vest06, shen08bhmass}. Some
  studies have suggested, however, that FWHM is not a good way to
  characterize line widths, particularly for noisy data
  \citep[e.g.,][]{from00bhmass, pete04bhmass, pete11bhmass}.}

Repeated measurements of individual systems suggests that variation in
the time lag and line widths result in constant $M_{\rm BH}$
estimates, supporting the validity of this technique
\citep[e.g.,][]{pete04bhmass, bent09revmap}. The local relation is
generally calibrated on the width of H$\alpha$ or H$\beta$; however,
at higher redshift we must use other lines (Mg {\sc ii} and C {\sc
  iv}), which have also been validated from local measurements
\citep[e.g.,][]{mclu02bhmass, vest02bhmass, vest06} but for which the
absolute calibrations are quite uncertain (as discussed below).

Despite these limitations, a number of authors have forged ahead to
produce $M_{\rm BH}$ estimates for many thousands of quasars from
large spectroscopic surveys, most prominently SDSS
\citep[e.g.,][]{shen08bhmass} but also with AGES \citep{koll06},
zCOSMOS \citep{merl10bhmass} and other surveys. Typical quasars have
relatively large BHs ($M_{\rm BH}>10^8$ \msun), with an effective
upper limit around $M_{\rm BH}\approx10^{10}$ \msun. Accurate BH mass
estimates allow for a reasonable measurement of the Eddington ratio
(see Eqn.~6 and Fig.~1). However, using optical spectra alone, these
calculations require some estimate of the bolometric correction (BC)
from the UV or optical continuum to obtain $L_{\rm bol}$, which can
depend on physical parameters (as discussed in \S~2.3) such as
luminosity, Eddington ratio, or BH mass \citep[e.g.,][]{marc04smbh,
  hopk07qlf, vasu07bolc, kell08qsox,davi11bolc, raim11bolc_aph}. The
Eddington ratio therefore becomes
\begin{equation}
\lambda \propto {\rm BC\;} {{L^{0.5}}\over{v^2}},
\end{equation}
meaning that the observed distribution in $\lambda$ is primarily
driven by the values of the broad emission line widths.

These estimates generally produce a distribution in $\lambda$ which
peaks around $\lambda = 0.1$--0.3, declining rapidly for $\lambda > 1$
and $\lambda < 0.01$ \citep[e.g.,][]{koll06, shen08bhmass}; contrast
with the Eddington ratios estimated for more typical X-ray AGNs (see
\S3.3.1). This confirms that optically-bright quasars do indeed
represent rapid growth phases, and so are in the regime of
optically-thick, geometrically-thin disk accretion as discussed in
\S2.3. The fact that few quasars are detected at $\lambda\lesssim0.01$
may result from the BH accretion flow switching to the optically-thin
mode at low Eddington ratios, but may also be the result of selection
effects that would prevent such a low-Eddington source from having the
blue continuum and broad lines characteristic of quasars
\citep{hopk09lowlum}.

Comparison of measurements at $z\approx$~6--7 to $z\lesssim 3$ suggest
that higher-redshift quasars are typically at lower $M_{\rm BH}$ but
accreting at higher $\lambda$ than their lower-redshift counterparts
\citep[e.g.,][]{trak11bhmass, dero11z6qso}. However, interpretation of
any observed Eddington ratios is complicated by selection effects that
can strongly skew the $\lambda$ distribution. Indeed, a full Bayesian
treatment of the observed distributions suggests that the typical {\em
  intrinsic} Eddington ratio for massive BHs at high $z$ is closer to
$\lambda\approx$~0.05 or lower \citep{kell10qsoedd}.

When interpreting these results it is crucial to determine the random
and systematic uncertainties in the estimates of $M_{\rm
  BH}$. Comparing $M_{\rm BH}$ measurements for nearby objects using
multiple techniques (including stellar and gas dynamics from
spectroscopy, analysis of megamasers, and reverberation mapping), the
error in virial $M_{\rm BH}$ estimators is found to be $\approx$0.4
dex \citep[e.g.,][]{onke04revmap,woo10revmap}. The latest measurements
of the radius-luminosity relationship suggest that its intrinsic
scatter is only $\approx$0.1 dex \citep{pete10bhmass}. Thus in
principle, radius-luminosity calibrators should be highly effective,
but there are important systematic uncertainties. One of these is the
overall normalization of the virial estimator, which is related to the
structure and kinematics of the broad-line emitting gas and is
calibrated by comparison to estimates from the $M_{\rm BH}-\sigma$
relation \citep[e.g.,][]{onke04revmap,woo10revmap}. This calibration
is performed using nearby low or moderate-luminosity AGN, but at
higher Eddington ratios radiation pressure can significantly change
the kinematics and geometry of the BLR \citep[e.g.,][]{marc08radpress,
  netz10bhmass}. Further issues are that low signal-to-noise
measurements may introduce large asymmetric uncertainties in $M_{\rm
  BH}$ \citep{denn09masserr}, and the value of luminosity included in
the virial estimator may include a color term accounting for the slope
of the continuum \citep{asse11bhmass}. These effects may account for
claimed discrepancies in the $M_{\rm BH}$ estimates using the UV and
Balmer lines.

Optically-bright spectroscopic quasars provide a wealth of information
on demographics, but the samples are strongly biased against AGN that
are obscured by dust and gas. Recent multiwavelength surveys have
shown that a significant fraction (likely a majority) of luminous AGN
at $z\gtrsim1$ are obscured \citep[e.g.,][]{tozz06, hick07abs,
  alex08compthick, pozz10obscagn, baue10obscagn, trei09obscagn,
  trei10obscagn}. A complete census requires a combination of
selection techniques, including searching for X-ray sources with
extremely low ratios of optical to X-ray flux
\citep[e.g.,][]{pozz10obscagn, baue10obscagn}, X-ray spectroscopy of
known quasars \citep[e.g.,][]{page04submm, coma11xmmdeep}, detection
of luminous, narrow emission lines \cite[e.g.,][]{zaka03,
  alex08compthick, reye08qso2, vign10qso2, gill10nev, june11agn}, or
through optical-to-mid-IR SEDs \citep[e.g.,][]{hick07abs,
  baue10obscagn}. Since most obscured quasars do not show bright
optical continua or broad emission lines, estimates of the bolometric
luminosity and $M_{\rm BH}$ are more difficult. However, given these
limitations, in general it appears that the luminosity and Eddington
ratio distributions are broadly similar to unobscured quasars detected
using similar techniques \citep[e.g.,][]{hick07abs, brus10cosmosagn,
  main11obsqso}. Even with the most powerful contemporary techniques
that employ a combination of deep IR, optical, and X-ray data, the
most heavily obscured quasars (Compton-thick sources with $N_{\rm H}
\gtrsim 10^{24}$ cm$^{-2}$) remain extremely difficult to identify
\citep[e.g.,][]{alex08compthick, trei10obscagn,
  alex11stack}. Therefore, estimates of the total obscured quasar
population are either lower limits or depend on assumptions of the
distribution of absorbing column densities ($N_{\rm H}$); however,
these studies robustly demonstrate that a large fraction of the rapid
BH growth is obscured. The nature of this obscured population may shed
light on the processes that fuel the growth of the most massive BHs,
as discussed below.

\begin{figure*}[tp]
\centering
\includegraphics[width=130mm]{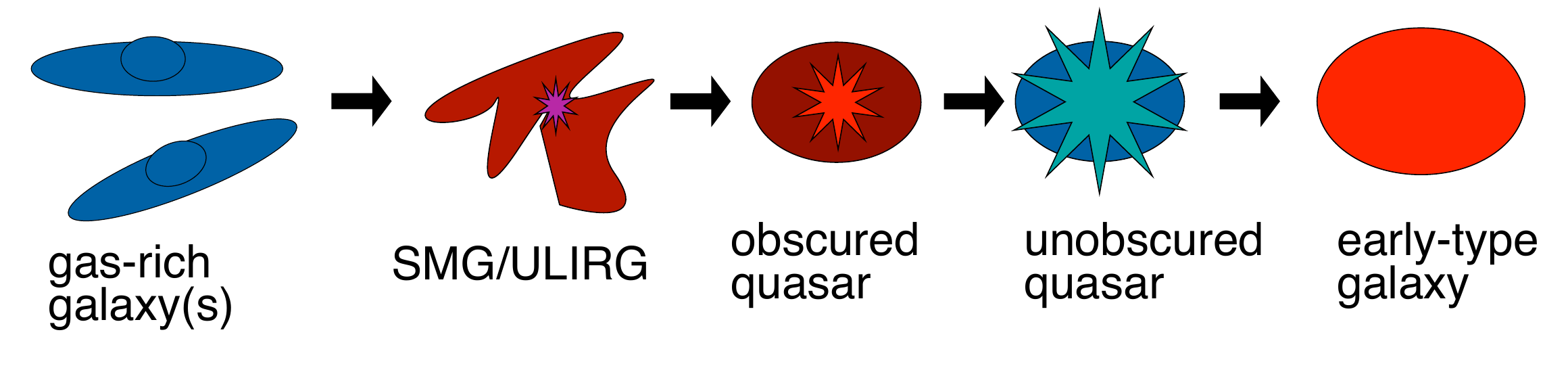}
\caption{Schematic diagram to illustrate the main components in the
  major-merger evolutionary scenario first proposed by \cite{sand88}.}
\end{figure*}

\subsection{Quasars, starbursts, mergers, and the evolutionary sequence}

Given a cosmological census of rapidly growing BHs, the next challenge
is to understand the physical processes responsible for driving the
accretion. The accretion rates of up-to $\approx$~10--100
\msun\ yr$^{-1}$ in quasars cannot easily be produced by secular
processes that could trigger lower-luminosity AGNs (as discussed in
\S3.2.3 \& \S3.3.2). Powerful quasars require higher gas inflow rates
and are more likely to be driven by gas-rich galaxy mergers
\citep[e.g.,][]{kauf00merge, spri05, hopk06merge}; however,
large-scale secular instabilities may also be effective in
particularly gas-rich distant galaxies \citep[e.g.,][]{mo98disk,
  bowe06gal, genz08ifsz2, bour11instab}. Support for major-merger
driven quasar activity comes from (1) the large fraction of systems
with disturbed morphologies (see \S3.3.2), and (2) the good agreement
between predictions for the merger rates from dark matter simulations
(for adopted empirical prescriptions the quasar fueling) and the
spatial clustering and space densities of distant quasars
\citep[e.g.,][]{hopk08frame1, trei10bhmerge}.

While mergers are favoured by a number of models
\citep[e.g.,][]{kauf00merge, spri05, hopk06merge, sija07agnfeed,
  dima08bhfeed}, {\em any} quasar triggering mechanism requires
relatively massive systems with large supplies of cold gas, which are
generally found in dark matter halos with $M_{\rm
  halo}\sim10^{12}$--$10^{13}$
\msun\ (\citealt[Figure~4][]{crot09qsohalo}; see also Fig.~5), just
below the ``maximal quenching" mass scales. Spatial clustering and
environment measurements of quasars \citep[e.g.,][]{ross09qsoclust,
  liet09qsoenviron, hick11qsoclust, carr11qsogroup} suggest that
quasars do indeed reside in halos of these masses at every redshift
(as discussed in \S3.4 and shown in Fig.~5). This implies that at high
redshift, quasars are found in the largest collapsed system in the
Universe (and are the progenitors of today's most massive early-type
galaxies) while in the local Universe quasars are found in much more
typical galaxy environments. Thus the mass of the dark matter halo may
itself be the key parameter in understanding the fuelling of quasars.

The rapid flow of cold gas that is necessary to fuel a quasar will
inevitably be expected to also result in high rates of star formation
(as discussed in \S~2.1--2.2). Robust evidence for links between
powerful starbursts and quasars come from studies of local powerful IR
galaxies ($L_{\rm IR} > 5\times10^{11}$ $L_{\odot}$). The vast
majority of such objects in the local Universe are major mergers of
galaxies, with higher luminosities found during late stages when the
galaxies are at small separations \citep[e.g.,][]{clem96ulirg,
  ishi04ulirg}. At higher $L_{\rm IR}$ the fraction of the luminosity
from the AGN increases, and the large masses of nuclear gas and dust
ensure that much of the BH growth is observed to be heavily obscured
\citep[e.g.,][]{tran01ulirg, yuan10lirg, iwas11lirgx, petr11lirg}. The
local results are broadly consistent with models in which mergers fuel
a rapid starburst and a phase of obscured BH growth, followed by an
unobscured phase after the gas is consumed or expelled from the galaxy
by stellar or quasar feedback \citep[e.g.,][]{sand88, dima05qso,
  hopk08frame1}; see Fig.~6. However, powerful starbursts are rare in
the local Universe, compared to higher redshift where they dominate
the star formation density \citep[e.g.,][]{lefl05irlf, rodi11irlf}. A
key question, then, is whether a similar starburst-quasar scenario is
the dominant process at high redshift, during the peak epoch of quasar
activity where the largest BHs accreted most of their mass. Testing
this picture is the subject of a number of recent studies.

One approach is to select high-redshift starburst galaxies based on
their IR or submmilimeter emission, and study the growth of BHs in
these systems. Useful observational tools are X-ray observations and
mid-IR spectroscopy, which can distinguish between dust heated by star
formation and the AGN. The most powerful starbursts at high redshift,
submillimeter galaxies (SMGs), have gas kinematics and morphologies
that are characteristic of mergers \citep[e.g.,][]{tacc08smg,
  enge10smg, reic11smgmerge}. A high fraction of these objects also
host AGN \citep[e.g.,][]{alex03submm, alex05, lair10submmx}, but they
generally have {\em Spitzer} IRS mid-IR spectra that are dominated by
star formation as indicated by luminous PAH emission features
\citep[e.g.,][]{vali07smg,pope08smgspitz,copp10irs}. Only 15\% of SMGs
are dominated in the mid-IR by steep AGN continua, and even these
powerful AGN generally do not produce the bulk of the bolometric
output, which is dominated at longer wavelengths by the cool dust from
star formation (as also found for star-forming galaxies detected at 70
$\mu$m; \citealt{syme10spitzer}). However, the presence of powerful
AGN in some starbursts is consistent with a ``transition'' phase
between powerful star formation and rapid BH growth
\citep[e.g.,][]{pope08smgspitz, copp10irs} as predicted by major
merger fueling models \citep[e.g.,][]{spri05, dima05qso,
  hopk08frame1}. In SMGs that host AGN activity, estimates of the
galaxy luminosities and BH masses \citep[as discussed in \S~5.2; e.g.,
][]{alex08bhmass, carr11qsogroup} indicate that the BH is undermassive
relative to the host galaxy and may therefore be in an early stage of
accretion that may precede the bright unobscured phase. These results
give broad support to a picture in which a rapid starburst is
associated with the triggering of accretion onto a BH which quickly
grows in mass.

Further exploration of this ``transition'' population comes from
studies of obscured quasars selected independently of any
star-formation signatures (such as far-IR or submillimeter emission).
In the unified AGN model, obscuration in quasars is purely due to
orientation and so the host galaxies and large-scale structures of
quasars would be independent of obscuration
\citep[e.g.,][]{anto93}. In contrast, evolutionary scenarios would
suggest significant differences in the host-galaxy properties
\citep[e.g.,][]{sand88, spri05, dima05qso, hopk08frame1}; see
Fig.~6. Some studies have focused on the SEDs of X-ray selected
obscured quasars, which can often be modeled with an extension of the
simple ``torus'' geometry, but in many cases also require a
contribution from the host galaxy at longer wavelengths indicating an
energetically important starburst \citep[e.g.,][]{main05qso2submm,
  vign09qso2, poll08obsc, pozz10obscagn}.

Additional tests of this evolutionary scenario come from clustering
analyses, as differences in host halo masses would be inconsistent
with a simple ``unified'' picture in which obscuration is purely due
to the orientation of an obscuring torus along the line of sight
\citep[e.g.,][]{anto93, urry95}. The first measurement of the spatial
correlation of IR-selected obscured quasars \citep{hick11qsoclust}
indicates that they are clustered at least as strongly as their
unobscured counterparts, with $M_{\rm halo}$ possibly above $10^{13}$
$h^{-1}$ \msun. This suggests the possibility that obscured quasars
represent an earlier evolutionary phase in which the BH is
undermassive relative to its host halo, such that obscured quasars are
found in more massive halos compared to unobscured quasars with
similar BH mass\footnotemark.

\footnotetext{We note that lower-luminosity X-ray selected AGN appear
  to show the opposite trend, with stronger clustering for unobscured
  compared to obscured sources \citep{alle11xclust}. This suggests
  that this evolutionary scenario may not hold for lower-luminosity
  systems, and provides strong motivation for more precise future
  clustering measurements that probe a wide range of luminosity and
  redshift.}

If obscured quasars represent an early phase of accretion then do we
see the shutdown of star formation upon the emergence of an unobscured
quasar? Evidence based on the submillimetre properties of quasars
suggests that unobscured quasars do coincide with a significant
decrease in star formation.  \cite{page01, page04submm} and
\cite{stev05submm} find a stark difference in the submillimetre
emission between X-ray obscured optical quasars and X-ray unobscured
optical quasars: the X-ray obscured optical quasars (which represent
$\approx$~15\% of the optical quasar population) have SFRs implied
from the submillimetre data that are about an order of magnitude
higher than the X-ray unobscured optical quasars.  Tentative evidence
for the catalyst of this decreased star formation is found from the
spectra of these systems. The X-ray spectra of the X-ray obscured
optical quasars indicate that the obscuration is due to an ionized
wind, providing evidence for a quasar-driven outflow that may be
associated with the short-lived transition from starburst to
AGN-dominated systems \citep{page11obsqso}. Near-IR IFU spectroscopy
of another X-ray obscured optical quasar shows evidence for an
energetic outflow on scales of $\approx$~4--8 kpc, which may be
starting to shut down the star formation across the whole galaxy
\citep{alex10outflow}.  At lower redshifts, massive and energetic
molecular outflows have been observed in powerful AGN
\citep{stur11qsoco} including the broad-absorption-line quasar Mrk 231
\citep{fisc10mrk231}, for which the outflow is extended on kpc scales
\citep{feru10outflow}. These results provide tentative evidence that
AGN-driven winds may indeed influence gas on the scale of the host
galaxy and help quench the formation of stars (as discussed in \S~5).

Multiwavelength studies of optical quasars also indicate a small
population ($\approx$~10\%) of objects that lack clear hot-dust
signatures \citep[e.g.,][]{jian10hdpqso, hao10hdpqso,
  hao11hdpqso}. There are a number of interpretations for the lack of
hot-dust emission in these objects,\footnotemark including the
possibility that some hot dust-poor quasars represent the end stage of
the quasar evolutionary sequence.

\footnotetext{In the highest-redshift ($z>6$) systems, there may
  simply not be enough dust in the Universe to form a significant
  dusty torus \citep{jian10hdpqso}.}

Taken together, the populations of the most-active systems at
$z\approx$~1--3 may be tentatively placed into a broad evolutionary
sequence ({SMGs/ULIRGs--obscured quasars--unobscured quasars--``hot
  dust-poor'' systems); see Fig.~6. However the precise relative
  number densities, duty cycles, and environments of these different
  objects remain poorly constrained, and it is therefore not clear how
  well the different populations can be accounted for by a single
  evolutionary scenario. Future observations with larger samples and
  improved diagnostics will provide the tools to verify or falsify
  this general picture for the rapid growth of distant massive BHs.

\subsection{Very high redshifts and the formation of ``seed'' black holes}

Through the preceeding sections we have discussed the significant
progress made in understanding the processes that trigger the rapid
growth of the most massive BHs. However, {\em any} scenario of quasar
fuelling requires an initial BH onto which accretion-driven growth can
occur, and the nature of how and when these initial seeds were formed
remains an important open question. In many cosmological simulations
that include BHs, these "seed" BHs are simply put in by hand with some
arbitrary mass \citep[e.g.,][]{bowe06gal, bowe08flip, boot10mhalo,
  fani11agn, fani12agn_aph}. The late-time properties of the BHs and
galaxies are generally insensitive to the mass of the seeds, as later
accretion ``erases'' any memory of the initial conditions.  However,
the precise nature of the seed BHs is still crucially important, as
the early formation and growth of BHs may in fact represent a key
component in the early formation of structure. Early BH accretion may
also produce a significant fraction of the total radiation background
responsible for reionization of the Universe at $z\simgt$~7
\citep[e.g.,][]{mada04bhreion, mira11highzbh}.

One primary challenge for any model of seed BH formation is posed by
the existence of luminous quasars with $M_{\rm BH}>10^9$ $M_{\odot}$
at $z>6$, when the Universe was $\lesssim$1 Gyr old
\citep[e.g.,][]{fan06, jian09z6qso, will10z6qso, dero11z6qso,
  mort11z7qso}. For a BH accreting with an Eddington ratio $\lambda$
and assuming a radiative efficiency $\epsilon$, the growth time of a
BH from an initial mass $M_{in}$ to a final mass $M_{fin}$ is given by
\citep[e.g.,][]{volo10bhform}:
\begin{equation}
t_{\rm growth} = 0.45 {\rm \; Gyr\;} \frac{\epsilon}{1-\epsilon} {{\lambda}^{-1}} \ln \left(\frac{M_{fin}}{M_{in}}\right)
\end{equation}
This formation time scale puts robust lower limits on the masses of
the seed BHs, for there must be sufficient time in the history of the
Universe for them to grow into the massive BHs seen in the high-$z$
quasars. For an initial BH formed at a high redshift (say $z=20$) to
reach $10^9$ $M_{\odot}$ $\approx$0.7 Gyr later at $z=6$, assuming
standard $\epsilon = 0.1$ and continuous accretion at the Eddington
limit, it must have started with $M_{in}\sim 100$ $M_{\odot}$ (with
the caveat that some additional growth can occur through BH-BH
mergers; e.g., \citealt{sesa07bhlisa, arun09bhlisa, sere11lisa}). For
a lower formation redshift or more stochastic accretion (and thus
lower average $\lambda$) intial BH masses would necessarily have been
higher. A successful model of seed formation must therefore produce
sufficiently massive BHs to satisfy these constraints.

There are currently three main candidate mechanisms for seed
formation: (1) remnants of massive population III stars (2) direct
collapse of primordial gas clouds, and (3) runaway collisions in dense
stellar clusters. Each model makes different predictions for the
masses and number densities of the BH seeds \citep[see][for a detailed
  review]{volo10bhform}.

The first candidate mechanism proposes that BH seeds are produced by
the deaths of Population III stars. Stellar models suggest that this
first generation of stars, which formed from very low-metallicity gas,
can have large masses of $\approx$~100--600 $M_{\odot}$
\citep{abel02firststar, brom09firststar}, and that the collapse of
stars with masses $\gtrsim260$ $M_{\odot}$ can produce $\sim$100
$M_{\odot}$ BH remnants \citep[e.g.,][]{fry01bhstar,
  mada01bhpop3}. This model has the advantage of following the
well-established mechanism for the formation of lower-mass BHs through
stellar processes, but is limited by the small mass of the
seeds. Radiative feedback limits the possible accretion rates onto
seed stellar-mass BHs \citep[e.g.,][]{john07firstbh, alva09seedbh},
making it difficult to construct mechanisms by which such BH seeds can
achive the high accretion rates required to produce the high-mass
quasars observed at $z\gtrsim 6$ \citep[e.g.,][]{haim01highzqso,
  volo05highzbh, volo06accr}. Therefore, stellar processes may be
responsible for a significant population of early BHs, but in the most
biased regions (where we find the high-$z$ quasars) another formation
mechanism is required. However, recent studies have challenged the
idea that the first stars are very massive, due to either
fragmentation or feedback effects \citep[e.g.,][]{glov08pop3,
  mcke08pop3, turk09pop3, tren09pop3, stac10pop3, grei11pop3}, casting
some doubt on the the role of Population III remnants as BH seeds.

The second candidate mechanism proposes that BHs form directly inside
the densest peaks in the matter distribution. In certain
circumstances, the BHs can be produced through the collapse of gas
clouds. In regions with radiation fields that can dissociate $H_2$
molecules, star formation is suppressed and gas cooling proceeds
through atomic processes. Detailed hydrodynamic simulations have
explored this process \citep[e.g.,][]{wise08protogal,rega09protogal,
  shan10seedbh, john11bhcollapse} and indicate that supermassive stars
at the centers of halos can, in principle, accrete up to a few
$\times10^5$ \msun\ over $\sim$2 Myr \citep{john11bhcollapse}. A
significant fraction of this mass may then go into a seed BH which can
continue to grow, although less rapidly because of the effects of
radiative feedback. A related possibility is that direct collapse does
not occur in an isolated halo, but is triggered by the mergers of
multiple halos, providing highly unstable conditions and large gas
inflow rates $>1$ \msun\ yr$^{-1}$ \citep[e.g.,][]{volo10quasistar,
  bege10quasistar}. Such conditions can produce ``quasistars'',
massive optically-thick structures in which the radiation pressure
from accretion is balanced by the ram pressure of infalling gas, in
which gas accretion can significantly exceed the Eddington limit for
the central BH \citep{bege06quasistar}. This process can produce BH
seeds with masses of up to $\sim$$10^5$ \msun\ in the most biased
regions of the early Universe.

The third candidate mechanism proposes that the seed BH formation is
preceded by the formation of dense stellar clusters, the centers of
which then undergo runaway merging due to stellar-dynamical processes
that produces a BH \citep[e.g.,][]{deve09bhcluster, deve10bhcluster,
  davi11bhcluster}. This process can proceed in regions with
significant molecular gas, and does not require extremely high inflow
rates as in direct gas-dynamical collapse, but generally produces
smaller BH seeds with masses $\sim$$10^3$ \msun.

Given these possible mechanisms for the formation and growth of seed
BHs, the challenge for observers is to distinguish between them. One
possibility is to trace the high-$z$ BH population through accurate
measurements of the number density of AGN at $z\gtrsim4$. The most
progress has been made through studies of high-$z$ optical quasars
\citep[e.g.,][]{fan06, jian09z6qso, will10z6qso, dero11z6qso,
  mort11z7qso}. However, these samples comprise only the most massive
end of the BH distribution, and these quasars appear to be growing
more rapidly than their lower-redshift counterparts
\citep{dero11z6qso} indicating that they are also biased to higher
Eddington ratios. In principle, X-ray observations allow us to explore
further down the mass distribution, and some success has been achieved
by searching for X-ray counterparts to high-redshift optical and IR
sources \citep{fior10agnevol}. In the future, a combination of
sensitive, high angular resolution X-ray surveys (e.g., {\em
  eROSITA}, \citealt{pred07erosita}; {\em Wide-Field X-ray
  Telescope}, \citealt{murr10wfxt, brus11wfxt}) over wide areas
covered by IR and optical observations will be required to accurately
measure the high-$z$ AGN luminosity function to faint limits.

Measurements of the AGN space density are sensitive not only to the
seed formation mechanisms, but also the rate of BH growth and BH--BH
mergers. In principle it may be possible to observe the seed BHs
directly, by searching for the signatures of accretion onto
$\lesssim10^5$ \msun\ BHs. Even for such BHs, the hard radiation field
may produce characteristic nebular line emission (in particular, He
{\sc II} $\lambda$1640) that could be detected with NIRSpec on the
{\em James Webb Space Telescope} \citep{john11bhcollapse}, while
quasistars could produce a significant population of sources
observable at mid-IR wavelengths with {\em JWST}
\citep{volo10quasistar}. In the more distant future, these BHs may
also be detectable through gravitational wave signatures when they
merge \citep[e.g.,][]{sesa07bhlisa, arun09bhlisa, sere11lisa}.

Finally, a complementary strategy for understanding seed BHs is to
look for residual signatures at lower redshifts. For the most massive
systems, the properties of the seeds have been ``washed out'' by
subsequent mass accretion events. However, some of the lower-mass
($\lesssim10^5$ \msun) BHs in the local Universe may represent relics
of the initial epoch of seed formation \citep[e.g.,
][]{vanw10bhrelic}. Therefore the formation process for seed BHs may
be reflected in the distribution of low-mass BHs within dark matter
halos at low redshift, which can be probed using accurate measurements
of the relation between BH masses and galaxy properties \citep[e.g.,
][]{volo08bhseed}.  By studying seed BHs in concert with the
subsequent fueling and evolution of powerful quasars, we can aim to
produce a coherent picture of the growth of massive BHs from their
early origins to the present epoch.

\section{What is the detailed nature of AGN feedback and its effects on black-hole fuelling and star formation?}

A recurring theme in any discussion of BH growth is the impact that
energy released by accretion has on the surrounding gas in the
system. In principle, a growing BH releases plenty of energy to impact
its surroundings: the thermal energy of the hot gas atmosphere in a
$10^{13}$ \msun\ dark-matter halo is $\approx$$10^{61}$ erg, while the
total accretion energy of a $10^9$ \msun\ BH is
$\approx2\times10^{62}$ erg! Therefore, given a sufficiently strong
coupling between the radiative or mechanical output of the BH and the
surrounding gas, the AGN should be able to disrupt its environment and
potentially regulate its own growth and star formation in the host
galaxy \citep[e.g.][]{silk98, bowe06gal}. In this section we explore
the various physical processes, such as winds and jets, by which
energy from the BH can couple to the surrounding gas. We also address
the observed consequences of this feedback, with emphasis on the
relations between BHs and their host stellar bulges, as well as the
gas content of dark-matter halos.

\subsection{Driving feedback through winds and jets}

Any feedback process from a growing BH requires coupling between the
energy released by the BH and the surrounding matter. In general these
can take two forms: (1) ``winds'' (often referred to as superwind-mode), 
which comprise wide-angle,
sub-relativistic outflows and tend to be driven by the radiative
output of the AGN, and (2) ``jets'' (often referrred to as radio-mode), 
which are relativistic outflows
with narrow opening angles that are launched directly from the
accretion flow itself; see Fig.~7. As described in \S3.32, \S3.3.2, \&
\S4.2 the radiatively-dominated AGN that drive winds are expected to
be relatively high-Eddington ratio systems, while jets are most commonly
produced (except for the highest-power sources) by lower-Eddington ratio
accretion flows.

\subsubsection{Radiatively-driven winds}

In recent years, a wide range of observational efforts have been aimed
at searching for evidence for AGN winds, largely through the presence
of highly blueshifted absorption and emission lines. Much work has
focused on bright, nearby AGN, for which high signal-to-noise
observations readily allow the detection of the relevant line
features. As discussed in \S2.1, IFU observations of local Seyferts
show evidence for outflowing gas on scales of $\sim$10-100 pc
\citep[e.g.,][]{stor07ifu, davi09ifu, stor10ngc4151, schn11ifu}, while
spectropolarimetry of a low-redshift quasar shows a high-velocity
($\sim$4,000 km s$^{-1}$) outflow within the nuclear torus itself,
close to the accretion disk \citep{youn07qsowind}.  High-velocity
winds are also commonly observed using X-ray spectroscopy, which can
trace atomic transitions in ionized gas. The presence of highly
blueshifted absorption and emission lines in a number of quasars and
Seyferts (e.g., PDS 456, PG 1211+143, NGC 4051, 3C 455, MR2251-178)
indicate outflows of $v\approx$~0.1--0.3$c$ \citep{poun03ngc5548,
  reev03outflow, tomb10outflow, lobb11ngc4051, goff11xray}, as
expected for models of momentum-driven winds, as discussed below.  A
lower velocity outflow ($\sim$500 km s$^{-1}$) is observed in the
local Seyfert 2 galaxy NGC 1068 (although it is confined to the
nuclear regions rather than extending on host galaxy scales). The
high-resolution X-ray spectrum for NGC 1068 is characteristic of
photoionization rather than collisional ionization, suggesting that
the outflow is driven by a radiatively efficient wind rather than a
jet \citep{evan10ngc1068}.

\begin{figure}[tp]
\centering
\includegraphics[width=75mm]{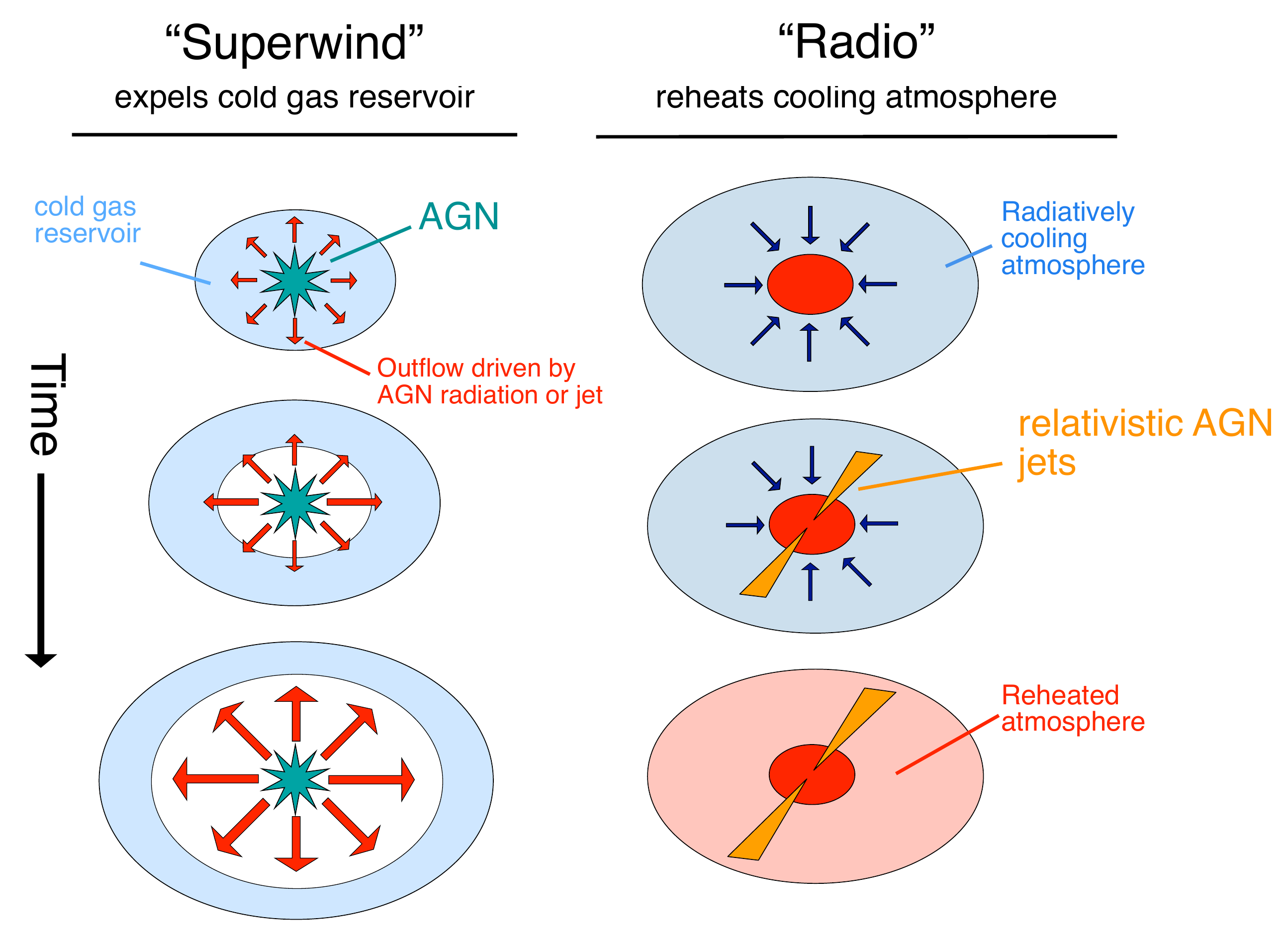}
\caption{Schematic diagrams to illustrate the two main modes of AGN
  outflows: ``superwind''-mode outflows such as those found in
  luminous AGNs and ``radio''-mode outflows such as those found in
  low-excitation radio-loud AGNs.}
\end{figure}

Similar statistical analyses show evidence for powerful outflows in
distant quasars. In the spectra of high-$z$ quasars from SDSS, the
peak of the broad C {\sc iv} emission is typically blueshifted,
indicating outflows with $\sim$1000 km s$^{-1}$
\citep[e.g.,][]{rich02blueshift, rich11blueshift}. Quasars with larger
blueshifts show lower equivalent widths of C {\sc iv} as well as
weaker observed X-ray emission. A similar trend with X-ray emission is
found for broad absorption line (BAL) quasars, which are believed to
represent the $\approx$~20--40\% of quasars (with evidence for a
redshift dependence; e.g.,\ \citealt{all11balqso}) for which the
nuclear continuum is viewed through the outflow (there is evidence
that maybe all quasars host energetic winds;
\citealt{gang08winds}). The maximum blueshift in BAL quasars
(corresponding to the terminal velocity of the wind) is higher for
sources with weaker X-ray emission \citep[e.g.,][]{gall06balqso,
  gibs09balqso}. Both these and the results on C {\sc iv} blueshifts
can be understood in terms of radiation driving of the quasar wind;
with a weaker X-ray continuum (possibly associated with an absorber
near the midplane), the gas close to the BH is less highly ionized and
so more easily driven by the UV continuum radiation from the quasars.

Evidence for outflows also comes from statistical studies of large
numbers of moderate-luminosity AGNs at $z\simlt$~0.4.  A number of
narrow-line Seyfert galaxies show complex [O {\sc iii}] emission with
a broad blue wing in addition to the main narrow component
\citep[e.g.,][]{veil91seyfline, boro05blueseyf}, and statistical
studies suggest these may be ubiquitous (Mullaney et al., in
prep). The width and luminosity of the broad component increase with
$L_{\rm [OIII]}$ but are independent of radio-loudness, possibly
indicating that the winds are driven by the radiative output of the
AGN rather than by relativistic jets.

These studies demonstrate that energetic winds are common in the AGN
population.  However, they cannot establish on the basis of this data
whether these winds generally have significant effects on the scale of
the host galaxy since, in the majority of the cases the winds are
observed only along the line of sight and there are no direct
constraints on the spatial distribution of the outflowing gas
\citep[e.g.,][]{trem07psb, arav08outflow, dunn10outflow}. To fully
understand the mass distribution, velocity structure, and energetics
of these winds requires improved constraints on the covering fraction
and clumpiness the outflows, which are difficult or impossible to
determine from spectroscopy along a single line of sight.

Recent observations have begun to overcome some of these challenges
through spatially resolved spectroscopy (IFU observations). For
example, a quasar hosted by a $z\sim 2$ SMG shows broad blue wings to
[O {\sc iii}] lines extend out to several kpc
\citep{alex10outflow}. The wind velocity in this case is $\approx$~300
km s$^{-1}$, which suggests that it could possibly be driven by the
powerful starburst instead of the AGN. In addition, the observed
outflow is sufficient to strongly disturb gas in the galaxy, but is
unlikely to completely unbind the material from the system (see
\S5.1.2 for IFU observations of several high-z radio-loud quasars).
At lower redshifts, long-slit spectroscopy of luminous narrow-line
quasars at $z<0.5$ has revealed ubiquitous galaxy-scale ionized clouds
with high velocity dispersions, which appear to be powered by the AGN
and may be associated with outflows \citep{gree11obsqso}. An extreme
example of an outflow has also been observed with CO line observations
of the nearby quasar Mrk 231, in which broad wings to the observed
lines indicate the presence of a high-velocity ($\sim$750 km s$^{-1}$)
wind that is resolved on kpc scales and contains $\sim$700
\msun\ yr$^{-1}$ of molecular gas \citep{feru10outflow}. This outflow
may be expected to evacuate the cold gas content of the galaxy within
$\sim$10 Myr. Observations are revealing more and more examples of
powerful feedback from radiatively driven winds, but the overall
prevalence of these outflows and their overall impact on the galaxy
population remains to be determined.

\subsubsection{Relativistic jets}

In contrast to the picture for radiative winds, relativistic jets from
AGN are commonly observed to influence gas on the scales of the host
galaxy or even its parent dark matter halo. Indeed, the brightest
structures observed in radio AGN are often on $\gtrsim$\,kpc scales
and are produced by the coupling of the AGN outflow to its
environment. The physical parameters that produce AGN jets (low
accretion rates, fueling from hot atmospheres, BH spin) remain poorly
understood; however, the impact of the jets on the hot gas in galaxies
and dark-matter halos is relatively well established.

In massive systems with hot ionized atmospheres, AGN jets interact
most strongly with the gas by inflating bubbles of relativistic plasma
in the ambient medium. In the classical picture for radio lobe
formation \citep{sche74radio}, the bubbles are strongly overpressured
and so produce shocks that can directly heat the atmosphere. However,
{\em Chandra} observations of hot gas around AGN jets yield only a few
clear examples of large-scale shocks, and these are predominantly in
low-luminosity AGN with FR~I morphologies \citep[e.g.,][]{form07m87,
  cros09cena}, as opposed to the powerful FR~II sources that are
expected to drive the strongest shocks. Therefore most of the
mechanical energy from the jets is transferred in the form of $P{\rm
  d}V$ work by the inflating bubbles \citep[e.g.,][]{birz04,
  birz08jet, cava10jet} rather than through strong shocks. The bubbles
are observed to carve out cavities in the hot gas, so there is
expected to be little mixing between the hot gas and the relativistic
plasma. In this case the pressure is simply the external pressure
$P_{\rm ext}$ (which can be measured from the X-ray images), so that
$\Delta E = P_{\rm ext} V_{\rm lobe}$. Assuming a timescale for the
lifetime of the radio source (for example from the buoyant rise time
of the inflated bubble), we can then estimate the total mechanical
power $P_{\rm mech}$ output by the AGN \citep[e.g.,][]{birz04,
  dunn05bubbles, dunn08bubbles, birz08jet, cava10jet}.

Such estimates show that for most radio-loud AGN, the mechanical power
in the relativistic outflows vastly exceeds (by as much as $\sim$1000
times) the radiative output of the AGN \citep[e.g.,][]{birz08jet,
  cava10jet}.  These studies also find correlations between the radio
luminosity and (much larger) mechanical power, which can be used to
estimate overall distribution of mechanical energy input in the
Universe \citep[e.g.,][]{hein07klf, merl08agnsynth, lafa10radio}; see
\S3.3.1. Analysis of the cavities can also be used to explore whether
the mechanical output of the AGN is sufficient to balance the energy
lost in the atmosphere through radiative cooling. While limited mainly
to bright, massive clusters (where the cavities are most easily
identified), the $P{\rm d}V$ work done by the cavities can usually
offset radiative losses \citep[e.g.,][]{birz04,
  dunn08bubbles}. Indeed, the heating from intermittent AGN activity
and the radiative cooling are often in relatively close balance
\citep{raff08feedback}.  These results suggest a limit cycle in which
the cooling of hot gas onto the nucleus of the galaxy fuels
intermittent AGN outbursts, which in turn provide sufficient heating
to slow down or stop the accretion flow \citep[e.g.,][]{best05,
  pope11feedback}. This scenario represents ``feedback'' in the true
sense of the term, in that the energy output by the AGN has a direct
impact on its own fuel supply.


While most systems are roughly consistent with obeying such a limit
cycle, there are some examples in which the radio jets impart far more
power than is necessary to simply reheat the gas, but instead may be
sufficient to unbind much of the hot atmosphere. Such sources are
common among galaxy groups \citep{giod10groups}, but the most extreme
case is in the massive cluster MS0735.6+7421
\citep[e.g.,][]{mcna05nature, mcna09ms07}; see also
\cite{siem10qso}. This object exhibits enormous cavities of $\sim$200
kpc in diameter, for which the $P{\rm d}V$ work required is
$\sim$$10^{62}$ erg, or close to the entire accretion energy of a
$10^9$ \msun\ BH. The AGN in this source may be driven by the
accretion of cold gas rather than fueling directly from the hot
atmosphere, and may be an example of a radio-loud quasar (that is, a
source with both powerful radiative {\em and} mechanical output),
rather than lower-Eddington, optically faint AGN.  In support of this
picture, high-excitation (radiatively efficient) radio AGN are
generally found in blue star-forming galaxies that are expected to be
rich in cold gas, while low-excitation (radiatively inefficient)
sources are found in red, passive galaxies with low cold gas content
\citep[e.g.,][]{smol09radio, herb10radio, smol11radiomol}. In the case
of MS0735.6+7421, the enormous mechanical power needed to produce the
cavities has prompted a more extreme possibility, that the outflow
must be powered not only by accretion energy but also by the
extraction of BH spin energy by the relativistic jet
\citep{mcna09ms07, mcna11spin}. Whatever the ultimate power source,
these extremely powerful mechanical outflows may be responsible for
expelling significant gas from the centers of halos, as discussed in
\S~\ref{sec.feedbackgas}.

Finally, there are examples of relativistic jets interacting not only
with hot, virialized plasma, but also with cooler gas in the host
galaxy. IFU observations of powerful ($\simgt$$10^{26}$ W Hz$^{-1}$ at
1.4 GHz) radio-loud quasars have revealed outflows in [O {\sc iii}]
that are aligned with the radio jet axis over scales of $\sim$10 kpc,
with $v\sim1000$ km s$^{-1}$ and ${\rm FWHM} \sim 1000$ km s$^{-1}$
\citep[e.g.][]{nesv06outflow, nesv07outflow, nesv08outflow}. These
results suggest that relativistic jets can ionize and expel
significant amounts of cool ($\sim$$10^4$ K) gas, in addition to
strong interactions with the virialized hot atmosphere.

\subsection{Observational consequences of feedback}

The discovery of the ubiquity of massive BHs in galactic centers and
the apparently tight relationship between the mass of a central BH and
the properties of its host stellar bulge has motivated a great deal of
interest in the physical consequences of BH growth. These results
provide compelling (although indirect) evidence for interplay between
the growth of the BH and the formation of the host galaxy, despite
enormous differences in mass ($\approx$~3 orders of magnitude) and
linear scale ($\approx$~8--9 orders of magnitude). Given the energetic
importance of BH accretion in the context of the galaxy, feedback from
AGN was immediately recognized as a possible mechanism to produce
these observed correlations \citep[e.g.,][]{silk98, fabian99feedback,
  king03msigma}

\subsubsection{Correlations between BHs and galaxy spheroids}

The first step in understanding BH-spheroid relationships is to
measure accurately the nature and intrinsic scatter in the relations,
as well as their evolution with redshift. A wide range of studies have
explored correlations of $M_{\rm BH}$ with bulge mass, luminosity,
velocity dispersion, and gravitational potential, and found a
remarkably small intrinsic scatter for massive galaxies in the local
Universe \citep[e.g.,][]{mago98, gebh00, ferr00, gult09msigma}. Recent
studies of lower-mass galaxies suggest a change in relationship at the
low end, with an increase in scatter \citep[e.g.,][]{gree08lowmass,
  gree10maser}, although some of this may be driven by the prevalence
of pseudobulges in disk galaxies rather than classical bulges, where
the correlations are much weaker or non existent
\citep[e.g.,][]{korm11halo, jian11lowmass}.

To understand the physical origin of these correlations, important
clues come from the {\em evolution} of the BH--spheroid mass
relationship with cosmic time, which can now be explored
observationally. The difficulty in measuring BH masses for distant
galaxies restricts the current studies to systems hosting AGN
activity. BH mass estimates are computed from either the virial method
(for broad-line sources; e.g., \citealt{woo08qsombh, kim08msigma,
  deca10mbhevol, benn10bhmass, merl10bhmass}; see \S4.1) or via some
assumptions about the relationship between obscured and unobscured
AGNs (as for SMGs; \citealt{alex08bhmass}), each of which have
significant uncertainty. The presence of an AGN also presents the
additional challenge of separating the light from the host galaxy to
that from the accreting BH, and difficulties in separating the mass or
potential of the stellar bulge from the rest of the galaxy
\citep[e.g.,][]{deca10mbhmethod, merl10bhmass}.

Despite these challenges, a number of authors have explored the
evolution in $M_{\rm BH}$--$M_{\rm bulge}$ for distant AGNs. Studies
of broad-line quasars consistently find that the BH is {\em
  overmassive} relative to its host galaxy compared to the local
relation \citep[e.g.,][]{walt04qsoco, woo08qsombh, deca10mbhevol,
  benn10bhmass, merl10bhmass}. In most cases comparisons are made
between $M_{\rm BH}$ and the total galaxy mass, rather than the
spheroid properties, which are challenging to determine with current
observations. These results have prompted speculation that BH growth
precedes that of the host galaxy. However the observed evolution may
be due in part to selection effects that cause more massive BHs to be
more readily detected \citep[e.g.,][]{laue07mbhbias, shen10mbhbias},
and some observational results have shown little redshift evolution,
at least in the relation between BH and total galaxy mass
\citep[e.g.,][]{jahn09bhgal}. In contrast, corresponding analyses of
X-ray selected AGN in powerful starbursts (SMGs) indicate that the BHs
lie up-to an order of magnitude {\em below} the relationship for
$z\approx$~2 quasars, suggesting that these BHs are in the processes
of ``catching up'' to their final mass \citep{alex08bhmass,
  carr11qsogroup}; however, SMGs may represent an earlier stage in the
evolution of quasars (see \S4.2 and Fig.~6). At present, the
significant uncertainties in the current approaches, as well as the
biases inherent in limiting studies to galaxies hosting AGN, make it
difficult to draw robust conclusions about the evolution of
BH-spheroid mass relationships. Improved observations and larger,
well-characterized galaxy and AGN samples should provide better
constraints on this evolution in the future.

\subsubsection{Physical drivers of BH-spheroid relationships}

Redshift evolution notwithstanding, a great deal of theoretical effort
has attempted to explain the observed BH-spheroid relationships in the
local Universe, with a particular focus on the effects of outflows
from AGN. As one example of the vast array of analytical models,
\citet{king11outflow} consider roughly Eddington-limited accretion in
which the optical depth to photon scattering is $\sim$1. In this case,
the outflow momentum is comparable to the photon momentum, which
yields a typical outflow velocity $v \simeq {\epsilon}/{\lambda} c
\sim 0.1 c$, as observed in the X-ray spectra of many AGN (discussed
below). Compton cooling by the AGN radiation field allows this outflow
to cool rapidly inside a radius of $\sim$1 kpc, reducing the thermal
energy of the flow but conserving its momentum. Setting the thrust
from such a ``momentum-driven'' flow to balance the gravitational
force on the bulge gas naturally produces a BH-spheroid mass
relationship with $M_{\rm BH} \propto \sigma^4$, as observed
\citep{king05}. This model also predicts that BHs in AGN should in
general be accreting near the Eddington limit, and growing up toward
the $M_{\rm BH}$--$\sigma$ relation that is observed for passive
systems \citep{king10bhedd}.

Such analytical models provide a clear physical picture for simple
generalized conditions (spherical symmetry, smooth matter
distributions, and constant accretion rates). An alternative approach
is to study the growth of BHs in the context of large cosmological
hydrodynamic simulations, which provide more realistic physical
conditions but have the limitation that they do not directly resolve
the sphere of influence of the BH. Instead, a common approach is to
assume that some fraction of the gas inflowing into the central
$\sim$kpc (through mergers or other processes) accretes onto the BH,
up to the Eddington limit \citep[e.g.,][]{spri05, dima05qso,
  dima08bhfeed, boot10mhalo}. This prescription is broadly valid if
the transport of gas from $\sim$kpc scales down to the BH is indeed
efficient, as discussed in \S2.1--2.2. Feedback is modeled not as a
full description of the outflow, but by transferring a fraction of the
radiative output of the BH to the thermal energy of the surrounding
gas. The feedback energy output is thus
\begin{equation}
\dot{E}_{\rm feed} = \epsilon_f \epsilon_r \dot{m} c^2,
\end{equation} 
where $\epsilon_r\sim 0.1$ is the standard radiative efficiency, and
$\epsilon_f$ is the fraction of the radiative output that couples to
the gas. Such models naturally yield a relationship between $M_{\rm
  BH}$ and $M_{\rm sph}$, and can reproduce the evolution with
redshift as observed from broad-line quasar samples
\citep[e.g.,][]{boot11bhscale}. To reproduce the observed
normalization for these relations requires $\epsilon_f
\approx0.05$--$0.15$ (depending on the specific subgrid prescriptions
for feedback; \citealt{spri05, dima08bhfeed, boot09bhsim,
  teys11agnfeed}) suggesting relatively efficient coupling between the
BH luminosity and the surrounding gas. A detailed study of the output
of one model \citep{boot10mhalo} suggests that the most fundamental
relationship is between $M_{\rm BH}$ and the binding energy of the
host dark-matter halo. In this model, the BH grows until it is massive
enough to unbind gas not only from its host galaxy but from the entire
surrounding halo, and it is this feedback (along with correlations
between the galaxy and $M_{\rm halo}$) that give rise to the observed
BH-spheroid correlations.  Recent observational results suggest that
such a $M_{\rm BH}$--$M_{\rm halo}$ relation is not fundamental for
lower-mass AGN \citep[e.g.,][]{gree08lowmass, gree10maser, korm11halo}
but it remains a valid possibility for higher-mass objects $M_{\rm BH}
\gtrsim 10^8$ \msun\ for which BH, galaxy spheroid, and dark-matter
halo correlations are more difficult to disentangle
\citep{volo11mbhmhalo}.

While feedback has naturally occupied a great deal of attention in
understanding BH-spheroid relationships, recent work has also proposed
that these relations do not imply a causal link between BHs and their
host galaxies or halos, but are simply a consequence of the
hierarchical growth of structure \citep[e.g.,][]{peng07mbh,
  jahn11mbh}. If mergers of galaxies are accompanied by the
corresponding mergers of their central BHs, then a succession of
mergers will tend to drive the resulting galaxies toward the
``average'' relationship between BH and spheroid mass, even if the
initial galaxies show little or no correlation between BH and spheroid
properties. These processes may also help explain the increased
scatter at the low-mass end of the BH-spheroid relationship, for which
the galaxies may have experienced fewer mergers. While in this picture
mergers play the predominant role in producing the BH-spheroid
relationships, AGN feedback may be important in setting the relative
rates of BH and galaxy growth over the history of the Universe.

\subsubsection{The impact of feedback on gas in galaxies and halos}

\label{sec.feedbackgas}

The energy released by growing BHs can impact not only on the stellar
content of their host galaxies (as suggested for feedback models of
the BH-spheroid relationships) but also the diffuse gas in their host
dark matter halos. In massive ($M_{\rm halo}\gtrsim10^{13}$ \msun)
halos with hot, ionized atmospheres, heating from relativistic jets
(as discussed above) can regulate the temperature of the halos and
prevent cooling to form new stars, as required in order to produce the
observed stellar mass and luminosity functions of galaxies
\citep[e.g.,][]{bowe06gal, crot06,bowe08flip}. Mechanically dominated
AGN activity is observed to be most prevalent in passive systems in
such massive halos \citep[e.g.,][]{best05, hick09corr, wake08radio,
  mand09agnclust, smol09radio}; see \S3.4 \& Fig.~5. In contrast,
moderate-luminosity radio-quiet AGN appear to have very little impact
on the stellar populations or star formation rates of their host
galaxies \citep[e.g.,][]{hick09corr, card10xhost, xue10xhost,
  lutz10agnsf, shao10agnsf, ammo11xhost, mull11agnsf}, although
outflows in such systems may still play a significant role in
moderating the growth of the BH (e.g., \citealt{king10outflow},
Mullaney et al., in prep).

Some of the most compelling evidence for AGN feedback comes from
observations of the hot gas itself, which represents the majority of
the baryonic content of group- and cluster-scale systems. Of
particular interest is the luminosity-temperature ($L_X$--$T_X$)
relation, which is broadly related to the baryon fraction in a halo as
a function of $M_{\rm halo}$. Observations show that the $L_X$--$T_X$
relation is remarkably steep at the low-temperature end, suggesting
that small to moderate-mass groups have lower baryon fractions in the
centers of their dark-matter halos than higher-mass systems
\citep[e.g.,][]{sun09groups, giod09baryons}. Outflows must have
removed some of the low-entropy gas from the centers of these halos,
and simulations indicate that winds from stellar processes (stellar
winds and supernovae) are insufficient to produce this
deficit. However, models suggest that feedback from growing massive
BHs (indeed, the same outflows that are invoked to produce the
BH-spheroid relation) can naturally produce the $L_X$--$T_X$ relations
similar to those observed \citep[e.g.,][]{bowe08flip, boot10mhalo,
  mcca10feedback, mcca11feedback}. Unlike models for the galaxy
luminosity function or BH-spheroid relationships, these predictions
are independent of assumptions about the star formation processes
within the host galaxy and depend only on the coupling of the AGN to
the surrounding gas. As such, they represent one of the most promising
avenues for quantifying the impact of growing BHs on their host
galaxies and halos.

\begin{figure*}[tp]
\centering
\includegraphics[width=140mm]{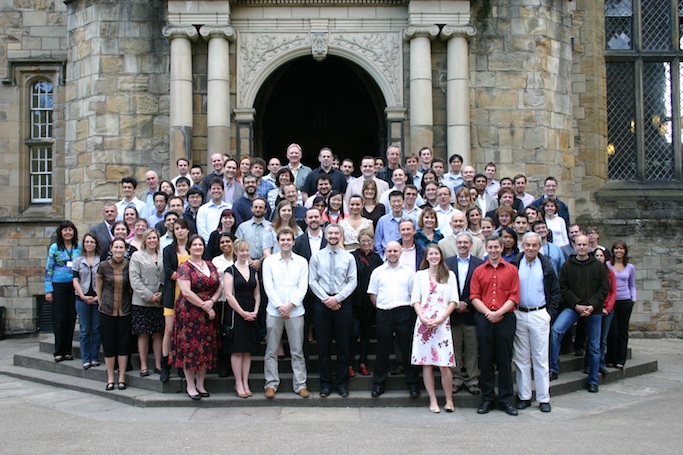}
\caption{The workshop participants before dinner at Durham Castle. Photograph taken by Sarah Noble.}
\end{figure*}

\section{Concluding remarks}

In this review we have discussed the processes and mechanics of BH
growth and explored the connections between the growth of BHs and star
formation, host-galaxy properties, signatures of interactions/mergers,
and environment. We have explored these processes in a very broad
range of systems, from the closest AGNs to AGNs at the highest
redshifts, and also the formation of seed BHs. Our overarching aim has
been to address the question: {\it What drives the growth of black
  holes?}

Surprisingly, there are often few external morphological and
kinematical clues that reveal when significant BH growth has been
triggered, with AGN host galaxies showing similar overall properties
to inactive galaxies of comparable stellar mass. This appears to be
true even down to small $\approx$~10--100 pc scales. However, the
accurate identification of gas inflow on these scales is mostly
restricted to low-luminosity AGNs in the local Universe, and
significant differences may be present in the central
$\approx$~10--100~pc regions of luminous AGNs. Comparable linear
spatial scales should be achievable for distant and high-luminosity
AGNs over the coming years with the final configuration of ALMA at
submm-mm wavelengths, which will be provide direct insight into the
properties and extent of the molecular gas and dust-continuum emission
on $\approx$~100~pc scales. A clear complication in identifying any
large-scale catalysts that may have triggered AGN activity is the
timescale required for the gas to flow from the host galaxy down to
the BH, which can be $\gg10^8$~yrs. Over such long timescales any
clear morphological or kinematical signatures of the triggering event
may be lost and therefore care needs to be taken when assessing the
role of galaxy major mergers or interactions in the fueling of AGN
activity.

A clear dependence between the SFR and luminous AGN fraction is found
in both distant and local galaxies, implying a connection between BH
and galaxy growth. This finding is perhaps not unsurprising since both
star formation and AGN activity are predominantly driven by a cold-gas
supply, either from the host galaxy and larger-scale environment, or
from stellar winds and circumnuclear supernovae events. However, these
results may also indicate a more direct and less coincidental
connection between star formation and AGN activity in at least some
sytems (such as in the common evolutionary scenario shown in
Fig.~6). High-resolution observations with ALMA and {\it JWST} of
well-selected galaxy samples could directly test this evolutionary
scenario by simultaneously providing SFRs, gas-inflow rates, and BH
mass accretion rates as a function of stellar age.

Recent results have shown that environment plays a major role in the
triggering of AGN activity. The global peak in AGN activity in
overdense regions occured at higher redshifts than in the field, with
a corresponding more rapid decline to the present day. The most rapid
BH growth also appears to be found in dark-matter halos of
$\approx10^{12}$--$10^{13}$~$M_{\odot}$ (see Fig.~5), with more
massive dark-matter halos found to harbour low-accretion rate
radio-loud AGNs hosted by relatively quiescent galaxies. This mass
dependence on AGN activity may be due to ``quenching'' of the cold gas
in massive dark-matter halos, which could be largely assisted by
large-scale mechanically powerful (but often radiatively weak; see
Fig.~7) outflows, such as those seen in nearby galaxy clusters. The
strong dependence on the fraction of galaxies that host radio-loud AGN
with galaxy mass provides further support for this picture.

However, on the basis of the current results it is not immediately
apparent why there exists a tight correlation between the mass of the
BH and the mass of the spheroid in nearby galaxies. Many models argue
that energetic radiatively driven outflows (such as those illustrated
in Fig.~7) regulate star formation in the host galaxy and ``forge''
the BH--spheroid mass relationship. Energetic radiatively driven
outflows are clearly seen in many luminous AGNs and may be a
ubiquitious feature of the AGN population but it is not yet clear what
impact these outflows have on star formation in the host
galaxy. Spatially resolved spectroscopy of AGNs with IFU observations
may provide important insight on this issue since they can reveal the
presence of large-scale outflowing gas and address when and where
energetic outflows are found. However, given the relatively modest BH
mass accretion and SFRs for the majority of the AGN population
(e.g.,\ see Fig.~4 for the AGN fraction as a function of different AGN
selection techniques), a ``regulatory'' mechanism may not be necessary
every time the BH and galaxy grows. It is presumably only the rapidly
evolving and gas-rich systems where the BH and galaxy growth can be so
high that they can move substantially around the BH--spheroid mass
plane on short timescales. These rapdily evolving systems are
therefore the most promising sites to search for the evidence of
large-scale energetic outflows. We should also bear in mind that the
tight BH--spheroid mass relationship is only applicable for galaxies
hosting classical bulges -- there are no significant BH--galaxy
correlations for systems hosted in pseudo bulges. Since galaxy major
mergers are likely to be the catalyst for the formation of classical
spheroids, this suggests that galaxy major mergers played a major role
in forging the BH--spheroid mass relationship.

Although rare, luminous AGNs such as the quasars selected in optical
surveys may play a much more important role in the evolution and
growth of BHs and galaxies than would be apparent from their
relatively low space density. An example of this is the finding that
the most luminous AGNs in the local Universe are responsible for
$\approx$~50\% of the integrated BH growth, despite only comprising
$\approx$~0.2\% of the optical AGN population. However, since AGNs can
be extremely variable, any measurement of the BH mass-accretion rate
only provides a snapshot of its growth path (e.g.,\ see Fig.~5 of
Hopkins \& Quataert 2010, which shows factors of $\approx$~100
variation in the predicted gas inflow rate over Myr timescales). This
adds significant complications to the interpretation of individual
objects, which can be overcome to some extent from studying AGN
populations as a function of a more stable property than mass
accretion rate such as BH mass, stellar mass, and star-formation
rate. This should be possible over the coming years with the advent of
sensitive large facilities, particularly those that combine depth with
breadth and yield large source densities (e.g.,\ BigBOSS, LSST, and
{\it e-ROSITA}).

The current theoretical models focus on either small-scale processes
at high resolution, with limited constraints on larger-scale
processes, or explore the large-scale growth of BHs and galaxies in
cosmological simulations using sub-grid models to account for the
smaller-scale physics. Both approaches have clear strengths but
evident weaknesses (i.e.,\ either limited spatial and time resolution
or limited cosmological volume). Since AGNs are gas-driven systems fed
from gas inflow down to $<10$~pc scales from typically $>1$~kpc
scales, high-resolution hydrodynamical simulations over a broad range
of size scales would be particularly beneficial for a detailed
understanding of AGN fueling and feedback processes. More detailed
theoretical insight may also come from improvements in the sub-grid
models using the physics learnt from yet higher-resolution small-scale
models to produce more realistic fueling and feedback prescriptions
for the cosmological simulations.

The next $\approx$~5--10~years therefore promise huge advances in our
understanding of ``What drives the growth of black holes?''. We look
forward to hosting another exciting workshop and reviewing our
progress on this issue over the next decade!

\vspace{0.2in}
\noindent {\bf Ackowledgements:}
\vspace{0.2in}

The original date of the workshop was 19$^{\rm th}$--22$^{\rm nd}$
April 2010 but the eruption of Icelandic volcano Eyjafjallajokull a
few days before caused the majority of airspace in Europe to be shut
down, and the workshop had to be rescheduled for 26$^{\rm
  th}$--29$^{\rm th}$ July 2010. We would like to thank the workshop
participants for being able to rapidly change their travel plans
($>90$\% of the original participates managed to attend the
re-scheduled workshop) and for providing 3.5 days of excellent
presentations and enthusiastic scientific discussions: {\bf SOC:}
P.~Best, R.~Davies, T.~di Matteo, A.~Fabian, J.~Greene, M.~Volonteri;
{\bf LOC:} L.~Borrero, K.~Coppin, A.~Danielson, J.~Geach
% ...

We thank the New Astronomy Review editors and referees for their time
and assistance in completing this review. We would also like to thank
Sera Markoff and many workshop participants for providing feedback on
earlier drafts of this review, and Sarah Noble for taking the workshop
photograph. Special thanks to Darren Croton for the inspiration for
the dark-matter halo figure and for sharing his original version with
us, to Chris Done for giving us permission to use a revised version of
the BH accretion mode figure originally published in Done, Gerlinkski,
\& Kubota (2007), to Philip Best, Andy Goulding, and Nic Ross for
providing us with the data used in various figures, and to James
Mullaney for producing an earlier version of the AGN outflow
figure. We gratefully acknowledge the Science and Technology
Facilities Countil (STFC), Royal Society, and Leverhulme Trust for
financial support.

\bibliographystyle{apj} \bibliography{research}

\begin{thebibliography}{690}
\expandafter\ifx\csname natexlab\endcsname\relax\def\natexlab#1{#1}\fi

\bibitem[{{Abazajian} {et~al.}(2009){Abazajian}, {Adelman-McCarthy},
  {Ag{\"u}eros}, {Allam}, {Allende Prieto}, {An}, {Anderson}, {Anderson},
  {Annis}, {Bahcall}, \& et~al.}]{abaz09sdss}
{Abazajian}, K.~N., et~al.\ 2009, \apjs, 182, 543

\bibitem[{{Abel} {et~al.}(2002){Abel}, {Bryan}, \& {Norman}}]{abel02firststar}
{Abel}, T., {Bryan}, G.~L., \& {Norman}, M.~L. 2002, Science, 295, 93

\bibitem[{{Agol} \& {Krolik}(2000)}]{agol00disk}
{Agol}, E. \& {Krolik}, J.~H. 2000, \apj, 528, 161

\bibitem[{{Aird} {et~al.}(2010){Aird}, {Nandra}, {Laird}, {Georgakakis},
  {Ashby}, {Barmby}, {Coil}, {Huang}, {Koekemoer}, {Steidel}, \&
  {Willmer}}]{aird10xlf}
{Aird}, J., et~al.\ 2010, \mnras, 401, 2531

\bibitem[{{Akiyama}(2005)}]{akiy05xhost}
{Akiyama}, M. 2005, \apj, 629, 72

\bibitem[{{Akylas} {et~al.}(2006){Akylas}, {Georgantopoulos}, {Georgakakis},
  {Kitsionas}, \& {Hatziminaoglou}}]{akyl06}
{Akylas}, A., {Georgantopoulos}, I., {Georgakakis}, A., {Kitsionas}, S., \&
  {Hatziminaoglou}, E. 2006, \aap, 459, 693

\bibitem[{{Alexander} {et~al.}(2011){Alexander}, {Bauer}, {Brandt}, {Daddi},
  {Hickox}, {Lehmer}, {Luo}, {Xue}, {Young}, {Comastri}, {Del Moro}, {Fabian},
  {Gilli}, {Goulding}, {Mainieri}, {Mullaney}, {Paolillo}, {Rafferty},
  {Schneider}, {Shemmer}, \& {Vignali}}]{alex11stack}
{Alexander}, D.~M., et~al.\ 2011, \apj, 738, 44

\bibitem[{{Alexander} {et~al.}(2003{\natexlab{a}}){Alexander}, {Bauer},
  {Brandt}, {Hornschemeier}, {Vignali}, {Garmire}, {Schneider}, {Chartas}, \&
  {Gallagher}}]{alex03submm}
{Alexander}, D.~M., et~al.\ 2003{\natexlab{a}}, \aj, 125, 383

\bibitem[{{Alexander} {et~al.}(2003{\natexlab{b}}){Alexander}, {Bauer},
  {Brandt}, {Schneider}, {Hornschemeier}, {Vignali}, {Barger}, {Broos},
  {Cowie}, {Garmire}, {Townsley}, {Bautz}, {Chartas}, \& {Sargent}}]{alex03}
{Alexander}, D.~M., et~al.\ 2003{\natexlab{b}}, \aj, 126, 539

\bibitem[{{Alexander} {et~al.}(2005){Alexander}, {Bauer}, {Chapman}, {Smail},
  {Blain}, {Brandt}, \& {Ivison}}]{alex05}
{Alexander}, D.~M., {Bauer}, F.~E., {Chapman}, S.~C., {Smail}, I., {Blain},
  A.~W., {Brandt}, W.~N., \& {Ivison}, R.~J. 2005, \apj, 632, 736

\bibitem[{{Alexander} {et~al.}(2001){Alexander}, {Brandt}, {Hornschemeier},
  {Garmire}, {Schneider}, {Bauer}, \& {Griffiths}}]{alex01xfaint}
{Alexander}, D.~M., {Brandt}, W.~N., {Hornschemeier}, A.~E., {Garmire}, G.~P.,
  {Schneider}, D.~P., {Bauer}, F.~E., \& {Griffiths}, R.~E. 2001, \aj, 122,
  2156

\bibitem[{{Alexander} {et~al.}(2008{\natexlab{a}}){Alexander}, {Brandt},
  {Smail}, {Swinbank}, {Bauer}, {Blain}, {Chapman}, {Coppin}, {Ivison}, \&
  {Men{\'e}ndez-Delmestre}}]{alex08bhmass}
{Alexander}, D.~M., et~al.\ 2008{\natexlab{a}}, \aj, 135, 1968

\bibitem[{{Alexander} {et~al.}(2008{\natexlab{b}}){Alexander}, {Chary}, {Pope},
  {Bauer}, {Brandt}, {Daddi}, {Dickinson}, {Elbaz}, \&
  {Reddy}}]{alex08compthick}
{Alexander}, D.~M., et~al.\ 2008{\natexlab{b}}, \apj, 687, 835

\bibitem[{{Alexander} {et~al.}(2010){Alexander}, {Swinbank}, {Smail},
  {McDermid}, \& {Nesvadba}}]{alex10outflow}
{Alexander}, D.~M., {Swinbank}, A.~M., {Smail}, I., {McDermid}, R., \&
  {Nesvadba}, N.~P.~H. 2010, \mnras, 402, 2211

\bibitem[{{Allen} {et~al.}(2011){Allen}, {Hewett}, {Maddox}, {Richards}, \&
  {Belokurov}}]{all11balqso}
{Allen}, J.~T., {Hewett}, P.~C., {Maddox}, N., {Richards}, G.~T., \&
  {Belokurov}, V. 2011, \mnras, 410, 860

\bibitem[{{Allevato} {et~al.}(2011){Allevato}, {Finoguenov}, {Cappelluti},
  {Miyaji}, {Hasinger}, {Salvato}, {Brusa}, {Gilli}, {Zamorani}, {Shankar},
  {James}, {McCracken}, {Bongiorno}, {Merloni}, {Peacock}, {Silverman}, \&
  {Comastri}}]{alle11xclust}
{Allevato}, V., et~al.\ 2011, \apj, 736, 99

\bibitem[{{Alonso-Herrero} {et~al.}(2011){Alonso-Herrero}, {Pereira-Santaella},
  {Rieke}, \& {Rigopoulou}}]{alon11lirg}
{Alonso-Herrero}, A., {Pereira-Santaella}, M., {Rieke}, G.~H., \& {Rigopoulou},
  D. 2011, \apj\ in press (arXiv:1109.1372)

\bibitem[{{Alonso-Herrero} {et~al.}(2006){Alonso-Herrero}, {P{\'e}rez-Gonz
  {\'a}lez}, {Alexander}, {Rieke}, {Rigopoulou}, {Le Floc'h}, {Barmby},
  {Papovich}, {Rigby}, {Bauer}, {Brandt}, {Egami}, {Willner}, {Dole}, \&
  {Huang}}]{alon06}
{Alonso-Herrero}, A., et~al.\ 2006, \apj, 640, 167

\bibitem[{{Alonso-Herrero} {et~al.}(2008){Alonso-Herrero},
  {P{\'e}rez-Gonz{\'a}lez}, {Rieke}, {Alexander}, {Rigby}, {Papovich},
  {Donley}, \& {Rigopoulou}}]{alon08agn}
{Alonso-Herrero}, A., et~al.\ 2008, \apj, 677, 127

\bibitem[{{Alvarez} {et~al.}(2009){Alvarez}, {Wise}, \& {Abel}}]{alva09seedbh}
{Alvarez}, M.~A., {Wise}, J.~H., \& {Abel}, T. 2009, \apjl, 701, L133

\bibitem[{{Ammons} {et~al.}(2011){Ammons}, {Rosario}, {Koo}, {Dutton},
  {Melbourne}, {Max}, {Mozena}, {Kocevski}, {McGrath}, {Bouwens}, \&
  {Magee}}]{ammo11xhost}
{Ammons}, S.~M., et~al.\ 2011, \apj, 740, 3

\bibitem[{{Antonucci}(1993)}]{anto93}
{Antonucci}, R. 1993, \araa, 31, 473

\bibitem[{{Arav} {et~al.}(2008){Arav}, {Moe}, {Costantini}, {Korista}, {Benn},
  \& {Ellison}}]{arav08outflow}
{Arav}, N., {Moe}, M., {Costantini}, E., {Korista}, K.~T., {Benn}, C., \&
  {Ellison}, S. 2008, \apj, 681, 954

\bibitem[{{Archibald} {et~al.}(2001){Archibald}, {Dunlop}, {Hughes},
  {Rawlings}, {Eales}, \& {Ivison}}]{arch01agnsubmm}
{Archibald}, E.~N., {Dunlop}, J.~S., {Hughes}, D.~H., {Rawlings}, S., {Eales},
  S.~A., \& {Ivison}, R.~J. 2001, \mnras, 323, 417

\bibitem[{{Arun} {et~al.}(2009){Arun}, {Babak}, {Berti}, {Cornish}, {Cutler},
  {Gair}, {Hughes}, {Iyer}, {Lang}, {Mandel}, {Porter}, {Sathyaprakash},
  {Sinha}, {Sintes}, {Trias}, {Van Den Broeck}, \& {Volonteri}}]{arun09bhlisa}
{Arun}, K.~G., et~al.\ 2009, Classical and Quantum Gravity,  26, 094027

\bibitem[{{Assef} {et~al.}(2011){Assef}, {Denney}, {Kochanek}, {Peterson},
  {Koz{\l}owski}, {Ageorges}, {Barrows}, {Buschkamp}, {Dietrich}, {Falco},
  {Feiz}, {Gemperlein}, {Germeroth}, {Grier}, {Hofmann}, {Juette}, {Khan},
  {Kilic}, {Knierim}, {Laun}, {Lederer}, {Lehmitz}, {Lenzen}, {Mall}, {Madsen},
  {Mandel}, {Martini}, {Mathur}, {Mogren}, {Mueller}, {Naranjo}, {Pasquali},
  {Polsterer}, {Pogge}, {Quirrenbach}, {Seifert}, {Stern}, {Shappee}, {Storz},
  {Van Saders}, {Weiser}, \& {Zhang}}]{asse11bhmass}
{Assef}, R.~J., et~al.\ 2011, \apj, 742,  93

\bibitem[{{Atlee} {et~al.}(2011){Atlee}, {Martini}, {Assef}, {Kelson}, \&
  {Mulchaey}}]{atle11clustagn}
{Atlee}, D.~W., {Martini}, P., {Assef}, R.~J., {Kelson}, D.~D., \& {Mulchaey},
  J.~S. 2011, \apj, 729, 22

\bibitem[{{B{\^ i}rzan} {et~al.}(2004){B{\^ i}rzan}, {Rafferty}, {McNamara},
  {Wise}, \& {Nulsen}}]{birz04}
{B{\^ i}rzan}, L., {Rafferty}, D.~A., {McNamara}, B.~R., {Wise}, M.~W., \&
  {Nulsen}, P.~E.~J. 2004, \apj, 607, 800

\bibitem[{{Babi{\'c}} {et~al.}(2007){Babi{\'c}}, {Miller}, {Jarvis}, {Turner},
  {Alexander}, \& {Croom}}]{babi07edd}
{Babi{\'c}}, A., {Miller}, L., {Jarvis}, M.~J., {Turner}, T.~J., {Alexander},
  D.~M., \& {Croom}, S.~M. 2007, \aap, 474, 755

\bibitem[{{Baganoff} {et~al.}(2001){Baganoff}, {Bautz}, {Brandt}, {Chartas},
  {Feigelson}, {Garmire}, {Maeda}, {Morris}, {Ricker}, {Townsley}, \&
  {Walter}}]{baga01sgra}
{Baganoff}, F.~K., et~al.\ 2001, \nat, 413, 45

\bibitem[{{Baganoff} {et~al.}(2003){Baganoff}, {Maeda}, {Morris}, {Bautz},
  {Brandt}, {Cui}, {Doty}, {Feigelson}, {Garmire}, {Pravdo}, {Ricker}, \&
  {Townsley}}]{baga03sgra}
{Baganoff}, F.~K., et~al.\ 2003, \apj, 591, 891

\bibitem[{{Bahcall} {et~al.}(1997){Bahcall}, {Kirhakos}, {Saxe}, \&
  {Schneider}}]{bahc97qsohost}
{Bahcall}, J.~N., {Kirhakos}, S., {Saxe}, D.~H., \& {Schneider}, D.~P. 1997,
  \apj, 479, 642

\bibitem[{{Balbus}(2003)}]{balb03disk}
{Balbus}, S.~A. 2003, \araa, 41, 555

\bibitem[{{Balbus} \& {Hawley}(1991)}]{balb91merge}
{Balbus}, S.~A. \& {Hawley}, J.~F. 1991, \apj, 376, 214

\bibitem[{{Balbus} \& {Hawley}(1998)}]{balb98}
---. 1998, Reviews of Modern Physics, 70, 1

\bibitem[{{Baldi} \& {Capetti}(2008)}]{bald08radiohost}
{Baldi}, R.~D. \& {Capetti}, A. 2008, \aap, 489, 989

\bibitem[{{Baldry} {et~al.}(2004){Baldry}, {Glazebrook}, {Brinkmann},
  {Ivezi{\'c}}, {Lupton}, {Nichol}, \& {Szalay}}]{bald04color}
{Baldry}, I.~K., {Glazebrook}, K., {Brinkmann}, J., {Ivezi{\'c}}, {\v Z}.,
  {Lupton}, R.~H., {Nichol}, R.~C., \& {Szalay}, A.~S. 2004, \apj, 600, 681

\bibitem[{{Baldwin} {et~al.}(1981){Baldwin}, {Phillips}, \&
  {Terlevich}}]{bald81bpt}
{Baldwin}, J.~A., {Phillips}, M.~M., \& {Terlevich}, R. 1981, \pasp, 93, 5

\bibitem[{{Ballantyne} {et~al.}(2006){Ballantyne}, {Everett}, \&
  {Terlevich}}]{ball06}
{Ballantyne}, D.~R., {Everett}, J.~E., \& {Murray}, N. 2006, \apj, 639, 740

\bibitem[{{Ballo} {et~al.}(2007){Ballo}, {Cristiani}, {Fasano}, {Fontanot},
  {Monaco}, {Nonino}, {Pignatelli}, {Tozzi}, {Vanzella}, {Fontana},
  {Giallongo}, {Grazian}, \& {Danese}}]{ball07edd}
{Ballo}, L., et~al.\ 2007, \apj, 667, 97

\bibitem[{{Bardelli} {et~al.}(2010){Bardelli}, {Schinnerer}, {Smol{\v c}ic},
  {Zamorani}, {Zucca}, {Mignoli}, {Halliday}, {Kova{\v c}}, {Ciliegi},
  {Caputi}, {Koekemoer}, {Bongiorno}, {Bondi}, {Bolzonella}, {Vergani},
  {Pozzetti}, {Carollo}, {Contini}, {Kneib}, {Le F{\`e}vre}, {Lilly},
  {Mainieri}, {Renzini}, {Scodeggio}, {Coppa}, {Cucciati}, {de la Torre}, {de
  Ravel}, {Franzetti}, {Garilli}, {Iovino}, {Kampczyk}, {Knobel}, {Lamareille},
  {Le Borgne}, {Le Brun}, {Maier}, {Pell{\`o}}, {Peng}, {Perez-Montero},
  {Ricciardelli}, {Silverman}, {Tanaka}, {Tasca}, {Tresse}, {Abbas}, {Bottini},
  {Cappi}, {Cassata}, {Cimatti}, {Guzzo}, {Leauthaud}, {Maccagni}, {Marinoni},
  {McCracken}, {Memeo}, {Meneux}, {Oesch}, {Porciani}, {Scaramella}, {Capak},
  {Sanders}, {Scoville}, {Taniguchi}, \& {Jahnke}}]{bard10radio}
{Bardelli}, S., et~al.\ 2010, \aap,  511, A1+

\bibitem[{{Barger} {et~al.}(2003{\natexlab{a}}){Barger}, {Cowie}, {Capak},
  {Alexander}, {Bauer}, {Brandt}, {Garmire}, \& {Hornschemeier}}]{barg03highz}
{Barger}, A.~J., et~al.\ 2003{\natexlab{a}}, \apjl, 584, L61

\bibitem[{{Barger} {et~al.}(2003{\natexlab{b}}){Barger}, {Cowie}, {Capak},
  {Alexander}, {Bauer}, {Fernandez}, {Brandt}, {Garmire}, \&
  {Hornschemeier}}]{barg03}
{Barger}, A.~J., et~al.\ 2003{\natexlab{b}}, \aj, 126, 632

\bibitem[{{Barger} {et~al.}(2005){Barger}, {Cowie}, {Mushotzky}, {Yang},
  {Wang}, {Steffen}, \& {Capak}}]{barg05}
{Barger}, A.~J., {Cowie}, L.~L., {Mushotzky}, R.~F., {Yang}, Y., {Wang}, W.-H.,
  {Steffen}, A.~T., \& {Capak}, P. 2005, \aj, 129, 578

\bibitem[{{Barnes} \& {Hernquist}(1992)}]{barn92merge}
{Barnes}, J.~E. \& {Hernquist}, L. 1992, \araa, 30, 705

\bibitem[{{Barnes} \& {Hernquist}(1996)}]{barn96merge}
---. 1996, \apj, 471, 115

\bibitem[{{Barrows} {et~al.}(2011){Barrows}, {Stern}, {Madsen}, {Harrison},
  {Assef}, {Comerford}, {Cushing}, {Fassnacht}, {Gonzalez}, {Griffith},
  {Hickox}, {Kirkpatrick}, \& {Lagattuta}}]{barr11dualagn}
{Barrows}, R.~S., et~al.\ 2011, \apj\ in press (arXiv:1109.3469)

\bibitem[{{Bauer} {et~al.}(2004){Bauer}, {Alexander}, {Brandt}, {Schneider},
  {Treister}, {Hornschemeier}, \& {Garmire}}]{baue04}
{Bauer}, F.~E., {Alexander}, D.~M., {Brandt}, W.~N., {Schneider}, D.~P.,
  {Treister}, E., {Hornschemeier}, A.~E., \& {Garmire}, G.~P. 2004, \aj, 128,
  2048

\bibitem[{{Bauer} {et~al.}(2010){Bauer}, {Yan}, {Sajina}, \&
  {Alexander}}]{baue10obscagn}
{Bauer}, F.~E., {Yan}, L., {Sajina}, A., \& {Alexander}, D.~M. 2010, \apj, 710,
  212

\bibitem[{{Baum} {et~al.}(2010){Baum}, {Gallimore}, {O'Dea}, {Buchanan},
  {Noel-Storr}, {Axon}, {Robinson}, {Elitzur}, {Dorn}, \&
  {Staudaher}}]{baum10agnsf}
{Baum}, S.~A., et~al.\ 2010, \apj, 710, 289

\bibitem[{{Begelman}(2002)}]{bege02super}
{Begelman}, M.~C. 2002, \apjl, 568, L97

\bibitem[{{Begelman}(2010)}]{bege10quasistar}
---. 2010, \mnras, 402, 673

\bibitem[{{Begelman} {et~al.}(2006){Begelman}, {Volonteri}, \&
  {Rees}}]{bege06quasistar}
{Begelman}, M.~C., {Volonteri}, M., \& {Rees}, M.~J. 2006, \mnras, 370, 289

\bibitem[{{Bell} {et~al.}(2005){Bell}, {Papovich}, {Wolf}, {Le Floc'h},
  {Caldwell}, {Barden}, {Egami}, {McIntosh}, {Meisenheimer},
  {P{\'e}rez-Gonz{\'a}lez}, {Rieke}, {Rieke}, {Rigby}, \& {Rix}}]{bell05sfevol}
{Bell}, E.~F., et~al.\ 2005, \apj, 625, 23

\bibitem[{{Bender} {et~al.}(2005){Bender}, {Kormendy}, {Bower}, {Green},
  {Thomas}, {Danks}, {Gull}, {Hutchings}, {Joseph}, {Kaiser}, {Lauer},
  {Nelson}, {Richstone}, {Weistrop}, \& {Woodgate}}]{bend05m31}
{Bender}, R., et~al.\ 2005, \apj, 631, 280

\bibitem[{{Bennert} {et~al.}(2008){Bennert}, {Canalizo}, {Jungwiert},
  {Stockton}, {Schweizer}, {Peng}, \& {Lacy}}]{benn08qsohost}
{Bennert}, N., {Canalizo}, G., {Jungwiert}, B., {Stockton}, A., {Schweizer},
  F., {Peng}, C.~Y., \& {Lacy}, M. 2008, \apj, 677, 846

\bibitem[{{Bennert} {et~al.}(2010){Bennert}, {Treu}, {Woo}, {Malkan}, {Le
  Bris}, {Auger}, {Gallagher}, \& {Blandford}}]{benn10bhmass}
{Bennert}, V.~N., {Treu}, T., {Woo}, J.-H., {Malkan}, M.~A., {Le Bris}, A.,
  {Auger}, M.~W., {Gallagher}, S., \& {Blandford}, R.~D. 2010, \apj, 708, 1507

\bibitem[{{Bentz} {et~al.}(2009){Bentz}, {Peterson}, {Netzer}, {Pogge}, \&
  {Vestergaard}}]{bent09revmap}
{Bentz}, M.~C., {Peterson}, B.~M., {Netzer}, H., {Pogge}, R.~W., \&
  {Vestergaard}, M. 2009, \apj, 697, 160

\bibitem[{{Best}(2004)}]{best04radio}
{Best}, P.~N. 2004, \mnras, 351, 70

\bibitem[{{Best} {et~al.}(2006){Best}, {Kaiser}, {Heckman}, \&
  {Kauffmann}}]{best06feed}
{Best}, P.~N., {Kaiser}, C.~R., {Heckman}, T.~M., \& {Kauffmann}, G. 2006,
  \mnras, 368, L67

\bibitem[{{Best} {et~al.}(2005){Best}, {Kauffmann}, {Heckman}, {Brinchmann},
  {Charlot}, {Ivezi{\'c}}, \& {White}}]{best05}
{Best}, P.~N., {Kauffmann}, G., {Heckman}, T.~M., {Brinchmann}, J., {Charlot},
  S., {Ivezi{\'c}}, {\v Z}., \& {White}, S.~D.~M. 2005, \mnras, 362, 25

\bibitem[{{Binette} {et~al.}(1994){Binette}, {Magris}, {Stasi{\'n}ska}, \&
  {Bruzual}}]{bine94photoion}
{Binette}, L., {Magris}, C.~G., {Stasi{\'n}ska}, G., \& {Bruzual}, A.~G. 1994,
  \aap, 292, 13

\bibitem[{{B{\^i}rzan} {et~al.}(2008){B{\^i}rzan}, {McNamara}, {Nulsen},
  {Carilli}, \& {Wise}}]{birz08jet}
{B{\^i}rzan}, L., {McNamara}, B.~R., {Nulsen}, P.~E.~J., {Carilli}, C.~L., \&
  {Wise}, M.~W. 2008, \apj, 686, 859

\bibitem[{{Blandford} \& {Begelman}(1999)}]{blan99fate}
{Blandford}, R.~D. \& {Begelman}, M.~C. 1999, \mnras, 303, L1

\bibitem[{{Bondi}(1952)}]{bond52}
{Bondi}, H. 1952, \mnras, 112, 195

\bibitem[{{Bondi} \& {Hoyle}(1944)}]{bond44}
{Bondi}, H. \& {Hoyle}, F. 1944, \mnras, 104, 273

\bibitem[{{Bongiorno} {et~al.}(2007){Bongiorno}, {Zamorani}, {Gavignaud},
  {Marano}, {Paltani}, {Mathez}, {M{\o}ller}, {Picat}, {Cirasuolo},
  {Lamareille}, {Bottini}, {Garilli}, {Le Brun}, {Le F{\`e}vre}, {Maccagni},
  {Scaramella}, {Scodeggio}, {Tresse}, {Vettolani}, {Zanichelli}, {Adami},
  {Arnouts}, {Bardelli}, {Bolzonella}, {Cappi}, {Charlot}, {Ciliegi},
  {Contini}, {Foucaud}, {Franzetti}, {Guzzo}, {Ilbert}, {Iovino}, {McCracken},
  {Marinoni}, {Mazure}, {Meneux}, {Merighi}, {Pell{\`o}}, {Pollo}, {Pozzetti},
  {Radovich}, {Zucca}, {Hatziminaoglou}, {Polletta}, {Bondi}, {Brinchmann},
  {Cucciati}, {de la Torre}, {Gregorini}, {Mellier}, {Merluzzi}, {Temporin},
  {Vergani}, \& {Walcher}}]{bong07qlf}
{Bongiorno}, A., et~al.\ 2007,  \aap, 472, 443

\bibitem[{{Booth} \& {Schaye}(2009)}]{boot09bhsim}
{Booth}, C.~M. \& {Schaye}, J. 2009, \mnras, 398, 53

\bibitem[{{Booth} \& {Schaye}(2010)}]{boot10mhalo}
---. 2010, \mnras, 405, L1

\bibitem[{{Booth} \& {Schaye}(2011)}]{boot11bhscale}
---. 2011, \mnras, 413, 1158

\bibitem[{{Boroson}(2005)}]{boro05blueseyf}
{Boroson}, T. 2005, \aj, 130, 381

\bibitem[{{Bournaud} \& {Combes}(2002)}]{bour02accr}
{Bournaud}, F. \& {Combes}, F. 2002, \aap, 392, 83

\bibitem[{{Bournaud} {et~al.}(2011){Bournaud}, {Dekel}, {Teyssier}, {Cacciato},
  {Daddi}, {Juneau}, \& {Shankar}}]{bour11instab}
{Bournaud}, F., {Dekel}, A., {Teyssier}, R., {Cacciato}, M., {Daddi}, E.,
  {Juneau}, S., \& {Shankar}, F. 2011, \apjl, 741, L33

\bibitem[{{Bournaud} {et~al.}(2005){Bournaud}, {Jog}, \&
  {Combes}}]{bour05merge}
{Bournaud}, F., {Jog}, C.~J., \& {Combes}, F. 2005, \aap, 437, 69

\bibitem[{{Bower} {et~al.}(2006){Bower}, {Benson}, {Malbon}, {Helly}, {Frenk},
  {Baugh}, {Cole}, \& {Lacey}}]{bowe06gal}
{Bower}, R.~G., {Benson}, A.~J., {Malbon}, R., {Helly}, J.~C., {Frenk}, C.~S.,
  {Baugh}, C.~M., {Cole}, S., \& {Lacey}, C.~G. 2006, \mnras, 370, 645

\bibitem[{{Bower} {et~al.}(2008){Bower}, {McCarthy}, \& {Benson}}]{bowe08flip}
{Bower}, R.~G., {McCarthy}, I.~G., \& {Benson}, A.~J. 2008, \mnras, 390, 1399

\bibitem[{{Boyle} {et~al.}(2000){Boyle}, {Shanks}, {Croom}, {Smith}, {Miller},
  {Loaring}, \& {Heymans}}]{boyl00qlf}
{Boyle}, B.~J., {Shanks}, T., {Croom}, S.~M., {Smith}, R.~J., {Miller}, L.,
  {Loaring}, N., \& {Heymans}, C. 2000, \mnras, 317, 1014

\bibitem[{{Boyle} \& {Terlevich}(1998)}]{boyl98qsosf}
{Boyle}, B.~J. \& {Terlevich}, R.~J. 1998, \mnras, 293, L49

\bibitem[{{Bradshaw} {et~al.}(2011){Bradshaw}, {Almaini}, {Hartley}, {Chuter},
  {Simpson}, {Conselice}, {Dunlop}, {McLure}, \& {Cirasuolo}}]{brad11agnenv}
{Bradshaw}, E.~J., et~al.\ 2011, \mnras, 415, 2626

\bibitem[{{Braito} {et~al.}(2004){Braito}, {Della Ceca}, {Piconcelli},
  {Severgnini}, {Bassani}, {Cappi}, {Franceschini}, {Iwasawa}, {Malaguti},
  {Marziani}, {Palumbo}, {Persic}, {Risaliti}, \& {Salvati}}]{brai04mrk231}
{Braito}, V., et~al.\ 2004, \aap, 420, 79

\bibitem[{{Brandt} \& {Alexander}(2010)}]{bran10}
{Brandt}, W.~N. \& {Alexander}, D.~M. 2010, Proceedings of the National Academy
  of Science, 107, 7184

\bibitem[{{Brandt} {et~al.}(2001){Brandt}, {Alexander}, {Hornschemeier},
  {Garmire}, {Schneider}, {Barger}, {Bauer}, {Broos}, {Cowie}, {Townsley},
  {Burrows}, {Chartas}, {Feigelson}, {Griffiths}, {Nousek}, \&
  {Sargent}}]{bran01b}
{Brandt}, W.~N., et~al.\ 2001, \aj,  122, 2810

\bibitem[{{Brandt} \& {Hasinger}(2005)}]{bran05}
{Brandt}, W.~N. \& {Hasinger}, G. 2005, \araa, 43, 827

\bibitem[{{Bregman} \& {Alexander}(2009)}]{breg09warp}
{Bregman}, M. \& {Alexander}, T. 2009, \apjl, 700, L192

\bibitem[{{Bromm} {et~al.}(2009){Bromm}, {Yoshida}, {Hernquist}, \&
  {McKee}}]{brom09firststar}
{Bromm}, V., {Yoshida}, N., {Hernquist}, L., \& {McKee}, C.~F. 2009, \nat, 459,
  49

\bibitem[{{Brusa} {et~al.}(2010){Brusa}, {Civano}, {Comastri}, {Miyaji},
  {Salvato}, {Zamorani}, {Cappelluti}, {Fiore}, {Hasinger}, {Mainieri},
  {Merloni}, {Bongiorno}, {Capak}, {Elvis}, {Gilli}, {Hao}, {Jahnke},
  {Koekemoer}, {Ilbert}, {Le Floc'h}, {Lusso}, {Mignoli}, {Schinnerer},
  {Silverman}, {Treister}, {Trump}, {Vignali}, {Zamojski}, {Aldcroft},
  {Aussel}, {Bardelli}, {Bolzonella}, {Cappi}, {Caputi}, {Contini},
  {Finoguenov}, {Fruscione}, {Garilli}, {Impey}, {Iovino}, {Iwasawa},
  {Kampczyk}, {Kartaltepe}, {Kneib}, {Knobel}, {Kovac}, {Lamareille},
  {Leborgne}, {Le Brun}, {Le Fevre}, {Lilly}, {Maier}, {McCracken}, {Pello},
  {Peng}, {Perez-Montero}, {de Ravel}, {Sanders}, {Scodeggio}, {Scoville},
  {Tanaka}, {Taniguchi}, {Tasca}, {de la Torre}, {Tresse}, {Vergani}, \&
  {Zucca}}]{brus10cosmosagn}
{Brusa}, M., et~al.\ 2010, \apj, 716, 348

\bibitem[{{Brusa} {et~al.}(2009{\natexlab{a}}){Brusa}, {Comastri}, {Gilli},
  {Hasinger}, {Iwasawa}, {Mainieri}, {Mignoli}, {Salvato}, {Zamorani},
  {Bongiorno}, {Cappelluti}, {Civano}, {Fiore}, {Merloni}, {Silverman},
  {Trump}, {Vignali}, {Capak}, {Elvis}, {Ilbert}, {Impey}, \&
  {Lilly}}]{brus09highz}
{Brusa}, M., et~al.\ 2009{\natexlab{a}}, \apj, 693, 8

\bibitem[{{Brusa} {et~al.}(2009{\natexlab{b}}){Brusa}, {Fiore}, {Santini},
  {Grazian}, {Comastri}, {Zamorani}, {Hasinger}, {Merloni}, {Civano},
  {Fontana}, \& {Mainieri}}]{brus09xhost}
{Brusa}, M., et~al.\ 2009{\natexlab{b}}, \aap, 507, 1277

\bibitem[{{Brusa} {et~al.}(2011){Brusa}, {Gilli}, {Civano}, {Comastri},
  {Fiore}, \& {Vignali}}]{brus11wfxt}
{Brusa}, M., {Gilli}, R., {Civano}, F., {Comastri}, A., {Fiore}, R., \&
  {Vignali}, C. 2011, Memorie della Societa Astronomica Italiana Supplementi,
  17, 106

\bibitem[{{Bundy} {et~al.}(2006){Bundy}, {Ellis}, {Conselice}, {Taylor},
  {Cooper}, {Willmer}, {Weiner}, {Coil}, {Noeske}, \&
  {Eisenhardt}}]{bund06sfevol}
{Bundy}, K., et~al.\ 2006, \apj, 651, 120

\bibitem[{{Bundy} {et~al.}(2008){Bundy}, {Georgakakis}, {Nandra}, {Ellis},
  {Conselice}, {Laird}, {Coil}, {Cooper}, {Faber}, {Newman}, {Pierce},
  {Primack}, \& {Yan}}]{bund08quench}
{Bundy}, K., et~al.\ 2008, \apj, 681, 931

\bibitem[{{Burlon} {et~al.}(2011){Burlon}, {Ajello}, {Greiner}, {Comastri},
  {Merloni}, \& {Gehrels}}]{burl11batagn}
{Burlon}, D., {Ajello}, M., {Greiner}, J., {Comastri}, A., {Merloni}, A., \&
  {Gehrels}, N. 2011, \apj, 728, 58

\bibitem[{{Canalizo} \& {Stockton}(2001)}]{cana01qso}
{Canalizo}, G. \& {Stockton}, A. 2001, \apj, 555, 719

\bibitem[{{Capetti} \& {Baldi}(2011)}]{cape11liner}
{Capetti}, A. \& {Baldi}, R.~D. 2011, \aap, 529, A126+

\bibitem[{{Cappi} {et~al.}(2006){Cappi}, {Panessa}, {Bassani}, {Dadina}, {Di
  Cocco}, {Comastri}, {della Ceca}, {Filippenko}, {Gianotti}, {Ho}, {Malaguti},
  {Mulchaey}, {Palumbo}, {Piconcelli}, {Sargent}, {Stephen}, {Trifoglio}, \&
  {Weaver}}]{capp06seyfx}
{Cappi}, M., et~al.\ 2006, \aap, 446, 459

\bibitem[{{Cardamone} {et~al.}(2010){Cardamone}, {Urry}, {Schawinski},
  {Treister}, {Brammer}, \& {Gawiser}}]{card10xhost}
{Cardamone}, C.~N., {Urry}, C.~M., {Schawinski}, K., {Treister}, E., {Brammer},
  G., \& {Gawiser}, E. 2010, \apjl, 721, L38

\bibitem[{{Carrera} {et~al.}(2011){Carrera}, {Page}, {Stevens}, {Ivison},
  {Dwelly}, {Ebrero}, \& {Falocco}}]{carr11qsogroup}
{Carrera}, F.~J., {Page}, M.~J., {Stevens}, J.~A., {Ivison}, R.~J., {Dwelly},
  T., {Ebrero}, J., \& {Falocco}, S. 2011, \mnras, 413, 2791

\bibitem[{{Cattaneo} {et~al.}(2006){Cattaneo}, {Dekel}, {Devriendt},
  {Guiderdoni}, \& {Blaizot}}]{catt06crit}
{Cattaneo}, A., {Dekel}, A., {Devriendt}, J., {Guiderdoni}, B., \& {Blaizot},
  J. 2006, \mnras, 370, 1651

\bibitem[{{Cattaneo} {et~al.}(2008){Cattaneo}, {Dekel}, {Faber}, \&
  {Guiderdoni}}]{catt08shutdown}
{Cattaneo}, A., {Dekel}, A., {Faber}, S.~M., \& {Guiderdoni}, B. 2008, \mnras,
  389, 567

\bibitem[{{Cattaneo} {et~al.}(2009){Cattaneo}, {Faber}, {Binney}, {Dekel},
  {Kormendy}, {Mushotzky}, {Babul}, {Best}, {Br{\"u}ggen}, {Fabian}, {Frenk},
  {Khalatyan}, {Netzer}, {Mahdavi}, {Silk}, {Steinmetz}, \&
  {Wisotzki}}]{catt09bhgal}
{Cattaneo}, A., et~al.\ 2009, \nat, 460, 213

\bibitem[{{Cavagnolo} {et~al.}(2010){Cavagnolo}, {McNamara}, {Nulsen},
  {Carilli}, {Jones}, \& {B{\^i}rzan}}]{cava10jet}
{Cavagnolo}, K.~W., {McNamara}, B.~R., {Nulsen}, P.~E.~J., {Carilli}, C.~L.,
  {Jones}, C., \& {B{\^i}rzan}, L. 2010, \apj, 720, 1066

\bibitem[{{Chapman} {et~al.}(2005){Chapman}, {Blain}, {Smail}, \&
  {Ivison}}]{chap05smg}
{Chapman}, S.~C., {Blain}, A.~W., {Smail}, I., \& {Ivison}, R.~J. 2005, \apj,
  622, 772

\bibitem[{{Chiaberge} {et~al.}(2005){Chiaberge}, {Capetti}, \&
  {Macchetto}}]{chia05liner}
{Chiaberge}, M., {Capetti}, A., \& {Macchetto}, F.~D. 2005, \apj, 625, 716

\bibitem[{{Cid Fernandes} {et~al.}(2004){Cid Fernandes}, {Gonz{\'a}lez
  Delgado}, {Schmitt}, {Storchi-Bergmann}, {Martins}, {P{\'e}rez}, {Heckman},
  {Leitherer}, \& {Schaerer}}]{cidf04llagn}
{Cid Fernandes}, R., et~al.\ 2004, \apj, 605, 105

\bibitem[{{Cisternas} {et~al.}(2011){Cisternas}, {Jahnke}, {Inskip},
  {Kartaltepe}, {Koekemoer}, {Lisker}, {Robaina}, {Scodeggio}, {Sheth},
  {Trump}, {Andrae}, {Miyaji}, {Lusso}, {Brusa}, {Capak}, {Cappelluti},
  {Civano}, {Ilbert}, {Impey}, {Leauthaud}, {Lilly}, {Salvato}, {Scoville}, \&
  {Taniguchi}}]{cist11agn}
{Cisternas}, M., et~al.\ 2011, \apj, 726, 57

\bibitem[{{Civano} {et~al.}(2010){Civano}, {Elvis}, {Lanzuisi}, {Jahnke},
  {Zamorani}, {Blecha}, {Bongiorno}, {Brusa}, {Comastri}, {Hao}, {Leauthaud},
  {Loeb}, {Mainieri}, {Piconcelli}, {Salvato}, {Scoville}, {Trump}, {Vignali},
  {Aldcroft}, {Bolzonella}, {Bressert}, {Finoguenov}, {Fruscione}, {Koekemoer},
  {Cappelluti}, {Fiore}, {Giodini}, {Gilli}, {Impey}, {Lilly}, {Lusso},
  {Puccetti}, {Silverman}, {Aussel}, {Capak}, {Frayer}, {Le Floch},
  {McCracken}, {Sanders}, {Schiminovich}, \& {Taniguchi}}]{civa10recoil}
{Civano}, F., et~al.\ 2010, \apj, 717, 209

\bibitem[{{Clements} {et~al.}(2002){Clements}, {McDowell}, {Shaked}, {Baker},
  {Borne}, {Colina}, {Lamb}, \& {Mundell}}]{clem02arp220}
{Clements}, D.~L., {McDowell}, J.~C., {Shaked}, S., {Baker}, A.~C., {Borne},
  K., {Colina}, L., {Lamb}, S.~A., \& {Mundell}, C. 2002, \apj, 581, 974

\bibitem[{{Clements} {et~al.}(1996){Clements}, {Sutherland}, {McMahon}, \&
  {Saunders}}]{clem96ulirg}
{Clements}, D.~L., {Sutherland}, W.~J., {McMahon}, R.~G., \& {Saunders}, W.
  1996, \mnras, 279, 477

\bibitem[{{Coil} {et~al.}(2009){Coil}, {Georgakakis}, {Newman}, {Cooper},
  {Croton}, {Davis}, {Koo}, {Laird}, {Nandra}, {Weiner}, {Willmer}, \&
  {Yan}}]{coil09xclust}
{Coil}, A.~L., et~al.\ 2009, \apj, 701, 1484

\bibitem[{{Colpi} \& {Dotti}(2009)}]{colp09binary_aph}
{Colpi}, M. \& {Dotti}, M. 2009, Review in Advanced Science Letters
  (arXiv:0906.4339)

\bibitem[{{Comastri} {et~al.}(2011){Comastri}, {Ranalli}, {Iwasawa}, {Vignali},
  {Gilli}, {Georgantopoulos}, {Barcons}, {Brandt}, {Brunner}, {Brusa},
  {Cappelluti}, {Carrera}, {Civano}, {Fiore}, {Hasinger}, {Mainieri},
  {Merloni}, {Nicastro}, {Paolillo}, {Puccetti}, {Rosati}, {Silverman},
  {Tozzi}, {Zamorani}, {Balestra}, {Bauer}, {Luo}, \& {Xue}}]{coma11xmmdeep}
{Comastri}, A., et~al.\ 2011, \aap, 526, L9+

\bibitem[{{Comerford} {et~al.}(2009){Comerford}, {Gerke}, {Newman}, {Davis},
  {Yan}, {Cooper}, {Faber}, {Koo}, {Coil}, {Rosario}, \&
  {Dutton}}]{come09binary}
{Comerford}, J.~M., et~al.\ 2009, \apj, 698, 956

\bibitem[{{Coppin} {et~al.}(2010){Coppin}, {Pope}, {Men{\'e}ndez-Delmestre},
  {Alexander}, {Dunlop}, {Egami}, {Gabor}, {Ibar}, {Ivison}, {Austermann},
  {Blain}, {Chapman}, {Clements}, {Dunne}, {Dye}, {Farrah}, {Hughes},
  {Mortier}, {Page}, {Rowan-Robinson}, {Scott}, {Simpson}, {Smail}, {Swinbank},
  {Vaccari}, \& {Yun}}]{copp10irs}
{Coppin}, K., et~al.\ 2010, \apj, 713, 503

\bibitem[{{Cowie} {et~al.}(1996){Cowie}, {Songaila}, {Hu}, \&
  {Cohen}}]{cowi96sfev}
{Cowie}, L.~L., {Songaila}, A., {Hu}, E.~M., \& {Cohen}, J.~G. 1996, \aj, 112,
  839

\bibitem[{{Crenshaw} {et~al.}(2003){Crenshaw}, {Kraemer}, \&
  {Gabel}}]{cren03nls1}
{Crenshaw}, D.~M., {Kraemer}, S.~B., \& {Gabel}, J.~R. 2003, \aj, 126, 1690

\bibitem[{{Croom} {et~al.}(2005){Croom}, {Boyle}, {Shanks}, {Smith}, {Miller},
  {Outram}, {Loaring}, {Hoyle}, \& {da {\^A}ngela}}]{croo05}
{Croom}, S.~M., et~al.\ 2005,  \mnras, 356, 415

\bibitem[{{Croom} {et~al.}(2009){Croom}, {Richards}, {Shanks}, {Boyle},
  {Strauss}, {Myers}, {Nichol}, {Pimbblet}, {Ross}, {Schneider}, {Sharp}, \&
  {Wake}}]{croo09twodfqz}
{Croom}, S.~M., et~al.\ 2009, \mnras, 399, 1755

\bibitem[{{Croston} {et~al.}(2009){Croston}, {Kraft}, {Hardcastle},
  {Birkinshaw}, {Worrall}, {Nulsen}, {Penna}, {Sivakoff}, {Jord{\'a}n},
  {Brassington}, {Evans}, {Forman}, {Gilfanov}, {Goodger}, {Harris}, {Jones},
  {Juett}, {Murray}, {Raychaudhury}, {Sarazin}, {Voss}, \&
  {Woodley}}]{cros09cena}
{Croston}, J.~H., et~al.\ 2009, \mnras, 395, 1999

\bibitem[{{Croton}(2009)}]{crot09qsohalo}
{Croton}, D.~J. 2009, \mnras, 394, 1109

\bibitem[{{Croton} {et~al.}(2006){Croton}, {Springel}, {White}, {De Lucia},
  {Frenk}, {Gao}, {Jenkins}, {Kauffmann}, {Navarro}, \& {Yoshida}}]{crot06}
{Croton}, D.~J., et~al.\ 2006, \mnras, 365, 11

\bibitem[{{Cuadra} {et~al.}(2008){Cuadra}, {Nayakshin}, \&
  {Martins}}]{cuad08sgrawind}
{Cuadra}, J., {Nayakshin}, S., \& {Martins}, F. 2008, \mnras, 383, 458

\bibitem[{{Cuadra} {et~al.}(2006){Cuadra}, {Nayakshin}, {Springel}, \& {Di
  Matteo}}]{cuad06sgrawind}
{Cuadra}, J., {Nayakshin}, S., {Springel}, V., \& {Di Matteo}, T. 2006, \mnras,
  366, 358

\bibitem[{{da {\^A}ngela} {et~al.}(2008){da {\^A}ngela}, {Shanks}, {Croom},
  {Weilbacher}, {Brunner}, {Couch}, {Miller}, {Myers}, {Nichol}, {Pimbblet},
  {de Propris}, {Richards}, {Ross}, {Schneider}, \& {Wake}}]{daan08clust}
{da {\^A}ngela}, J., et~al.\ 2008, \mnras, 383, 565

\bibitem[{{Daddi} {et~al.}(2007{\natexlab{a}}){Daddi}, {Alexander},
  {Dickinson}, {Gilli}, {Renzini}, {Elbaz}, {Cimatti}, {Chary}, {Frayer},
  {Bauer}, {Brandt}, {Giavalisco}, {Grogin}, {Huynh}, {Kurk}, {Mignoli},
  {Morrison}, {Pope}, \& {Ravindranath}}]{dadd07comp}
{Daddi}, E., et~al.\ 2007{\natexlab{a}}, \apj, 670, 173

\bibitem[{{Daddi} {et~al.}(2010){Daddi}, {Bournaud}, {Walter}, {Dannerbauer},
  {Carilli}, {Dickinson}, {Elbaz}, {Morrison}, {Riechers}, {Onodera}, {Salmi},
  {Krips}, \& {Stern}}]{dadd10fgas}
{Daddi}, E., et~al.\ 2010, \apj, 713, 686

\bibitem[{{Daddi} {et~al.}(2007{\natexlab{b}}){Daddi}, {Dickinson}, {Morrison},
  {Chary}, {Cimatti}, {Elbaz}, {Frayer}, {Renzini}, {Pope}, {Alexander},
  {Bauer}, {Giavalisco}, {Huynh}, {Kurk}, \& {Mignoli}}]{dadd07sf}
{Daddi}, E., et~al.\ 2007{\natexlab{b}}, \apj, 670, 156

\bibitem[{{Darg} {et~al.}(2010){Darg}, {Kaviraj}, {Lintott}, {Schawinski},
  {Sarzi}, {Bamford}, {Silk}, {Andreescu}, {Murray}, {Nichol}, {Raddick},
  {Slosar}, {Szalay}, {Thomas}, \& {Vandenberg}}]{darg10mergeanal}
{Darg}, D.~W., et~al.\ 2010, \mnras, 401, 1552

\bibitem[{{Davies} {et~al.}(2011){Davies}, {Miller}, \&
  {Bellovary}}]{davi11bhcluster}
{Davies}, M.~B., {Miller}, M.~C., \& {Bellovary}, J.~M. 2011, \apjl, 740, L42

\bibitem[{{Davies} {et~al.}(2009){Davies}, {Maciejewski}, {Hicks}, {Tacconi},
  {Genzel}, \& {Engel}}]{davi09ifu}
{Davies}, R.~I., {Maciejewski}, W., {Hicks}, E.~K.~S., {Tacconi}, L.~J.,
  {Genzel}, R., \& {Engel}, H. 2009, \apj, 702, 114

\bibitem[{{Davies} {et~al.}(2007){Davies}, {M{\"u}ller S{\'a}nchez}, {Genzel},
  {Tacconi}, {Hicks}, {Friedrich}, \& {Sternberg}}]{davi07agnsf}
{Davies}, R.~I., {M{\"u}ller S{\'a}nchez}, F., {Genzel}, R., {Tacconi}, L.~J.,
  {Hicks}, E.~K.~S., {Friedrich}, S., \& {Sternberg}, A. 2007, \apj, 671, 1388

\bibitem[{{Davis} \& {Laor}(2011)}]{davi11bolc}
{Davis}, S.~W. \& {Laor}, A. 2011, \apj, 728, 98

\bibitem[{{de Koff} {et~al.}(1996){de Koff}, {Baum}, {Sparks}, {Biretta},
  {Golombek}, {Macchetto}, {McCarthy}, \& {Miley}}]{deko96radiohost}
{de Koff}, S., {Baum}, S.~A., {Sparks}, W.~B., {Biretta}, J., {Golombek}, D.,
  {Macchetto}, F., {McCarthy}, P., \& {Miley}, G.~K. 1996, \apjs, 107, 621

\bibitem[{{De Lucia} {et~al.}(2006){De Lucia}, {Springel}, {White}, {Croton},
  \& {Kauffmann}}]{delu06ell}
{De Lucia}, G., {Springel}, V., {White}, S.~D.~M., {Croton}, D., \&
  {Kauffmann}, G. 2006, \mnras, 366, 499

\bibitem[{{De Rosa} {et~al.}(2011){De Rosa}, {Decarli}, {Walter}, {Fan},
  {Jiang}, {Kurk}, {Pasquali}, \& {Rix}}]{dero11z6qso}
{De Rosa}, G., {Decarli}, R., {Walter}, F., {Fan}, X., {Jiang}, L., {Kurk}, J.,
  {Pasquali}, A., \& {Rix}, H.~W. 2011, \apj, 739, 56

\bibitem[{{Decarli} {et~al.}(2010{\natexlab{a}}){Decarli}, {Falomo}, {Treves},
  {Kotilainen}, {Labita}, \& {Scarpa}}]{deca10mbhmethod}
{Decarli}, R., {Falomo}, R., {Treves}, A., {Kotilainen}, J.~K., {Labita}, M.,
  \& {Scarpa}, R. 2010{\natexlab{a}}, \mnras, 402, 2441

\bibitem[{{Decarli} {et~al.}(2010{\natexlab{b}}){Decarli}, {Falomo}, {Treves},
  {Labita}, {Kotilainen}, \& {Scarpa}}]{deca10mbhevol}
{Decarli}, R., {Falomo}, R., {Treves}, A., {Labita}, M., {Kotilainen}, J.~K.,
  \& {Scarpa}, R. 2010{\natexlab{b}}, \mnras, 402, 2453

\bibitem[{{Dekel} \& {Birnboim}(2006)}]{deke06shock}
{Dekel}, A. \& {Birnboim}, Y. 2006, \mnras, 368, 2

\bibitem[{{Denney} {et~al.}(2009){Denney}, {Peterson}, {Dietrich},
  {Vestergaard}, \& {Bentz}}]{denn09masserr}
{Denney}, K.~D., {Peterson}, B.~M., {Dietrich}, M., {Vestergaard}, M., \&
  {Bentz}, M.~C. 2009, \apj, 692, 246

\bibitem[{{Devecchi} \& {Volonteri}(2009)}]{deve09bhcluster}
{Devecchi}, B. \& {Volonteri}, M. 2009, \apj, 694, 302

\bibitem[{{Devecchi} {et~al.}(2010){Devecchi}, {Volonteri}, {Colpi}, \&
  {Haardt}}]{deve10bhcluster}
{Devecchi}, B., {Volonteri}, M., {Colpi}, M., \& {Haardt}, F. 2010, \mnras,
  409, 1057

\bibitem[{{Di Matteo} {et~al.}(2003){Di Matteo}, {Allen}, {Fabian}, {Wilson},
  \& {Young}}]{dima03m87}
{Di Matteo}, T., {Allen}, S.~W., {Fabian}, A.~C., {Wilson}, A.~S., \& {Young},
  A.~J. 2003, \apj, 582, 133

\bibitem[{{Di Matteo} {et~al.}(2008){Di Matteo}, {Colberg}, {Springel},
  {Hernquist}, \& {Sijacki}}]{dima08bhfeed}
{Di Matteo}, T., {Colberg}, J., {Springel}, V., {Hernquist}, L., \& {Sijacki},
  D. 2008, \apj, 676, 33

\bibitem[{{Di Matteo} {et~al.}(2005){Di Matteo}, {Springel}, \&
  {Hernquist}}]{dima05qso}
{Di Matteo}, T., {Springel}, V., \& {Hernquist}, L. 2005, \nat, 433, 604

\bibitem[{{Diamond-Stanic} \& {Rieke}(2011)}]{diam11agnsf_aph}
{Diamond-Stanic}, A.~M. \& {Rieke}, G.~H. 2011, \apj\ submitted
  (arXiv:1106.3565)

\bibitem[{{Digby-North} {et~al.}(2010){Digby-North}, {Nandra}, {Laird},
  {Steidel}, {Georgakakis}, {Bogosavljevi{\'c}}, {Erb}, {Shapley}, {Reddy}, \&
  {Aird}}]{digb10proto}
{Digby-North}, J.~A., et~al.\ 2010, \mnras, 407, 846

\bibitem[{{Dodds-Eden} {et~al.}(2011){Dodds-Eden}, {Gillessen}, {Fritz},
  {Eisenhauer}, {Trippe}, {Genzel}, {Ott}, {Bartko}, {Pfuhl}, {Bower},
  {Goldwurm}, {Porquet}, {Trap}, \& {Yusef-Zadeh}}]{dodd11sgra}
{Dodds-Eden}, K., et~al.\ 2011, \apj,  728, 37

\bibitem[{{Dodds-Eden} {et~al.}(2010){Dodds-Eden}, {Sharma}, {Quataert},
  {Genzel}, {Gillessen}, {Eisenhauer}, \& {Porquet}}]{dodd10flare}
{Dodds-Eden}, K., {Sharma}, P., {Quataert}, E., {Genzel}, R., {Gillessen}, S.,
  {Eisenhauer}, F., \& {Porquet}, D. 2010, \apj, 725, 450

\bibitem[{{Done} {et~al.}(2007){Done}, {Gierli{\'n}ski}, \&
  {Kubota}}]{done07disk}
{Done}, C., {Gierli{\'n}ski}, M., \& {Kubota}, A. 2007, \aapr, 15, 1

\bibitem[{{Donley} {et~al.}(2008){Donley}, {Rieke}, {P{\'e}rez-Gonz{\'a}lez},
  \& {Barro}}]{donl08spitz}
{Donley}, J.~L., {Rieke}, G.~H., {P{\'e}rez-Gonz{\'a}lez}, P.~G., \& {Barro},
  G. 2008, \apj, 687, 111

\bibitem[{{Donley} {et~al.}(2005){Donley}, {Rieke}, {Rigby}, \&
  {P{\'e}rez-Gonz{\'a}lez}}]{donl05radio}
{Donley}, J.~L., {Rieke}, G.~H., {Rigby}, J.~R., \& {P{\'e}rez-Gonz{\'a}lez},
  P.~G. 2005, \apj, 634, 169

\bibitem[{{Donoso} {et~al.}(2010){Donoso}, {Li}, {Kauffmann}, {Best}, \&
  {Heckman}}]{dono10clust}
{Donoso}, E., {Li}, C., {Kauffmann}, G., {Best}, P.~N., \& {Heckman}, T.~M.
  2010, \mnras, 407, 1078

\bibitem[{{Dotti} {et~al.}(2009){Dotti}, {Montuori}, {Decarli}, {Volonteri},
  {Colpi}, \& {Haardt}}]{dott09binary}
{Dotti}, M., {Montuori}, C., {Decarli}, R., {Volonteri}, M., {Colpi}, M., \&
  {Haardt}, F. 2009, \mnras, 398, L73

\bibitem[{{Dotti} \& {Ruszkowski}(2010)}]{dott10agnpair}
{Dotti}, M. \& {Ruszkowski}, M. 2010, \apjl, 713, L37

\bibitem[{{Down} {et~al.}(2010){Down}, {Rawlings}, {Sivia}, \&
  {Baker}}]{down10qsodisk}
{Down}, E.~J., {Rawlings}, S., {Sivia}, D.~S., \& {Baker}, J.~C. 2010, \mnras,
  401, 633

\bibitem[{{Downes} \& {Eckart}(2007)}]{down07arp220}
{Downes}, D. \& {Eckart}, A. 2007, \aap, 468, L57

\bibitem[{{Dumas} {et~al.}(2007){Dumas}, {Mundell}, {Emsellem}, \&
  {Nagar}}]{duma07ifu}
{Dumas}, G., {Mundell}, C.~G., {Emsellem}, E., \& {Nagar}, N.~M. 2007, \mnras,
  379, 1249

\bibitem[{{Dunlop} {et~al.}(2003){Dunlop}, {McLure}, {Kukula}, {Baum}, {O'Dea},
  \& {Hughes}}]{dunl03qsohost}
{Dunlop}, J.~S., {McLure}, R.~J., {Kukula}, M.~J., {Baum}, S.~A., {O'Dea},
  C.~P., \& {Hughes}, D.~H. 2003, \mnras, 340, 1095

\bibitem[{{Dunn} {et~al.}(2010){Dunn}, {Bautista}, {Arav}, {Moe}, {Korista},
  {Costantini}, {Benn}, {Ellison}, \& {Edmonds}}]{dunn10outflow}
{Dunn}, J.~P., et~al.\ 2010, \apj,  709, 611

\bibitem[{{Dunn} \& {Fabian}(2004)}]{dunn04bubbles}
{Dunn}, R.~J.~H. \& {Fabian}, A.~C. 2004, \mnras, 355, 862

\bibitem[{{Dunn} \& {Fabian}(2008)}]{dunn08bubbles}
---. 2008, \mnras, 385, 757

\bibitem[{{Dunn} {et~al.}(2005){Dunn}, {Fabian}, \& {Taylor}}]{dunn05bubbles}
{Dunn}, R.~J.~H., {Fabian}, A.~C., \& {Taylor}, G.~B. 2005, \mnras, 364, 1343

\bibitem[{{Dwelly} \& {Page}(2006)}]{dwel06obsc}
{Dwelly}, T. \& {Page}, M.~J. 2006, \mnras, 372, 1755

\bibitem[{{Eastman} {et~al.}(2007){Eastman}, {Martini}, {Sivakoff}, {Kelson},
  {Mulchaey}, \& {Tran}}]{east07clustagn}
{Eastman}, J., {Martini}, P., {Sivakoff}, G., {Kelson}, D.~D., {Mulchaey},
  J.~S., \& {Tran}, K.-V. 2007, \apjl, 664, L9

\bibitem[{{Eckart} {et~al.}(2009){Eckart}, {Baganoff}, {Morris}, {Kunneriath},
  {Zamaninasab}, {Witzel}, {Sch{\"o}del}, {Garc{\'{\i}}a-Mar{\'{\i}}n},
  {Meyer}, {Bower}, {Marrone}, {Bautz}, {Brandt}, {Garmire}, {Ricker},
  {Straubmeier}, {Roberts}, {Muzic}, {Mauerhan}, \& {Zensus}}]{ecka09sgra}
{Eckart}, A., et~al.\ 2009, \aap, 500, 935

\bibitem[{{Eckart} {et~al.}(2006){Eckart}, {Baganoff}, {Sch{\"o}del}, {Morris},
  {Genzel}, {Bower}, {Marrone}, {Moran}, {Viehmann}, {Bautz}, {Brandt},
  {Garmire}, {Ott}, {Trippe}, {Ricker}, {Straubmeier}, {Roberts},
  {Yusef-Zadeh}, {Zhao}, \& {Rao}}]{ecka06sgra}
{Eckart}, A., et~al.\ 2006, \aap, 450, 535

\bibitem[{{Efstathiou} {et~al.}(1995){Efstathiou}, {Hough}, \&
  {Young}}]{efst95ngc1068}
{Efstathiou}, A., {Hough}, J.~H., \& {Young}, S. 1995, \mnras, 277, 1134

\bibitem[{{Efstathiou} \& {Rowan-Robinson}(1995)}]{efst95disk}
{Efstathiou}, A. \& {Rowan-Robinson}, M. 1995, \mnras, 273, 649

\bibitem[{{Elbaz} {et~al.}(2011){Elbaz}, {Dickinson}, {Hwang},
  {D{\'{\i}}az-Santos}, {Magdis}, {Magnelli}, {Le Borgne}, {Galliano},
  {Pannella}, {Chanial}, {Armus}, {Charmandaris}, {Daddi}, {Aussel}, {Popesso},
  {Kartaltepe}, {Altieri}, {Valtchanov}, {Coia}, {Dannerbauer}, {Dasyra},
  {Leiton}, {Mazzarella}, {Alexander}, {Buat}, {Burgarella}, {Chary}, {Gilli},
  {Ivison}, {Juneau}, {Le Floc'h}, {Lutz}, {Morrison}, {Mullaney}, {Murphy},
  {Pope}, {Scott}, {Brodwin}, {Calzetti}, {Cesarsky}, {Charlot}, {Dole},
  {Eisenhardt}, {Ferguson}, {F{\"o}rster Schreiber}, {Frayer}, {Giavalisco},
  {Huynh}, {Koekemoer}, {Papovich}, {Reddy}, {Surace}, {Teplitz}, {Yun}, \&
  {Wilson}}]{elba11ms}
{Elbaz}, D., et~al.\ 2011, \aap,  533, A119

\bibitem[{{Elitzur} \& {Shlosman}(2006)}]{elit06torus}
{Elitzur}, M. \& {Shlosman}, I. 2006, \apjl, 648, L101

\bibitem[{{Ellison} {et~al.}(2011){Ellison}, {Patton}, {Mendel}, \&
  {Scudder}}]{elli11agninter}
{Ellison}, S.~L., {Patton}, D.~R., {Mendel}, J.~T., \& {Scudder}, J.~M. 2011,
  \mnras, 1541

\bibitem[{{Elvis} {et~al.}(1994){Elvis}, {Wilkes}, {McDowell}, {Green},
  {Bechtold}, {Willner}, {Oey}, {Polomski}, \& {Cutri}}]{elvi94}
{Elvis}, M., et~al.\ 1994,  \apjs, 95, 1

\bibitem[{{Engel} {et~al.}(2010){Engel}, {Tacconi}, {Davies}, {Neri}, {Smail},
  {Chapman}, {Genzel}, {Cox}, {Greve}, {Ivison}, {Blain}, {Bertoldi}, \&
  {Omont}}]{enge10smg}
{Engel}, H., et~al.\ 2010, \apj, 724, 233

\bibitem[{{Englmaier} \& {Shlosman}(2004)}]{engl04bars}
{Englmaier}, P. \& {Shlosman}, I. 2004, \apjl, 617, L115

\bibitem[{{Eracleous} {et~al.}(2002){Eracleous}, {Shields}, {Chartas}, \&
  {Moran}}]{erac02liner}
{Eracleous}, M., {Shields}, J.~C., {Chartas}, G., \& {Moran}, E.~C. 2002, \apj,
  565, 108

\bibitem[{{Esin} {et~al.}(1997){Esin}, {McClintock}, \& {Narayan}}]{esin97nova}
{Esin}, A.~A., {McClintock}, J.~E., \& {Narayan}, R. 1997, \apj, 489, 865

\bibitem[{{Evans} {et~al.}(2010){Evans}, {Ogle}, {Marshall}, {Nowak},
  {Bianchi}, {Guainazzi}, {Longinotti}, {Dewey}, {Schulz}, {Noble}, {Houck}, \&
  {Canizares}}]{evan10ngc1068}
{Evans}, D.~A., et~al.\ 2010, in Astronomical  Society of the Pacific Conference Series, Vol. 427, Accretion and Ejection in  AGN: a Global View, ed. {L.~Maraschi, G.~Ghisellini, R.~Della Ceca, \&  F.~Tavecchio}, 97--+

\bibitem[{{Evans} {et~al.}(2006){Evans}, {Worrall}, {Hardcastle}, {Kraft}, \&
  {Birkinshaw}}]{evans2006}
{Evans}, D.~A., {Worrall}, D.~M., {Hardcastle}, M.~J., {Kraft}, R.~P., \&
  {Birkinshaw}, M. 2006, \apj, 642, 96

\bibitem[{{Fabian}(1999)}]{fabian99feedback}
{Fabian}, A.~C. 1999, \mnras, 308, L39

\bibitem[{{Fakhouri} {et~al.}(2010){Fakhouri}, {Ma}, \&
  {Boylan-Kolchin}}]{fakh10halorate}
{Fakhouri}, O., {Ma}, C.-P., \& {Boylan-Kolchin}, M. 2010, \mnras, 406, 2267

\bibitem[{{Falcke} {et~al.}(2004){Falcke}, {K{\"o}rding}, \&
  {Markoff}}]{falc04bhplane}
{Falcke}, H., {K{\"o}rding}, E., \& {Markoff}, S. 2004, \aap, 414, 895

\bibitem[{{Fan} {et~al.}(2001){Fan}, {Narayanan}, {Lupton}, {Strauss}, {Knapp},
  {Becker}, {White}, {Pentericci}, {Leggett}, {Haiman}, {Gunn}, {Ivezi{\'c}},
  {Schneider}, {Anderson}, {Brinkmann}, {Bahcall}, {Connolly}, {Csabai}, {Doi},
  {Fukugita}, {Geballe}, {Grebel}, {Harbeck}, {Hennessy}, {Lamb}, {Miknaitis},
  {Munn}, {Nichol}, {Okamura}, {Pier}, {Prada}, {Richards}, {Szalay}, \&
  {York}}]{fan01z6qso}
{Fan}, X., et~al.\ 2001, \aj, 122, 2833

\bibitem[{{Fan} {et~al.}(2006){Fan}, {Strauss}, {Richards}, {Hennawi},
  {Becker}, {White}, {Diamond-Stanic}, {Donley}, {Jiang}, {Kim}, {Vestergaard},
  {Young}, {Gunn}, {Lupton}, {Knapp}, {Schneider}, {Brandt}, {Bahcall},
  {Barentine}, {Brinkmann}, {Brewington}, {Fukugita}, {Harvanek}, {Kleinman},
  {Krzesinski}, {Long}, {Neilsen}, {Nitta}, {Snedden}, \& {Voges}}]{fan06}
{Fan}, X., et~al.\ 2006, \aj, 131,  1203

\bibitem[{{Fanaroff} \& {Riley}(1974)}]{fana74}
{Fanaroff}, B.~L. \& {Riley}, J.~M. 1974, \mnras, 167, 31P

\bibitem[{{Fanidakis} {et~al.}(2011){Fanidakis}, {Baugh}, {Benson}, {Bower},
  {Cole}, {Done}, \& {Frenk}}]{fani11agn}
{Fanidakis}, N., {Baugh}, C.~M., {Benson}, A.~J., {Bower}, R.~G., {Cole}, S.,
  {Done}, C., \& {Frenk}, C.~S. 2011, \mnras, 410, 53

\bibitem[{{Fanidakis} {et~al.}(2012){Fanidakis}, {Baugh}, {Benson}, {Bower},
  {Cole}, {Done}, {Frenk}, {Hickox}, {Lacey}, \& {Lagos}}]{fani12agn_aph}
{Fanidakis}, N., et~al.\ 2012, \mnras\ in press

\bibitem[{{Fathi} {et~al.}(2006){Fathi}, {Storchi-Bergmann}, {Riffel}, {Winge},
  {Axon}, {Robinson}, {Capetti}, \& {Marconi}}]{fath06ifu}
{Fathi}, K., {Storchi-Bergmann}, T., {Riffel}, R.~A., {Winge}, C., {Axon},
  D.~J., {Robinson}, A., {Capetti}, A., \& {Marconi}, A. 2006, \apjl, 641, L25

\bibitem[{{Fender} {et~al.}(2004){Fender}, {Belloni}, \& {Gallo}}]{fend04jet}
{Fender}, R.~P., {Belloni}, T.~M., \& {Gallo}, E. 2004, \mnras, 355, 1105

\bibitem[{{Ferrarese} \& {Merritt}(2000)}]{ferr00}
{Ferrarese}, L. \& {Merritt}, D. 2000, \apjl, 539, L9

\bibitem[{{Feruglio} {et~al.}(2011){Feruglio}, {Daddi}, {Fiore}, {Alexander},
  {Piconcelli}, \& {Malacaria}}]{feru11kalpha}
{Feruglio}, C., {Daddi}, E., {Fiore}, F., {Alexander}, D.~M., {Piconcelli}, E.,
  \& {Malacaria}, C. 2011, \apjl, 729, L4+

\bibitem[{{Feruglio} {et~al.}(2010){Feruglio}, {Maiolino}, {Piconcelli},
  {Menci}, {Aussel}, {Lamastra}, \& {Fiore}}]{feru10outflow}
{Feruglio}, C., {Maiolino}, R., {Piconcelli}, E., {Menci}, N., {Aussel}, H.,
  {Lamastra}, A., \& {Fiore}, F. 2010, \aap, 518, L155+

\bibitem[{{Filho} {et~al.}(2006){Filho}, {Barthel}, \& {Ho}}]{filh06radio}
{Filho}, M.~E., {Barthel}, P.~D., \& {Ho}, L.~C. 2006, \aap, 451, 71

\bibitem[{{Fine} {et~al.}(2011){Fine}, {Shanks}, {Nikoloudakis}, \&
  {Sawangwit}}]{fine11radio}
{Fine}, S., {Shanks}, T., {Nikoloudakis}, N., \& {Sawangwit}, U. 2011, \mnras\
  in press (arXiv:1107.5666)

\bibitem[{{Finn} {et~al.}(2010){Finn}, {Desai}, {Rudnick}, {Poggianti}, {Bell},
  {Hinz}, {Jablonka}, {Milvang-Jensen}, {Moustakas}, {Rines}, \&
  {Zaritsky}}]{finn10clustsf}
{Finn}, R.~A., et~al.\ 2010, \apj, 720, 87

\bibitem[{{Fiore}(2010)}]{fior10agnevol}
{Fiore}, F. 2010, in American Institute of Physics Conference Series, Vol.
  1248, American Institute of Physics Conference Series, ed. {A.~Comastri,
  L.~Angelini, \& M.~Cappi}, 373--380

\bibitem[{{Fiore} {et~al.}(2003){Fiore}, {Brusa}, {Cocchia}, {Baldi},
  {Carangelo}, {Ciliegi}, {Comastri}, {La Franca}, {Maiolino}, {Matt},
  {Molendi}, {Mignoli}, {Perola}, {Severgnini}, \& {Vignali}}]{fior03xevol}
{Fiore}, F., et~al.\ 2003, \aap, 409, 79

\bibitem[{{Fiore} {et~al.}(2008){Fiore}, {Grazian}, {Santini}, {Puccetti},
  {Brusa}, {Feruglio}, {Fontana}, {Giallongo}, {Comastri}, {Gruppioni},
  {Pozzi}, {Zamorani}, \& {Vignali}}]{fior08agn}
{Fiore}, F., et~al.\ 2008, \apj, 672, 94

\bibitem[{{Fiore} {et~al.}(2009){Fiore}, {Puccetti}, {Brusa}, {Salvato},
  {Zamorani}, {Aldcroft}, {Aussel}, {Brunner}, {Capak}, {Cappelluti}, {Civano},
  {Comastri}, {Elvis}, {Feruglio}, {Finoguenov}, {Fruscione}, {Gilli},
  {Hasinger}, {Koekemoer}, {Kartaltepe}, {Ilbert}, {Impey}, {Le Floc'h},
  {Lilly}, {Mainieri}, {Martinez-Sansigre}, {McCracken}, {Menci}, {Merloni},
  {Miyaji}, {Sanders}, {Sargent}, {Schinnerer}, {Scoville}, {Silverman},
  {Smolcic}, {Steffen}, {Santini}, {Taniguchi}, {Thompson}, {Trump}, {Vignali},
  {Urry}, \& {Yan}}]{fior09obsc}
{Fiore}, F., et~al.\ 2009, \apj, 693, 447

\bibitem[{{Fischer} {et~al.}(2010){Fischer}, {Sturm}, {Gonz{\'a}lez-Alfonso},
  {Graci{\'a}-Carpio}, {Hailey-Dunsheath}, {Poglitsch}, {Contursi}, {Lutz},
  {Genzel}, {Sternberg}, {Verma}, \& {Tacconi}}]{fisc10mrk231}
{Fischer}, J., et~al.\ 2010, \aap, 518,  L41

\bibitem[{{Floyd} {et~al.}(2010){Floyd}, {Axon}, {Baum}, {Capetti},
  {Chiaberge}, {Madrid}, {O'Dea}, {Perlman}, \& {Sparks}}]{floy10radio}
{Floyd}, D.~J.~E., et~al.\ 2010, \apj, 713,  66

\bibitem[{{Floyd} {et~al.}(2009){Floyd}, {Bate}, \& {Webster}}]{floy09qsolens}
{Floyd}, D.~J.~E., {Bate}, N.~F., \& {Webster}, R.~L. 2009, \mnras, 398, 233

\bibitem[{{Floyd} {et~al.}(2004){Floyd}, {Kukula}, {Dunlop}, {McLure},
  {Miller}, {Percival}, {Baum}, \& {O'Dea}}]{floy04qsohost}
{Floyd}, D.~J.~E., {Kukula}, M.~J., {Dunlop}, J.~S., {McLure}, R.~J., {Miller},
  L., {Percival}, W.~J., {Baum}, S.~A., \& {O'Dea}, C.~P. 2004, \mnras, 355,
  196

\bibitem[{{Forman} {et~al.}(2007){Forman}, {Jones}, {Churazov}, {Markevitch},
  {Nulsen}, {Vikhlinin}, {Begelman}, {B{\"o}hringer}, {Eilek}, {Heinz},
  {Kraft}, {Owen}, \& {Pahre}}]{form07m87}
{Forman}, W., et~al.\ 2007, \apj, 665, 1057

\bibitem[{{Fromerth} \& {Melia}(2000)}]{from00bhmass}
{Fromerth}, M.~J. \& {Melia}, F. 2000, \apj, 533, 172

\bibitem[{{Fryer} \& {Kalogera}(2001)}]{fry01bhstar}
{Fryer}, C.~L. \& {Kalogera}, V. 2001, \apj, 554, 548

\bibitem[{{Gabor} {et~al.}(2009){Gabor}, {Impey}, {Jahnke}, {Simmons}, {Trump},
  {Koekemoer}, {Brusa}, {Cappelluti}, {Schinnerer}, {Smol{\v c}i{\'c}},
  {Salvato}, {Rhodes}, {Mobasher}, {Capak}, {Massey}, {Leauthaud}, \&
  {Scoville}}]{gabo09xmorph}
{Gabor}, J.~M., et~al.\ 2009, \apj, 691, 705

\bibitem[{{Gadotti} \& {Kauffmann}(2009)}]{gado09bhbulge}
{Gadotti}, D.~A. \& {Kauffmann}, G. 2009, \mnras, 399, 621

\bibitem[{{Gallagher} {et~al.}(2006){Gallagher}, {Brandt}, {Chartas},
  {Priddey}, {Garmire}, \& {Sambruna}}]{gall06balqso}
{Gallagher}, S.~C., {Brandt}, W.~N., {Chartas}, G., {Priddey}, R., {Garmire},
  G.~P., \& {Sambruna}, R.~M. 2006, \apj, 644, 709

\bibitem[{{Gallo} {et~al.}(2003){Gallo}, {Fender}, \& {Pooley}}]{gall03radiox}
{Gallo}, E., {Fender}, R.~P., \& {Pooley}, G.~G. 2003, \mnras, 344, 60

\bibitem[{{Gandhi} {et~al.}(2009){Gandhi}, {Horst}, {Smette}, {H{\"o}nig},
  {Comastri}, {Gilli}, {Vignali}, \& {Duschl}}]{gand09seyfir}
{Gandhi}, P., {Horst}, H., {Smette}, A., {H{\"o}nig}, S., {Comastri}, A.,
  {Gilli}, R., {Vignali}, C., \& {Duschl}, W. 2009, \aap, 502, 457

\bibitem[{{Ganguly} \& {Brotherton}(2008)}]{gang08winds}
{Ganguly}, R. \& {Brotherton}, M.~S. 2008, \apj, 672, 102

\bibitem[{{Garc{\'{\i}}a-Burillo} {et~al.}(2005){Garc{\'{\i}}a-Burillo},
  {Combes}, {Schinnerer}, {Boone}, \& {Hunt}}]{garc05agnmol}
{Garc{\'{\i}}a-Burillo}, S., {Combes}, F., {Schinnerer}, E., {Boone}, F., \&
  {Hunt}, L.~K. 2005, \aap, 441, 1011

\bibitem[{{Geach} {et~al.}(2011){Geach}, {Smail}, {Moran}, {MacArthur},
  {Lagos}, \& {Edge}}]{geac11fgas}
{Geach}, J.~E., {Smail}, I., {Moran}, S.~M., {MacArthur}, L.~A., {Lagos},
  C.~d.~P., \& {Edge}, A.~C. 2011, \apjl, 730, L19+

\bibitem[{{Gebhardt} {et~al.}(2000){Gebhardt}, {Bender}, {Bower}, {Dressler},
  {Faber}, {Filippenko}, {Green}, {Grillmair}, {Ho}, {Kormendy}, {Lauer},
  {Magorrian}, {Pinkney}, {Richstone}, \& {Tremaine}}]{gebh00}
{Gebhardt}, K., et~al.\ 2000, \apjl, 539, L13

\bibitem[{{Genzel} {et~al.}(2008){Genzel}, {Burkert}, {Bouch{\'e}}, {Cresci},
  {F{\"o}rster Schreiber}, {Shapley}, {Shapiro}, {Tacconi}, {Buschkamp},
  {Cimatti}, {Daddi}, {Davies}, {Eisenhauer}, {Erb}, {Genel}, {Gerhard},
  {Hicks}, {Lutz}, {Naab}, {Ott}, {Rabien}, {Renzini}, {Steidel}, {Sternberg},
  \& {Lilly}}]{genz08ifsz2}
{Genzel}, R., et~al.\ 2008, \apj, 687, 59

\bibitem[{{Genzel} {et~al.}(2003){Genzel}, {Sch{\"o}del}, {Ott}, {Eckart},
  {Alexander}, {Lacombe}, {Rouan}, \& {Aschenbach}}]{genz03sgra}
{Genzel}, R., {Sch{\"o}del}, R., {Ott}, T., {Eckart}, A., {Alexander}, T.,
  {Lacombe}, F., {Rouan}, D., \& {Aschenbach}, B. 2003, \nat, 425, 934

\bibitem[{{Georgakakis} {et~al.}(2009){Georgakakis}, {Coil}, {Laird},
  {Griffith}, {Nandra}, {Lotz}, {Pierce}, {Cooper}, {Newman}, \&
  {Koekemoer}}]{geor09xmorph}
{Georgakakis}, A., et~al.\ 2009, \mnras, 397, 623

\bibitem[{{Georgakakis} {et~al.}(2011){Georgakakis}, {Coil}, {Willmer},
  {Nandra}, {Kocevski}, {Cooper}, {Rosario}, {Koo}, {Trump}, \&
  {Juneau}}]{geor11xagn}
{Georgakakis}, A., et~al.\ 2011, \mnras, 1658

\bibitem[{{Georgakakis} {et~al.}(2008){Georgakakis}, {Nandra}, {Yan},
  {Willner}, {Lotz}, {Pierce}, {Cooper}, {Laird}, {Koo}, {Barmby}, {Newman},
  {Primack}, \& {Coil}}]{geor08agn}
{Georgakakis}, A., et~al.\ 2008, \mnras, 385, 2049

\bibitem[{{Georgantopoulos} {et~al.}(2011){Georgantopoulos}, {Rovilos}, \&
  {Comastri}}]{geor11submmx}
{Georgantopoulos}, I., {Rovilos}, E., \& {Comastri}, A. 2011, \aap, 526, A46+

\bibitem[{{Ghez} {et~al.}(2008){Ghez}, {Salim}, {Weinberg}, {Lu}, {Do}, {Dunn},
  {Matthews}, {Morris}, {Yelda}, {Becklin}, {Kremenek}, {Milosavljevic}, \&
  {Naiman}}]{ghez08sgra}
{Ghez}, A.~M., et~al.\ 2008, \apj, 689, 1044

\bibitem[{{Ghez} {et~al.}(2004){Ghez}, {Wright}, {Matthews}, {Thompson}, {Le
  Mignant}, {Tanner}, {Hornstein}, {Morris}, {Becklin}, \&
  {Soifer}}]{ghez04flare}
{Ghez}, A.~M., et~al.\ 2004, \apjl, 601, L159

\bibitem[{{Giacconi} {et~al.}(2002){Giacconi}, {Zirm}, {Wang}, {Rosati},
  {Nonino}, {Tozzi}, {Gilli}, {Mainieri}, {Hasinger}, {Kewley}, {Bergeron},
  {Borgani}, {Gilmozzi}, {Grogin}, {Koekemoer}, {Schreier}, {Zheng}, \&
  {Norman}}]{giac02}
{Giacconi}, R., et~al.\ 2002, \apjs, 139, 369

\bibitem[{{Gibson} {et~al.}(2009){Gibson}, {Brandt}, {Gallagher}, \&
  {Schneider}}]{gibs09balqso}
{Gibson}, R.~R., {Brandt}, W.~N., {Gallagher}, S.~C., \& {Schneider}, D.~P.
  2009, \apj, 696, 924

\bibitem[{{Gillessen} {et~al.}(2006){Gillessen}, {Eisenhauer}, {Quataert},
  {Genzel}, {Paumard}, {Trippe}, {Ott}, {Abuter}, {Eckart}, {Lagage},
  {Lehnert}, {Tacconi}, \& {Martins}}]{gill06sgra}
{Gillessen}, S., et~al.\ 2006, \apjl, 640, L163

\bibitem[{{Gillessen} {et~al.}(2009){Gillessen}, {Eisenhauer}, {Trippe},
  {Alexander}, {Genzel}, {Martins}, \& {Ott}}]{gill09sgra}
{Gillessen}, S., {Eisenhauer}, F., {Trippe}, S., {Alexander}, T., {Genzel}, R.,
  {Martins}, F., \& {Ott}, T. 2009, \apj, 692, 1075

\bibitem[{{Gilli} {et~al.}(2007){Gilli}, {Comastri}, \& {Hasinger}}]{gill07cxb}
{Gilli}, R., {Comastri}, A., \& {Hasinger}, G. 2007, \aap, 463, 79

\bibitem[{{Gilli} {et~al.}(2010){Gilli}, {Vignali}, {Mignoli}, {Iwasawa},
  {Comastri}, \& {Zamorani}}]{gill10nev}
{Gilli}, R., {Vignali}, C., {Mignoli}, M., {Iwasawa}, K., {Comastri}, A., \&
  {Zamorani}, G. 2010, \aap, 519, A92+

\bibitem[{{Gilli} {et~al.}(2009){Gilli}, {Zamorani}, {Miyaji}, {Silverman},
  {Brusa}, {Mainieri}, {Cappelluti}, {Daddi}, {Porciani}, {Pozzetti}, {Civano},
  {Comastri}, {Finoguenov}, {Fiore}, {Salvato}, {Vignali}, {Hasinger}, {Lilly},
  {Impey}, {Trump}, {Capak}, {McCracken}, {Scoville}, {Taniguchi}, {Carollo},
  {Contini}, {Kneib}, {Le Fevre}, {Renzini}, {Scodeggio}, {Bardelli},
  {Bolzonella}, {Bongiorno}, {Caputi}, {Cimatti}, {Coppa}, {Cucciati}, {de La
  Torre}, {de Ravel}, {Franzetti}, {Garilli}, {Iovino}, {Kampczyk}, {Knobel},
  {Kova{\v c}}, {Lamareille}, {Le Borgne}, {Le Brun}, {Maier}, {Mignoli},
  {Pell{\`o}}, {Peng}, {Perez Montero}, {Ricciardelli}, {Tanaka}, {Tasca},
  {Tresse}, {Vergani}, {Zucca}, {Abbas}, {Bottini}, {Cappi}, {Cassata},
  {Fumana}, {Guzzo}, {Leauthaud}, {Maccagni}, {Marinoni}, {Memeo}, {Meneux},
  {Oesch}, {Scaramella}, \& {Walcher}}]{gill09xclust}
{Gilli}, R., et~al.\ 2009, \aap, 494, 33

\bibitem[{{Gilmour} {et~al.}(2007){Gilmour}, {Gray}, {Almaini}, {Best}, {Wolf},
  {Meisenheimer}, {Papovich}, \& {Bell}}]{gilm07agnsuperclust}
{Gilmour}, R., {Gray}, M.~E., {Almaini}, O., {Best}, P., {Wolf}, C.,
  {Meisenheimer}, K., {Papovich}, C., \& {Bell}, E. 2007, \mnras, 380, 1467

\bibitem[{{Giodini} {et~al.}(2009){Giodini}, {Pierini}, {Finoguenov}, {Pratt},
  {Boehringer}, {Leauthaud}, {Guzzo}, {Aussel}, {Bolzonella}, {Capak}, {Elvis},
  {Hasinger}, {Ilbert}, {Kartaltepe}, {Koekemoer}, {Lilly}, {Massey},
  {McCracken}, {Rhodes}, {Salvato}, {Sanders}, {Scoville}, {Sasaki}, {Smolcic},
  {Taniguchi}, {Thompson}, \& {the COSMOS Collaboration}}]{giod09baryons}
{Giodini}, S., et~al.\ 2009, \apj, 703, 982

\bibitem[{{Giodini} {et~al.}(2010){Giodini}, {Smol{\v c}i{\'c}}, {Finoguenov},
  {Boehringer}, {B{\^i}rzan}, {Zamorani}, {Oklop{\v c}i{\'c}}, {Pierini},
  {Pratt}, {Schinnerer}, {Massey}, {Koekemoer}, {Salvato}, {Sanders},
  {Kartaltepe}, \& {Thompson}}]{giod10groups}
{Giodini}, S., et~al.\ 2010,  \apj, 714, 218

\bibitem[{{Gladstone} {et~al.}(2009){Gladstone}, {Roberts}, \&
  {Done}}]{glad09ulx}
{Gladstone}, J.~C., {Roberts}, T.~P., \& {Done}, C. 2009, \mnras, 397, 1836

\bibitem[{{Glover} {et~al.}(2008){Glover}, {Clark}, {Greif}, {Johnson},
  {Bromm}, {Klessen}, \& {Stacy}}]{glov08pop3}
{Glover}, S.~C.~O., {Clark}, P.~C., {Greif}, T.~H., {Johnson}, J.~L., {Bromm},
  V., {Klessen}, R.~S., \& {Stacy}, A. 2008, in IAU Symposium, Vol. 255, IAU
  Symposium, ed. {L.~K.~Hunt, S.~Madden, \& R.~Schneider}, 3--17

\bibitem[{{Gofford} {et~al.}(2011){Gofford}, {Reeves}, {Turner}, {Tombesi},
  {Braito}, {Porquet}, {Miller}, {Kraemer}, \& {Fukazawa}}]{goff11xray}
{Gofford}, J., et~al.\ 2011,  \mnras, 414, 3307

\bibitem[{{Gonz{\'a}lez Delgado} {et~al.}(2001){Gonz{\'a}lez Delgado},
  {Heckman}, \& {Leitherer}}]{gonz01agnsf}
{Gonz{\'a}lez Delgado}, R.~M., {Heckman}, T., \& {Leitherer}, C. 2001, \apj,
  546, 845

\bibitem[{{Gonz{\'a}lez-Mart{\'{\i}}n}
  {et~al.}(2009){Gonz{\'a}lez-Mart{\'{\i}}n}, {Masegosa}, {M{\'a}rquez},
  {Guainazzi}, \& {Jim{\'e}nez-Bail{\'o}n}}]{gonz09linerx}
{Gonz{\'a}lez-Mart{\'{\i}}n}, O., {Masegosa}, J., {M{\'a}rquez}, I.,
  {Guainazzi}, M., \& {Jim{\'e}nez-Bail{\'o}n}, E. 2009, \aap, 506, 1107

\bibitem[{{Goodman}(2003)}]{good03disc}
{Goodman}, J. 2003, \mnras, 339, 937

\bibitem[{{Goulding} \& {Alexander}(2009)}]{goul09irs}
{Goulding}, A.~D. \& {Alexander}, D.~M. 2009, \mnras, 398, 1165

\bibitem[{{Goulding} {et~al.}(2010){Goulding}, {Alexander}, {Lehmer}, \&
  {Mullaney}}]{goul10census}
{Goulding}, A.~D., {Alexander}, D.~M., {Lehmer}, B.~D., \& {Mullaney}, J.~R.
  2010, \mnras, 406, 597

\bibitem[{{Granato} \& {Danese}(1994)}]{gran94torus}
{Granato}, G.~L. \& {Danese}, L. 1994, \mnras, 268, 235

\bibitem[{{Greene} \& {Ho}(2007)}]{gree07mfunc}
{Greene}, J.~E. \& {Ho}, L.~C. 2007, \apj, 667, 131

\bibitem[{{Greene} {et~al.}(2008){Greene}, {Ho}, \& {Barth}}]{gree08lowmass}
{Greene}, J.~E., {Ho}, L.~C., \& {Barth}, A.~J. 2008, \apj, 688, 159

\bibitem[{{Greene} {et~al.}(2010){Greene}, {Peng}, {Kim}, {Kuo}, {Braatz},
  {Violette Impellizzeri}, {Condon}, {Lo}, {Henkel}, \& {Reid}}]{gree10maser}
{Greene}, J.~E., et~al.\ 2010, \apj, 721, 26

\bibitem[{{Greene} {et~al.}(2011){Greene}, {Zakamska}, {Ho}, \&
  {Barth}}]{gree11obsqso}
{Greene}, J.~E., {Zakamska}, N.~L., {Ho}, L.~C., \& {Barth}, A.~J. 2011, \apj,
  732, 9

\bibitem[{{Greenhill} \& {Gwinn}(1997)}]{gree97ngc1068}
{Greenhill}, L.~J. \& {Gwinn}, C.~R. 1997, \apss, 248, 261

\bibitem[{{Greif} {et~al.}(2011){Greif}, {Springel}, {White}, {Glover},
  {Clark}, {Smith}, {Klessen}, \& {Bromm}}]{grei11pop3}
{Greif}, T.~H., {Springel}, V., {White}, S.~D.~M., {Glover}, S.~C.~O., {Clark},
  P.~C., {Smith}, R.~J., {Klessen}, R.~S., \& {Bromm}, V. 2011, \apj, 737, 75

\bibitem[{{Grogin} {et~al.}(2005){Grogin}, {Conselice}, {Chatzichristou},
  {Alexander}, {Bauer}, {Hornschemeier}, {Jogee}, {Koekemoer}, {Laidler},
  {Livio}, {Lucas}, {Paolillo}, {Ravindranath}, {Schreier}, {Simmons}, \&
  {Urry}}]{grog05xhost}
{Grogin}, N.~A., et~al.\ 2005, \apjl, 627, L97

\bibitem[{{Guainazzi} {et~al.}(2005){Guainazzi}, {Matt}, \& {Perola}}]{guai05}
{Guainazzi}, M., {Matt}, G., \& {Perola}, G.~C. 2005, \aap, 444, 119

\bibitem[{{G{\"u}ltekin} {et~al.}(2009){G{\"u}ltekin}, {Richstone}, {Gebhardt},
  {Lauer}, {Tremaine}, {Aller}, {Bender}, {Dressler}, {Faber}, {Filippenko},
  {Green}, {Ho}, {Kormendy}, {Magorrian}, {Pinkney}, \&
  {Siopis}}]{gult09msigma}
{G{\"u}ltekin}, K., et~al.\ 2009, \apj, 698, 198

\bibitem[{{Guyon} {et~al.}(2006){Guyon}, {Sanders}, \&
  {Stockton}}]{guyo06qsohost}
{Guyon}, O., {Sanders}, D.~B., \& {Stockton}, A. 2006, \apjs, 166, 89

\bibitem[{{Haiman} \& {Loeb}(2001)}]{haim01highzqso}
{Haiman}, Z. \& {Loeb}, A. 2001, \apj, 552, 459

\bibitem[{{Hao} {et~al.}(2010){Hao}, {Elvis}, {Civano}, {Lanzuisi}, {Brusa},
  {Lusso}, {Zamorani}, {Comastri}, {Bongiorno}, {Impey}, {Koekemoer}, {Le
  Floc'h}, {Salvato}, {Sanders}, {Trump}, \& {Vignali}}]{hao10hdpqso}
{Hao}, H., et~al.\ 2010, \apjl, 724, L59

\bibitem[{{Hao} {et~al.}(2011){Hao}, {Elvis}, {Civano}, \&
  {Lawrence}}]{hao11hdpqso}
{Hao}, H., {Elvis}, M., {Civano}, F., \& {Lawrence}, A. 2011, \apj, 733, 108

\bibitem[{{Hao} {et~al.}(2005){Hao}, {Strauss}, {Tremonti}, {Schlegel},
  {Heckman}, {Kauffmann}, {Blanton}, {Fan}, {Gunn}, {Hall}, {Ivezi{\'c}},
  {Knapp}, {Krolik}, {Lupton}, {Richards}, {Schneider}, {Strateva}, {Zakamska},
  {Brinkmann}, {Brunner}, \& {Szokoly}}]{hao05agn}
{Hao}, L., et~al.\ 2005, \aj, 129,  1783

\bibitem[{{Hardcastle} {et~al.}(2006){Hardcastle}, {Evans}, \&
  {Croston}}]{hard06radio}
{Hardcastle}, M.~J., {Evans}, D.~A., \& {Croston}, J.~H. 2006, \mnras, 370,
  1893

\bibitem[{{Hardcastle} {et~al.}(2007){Hardcastle}, {Evans}, \&
  {Croston}}]{hard07radio}
---. 2007, \mnras, 376, 1849

\bibitem[{{Hartwick} \& {Schade}(1990)}]{hart90qso}
{Hartwick}, F.~D.~A. \& {Schade}, D. 1990, \araa, 28, 437

\bibitem[{{Hasinger}(2008)}]{hasi08agn}
{Hasinger}, G. 2008, \aap, 490, 905

\bibitem[{{Hasinger} {et~al.}(2005){Hasinger}, {Miyaji}, \& {Schmidt}}]{hasi05}
{Hasinger}, G., {Miyaji}, T., \& {Schmidt}, M. 2005, \aap, 441, 417

\bibitem[{{Heckman}(1980)}]{heck80liner}
{Heckman}, T.~M. 1980, \aap, 87, 152

\bibitem[{{Heckman} {et~al.}(2004){Heckman}, {Kauffmann}, {Brinchmann},
  {Charlot}, {Tremonti}, \& {White}}]{heck04bh}
{Heckman}, T.~M., {Kauffmann}, G., {Brinchmann}, J., {Charlot}, S., {Tremonti},
  C., \& {White}, S.~D.~M. 2004, \apj, 613, 109

\bibitem[{{Heinz} {et~al.}(2007){Heinz}, {Merloni}, \& {Schwab}}]{hein07klf}
{Heinz}, S., {Merloni}, A., \& {Schwab}, J. 2007, \apjl, 658, L9

\bibitem[{{Herbert} {et~al.}(2010){Herbert}, {Jarvis}, {Willott}, {McLure},
  {Mitchell}, {Rawlings}, {Hill}, \& {Dunlop}}]{herb10radio}
{Herbert}, P.~D., {Jarvis}, M.~J., {Willott}, C.~J., {McLure}, R.~J.,
  {Mitchell}, E., {Rawlings}, S., {Hill}, G.~J., \& {Dunlop}, J.~S. 2010,
  \mnras, 406, 1841

\bibitem[{{Herbert} {et~al.}(2011){Herbert}, {Jarvis}, {Willott}, {McLure},
  {Mitchell}, {Rawlings}, {Hill}, \& {Dunlop}}]{herb11radio}
---. 2011, \mnras, 410, 1360

\bibitem[{{Hernquist} \& {Spergel}(1992)}]{hern92shell}
{Hernquist}, L. \& {Spergel}, D.~N. 1992, \apjl, 399, L117

\bibitem[{{Hickox} {et~al.}(2007){Hickox}, {Jones}, {Forman}, {Murray},
  {Brodwin}, {Brown}, {Eisenhardt}, {Stern}, {Kochanek}, {Eisenstein}, {Cool},
  {Jannuzi}, {Dey}, {Brand}, {Gorjian}, \& {Caldwell}}]{hick07abs}
{Hickox}, R.~C., et~al.\ 2007, \apj, 671, 1365

\bibitem[{{Hickox} {et~al.}(2009){Hickox}, {Jones}, {Forman}, {Murray},
  {Kochanek}, {Eisenstein}, {Jannuzi}, {Dey}, {Brown}, {Stern}, {Eisenhardt},
  {Gorjian}, {Brodwin}, {Narayan}, {Cool}, {Kenter}, {Caldwell}, \&
  {Anderson}}]{hick09corr}
{Hickox}, R.~C., et~al.\ 2009,  \apj, 696, 891

\bibitem[{{Hickox} {et~al.}(2011){Hickox}, {Myers}, {Brodwin}, {Alexander},
  {Forman}, {Jones}, {Murray}, {Brown}, {Cool}, {Kochanek}, {Dey}, {Jannuzi},
  {Eisenstein}, {Assef}, {Eisenhardt}, {Gorjian}, {Stern}, {Le Floc'h},
  {Caldwell}, {Goulding}, \& {Mullaney}}]{hick11qsoclust}
{Hickox}, R.~C., et~al.\ 2011, \apj, 731, 117

\bibitem[{{Hicks} {et~al.}(2009){Hicks}, {Davies}, {Malkan}, {Genzel},
  {Tacconi}, {M{\"u}ller S{\'a}nchez}, \& {Sternberg}}]{hick09agnmol}
{Hicks}, E.~K.~S., {Davies}, R.~I., {Malkan}, M.~A., {Genzel}, R., {Tacconi},
  L.~J., {M{\"u}ller S{\'a}nchez}, F., \& {Sternberg}, A. 2009, \apj, 696, 448

\bibitem[{{Ho}(2008)}]{ho08llagn}
{Ho}, L.~C. 2008, \araa, 46, 475

\bibitem[{{Ho} {et~al.}(2001){Ho}, {Feigelson}, {Townsley}, {Sambruna},
  {Garmire}, {Brandt}, {Filippenko}, {Griffiths}, {Ptak}, \&
  {Sargent}}]{ho01agnx}
{Ho}, L.~C., et~al.\ 2001, \apjl, 549, L51

\bibitem[{{Ho} {et~al.}(1997{\natexlab{a}}){Ho}, {Filippenko}, \&
  {Sargent}}]{ho97agnhost}
{Ho}, L.~C., {Filippenko}, A.~V., \& {Sargent}, W.~L.~W. 1997{\natexlab{a}},
  \apjs, 112, 315

\bibitem[{{Ho} {et~al.}(1997{\natexlab{b}}){Ho}, {Filippenko}, \&
  {Sargent}}]{ho97agn}
---. 1997{\natexlab{b}}, \apj, 487, 568

\bibitem[{{Ho} {et~al.}(2003){Ho}, {Filippenko}, \& {Sargent}}]{ho03line}
---. 2003, \apj, 583, 159

\bibitem[{{Ho} {et~al.}(1997{\natexlab{c}}){Ho}, {Filippenko}, {Sargent}, \&
  {Peng}}]{ho97agnbroad}
{Ho}, L.~C., {Filippenko}, A.~V., {Sargent}, W.~L.~W., \& {Peng}, C.~Y.
  1997{\natexlab{c}}, \apjs, 112, 391

\bibitem[{{Hobbs} {et~al.}(2011){Hobbs}, {Nayakshin}, {Power}, \&
  {King}}]{hobb11bhfeed}
{Hobbs}, A., {Nayakshin}, S., {Power}, C., \& {King}, A. 2011, \mnras, 413,
  2633

\bibitem[{{H{\"o}nig} {et~al.}(2006){H{\"o}nig}, {Beckert}, {Ohnaka}, \&
  {Weigelt}}]{honi06clumpy}
{H{\"o}nig}, S.~F., {Beckert}, T., {Ohnaka}, K., \& {Weigelt}, G. 2006, \aap,
  452, 459

\bibitem[{{H{\"o}nig} \& {Kishimoto}(2010)}]{honi10clumpy}
{H{\"o}nig}, S.~F. \& {Kishimoto}, M. 2010, \aap, 523, A27

\bibitem[{{Hopkins}(2011)}]{hopk11delay}
{Hopkins}, P.~F. 2011, \mnras\ in press (arXiv:1101.4230)

\bibitem[{{Hopkins} {et~al.}(2010){Hopkins}, {Bundy}, {Croton}, {Hernquist},
  {Keres}, {Khochfar}, {Stewart}, {Wetzel}, \& {Younger}}]{hopk10merge}
{Hopkins}, P.~F., et~al.\ 2010, \apj,  715, 202

\bibitem[{{Hopkins} \& {Hernquist}(2009)}]{hopk09fueling}
{Hopkins}, P.~F. \& {Hernquist}, L. 2009, \apj, 694, 599

\bibitem[{{Hopkins} {et~al.}(2006{\natexlab{a}}){Hopkins}, {Hernquist}, {Cox},
  {Di Matteo}, {Robertson}, \& {Springel}}]{hopk06apjs}
{Hopkins}, P.~F., {Hernquist}, L., {Cox}, T.~J., {Di Matteo}, T., {Robertson},
  B., \& {Springel}, V. 2006{\natexlab{a}}, \apjs, 163, 1

\bibitem[{{Hopkins} {et~al.}(2008){Hopkins}, {Hernquist}, {Cox}, \& {Kere{\v
  s}}}]{hopk08frame1}
{Hopkins}, P.~F., {Hernquist}, L., {Cox}, T.~J., \& {Kere{\v s}}, D. 2008,
  \apjs, 175, 356

\bibitem[{{Hopkins} {et~al.}(2009){Hopkins}, {Hickox}, {Quataert}, \&
  {Hernquist}}]{hopk09lowlum}
{Hopkins}, P.~F., {Hickox}, R., {Quataert}, E., \& {Hernquist}, L. 2009,
  \mnras, 398, 333

\bibitem[{{Hopkins} \& {Quataert}(2010)}]{hopk10bhgas}
{Hopkins}, P.~F. \& {Quataert}, E. 2010, \mnras, 407, 1529

\bibitem[{{Hopkins} {et~al.}(2007){Hopkins}, {Richards}, \&
  {Hernquist}}]{hopk07qlf}
{Hopkins}, P.~F., {Richards}, G.~T., \& {Hernquist}, L. 2007, \apj, 654, 731

\bibitem[{{Hopkins} {et~al.}(2006{\natexlab{b}}){Hopkins}, {Somerville},
  {Hernquist}, {Cox}, {Robertson}, \& {Li}}]{hopk06merge}
{Hopkins}, P.~F., {Somerville}, R.~S., {Hernquist}, L., {Cox}, T.~J.,
  {Robertson}, B., \& {Li}, Y. 2006{\natexlab{b}}, \apj, 652, 864

\bibitem[{{Horst} {et~al.}(2009){Horst}, {Duschl}, {Gandhi}, \&
  {Smette}}]{hors09seyfir}
{Horst}, H., {Duschl}, W.~J., {Gandhi}, P., \& {Smette}, A. 2009, \aap, 495,
  137

\bibitem[{{Horst} {et~al.}(2008){Horst}, {Gandhi}, {Smette}, \&
  {Duschl}}]{hors08irx}
{Horst}, H., {Gandhi}, P., {Smette}, A., \& {Duschl}, W.~J. 2008, \aap, 479,
  389

\bibitem[{{Huchra} \& {Burg}(1992)}]{huch92agn}
{Huchra}, J. \& {Burg}, R. 1992, \apj, 393, 90

\bibitem[{{Hunt} \& {Malkan}(2004)}]{hunt04seyfhst}
{Hunt}, L.~K. \& {Malkan}, M.~A. 2004, \apj, 616, 707

\bibitem[{{Ishida}(2004)}]{ishi04ulirg}
{Ishida}, C.~M. 2004, PhD thesis, University of Hawai'i

\bibitem[{{Iwasawa} {et~al.}(2005){Iwasawa}, {Sanders}, {Evans}, {Trentham},
  {Miniutti}, \& {Spoon}}]{iwas05arp220}
{Iwasawa}, K., {Sanders}, D.~B., {Evans}, A.~S., {Trentham}, N., {Miniutti},
  G., \& {Spoon}, H.~W.~W. 2005, \mnras, 357, 565

\bibitem[{{Iwasawa} {et~al.}(2011){Iwasawa}, {Sanders}, {Teng}, {U}, {Armus},
  {Evans}, {Howell}, {Komossa}, {Mazzarella}, {Petric}, {Surace}, {Vavilkin},
  {Veilleux}, \& {Trentham}}]{iwas11lirgx}
{Iwasawa}, K., et~al.\ 2011,  \aap, 529, A106+

\bibitem[{{Jahnke} {et~al.}(2009){Jahnke}, {Bongiorno}, {Brusa}, {Capak},
  {Cappelluti}, {Cisternas}, {Civano}, {Colbert}, {Comastri}, {Elvis},
  {Hasinger}, {Ilbert}, {Impey}, {Inskip}, {Koekemoer}, {Lilly}, {Maier},
  {Merloni}, {Riechers}, {Salvato}, {Schinnerer}, {Scoville}, {Silverman},
  {Taniguchi}, {Trump}, \& {Yan}}]{jahn09bhgal}
{Jahnke}, K., et~al.\ 2009, \apjl, 706, L215

\bibitem[{{Jahnke} \& {Macci{\`o}}(2011)}]{jahn11mbh}
{Jahnke}, K. \& {Macci{\`o}}, A.~V. 2011, \apj, 734, 92

\bibitem[{{Jiang} {et~al.}(2009){Jiang}, {Fan}, {Bian}, {Annis}, {Chiu},
  {Jester}, {Lin}, {Lupton}, {Richards}, {Strauss}, {Malanushenko},
  {Malanushenko}, \& {Schneider}}]{jian09z6qso}
{Jiang}, L., et~al.\ 2009, \aj, 138,  305

\bibitem[{{Jiang} {et~al.}(2010){Jiang}, {Fan}, {Brandt}, {Carilli}, {Egami},
  {Hines}, {Kurk}, {Richards}, {Shen}, {Strauss}, {Vestergaard}, \&
  {Walter}}]{jian10hdpqso}
{Jiang}, L., et~al.\ 2010, \nat, 464, 380

\bibitem[{{Jiang} {et~al.}(2011){Jiang}, {Greene}, {Ho}, {Xiao}, \&
  {Barth}}]{jian11lowmass}
{Jiang}, Y.-F., {Greene}, J.~E., {Ho}, L.~C., {Xiao}, T., \& {Barth}, A.~J.
  2011, \apj, 742, 68

\bibitem[{{Jin} {et~al.}(2009){Jin}, {Done}, {Ward}, {Gierli{\'n}ski}, \&
  {Mullaney}}]{jin09super}
{Jin}, C., {Done}, C., {Ward}, M., {Gierli{\'n}ski}, M., \& {Mullaney}, J.
  2009, \mnras, 398, L16

\bibitem[{{Jogee}(2006)}]{joge06agn}
{Jogee}, S. 2006, in Lecture Notes in Physics, Berlin Springer Verlag, Vol.
  693, Physics of Active Galactic Nuclei at all Scales, ed. {D.~Alloin}, 143--+

\bibitem[{{Johnson} \& {Bromm}(2007)}]{john07firstbh}
{Johnson}, J.~L. \& {Bromm}, V. 2007, \mnras, 374, 1557

\bibitem[{{Johnson} {et~al.}(2011){Johnson}, {Khochfar}, {Greif}, \&
  {Durier}}]{john11bhcollapse}
{Johnson}, J.~L., {Khochfar}, S., {Greif}, T.~H., \& {Durier}, F. 2011, \mnras,
  410, 919

\bibitem[{{Juneau} {et~al.}(2011){Juneau}, {Dickinson}, {Alexander}, \&
  {Salim}}]{june11agn}
{Juneau}, S., {Dickinson}, M., {Alexander}, D.~M., \& {Salim}, S. 2011, \apj,
  736, 104

\bibitem[{{Juneau} {et~al.}(2005){Juneau}, {Glazebrook}, {Crampton},
  {McCarthy}, {Savaglio}, {Abraham}, {Carlberg}, {Chen}, {Le Borgne}, {Marzke},
  {Roth}, {J{\o}rgensen}, {Hook}, \& {Murowinski}}]{june05sfevol}
{Juneau}, S., et~al.\ 2005, \apjl, 619, L135

\bibitem[{{Jungwiert} {et~al.}(2001){Jungwiert}, {Combes}, \& {Palou{\v
  s}}}]{jung01loss}
{Jungwiert}, B., {Combes}, F., \& {Palou{\v s}}, J. 2001, \aap, 376, 85

\bibitem[{{Kaspi} {et~al.}(2005){Kaspi}, {Maoz}, {Netzer}, {Peterson},
  {Vestergaard}, \& {Jannuzi}}]{kasp05}
{Kaspi}, S., {Maoz}, D., {Netzer}, H., {Peterson}, B.~M., {Vestergaard}, M., \&
  {Jannuzi}, B.~T. 2005, \apj, 629, 61

\bibitem[{{Kauffmann}(1996)}]{kauf96ell}
{Kauffmann}, G. 1996, \mnras, 281, 487

\bibitem[{{Kauffmann} \& {Haehnelt}(2000)}]{kauf00merge}
{Kauffmann}, G. \& {Haehnelt}, M. 2000, \mnras, 311, 576

\bibitem[{{Kauffmann} \& {Heckman}(2009)}]{kauf09modes}
{Kauffmann}, G. \& {Heckman}, T.~M. 2009, \mnras, 397, 135

\bibitem[{{Kauffmann} {et~al.}(2008){Kauffmann}, {Heckman}, \&
  {Best}}]{kauf08radio}
{Kauffmann}, G., {Heckman}, T.~M., \& {Best}, P.~N. 2008, \mnras, 384, 953

\bibitem[{{Kauffmann} {et~al.}(2003){Kauffmann}, {Heckman}, {Tremonti},
  {Brinchmann}, {Charlot}, {White}, {Ridgway}, {Brinkmann}, {Fukugita}, {Hall},
  {Ivezi{\'c}}, {Richards}, \& {Schneider}}]{kauf03host}
{Kauffmann}, G., et~al.\ 2003, \mnras, 346, 1055

\bibitem[{{Kauffmann} {et~al.}(2004){Kauffmann}, {White}, {Heckman},
  {M{\'e}nard}, {Brinchmann}, {Charlot}, {Tremonti}, \& {Brinkmann}}]{kauf04}
{Kauffmann}, G., {White}, S.~D.~M., {Heckman}, T.~M., {M{\'e}nard}, B.,
  {Brinchmann}, J., {Charlot}, S., {Tremonti}, C., \& {Brinkmann}, J. 2004,
  \mnras, 353, 713

\bibitem[{{Kawakatu} {et~al.}(2007){Kawakatu}, {Imanishi}, \&
  {Nagao}}]{kawa07bhmass}
{Kawakatu}, N., {Imanishi}, M., \& {Nagao}, T. 2007, \apj, 661, 660

\bibitem[{{Kawakatu} \& {Wada}(2008)}]{kawa08disk}
{Kawakatu}, N. \& {Wada}, K. 2008, \apj, 681, 73

\bibitem[{{Kelly} {et~al.}(2008){Kelly}, {Bechtold}, {Trump}, {Vestergaard}, \&
  {Siemiginowska}}]{kell08qsox}
{Kelly}, B.~C., {Bechtold}, J., {Trump}, J.~R., {Vestergaard}, M., \&
  {Siemiginowska}, A. 2008, \apjs, 176, 355

\bibitem[{{Kelly} {et~al.}(2010){Kelly}, {Vestergaard}, {Fan}, {Hopkins},
  {Hernquist}, \& {Siemiginowska}}]{kell10qsoedd}
{Kelly}, B.~C., {Vestergaard}, M., {Fan}, X., {Hopkins}, P., {Hernquist}, L.,
  \& {Siemiginowska}, A. 2010, \apj, 719, 1315

\bibitem[{{Kennicutt}(1989)}]{kenn89sflaw}
{Kennicutt}, Jr., R.~C. 1989, \apj, 344, 685

\bibitem[{{Kere{\v s}} {et~al.}(2009){Kere{\v s}}, {Katz}, {Fardal},
  {Dav{\'e}}, \& {Weinberg}}]{kere09shock}
{Kere{\v s}}, D., {Katz}, N., {Fardal}, M., {Dav{\'e}}, R., \& {Weinberg},
  D.~H. 2009, \mnras, 395, 160

\bibitem[{{Kerr}(1963)}]{kerr63}
{Kerr}, R.~P. 1963, Physical Review Letters, 11, 237

\bibitem[{{Kewley} {et~al.}(2001){Kewley}, {Dopita}, {Sutherland}, {Heisler},
  \& {Trevena}}]{kewl01opt}
{Kewley}, L.~J., {Dopita}, M.~A., {Sutherland}, R.~S., {Heisler}, C.~A., \&
  {Trevena}, J. 2001, \apj, 556, 121

\bibitem[{{Kewley} {et~al.}(2006){Kewley}, {Groves}, {Kauffmann}, \&
  {Heckman}}]{kewl06agn}
{Kewley}, L.~J., {Groves}, B., {Kauffmann}, G., \& {Heckman}, T. 2006, \mnras,
  372, 961

\bibitem[{{Kim} {et~al.}(1998){Kim}, {Veilleux}, \& {Sanders}}]{kim98ulirg}
{Kim}, D.-C., {Veilleux}, S., \& {Sanders}, D.~B. 1998, \apj, 508, 627

\bibitem[{{Kim} {et~al.}(2008){Kim}, {Ho}, {Peng}, {Barth}, {Im}, {Martini}, \&
  {Nelson}}]{kim08msigma}
{Kim}, M., {Ho}, L.~C., {Peng}, C.~Y., {Barth}, A.~J., {Im}, M., {Martini}, P.,
  \& {Nelson}, C.~H. 2008, \apj, 687, 767

\bibitem[{{King}(2003)}]{king03msigma}
{King}, A. 2003, \apjl, 596, L27

\bibitem[{{King}(2005)}]{king05}
---. 2005, \apjl, 635, L121

\bibitem[{{King}(2010{\natexlab{a}})}]{king10bhedd}
{King}, A.~R. 2010{\natexlab{a}}, \mnras, 408, L95

\bibitem[{{King}(2010{\natexlab{b}})}]{king10outflow}
---. 2010{\natexlab{b}}, \mnras, 402, 1516

\bibitem[{{King} {et~al.}(2001){King}, {Davies}, {Ward}, {Fabbiano}, \&
  {Elvis}}]{king01ulx}
{King}, A.~R., {Davies}, M.~B., {Ward}, M.~J., {Fabbiano}, G., \& {Elvis}, M.
  2001, \apjl, 552, L109

\bibitem[{{King} \& {Pringle}(2007)}]{king07agnfuel}
{King}, A.~R. \& {Pringle}, J.~E. 2007, \mnras, 377, L25

\bibitem[{{King} {et~al.}(2011){King}, {Zubovas}, \& {Power}}]{king11outflow}
{King}, A.~R., {Zubovas}, K., \& {Power}, C. 2011, \mnras, L263+

\bibitem[{{Kishimoto} {et~al.}(2011){Kishimoto}, {H{\"o}nig}, {Antonucci},
  {Barvainis}, {Kotani}, {Tristram}, {Weigelt}, \& {Levin}}]{kish11torus}
{Kishimoto}, M., {H{\"o}nig}, S.~F., {Antonucci}, R., {Barvainis}, R.,
  {Kotani}, T., {Tristram}, K.~R.~W., {Weigelt}, G., \& {Levin}, K. 2011, \aap,
  527, A121+

\bibitem[{{Knapen} {et~al.}(2000){Knapen}, {Shlosman}, \&
  {Peletier}}]{knap00agn}
{Knapen}, J.~H., {Shlosman}, I., \& {Peletier}, R.~F. 2000, \apj, 529, 93

\bibitem[{{Kocevski} {et~al.}(2011){Kocevski}, {Faber}, {Mozena}, {Koekemoer},
  {Nandra}, {Rangel}, {Laird}, {Brusa}, {Wuyts}, {Trump}, {Koo}, {Somerville},
  {Bell}, {Lotz}, {Alexander}, {Bournaud}, {Conselice}, {Dahlen}, {Dekel},
  {Donley}, {Dunlop}, {Finoguenov}, {Georgakakis}, {Giavalisco}, {Guo},
  {Grogin}, {Hathi}, {Juneau}, {Kartaltepe}, {Lucas}, {McGrath}, {McIntosh},
  {Mobasher}, {Robaina}, {Rosario}, {Straughn}, {van der Wel}, \&
  {Villforth}}]{koce11xmorph_aph}
{Kocevski}, D.~D., et~al.\ 2011, \apj\ submitted (arXiv:1109.2588)

\bibitem[{{Kocevski} {et~al.}(2009){Kocevski}, {Lubin}, {Gal}, {Lemaux},
  {Fassnacht}, \& {Squires}}]{koce09xsuperclust}
{Kocevski}, D.~D., {Lubin}, L.~M., {Gal}, R., {Lemaux}, B.~C., {Fassnacht},
  C.~D., \& {Squires}, G.~K. 2009, \apj, 690, 295

\bibitem[{{Kollmeier} {et~al.}(2006){Kollmeier}, {Onken}, {Kochanek}, {Gould},
  {Weinberg}, {Dietrich}, {Cool}, {Dey}, {Eisenstein}, {Jannuzi}, {Le Floc'h},
  \& {Stern}}]{koll06}
{Kollmeier}, J.~A., et~al.\ 2006, \apj, 648, 128

\bibitem[{{Komossa} {et~al.}(2008){Komossa}, {Zhou}, \& {Lu}}]{komo08recoil}
{Komossa}, S., {Zhou}, H., \& {Lu}, H. 2008, \apjl, 678, L81

\bibitem[{{K{\"o}rding} {et~al.}(2008){K{\"o}rding}, {Jester}, \&
  {Fender}}]{kord08klf}
{K{\"o}rding}, E.~G., {Jester}, S., \& {Fender}, R. 2008, \mnras, 383, 277

\bibitem[{{Kormendy} \& {Bender}(2011)}]{korm11halo}
{Kormendy}, J. \& {Bender}, R. 2011, \nat, 469, 377

\bibitem[{{Kormendy} {et~al.}(2011){Kormendy}, {Bender}, \&
  {Cornell}}]{korm11pseudo}
{Kormendy}, J., {Bender}, R., \& {Cornell}, M.~E. 2011, \nat, 469, 374

\bibitem[{{Kormendy} \& {Kennicutt}(2004)}]{korm04pseudo}
{Kormendy}, J. \& {Kennicutt}, Jr., R.~C. 2004, \araa, 42, 603

\bibitem[{{Kormendy} \& {Richstone}(1995)}]{korm95}
{Kormendy}, J. \& {Richstone}, D. 1995, \araa, 33, 581

\bibitem[{{Koss} {et~al.}(2010){Koss}, {Mushotzky}, {Veilleux}, \&
  {Winter}}]{koss10batagn}
{Koss}, M., {Mushotzky}, R., {Veilleux}, S., \& {Winter}, L. 2010, \apjl, 716,
  L125

\bibitem[{{Koss} {et~al.}(2011){Koss}, {Mushotzky}, {Veilleux}, {Winter},
  {Baumgartner}, {Tueller}, {Gehrels}, \& {Valencic}}]{koss11bathost}
{Koss}, M., {Mushotzky}, R., {Veilleux}, S., {Winter}, L.~M., {Baumgartner},
  W., {Tueller}, J., {Gehrels}, N., \& {Valencic}, L. 2011, \apj, 739, 57

\bibitem[{{Kotilainen} {et~al.}(2009){Kotilainen}, {Falomo}, {Decarli},
  {Treves}, {Uslenghi}, \& {Scarpa}}]{koti09qsohost}
{Kotilainen}, J.~K., {Falomo}, R., {Decarli}, R., {Treves}, A., {Uslenghi}, M.,
  \& {Scarpa}, R. 2009, \apj, 703, 1663

\bibitem[{{Kotilainen} {et~al.}(2007){Kotilainen}, {Falomo}, {Labita},
  {Treves}, \& {Uslenghi}}]{koti07qsohost}
{Kotilainen}, J.~K., {Falomo}, R., {Labita}, M., {Treves}, A., \& {Uslenghi},
  M. 2007, \apj, 660, 1039

\bibitem[{{Krumpe} {et~al.}(2010){Krumpe}, {Miyaji}, \& {Coil}}]{krum10xclust}
{Krumpe}, M., {Miyaji}, T., \& {Coil}, A.~L. 2010, \apj, 713, 558

\bibitem[{{Kuo} {et~al.}(2008){Kuo}, {Lim}, {Tang}, \& {Ho}}]{kuo08tidal}
{Kuo}, C.-Y., {Lim}, J., {Tang}, Y.-W., \& {Ho}, P.~T.~P. 2008, \apj, 679, 1047

\bibitem[{{Kuraszkiewicz} {et~al.}(2009){Kuraszkiewicz}, {Wilkes}, {Schmidt},
  {Smith}, {Cutri}, \& {Czerny}}]{kura09redagn}
{Kuraszkiewicz}, J., {Wilkes}, B.~J., {Schmidt}, G., {Smith}, P.~S., {Cutri},
  R., \& {Czerny}, B. 2009, \apj, 692, 1180

\bibitem[{{La Franca} {et~al.}(2005){La Franca}, {Fiore}, {Comastri}, {Perola},
  {Sacchi}, {Brusa}, {Cocchia}, {Feruglio}, {Matt}, {Vignali}, {Carangelo},
  {Ciliegi}, {Lamastra}, {Maiolino}, {Mignoli}, {Molendi}, \&
  {Puccetti}}]{lafa05obsc}
{La Franca}, F., et~al.\ 2005, \apj, 635, 864

\bibitem[{{La Franca} {et~al.}(2010){La Franca}, {Melini}, \&
  {Fiore}}]{lafa10radio}
{La Franca}, F., {Melini}, G., \& {Fiore}, F. 2010, \apj, 718, 368

\bibitem[{{Lacy} {et~al.}(2004){Lacy}, {Storrie-Lombardi}, {Sajina},
  {Appleton}, {Armus}, {Chapman}, {Choi}, {Fadda}, {Fang}, {Frayer},
  {Heinrichsen}, {Helou}, {Im}, {Marleau}, {Masci}, {Shupe}, {Soifer},
  {Surace}, {Teplitz}, {Wilson}, \& {Yan}}]{lacy04}
{Lacy}, M., et~al.\ 2004, \apjs, 154, 166

\bibitem[{{Laine} {et~al.}(2002){Laine}, {Shlosman}, {Knapen}, \&
  {Peletier}}]{lain02bars}
{Laine}, S., {Shlosman}, I., {Knapen}, J.~H., \& {Peletier}, R.~F. 2002, \apj,
  567, 97

\bibitem[{{Laird} {et~al.}(2010){Laird}, {Nandra}, {Pope}, \&
  {Scott}}]{lair10submmx}
{Laird}, E.~S., {Nandra}, K., {Pope}, A., \& {Scott}, D. 2010, \mnras, 401,
  2763

\bibitem[{{Lauer} {et~al.}(2007){Lauer}, {Tremaine}, {Richstone}, \&
  {Faber}}]{laue07mbhbias}
{Lauer}, T.~R., {Tremaine}, S., {Richstone}, D., \& {Faber}, S.~M. 2007, \apj,
  670, 249

\bibitem[{{Lawrence}(1991)}]{lawr91unified}
{Lawrence}, A. 1991, \mnras, 252, 586

\bibitem[{{Lawrence} \& {Elvis}(2010)}]{lawr10disk}
{Lawrence}, A. \& {Elvis}, M. 2010, \apj, 714, 561

\bibitem[{{Le Floc'h} {et~al.}(2005){Le Floc'h}, {Papovich}, {Dole}, {Bell},
  {Lagache}, {Rieke}, {Egami}, {P{\'e}rez-Gonz{\'a}lez}, {Alonso-Herrero},
  {Rieke}, {Blaylock}, {Engelbracht}, {Gordon}, {Hines}, {Misselt}, {Morrison},
  \& {Mould}}]{lefl05irlf}
{Le Floc'h}, E., et~al.\ 2005, \apj, 632, 169

\bibitem[{{Lehmer} {et~al.}(2010){Lehmer}, {Alexander}, {Bauer}, {Brandt},
  {Goulding}, {Jenkins}, {Ptak}, \& {Roberts}}]{lehm10sfx}
{Lehmer}, B.~D., {Alexander}, D.~M., {Bauer}, F.~E., {Brandt}, W.~N.,
  {Goulding}, A.~D., {Jenkins}, L.~P., {Ptak}, A., \& {Roberts}, T.~P. 2010,
  \apj, 724, 559

\bibitem[{{Lehmer} {et~al.}(2009{\natexlab{a}}){Lehmer}, {Alexander},
  {Chapman}, {Smail}, {Bauer}, {Brandt}, {Geach}, {Matsuda}, {Mullaney}, \&
  {Swinbank}}]{lehm09catalog}
{Lehmer}, B.~D., et~al.\ 2009{\natexlab{a}}, \mnras, 400, 299

\bibitem[{{Lehmer} {et~al.}(2009{\natexlab{b}}){Lehmer}, {Alexander}, {Geach},
  {Smail}, {Basu-Zych}, {Bauer}, {Chapman}, {Matsuda}, {Scharf}, {Volonteri},
  \& {Yamada}}]{lehm09proto}
{Lehmer}, B.~D., et~al.\ 2009{\natexlab{b}}, \apj, 691, 687

\bibitem[{{Lehmer} {et~al.}(2007){Lehmer}, {Brandt}, {Alexander}, {Bell},
  {McIntosh}, {Bauer}, {Hasinger}, {Mainieri}, {Miyaji}, {Schneider}, \&
  {Steffen}}]{lehm07xevol}
{Lehmer}, B.~D., et~al.\ 2007, \apj, 657, 681

\bibitem[{{Li} {et~al.}(2006){Li}, {Kauffmann}, {Wang}, {White}, {Heckman}, \&
  {Jing}}]{li06agnclust}
{Li}, C., {Kauffmann}, G., {Wang}, L., {White}, S.~D.~M., {Heckman}, T.~M., \&
  {Jing}, Y.~P. 2006, \mnras, 373, 457

\bibitem[{{Lietzen} {et~al.}(2009){Lietzen}, {Hein{\"a}m{\"a}ki}, {Nurmi},
  {Tago}, {Saar}, {Liivam{\"a}gi}, {Tempel}, {Einasto}, {Einasto}, {Gramann},
  \& {Takalo}}]{liet09qsoenviron}
{Lietzen}, H., et~al.\ 2009, \aap, 501, 145

\bibitem[{{Lintott} {et~al.}(2011){Lintott}, {Schawinski}, {Bamford}, {Slosar},
  {Land}, {Thomas}, {Edmondson}, {Masters}, {Nichol}, {Raddick}, {Szalay},
  {Andreescu}, {Murray}, \& {Vandenberg}}]{lint11zoo}
{Lintott}, C., et~al.\ 2011,  \mnras, 410, 166

\bibitem[{{Liu} \& {Melia}(2002)}]{liu02sgra}
{Liu}, S. \& {Melia}, F. 2002, \apjl, 566, L77

\bibitem[{{Liu} {et~al.}(2011){Liu}, {Shen}, {Strauss}, \&
  {Hao}}]{liu11agnpair}
{Liu}, X., {Shen}, Y., {Strauss}, M.~A., \& {Hao}, L. 2011, \apj, 737, 101

\bibitem[{{Lobban} {et~al.}(2011){Lobban}, {Reeves}, {Miller}, {Turner},
  {Braito}, {Kraemer}, \& {Crenshaw}}]{lobb11ngc4051}
{Lobban}, A.~P., {Reeves}, J.~N., {Miller}, L., {Turner}, T.~J., {Braito}, V.,
  {Kraemer}, S.~B., \& {Crenshaw}, D.~M. 2011, \mnras, 544

\bibitem[{{Luo} {et~al.}(2008){Luo}, {Bauer}, {Brandt}, {Alexander}, {Lehmer},
  {Schneider}, {Brusa}, {Comastri}, {Fabian}, {Finoguenov}, {Gilli},
  {Hasinger}, {Hornschemeier}, {Koekemoer}, {Mainieri}, {Paolillo}, {Rosati},
  {Shemmer}, {Silverman}, {Smail}, {Steffen}, \& {Vignali}}]{luo08cdfs}
{Luo}, B., et~al.\ 2008,  \apjs, 179, 19

\bibitem[{{Luo} {et~al.}(2011){Luo}, {Brandt}, {Xue}, {Alexander}, {Brusa},
  {Bauer}, {Comastri}, {Fabian}, {Gilli}, {Lehmer}, {Rafferty}, {Schneider}, \&
  {Vignali}}]{luo11obsc}
{Luo}, B., et~al.\ 2011, \apj,  740, 37

\bibitem[{{Luo} {et~al.}(2010){Luo}, {Brandt}, {Xue}, {Brusa}, {Alexander},
  {Bauer}, {Comastri}, {Koekemoer}, {Lehmer}, {Mainieri}, {Rafferty},
  {Schneider}, {Silverman}, \& {Vignali}}]{luo10cdfs}
{Luo}, B., et~al.\ 2010, \apjs, 187, 560

\bibitem[{{Lutz} {et~al.}(2010){Lutz}, {Mainieri}, {Rafferty}, {Shao},
  {Hasinger}, {Wei{\ss}}, {Walter}, {Smail}, {Alexander}, {Brandt}, {Chapman},
  {Coppin}, {F{\"o}rster Schreiber}, {Gawiser}, {Genzel}, {Greve}, {Ivison},
  {Koekemoer}, {Kurczynski}, {Menten}, {Nordon}, {Popesso}, {Schinnerer},
  {Silverman}, {Wardlow}, \& {Xue}}]{lutz10agnsf}
{Lutz}, D., et~al.\ 2010,  \apj, 712, 1287

\bibitem[{{Lutz} {et~al.}(2004){Lutz}, {Maiolino}, {Spoon}, \&
  {Moorwood}}]{lutz04irx}
{Lutz}, D., {Maiolino}, R., {Spoon}, H.~W.~W., \& {Moorwood}, A.~F.~M. 2004,
  \aap, 418, 465

\bibitem[{{Lutz} {et~al.}(2008){Lutz}, {Sturm}, {Tacconi}, {Valiante},
  {Schweitzer}, {Netzer}, {Maiolino}, {Andreani}, {Shemmer}, \&
  {Veilleux}}]{lutz08qsosf}
{Lutz}, D., et~al.\ 2008, \apj, 684, 853

\bibitem[{{Lynden-Bell}(1969)}]{lyn69}
{Lynden-Bell}, D. 1969, \nat, 223, 690

\bibitem[{{Maccarone}(2003)}]{macc03bhb}
{Maccarone}, T.~J. 2003, \aap, 409, 697

\bibitem[{{Maciejewski}(2004)}]{maci04spiral}
{Maciejewski}, W. 2004, \mnras, 354, 892

\bibitem[{{Madau} \& {Rees}(2001)}]{mada01bhpop3}
{Madau}, P. \& {Rees}, M.~J. 2001, \apjl, 551, L27

\bibitem[{{Madau} {et~al.}(2004){Madau}, {Rees}, {Volonteri}, {Haardt}, \&
  {Oh}}]{mada04bhreion}
{Madau}, P., {Rees}, M.~J., {Volonteri}, M., {Haardt}, F., \& {Oh}, S.~P. 2004,
  \apj, 604, 484

\bibitem[{{Magorrian} {et~al.}(1998){Magorrian}, {Tremaine}, {Richstone},
  {Bender}, {Bower}, {Dressler}, {Faber}, {Gebhardt}, {Green}, {Grillmair},
  {Kormendy}, \& {Lauer}}]{mago98}
{Magorrian}, J., et~al.\ 1998, \aj, 115, 2285

\bibitem[{{Mainieri} {et~al.}(2011){Mainieri}, {Bongiorno}, {Merloni}, {Aller},
  {Carollo}, {Iwasawa}, {Koekemoer}, {Mignoli}, {Silverman}, {Bolzonella},
  {Brusa}, {Comastri}, {Gilli}, {Halliday}, {Ilbert}, {Lusso}, {Salvato},
  {Vignali}, {Zamorani}, {Contini}, {Kneib}, {Le F{\`e}vre}, {Lilly},
  {Renzini}, {Scodeggio}, {Balestra}, {Bardelli}, {Caputi}, {Coppa},
  {Cucciati}, {de la Torre}, {de Ravel}, {Franzetti}, {Garilli}, {Iovino},
  {Kampczyk}, {Knobel}, {Kova{\v c}}, {Lamareille}, {Le Borgne}, {Le Brun},
  {Maier}, {Nair}, {Pello}, {Peng}, {Perez Montero}, {Pozzetti},
  {Ricciardelli}, {Tanaka}, {Tasca}, {Tresse}, {Vergani}, {Zucca}, {Aussel},
  {Capak}, {Cappelluti}, {Elvis}, {Fiore}, {Hasinger}, {Impey}, {Le Floc'h},
  {Scoville}, {Taniguchi}, \& {Trump}}]{main11obsqso}
{Mainieri}, V., et~al.\ 2011, \aap, 535, A80

\bibitem[{{Mainieri} {et~al.}(2005{\natexlab{a}}){Mainieri}, {Rigopoulou},
  {Lehmann}, {Scott}, {Matute}, {Almaini}, {Tozzi}, {Hasinger}, \&
  {Dunlop}}]{main05qso2submm}
{Mainieri}, V., et~al.\ 2005{\natexlab{a}}, \mnras, 356, 1571

\bibitem[{{Mainieri} {et~al.}(2005{\natexlab{b}}){Mainieri}, {Rosati}, {Tozzi},
  {Bergeron}, {Gilli}, {Hasinger}, {Nonino}, {Lehmann}, {Alexander}, {Idzi},
  {Koekemoer}, {Norman}, {Szokoly}, \& {Zheng}}]{main05xfaint}
{Mainieri}, V., et~al.\ 2005{\natexlab{b}}, \aap, 437, 805

\bibitem[{{Maiolino} \& {Rieke}(1995)}]{maio95}
{Maiolino}, R. \& {Rieke}, G.~H. 1995, \apj, 454, 95

\bibitem[{{Malbon} {et~al.}(2007){Malbon}, {Baugh}, {Frenk}, \&
  {Lacey}}]{malb07smbh}
{Malbon}, R.~K., {Baugh}, C.~M., {Frenk}, C.~S., \& {Lacey}, C.~G. 2007,
  \mnras, 382, 1394

\bibitem[{{Malin} \& {Carter}(1983)}]{mali83shell}
{Malin}, D.~F. \& {Carter}, D. 1983, \apj, 274, 534

\bibitem[{{Malkan} {et~al.}(1998){Malkan}, {Gorjian}, \& {Tam}}]{malk98agn}
{Malkan}, M.~A., {Gorjian}, V., \& {Tam}, R. 1998, \apjs, 117, 25

\bibitem[{{Mandelbaum} {et~al.}(2009){Mandelbaum}, {Li}, {Kauffmann}, \&
  {White}}]{mand09agnclust}
{Mandelbaum}, R., {Li}, C., {Kauffmann}, G., \& {White}, S.~D.~M. 2009, \mnras,
  393, 377

\bibitem[{{Marconi} {et~al.}(2008){Marconi}, {Axon}, {Maiolino}, {Nagao},
  {Pastorini}, {Pietrini}, {Robinson}, \& {Torricelli}}]{marc08radpress}
{Marconi}, A., {Axon}, D.~J., {Maiolino}, R., {Nagao}, T., {Pastorini}, G.,
  {Pietrini}, P., {Robinson}, A., \& {Torricelli}, G. 2008, \apj, 678, 693

\bibitem[{{Marconi} \& {Hunt}(2003)}]{marc03}
{Marconi}, A. \& {Hunt}, L.~K. 2003, \apjl, 589, L21

\bibitem[{{Marconi} {et~al.}(2004){Marconi}, {Risaliti}, {Gilli}, {Hunt},
  {Maiolino}, \& {Salvati}}]{marc04smbh}
{Marconi}, A., {Risaliti}, G., {Gilli}, R., {Hunt}, L.~K., {Maiolino}, R., \&
  {Salvati}, M. 2004, \mnras, 351, 169

\bibitem[{{Markoff} {et~al.}(2001){Markoff}, {Falcke}, {Yuan}, \&
  {Biermann}}]{mark01flare}
{Markoff}, S., {Falcke}, H., {Yuan}, F., \& {Biermann}, P.~L. 2001, \aap, 379,
  L13

\bibitem[{{Marrone} {et~al.}(2008){Marrone}, {Baganoff}, {Morris}, {Moran},
  {Ghez}, {Hornstein}, {Dowell}, {Mu{\~n}oz}, {Bautz}, {Ricker}, {Brandt},
  {Garmire}, {Lu}, {Matthews}, {Zhao}, {Rao}, \& {Bower}}]{marr08sgra}
{Marrone}, D.~P., et~al.\ 2008, \apj, 682,  373

\bibitem[{{Martini} {et~al.}(2006){Martini}, {Kelson}, {Kim}, {Mulchaey}, \&
  {Athey}}]{mart06clustagn}
{Martini}, P., {Kelson}, D.~D., {Kim}, E., {Mulchaey}, J.~S., \& {Athey}, A.~A.
  2006, \apj, 644, 116

\bibitem[{{Martini} {et~al.}(2003){Martini}, {Regan}, {Mulchaey}, \&
  {Pogge}}]{mart03dust}
{Martini}, P., {Regan}, M.~W., {Mulchaey}, J.~S., \& {Pogge}, R.~W. 2003, \apj,
  589, 774

\bibitem[{{Martini} {et~al.}(2009){Martini}, {Sivakoff}, \&
  {Mulchaey}}]{mart09clustagn}
{Martini}, P., {Sivakoff}, G.~R., \& {Mulchaey}, J.~S. 2009, \apj, 701, 66

\bibitem[{{Mathur} {et~al.}(2011){Mathur}, {Fields}, {Peterson}, \&
  {Grupe}}]{math11nls1_aph}
{Mathur}, S., {Fields}, D., {Peterson}, B.~M., \& {Grupe}, D. 2011, \apj\
  submitted (1102.0537)

\bibitem[{{McCarthy} {et~al.}(2011){McCarthy}, {Schaye}, {Bower}, {Ponman},
  {Booth}, {Dalla Vecchia}, \& {Springel}}]{mcca11feedback}
{McCarthy}, I.~G., {Schaye}, J., {Bower}, R.~G., {Ponman}, T.~J., {Booth},
  C.~M., {Dalla Vecchia}, C., \& {Springel}, V. 2011, \mnras, 412, 1965

\bibitem[{{McCarthy} {et~al.}(2010){McCarthy}, {Schaye}, {Ponman}, {Bower},
  {Booth}, {Dalla Vecchia}, {Crain}, {Springel}, {Theuns}, \&
  {Wiersma}}]{mcca10feedback}
{McCarthy}, I.~G., et~al.\ 2010, \mnras, 406, 822

\bibitem[{{McHardy} {et~al.}(2006){McHardy}, {Koerding}, {Knigge}, {Uttley}, \&
  {Fender}}]{mcha06bhbagn}
{McHardy}, I.~M., {Koerding}, E., {Knigge}, C., {Uttley}, P., \& {Fender},
  R.~P. 2006, \nat, 444, 730

\bibitem[{{McKee} \& {Ostriker}(2007)}]{mcke07sf}
{McKee}, C.~F. \& {Ostriker}, E.~C. 2007, \araa, 45, 565

\bibitem[{{McKee} \& {Tan}(2008)}]{mcke08pop3}
{McKee}, C.~F. \& {Tan}, J.~C. 2008, \apj, 681, 771

\bibitem[{{McLure} \& {Dunlop}(2004)}]{mclu04mbh}
{McLure}, R.~J. \& {Dunlop}, J.~S. 2004, \mnras, 352, 1390

\bibitem[{{McLure} \& {Jarvis}(2002)}]{mclu02bhmass}
{McLure}, R.~J. \& {Jarvis}, M.~J. 2002, \mnras, 337, 109

\bibitem[{{McLure} {et~al.}(2004){McLure}, {Willott}, {Jarvis}, {Rawlings},
  {Hill}, {Mitchell}, {Dunlop}, \& {Wold}}]{mclu04radiohost}
{McLure}, R.~J., {Willott}, C.~J., {Jarvis}, M.~J., {Rawlings}, S., {Hill},
  G.~J., {Mitchell}, E., {Dunlop}, J.~S., \& {Wold}, M. 2004, \mnras, 351, 347

\bibitem[{{McNamara} {et~al.}(2009){McNamara}, {Kazemzadeh}, {Rafferty},
  {B{\^i}rzan}, {Nulsen}, {Kirkpatrick}, \& {Wise}}]{mcna09ms07}
{McNamara}, B.~R., {Kazemzadeh}, F., {Rafferty}, D.~A., {B{\^i}rzan}, L.,
  {Nulsen}, P.~E.~J., {Kirkpatrick}, C.~C., \& {Wise}, M.~W. 2009, \apj, 698,
  594

\bibitem[{{McNamara} \& {Nulsen}(2007)}]{mcna07araa}
{McNamara}, B.~R. \& {Nulsen}, P.~E.~J. 2007, \araa, 45, 117

\bibitem[{{McNamara} {et~al.}(2005){McNamara}, {Nulsen}, {Wise}, {Rafferty},
  {Carilli}, {Sarazin}, \& {Blanton}}]{mcna05nature}
{McNamara}, B.~R., {Nulsen}, P.~E.~J., {Wise}, M.~W., {Rafferty}, D.~A.,
  {Carilli}, C., {Sarazin}, C.~L., \& {Blanton}, E.~L. 2005, \nat, 433, 45

\bibitem[{{McNamara} {et~al.}(2011){McNamara}, {Rohanizadegan}, \&
  {Nulsen}}]{mcna11spin}
{McNamara}, B.~R., {Rohanizadegan}, M., \& {Nulsen}, P.~E.~J. 2011, \apj, 727,
  39

\bibitem[{{Melia} \& {Falcke}(2001)}]{meli01sgra}
{Melia}, F. \& {Falcke}, H. 2001, \araa, 39, 309

\bibitem[{{Merloni}(2004)}]{merl04smbh}
{Merloni}, A. 2004, \mnras, 353, 1035

\bibitem[{{Merloni} {et~al.}(2010){Merloni}, {Bongiorno}, {Bolzonella},
  {Brusa}, {Civano}, {Comastri}, {Elvis}, {Fiore}, {Gilli}, {Hao}, {Jahnke},
  {Koekemoer}, {Lusso}, {Mainieri}, {Mignoli}, {Miyaji}, {Renzini}, {Salvato},
  {Silverman}, {Trump}, {Vignali}, {Zamorani}, {Capak}, {Lilly}, {Sanders},
  {Taniguchi}, {Bardelli}, {Carollo}, {Caputi}, {Contini}, {Coppa}, {Cucciati},
  {de la Torre}, {de Ravel}, {Franzetti}, {Garilli}, {Hasinger}, {Impey},
  {Iovino}, {Iwasawa}, {Kampczyk}, {Kneib}, {Knobel}, {Kova{\v c}},
  {Lamareille}, {Le Borgne}, {Le Brun}, {Le F{\`e}vre}, {Maier}, {Pello},
  {Peng}, {Perez Montero}, {Ricciardelli}, {Scodeggio}, {Tanaka}, {Tasca},
  {Tresse}, {Vergani}, \& {Zucca}}]{merl10bhmass}
{Merloni}, A., et~al.\ 2010, \apj, 708, 137

\bibitem[{{Merloni} \& {Heinz}(2008)}]{merl08agnsynth}
{Merloni}, A. \& {Heinz}, S. 2008, \mnras, 388, 1011

\bibitem[{{Merloni} {et~al.}(2003){Merloni}, {Heinz}, \& {di
  Matteo}}]{merl03bhplane}
{Merloni}, A., {Heinz}, S., \& {di Matteo}, T. 2003, \mnras, 345, 1057

\bibitem[{{Merloni} {et~al.}(2004){Merloni}, {Rudnick}, \& {Di
  Matteo}}]{merl04bhsf}
{Merloni}, A., {Rudnick}, G., \& {Di Matteo}, T. 2004, \mnras, 354, L37

\bibitem[{{Micic} {et~al.}(2011){Micic}, {Holley-Bockelmann}, \&
  {Sigurdsson}}]{mici11mbh}
{Micic}, M., {Holley-Bockelmann}, K., \& {Sigurdsson}, S. 2011, \mnras, 414,
  1127

\bibitem[{{Mihos} \& {Hernquist}(1996)}]{miho96merge}
{Mihos}, J.~C. \& {Hernquist}, L. 1996, \apj, 464, 641

\bibitem[{{Mirabel} {et~al.}(2011){Mirabel}, {Dijkstra}, {Laurent}, {Loeb}, \&
  {Pritchard}}]{mira11highzbh}
{Mirabel}, I.~F., {Dijkstra}, M., {Laurent}, P., {Loeb}, A., \& {Pritchard},
  J.~R. 2011, \aap, 528, A149+

\bibitem[{{Mirabel} \& {Rodr{\'{\i}}guez}(1994)}]{mira94superlum}
{Mirabel}, I.~F. \& {Rodr{\'{\i}}guez}, L.~F. 1994, \nat, 371, 46

\bibitem[{{Mirabel} \& {Rodr{\'{\i}}guez}(1998)}]{mira98microqso}
---. 1998, \nat, 392, 673

\bibitem[{{Mirabel} {et~al.}(1992){Mirabel}, {Rodriguez}, {Cordier}, {Paul}, \&
  {Lebrun}}]{mira92jet}
{Mirabel}, I.~F., {Rodriguez}, L.~F., {Cordier}, B., {Paul}, J., \& {Lebrun},
  F. 1992, \nat, 358, 215

\bibitem[{{Mo} {et~al.}(1998){Mo}, {Mao}, \& {White}}]{mo98disk}
{Mo}, H.~J., {Mao}, S., \& {White}, S.~D.~M. 1998, \mnras, 295, 319

\bibitem[{{Mortlock} {et~al.}(2011){Mortlock}, {Warren}, {Venemans}, {Patel},
  {Hewett}, {McMahon}, {Simpson}, {Theuns}, {Gonz{\'a}les-Solares}, {Adamson},
  {Dye}, {Hambly}, {Hirst}, {Irwin}, {Kuiper}, {Lawrence}, \&
  {R{\"o}ttgering}}]{mort11z7qso}
{Mortlock}, D.~J., et~al.\ 2011,  \nat, 474, 616

\bibitem[{{Mulchaey} {et~al.}(1994){Mulchaey}, {Koratkar}, {Ward}, {Wilson},
  {Whittle}, {Antonucci}, {Kinney}, \& {Hurt}}]{mulc94torus}
{Mulchaey}, J.~S., {Koratkar}, A., {Ward}, M.~J., {Wilson}, A.~S., {Whittle},
  M., {Antonucci}, R.~R.~J., {Kinney}, A.~L., \& {Hurt}, T. 1994, \apj, 436,
  586

\bibitem[{{Mulchaey} \& {Regan}(1997)}]{mulc97bar}
{Mulchaey}, J.~S. \& {Regan}, M.~W. 1997, \apjl, 482, L135+

\bibitem[{{Mullaney} {et~al.}(2011{\natexlab{a}}){Mullaney}, {Alexander},
  {Goulding}, \& {Hickox}}]{mull11agnsed}
{Mullaney}, J.~R., {Alexander}, D.~M., {Goulding}, A.~D., \& {Hickox}, R.~C.
  2011{\natexlab{a}}, \mnras, 414, 1082

\bibitem[{{Mullaney} {et~al.}(2010){Mullaney}, {Alexander}, {Huynh},
  {Goulding}, \& {Frayer}}]{mull10agnsf}
{Mullaney}, J.~R., {Alexander}, D.~M., {Huynh}, M., {Goulding}, A.~D., \&
  {Frayer}, D. 2010, \mnras, 401, 995

\bibitem[{{Mullaney} {et~al.}(2011{\natexlab{b}}){Mullaney}, {Pannella},
  {Daddi}, {Alexander}, {Elbaz}, {Hickox}, {Bournaud}, {Altieri}, {Aussel},
  {Coia}, {Dannerbauer}, {Dasyra}, {Dickinson}, {Hwang}, {Kartaltepe},
  {Leiton}, {Magdis}, {Magnelli}, {Popesso}, {Valtchanov}, {Bauer}, {Brandt},
  {Del Moro}, {Hanish}, {Ivison}, {Juneau}, {Luo}, {Lutz}, {Sargent}, {Scott},
  \& {Xue}}]{mull11agnsf}
{Mullaney}, J.~R., et~al.\ 2011{\natexlab{b}}, \mnras, 1756

\bibitem[{{M{\"u}ller S{\'a}nchez} {et~al.}(2009){M{\"u}ller S{\'a}nchez},
  {Davies}, {Genzel}, {Tacconi}, {Eisenhauer}, {Hicks}, {Friedrich}, \&
  {Sternberg}}]{mull09ngc1068}
{M{\"u}ller S{\'a}nchez}, F., et~al.\ 2009, \apj, 691, 749

\bibitem[{{Muno} {et~al.}(2007){Muno}, {Baganoff}, {Brandt}, {Park}, \&
  {Morris}}]{muno07refl}
{Muno}, M.~P., {Baganoff}, F.~K., {Brandt}, W.~N., {Park}, S., \& {Morris},
  M.~R. 2007, \apjl, 656, L69

\bibitem[{{Murray} {et~al.}(2010){Murray}, {Giacconi}, {Ptak}, {Rosati},
  {Weisskopf}, {Borgani}, {Jones}, {Pareschi}, {Tozzi}, {Gilli}, {Campana},
  {Paolillo}, {Tagliaferri}, {Bautz}, {Vikhlinin}, {Hickox}, \&
  {Forman}}]{murr10wfxt}
{Murray}, S., et~al.\ 2010, in American Institute of Physics  Conference Series, Vol. 1248, American Institute of Physics Conference  Series, ed. {A.~Comastri, L.~Angelini, \& M.~Cappi}, 549--554

\bibitem[{{Myers} {et~al.}(2006){Myers}, {Brunner}, {Richards}, {Nichol},
  {Schneider}, {Vanden Berk}, {Scranton}, {Gray}, \& {Brinkmann}}]{myer06clust}
{Myers}, A.~D., et~al.\ 2006, \apj, 638, 622

\bibitem[{{Nandra} {et~al.}(2007){Nandra}, {Georgakakis}, {Willmer}, {Cooper},
  {Croton}, {Davis}, {Faber}, {Koo}, {Laird}, \& {Newman}}]{nand07host}
{Nandra}, K., et~al.\ 2007, \apjl, 660, L11

\bibitem[{{Narayan} \& {Yi}(1994)}]{nara94adaf}
{Narayan}, R. \& {Yi}, I. 1994, \apjl, 428, L13

\bibitem[{{Nardini} \& {Risaliti}(2011)}]{nard11comp}
{Nardini}, E. \& {Risaliti}, G. 2011, \mnras, 415, 619

\bibitem[{{Nardini} {et~al.}(2010){Nardini}, {Risaliti}, {Watabe}, {Salvati},
  \& {Sani}}]{nard10ulirg}
{Nardini}, E., {Risaliti}, G., {Watabe}, Y., {Salvati}, M., \& {Sani}, E. 2010,
  \mnras, 405, 2505

\bibitem[{{Nayakshin} {et~al.}(2007){Nayakshin}, {Cuadra}, \&
  {Springel}}]{naya07sgra}
{Nayakshin}, S., {Cuadra}, J., \& {Springel}, V. 2007, \mnras, 379, 21

\bibitem[{{Nayakshin} \& {Sunyaev}(2005)}]{naya05sfcen}
{Nayakshin}, S. \& {Sunyaev}, R. 2005, \mnras, 364, L23

\bibitem[{{Nayakshin} {et~al.}(2009){Nayakshin}, {Wilkinson}, \&
  {King}}]{naya09feed}
{Nayakshin}, S., {Wilkinson}, M.~I., \& {King}, A. 2009, \mnras, 398, L54

\bibitem[{{Nenkova} {et~al.}(2002){Nenkova}, {Ivezi{\'c}}, \&
  {Elitzur}}]{nenk02agndust}
{Nenkova}, M., {Ivezi{\'c}}, {\v Z}., \& {Elitzur}, M. 2002, \apjl, 570, L9

\bibitem[{{Nenkova} {et~al.}(2008){Nenkova}, {Sirocky}, {Nikutta},
  {Ivezi{\'c}}, \& {Elitzur}}]{nenk08torus}
{Nenkova}, M., {Sirocky}, M.~M., {Nikutta}, R., {Ivezi{\'c}}, {\v Z}., \&
  {Elitzur}, M. 2008, \apj, 685, 160

\bibitem[{{Nesvadba} {et~al.}(2007){Nesvadba}, {Lehnert}, {De Breuck},
  {Gilbert}, \& {van Breugel}}]{nesv07outflow}
{Nesvadba}, N.~P.~H., {Lehnert}, M.~D., {De Breuck}, C., {Gilbert}, A., \& {van
  Breugel}, W. 2007, \aap, 475, 145

\bibitem[{{Nesvadba} {et~al.}(2008){Nesvadba}, {Lehnert}, {De Breuck},
  {Gilbert}, \& {van Breugel}}]{nesv08outflow}
{Nesvadba}, N.~P.~H., {Lehnert}, M.~D., {De Breuck}, C., {Gilbert}, A.~M., \&
  {van Breugel}, W. 2008, \aap, 491, 407

\bibitem[{{Nesvadba} {et~al.}(2006){Nesvadba}, {Lehnert}, {Eisenhauer},
  {Gilbert}, {Tecza}, \& {Abuter}}]{nesv06outflow}
{Nesvadba}, N.~P.~H., {Lehnert}, M.~D., {Eisenhauer}, F., {Gilbert}, A.,
  {Tecza}, M., \& {Abuter}, R. 2006, \apj, 650, 693

\bibitem[{{Netzer} {et~al.}(2007){Netzer}, {Lutz}, {Schweitzer}, {Contursi},
  {Sturm}, {Tacconi}, {Veilleux}, {Kim}, {Rupke}, {Baker}, {Dasyra},
  {Mazzarella}, \& {Lord}}]{netz07qsosf}
{Netzer}, H., et~al.\ 2007, \apj, 666, 806

\bibitem[{{Netzer} \& {Marziani}(2010)}]{netz10bhmass}
{Netzer}, H. \& {Marziani}, P. 2010, \apj, 724, 318

\bibitem[{{Netzer} \& {Trakhtenbrot}(2007)}]{netz07agnevol}
{Netzer}, H. \& {Trakhtenbrot}, B. 2007, \apj, 654, 754

\bibitem[{{Noeske} {et~al.}(2007){Noeske}, {Weiner}, {Faber}, {Papovich},
  {Koo}, {Somerville}, {Bundy}, {Conselice}, {Newman}, {Schiminovich}, {Le
  Floc'h}, {Coil}, {Rieke}, {Lotz}, {Primack}, {Barmby}, {Cooper}, {Davis},
  {Ellis}, {Fazio}, {Guhathakurta}, {Huang}, {Kassin}, {Martin}, {Phillips},
  {Rich}, {Small}, {Willmer}, \& {Wilson}}]{noes07}
{Noeske}, K.~G., et~al.\ 2007, \apjl, 660, L43

\bibitem[{{Nowak}(1995)}]{nowa95bhb}
{Nowak}, M.~A. 1995, \pasp, 107, 1207

\bibitem[{{Ohsuga}(2007)}]{ohsu07agnfeed}
{Ohsuga}, K. 2007, \apj, 659, 205

\bibitem[{{Ohsuga} {et~al.}(2005){Ohsuga}, {Mori}, {Nakamoto}, \&
  {Mineshige}}]{ohsu05super}
{Ohsuga}, K., {Mori}, M., {Nakamoto}, T., \& {Mineshige}, S. 2005, \apj, 628,
  368

\bibitem[{{Onken} {et~al.}(2004){Onken}, {Ferrarese}, {Merritt}, {Peterson},
  {Pogge}, {Vestergaard}, \& {Wandel}}]{onke04revmap}
{Onken}, C.~A., {Ferrarese}, L., {Merritt}, D., {Peterson}, B.~M., {Pogge},
  R.~W., {Vestergaard}, M., \& {Wandel}, A. 2004, \apj, 615, 645

\bibitem[{{Orban de Xivry} {et~al.}(2011){Orban de Xivry}, {Davies},
  {Schartmann}, {Komossa}, {Marconi}, {Hicks}, {Engel}, \&
  {Tacconi}}]{orba11nls1}
{Orban de Xivry}, G., {Davies}, R., {Schartmann}, M., {Komossa}, S., {Marconi},
  A., {Hicks}, E., {Engel}, H., \& {Tacconi}, L. 2011, \mnras, 1359

\bibitem[{{Osterbrock} \& {Pogge}(1985)}]{oste85nls1}
{Osterbrock}, D.~E. \& {Pogge}, R.~W. 1985, \apj, 297, 166

\bibitem[{{Page} {et~al.}(2011){Page}, {Carrera}, {Stevens}, {Ebrero}, \&
  {Blustin}}]{page11obsqso}
{Page}, M.~J., {Carrera}, F.~J., {Stevens}, J.~A., {Ebrero}, J., \& {Blustin},
  A.~J. 2011, \mnras, 416, 2792

\bibitem[{{Page} {et~al.}(2004){Page}, {Stevens}, {Ivison}, \&
  {Carrera}}]{page04submm}
{Page}, M.~J., {Stevens}, J.~A., {Ivison}, R.~J., \& {Carrera}, F.~J. 2004,
  \apjl, 611, L85

\bibitem[{{Page} {et~al.}(2001){Page}, {Stevens}, {Mittaz}, \&
  {Carrera}}]{page01}
{Page}, M.~J., {Stevens}, J.~A., {Mittaz}, J.~P.~D., \& {Carrera}, F.~J. 2001,
  Science, 294, 2516

\bibitem[{{Pakull} {et~al.}(2010){Pakull}, {Soria}, \&
  {Motch}}]{paku10microqso}
{Pakull}, M.~W., {Soria}, R., \& {Motch}, C. 2010, \nat, 466, 209

\bibitem[{{Pannella} {et~al.}(2009){Pannella}, {Carilli}, {Daddi}, {McCracken},
  {Owen}, {Renzini}, {Strazzullo}, {Civano}, {Koekemoer}, {Schinnerer},
  {Scoville}, {Smol{\v c}i{\'c}}, {Taniguchi}, {Aussel}, {Kneib}, {Ilbert},
  {Mellier}, {Salvato}, {Thompson}, \& {Willott}}]{pann09sf}
{Pannella}, M., et~al.\ 2009, \apjl, 698, L116

\bibitem[{{Paumard} {et~al.}(2006){Paumard}, {Genzel}, {Martins}, {Nayakshin},
  {Beloborodov}, {Levin}, {Trippe}, {Eisenhauer}, {Ott}, {Gillessen}, {Abuter},
  {Cuadra}, {Alexander}, \& {Sternberg}}]{paum06disk}
{Paumard}, T., et~al.\ 2006,  \apj, 643, 1011

\bibitem[{{Peng}(2007)}]{peng07mbh}
{Peng}, C.~Y. 2007, \apj, 671, 1098

\bibitem[{{P{\'e}rez-Gonz{\'a}lez} {et~al.}(2005){P{\'e}rez-Gonz{\'a}lez},
  {Rieke}, {Egami}, {Alonso-Herrero}, {Dole}, {Papovich}, {Blaylock}, {Jones},
  {Rieke}, {Rigby}, {Barmby}, {Fazio}, {Huang}, \& {Martin}}]{pere05sfevol}
{P{\'e}rez-Gonz{\'a}lez}, P.~G., et~al.\ 2005,  \apj, 630, 82

\bibitem[{{Peterson}(2010)}]{pete10bhmass}
{Peterson}, B.~M. 2010, in IAU Symposium, Vol. 267, IAU Symposium, 151--160

\bibitem[{{Peterson}(2011)}]{pete11bhmass}
{Peterson}, B.~M. 2011, arXiv:1109.4181

\bibitem[{{Peterson} {et~al.}(2004){Peterson}, {Ferrarese}, {Gilbert}, {Kaspi},
  {Malkan}, {Maoz}, {Merritt}, {Netzer}, {Onken}, {Pogge}, {Vestergaard}, \&
  {Wandel}}]{pete04bhmass}
{Peterson}, B.~M., et~al.\ 2004, \apj, 613, 682

\bibitem[{{Petric} {et~al.}(2011{\natexlab{a}}){Petric}, {Armus}, {Howell},
  {Chan}, {Mazzarella}, {Evans}, {Surace}, {Sanders}, {Appleton},
  {Charmandaris}, {D{\'{\i}}az-Santos}, {Frayer}, {Haan}, {Inami}, {Iwasawa},
  {Kim}, {Madore}, {Marshall}, {Spoon}, {Stierwalt}, {Sturm}, {U}, {Vavilkin},
  \& {Veilleux}}]{petr11irs}
{Petric}, A.~O., et~al.\ 2011{\natexlab{a}}, \apj, 730, 28

\bibitem[{{Petric} {et~al.}(2011{\natexlab{b}}){Petric}, {Armus}, {Howell},
  {Chan}, {Mazzarella}, {Evans}, {Surace}, {Sanders}, {Appleton},
  {Charmandaris}, {D{\'{\i}}az-Santos}, {Frayer}, {Haan}, {Inami}, {Iwasawa},
  {Kim}, {Madore}, {Marshall}, {Spoon}, {Stierwalt}, {Sturm}, {U}, {Vavilkin},
  \& {Veilleux}}]{petr11lirg}
---. 2011{\natexlab{b}}, \apj, 730, 28

\bibitem[{{Pier} \& {Krolik}(1992)}]{pier92torus}
{Pier}, E.~A. \& {Krolik}, J.~H. 1992, \apj, 401, 99

\bibitem[{{Pierce} {et~al.}(2007){Pierce}, {Lotz}, {Laird}, {Lin}, {Nandra},
  {Primack}, {Faber}, {Barmby}, {Park}, {Willner}, {Gwyn}, {Koo}, {Coil},
  {Cooper}, {Georgakakis}, {Koekemoer}, {Noeske}, {Weiner}, \&
  {Willmer}}]{pier07morph}
{Pierce}, C.~M., et~al.\ 2007, \apjl, 660, L19

\bibitem[{{Pogge} \& {Martini}(2002)}]{pogg02nucl}
{Pogge}, R.~W. \& {Martini}, P. 2002, \apj, 569, 624

\bibitem[{{Polletta} {et~al.}(2007){Polletta}, {Tajer}, {Maraschi},
  {Trinchieri}, {Lonsdale}, {Chiappetti}, {Andreon}, {Pierre}, {Le F{\`e}vre},
  {Zamorani}, {Maccagni}, {Garcet}, {Surdej}, {Franceschini}, {Alloin},
  {Shupe}, {Surace}, {Fang}, {Rowan-Robinson}, {Smith}, \&
  {Tresse}}]{poll07agnsed}
{Polletta}, M., et~al.\ 2007, \apj, 663, 81

\bibitem[{{Polletta} {et~al.}(2008){Polletta}, {Weedman}, {H{\"o}nig},
  {Lonsdale}, {Smith}, \& {Houck}}]{poll08obsc}
{Polletta}, M., {Weedman}, D., {H{\"o}nig}, S., {Lonsdale}, C.~J., {Smith},
  H.~E., \& {Houck}, J. 2008, \apj, 675, 960

\bibitem[{{Polletta} {et~al.}(2006){Polletta}, {Wilkes}, {Siana}, {Lonsdale},
  {Kilgard}, {Smith}, {Kim}, {Owen}, {Efstathiou}, {Jarrett}, {Stacey},
  {Franceschini}, {Rowan-Robinson}, {Babbedge}, {Berta}, {Fang}, {Farrah},
  {Gonz{\'a}lez-Solares}, {Morrison}, {Surace}, \& {Shupe}}]{poll06}
{Polletta}, M.~d.~C., et~al.\ 2006, \apj, 642, 673

\bibitem[{{Ponti} {et~al.}(2010){Ponti}, {Terrier}, {Goldwurm}, {Belanger}, \&
  {Trap}}]{pont10refl}
{Ponti}, G., {Terrier}, R., {Goldwurm}, A., {Belanger}, G., \& {Trap}, G. 2010,
  \apj, 714, 732

\bibitem[{{Pooley} {et~al.}(2007){Pooley}, {Blackburne}, {Rappaport}, \&
  {Schechter}}]{pool07qsolens}
{Pooley}, D., {Blackburne}, J.~A., {Rappaport}, S., \& {Schechter}, P.~L. 2007,
  \apj, 661, 19

\bibitem[{{Pope} {et~al.}(2008){Pope}, {Chary}, {Alexander}, {Armus},
  {Dickinson}, {Elbaz}, {Frayer}, {Scott}, \& {Teplitz}}]{pope08smgspitz}
{Pope}, A., et~al.\ 2008, \apj, 675,  1171

\bibitem[{{Pope}(2011)}]{pope11feedback}
{Pope}, E.~C.~D. 2011, \mnras, 755

\bibitem[{{Pounds} {et~al.}(2003){Pounds}, {Reeves}, {Page}, {Edelson}, {Matt},
  \& {Perola}}]{poun03ngc5548}
{Pounds}, K.~A., {Reeves}, J.~N., {Page}, K.~L., {Edelson}, R., {Matt}, G., \&
  {Perola}, G.~C. 2003, \mnras, 341, 953

\bibitem[{{Power} {et~al.}(2011){Power}, {Nayakshin}, \& {King}}]{powe11bhsim}
{Power}, C., {Nayakshin}, S., \& {King}, A. 2011, \mnras, 412, 269

\bibitem[{{Pozzi} {et~al.}(2010){Pozzi}, {Vignali}, {Comastri}, {Bellocchi},
  {Fritz}, {Gruppioni}, {Mignoli}, {Maiolino}, {Pozzetti}, {Brusa}, {Fiore}, \&
  {Zamorani}}]{pozz10obscagn}
{Pozzi}, F., et~al.\ 2010, \aap, 517, A11+

\bibitem[{{Predehl} {et~al.}(2007){Predehl}, {Andritschke}, {Bornemann},
  {Br{\"a}uninger}, {Briel}, {Brunner}, {Burkert}, {Dennerl}, {Eder},
  {Freyberg}, {Friedrich}, {F{\"u}rmetz}, {Hartmann}, {Hartner}, {Hasinger},
  {Herrmann}, {Holl}, {Huber}, {Kendziorra}, {Kink}, {Meidinger}, {M{\"u}ller},
  {Pavlinsky}, {Pfeffermann}, {Roh{\'e}}, {Santangelo}, {Schmitt}, {Schwope},
  {Steinmetz}, {Str{\"u}der}, {Sunyaev}, {Tiedemann}, {Vongehr}, {Wilms},
  {Erhard}, {Gutruf}, {Jugler}, {Kampf}, {Graue}, {Citterio}, {Valsecci},
  {Vernani}, \& {Zimmerman}}]{pred07erosita}
{Predehl}, P., et~al.\ 2007, in Proceedings of the SPIE, Volume  6686, pp.\ 668617-668617-9, Vol. 6686

\bibitem[{{Prieto} {et~al.}(2005){Prieto}, {Maciejewski}, \&
  {Reunanen}}]{prie05ngc1097}
{Prieto}, M.~A., {Maciejewski}, W., \& {Reunanen}, J. 2005, \aj, 130, 1472

\bibitem[{{Pringle}(1981)}]{prin81disk}
{Pringle}, J.~E. 1981, \araa, 19, 137

\bibitem[{{Raban} {et~al.}(2009){Raban}, {Jaffe}, {R{\"o}ttgering},
  {Meisenheimer}, \& {Tristram}}]{raba09ngc1068}
{Raban}, D., {Jaffe}, W., {R{\"o}ttgering}, H., {Meisenheimer}, K., \&
  {Tristram}, K.~R.~W. 2009, \mnras, 394, 1325

\bibitem[{{Rafferty} {et~al.}(2011){Rafferty}, {Brandt}, {Alexander}, {Xue},
  {Bauer}, {Lehmer}, {Luo}, \& {Papovich}}]{raff11agnsf}
{Rafferty}, D.~A., {Brandt}, W.~N., {Alexander}, D.~M., {Xue}, Y.~Q., {Bauer},
  F.~E., {Lehmer}, B.~D., {Luo}, B., \& {Papovich}, C. 2011, \apj, 742, 3

\bibitem[{{Rafferty} {et~al.}(2008){Rafferty}, {McNamara}, \&
  {Nulsen}}]{raff08feedback}
{Rafferty}, D.~A., {McNamara}, B.~R., \& {Nulsen}, P.~E.~J. 2008, \apj, 687,
  899

\bibitem[{{Rafferty} {et~al.}(2006){Rafferty}, {McNamara}, {Nulsen}, \&
  {Wise}}]{raff06feedback}
{Rafferty}, D.~A., {McNamara}, B.~R., {Nulsen}, P.~E.~J., \& {Wise}, M.~W.
  2006, \apj, 652, 216

\bibitem[{{Raimundo} {et~al.}(2010){Raimundo}, {Fabian}, {Bauer}, {Alexander},
  {Brandt}, {Luo}, {Vasudevan}, \& {Xue}}]{raim10nh}
{Raimundo}, S.~I., {Fabian}, A.~C., {Bauer}, F.~E., {Alexander}, D.~M.,
  {Brandt}, W.~N., {Luo}, B., {Vasudevan}, R.~V., \& {Xue}, Y.~Q. 2010, \mnras,
  408, 1714

\bibitem[{{Raimundo} {et~al.}(2011){Raimundo}, {Fabian}, {Vasudevan}, {Gandhi},
  \& {Wu}}]{raim11bolc_aph}
{Raimundo}, S.~I., {Fabian}, A.~C., {Vasudevan}, R.~V., {Gandhi}, P., \& {Wu},
  J. 2011, \mnras\ in press (arXiv:1109.6225)

\bibitem[{{Ramos Almeida} {et~al.}(2011{\natexlab{a}}){Ramos Almeida},
  {Bessiere}, {Tadhunter}, {P{\'e}rez-Gonz{\'a}lez}, {Barro}, {Inskip},
  {Morganti}, {Holt}, \& {Dicken}}]{ramo11trigger}
{Ramos Almeida}, C., et~al.\ 2011{\natexlab{a}}, \mnras, 1702

\bibitem[{{Ramos Almeida} {et~al.}(2011{\natexlab{b}}){Ramos Almeida},
  {Tadhunter}, {Inskip}, {Morganti}, {Holt}, \& {Dicken}}]{ramo11radiohost}
{Ramos Almeida}, C., {Tadhunter}, C.~N., {Inskip}, K.~J., {Morganti}, R.,
  {Holt}, J., \& {Dicken}, D. 2011{\natexlab{b}}, \mnras, 410, 1550

\bibitem[{{Rees}(1984)}]{rees84agn}
{Rees}, M.~J. 1984, \araa, 22, 471

\bibitem[{{Reeves} {et~al.}(2003){Reeves}, {O'Brien}, \&
  {Ward}}]{reev03outflow}
{Reeves}, J.~N., {O'Brien}, P.~T., \& {Ward}, M.~J. 2003, \apjl, 593, L65

\bibitem[{{Regan} \& {Haehnelt}(2009)}]{rega09protogal}
{Regan}, J.~A. \& {Haehnelt}, M.~G. 2009, \mnras, 396, 343

\bibitem[{{Reid}(1993)}]{reid93sgra}
{Reid}, M.~J. 1993, \araa, 31, 345

\bibitem[{{Remillard} \& {McClintock}(2006)}]{remi06bhb}
{Remillard}, R.~A. \& {McClintock}, J.~E. 2006, \araa, 44, 49

\bibitem[{{Reyes} {et~al.}(2008){Reyes}, {Zakamska}, {Strauss}, {Green},
  {Krolik}, {Shen}, {Richards}, {Anderson}, \& {Schneider}}]{reye08qso2}
{Reyes}, R., et~al.\ 2008,  \aj, 136, 2373

\bibitem[{{Richards} {et~al.}(2005){Richards}, {Croom}, {Anderson},
  {Bland-Hawthorn}, {Boyle}, {De Propris}, {Drinkwater}, {Fan}, {Gunn},
  {Ivezi{\'c}}, {Jester}, {Loveday}, {Meiksin}, {Miller}, {Myers}, {Nichol},
  {Outram}, {Pimbblet}, {Roseboom}, {Ross}, {Schneider}, {Shanks}, {Sharp},
  {Stoughton}, {Strauss}, {Szalay}, {Vanden Berk}, \& {York}}]{rich05}
{Richards}, G.~T., et~al.\ 2005, \mnras, 360, 839

\bibitem[{{Richards} {et~al.}(2011){Richards}, {Kruczek}, {Gallagher}, {Hall},
  {Hewett}, {Leighly}, {Deo}, {Kratzer}, \& {Shen}}]{rich11blueshift}
{Richards}, G.~T., et~al.\ 2011, \aj, 141, 167

\bibitem[{{Richards} {et~al.}(2006){Richards}, {Strauss}, {Fan}, {Hall},
  {Jester}, {Schneider}, {Vanden Berk}, {Stoughton}, {Anderson}, {Brunner},
  {Gray}, {Gunn}, {Ivezi{\'c}}, {Kirkland}, {Knapp}, {Loveday}, {Meiksin},
  {Pope}, {Szalay}, {Thakar}, {Yanny}, {York}, {Barentine}, {Brewington},
  {Brinkmann}, {Fukugita}, {Harvanek}, {Kent}, {Kleinman}, {Krzesi{\'n}ski},
  {Long}, {Lupton}, {Nash}, {Neilsen}, {Nitta}, {Schlegel}, \&
  {Snedden}}]{rich06qlf}
{Richards}, G.~T., et~al.\ 2006, \aj, 131, 2766

\bibitem[{{Richards} {et~al.}(2002){Richards}, {Vanden Berk}, {Reichard},
  {Hall}, {Schneider}, {SubbaRao}, {Thakar}, \& {York}}]{rich02blueshift}
{Richards}, G.~T., {Vanden Berk}, D.~E., {Reichard}, T.~A., {Hall}, P.~B.,
  {Schneider}, D.~P., {SubbaRao}, M., {Thakar}, A.~R., \& {York}, D.~G. 2002,
  \aj, 124, 1

\bibitem[{{Riechers} {et~al.}(2011){Riechers}, {Carilli}, {Walter}, {Weiss},
  {Wagg}, {Bertoldi}, {Downes}, {Henkel}, \& {Hodge}}]{reic11smgmerge}
{Riechers}, D.~A., et~al.\ 2011, \apjl, 733,  L11+

\bibitem[{{Riffel} {et~al.}(2008){Riffel}, {Storchi-Bergmann}, {Winge},
  {McGregor}, {Beck}, \& {Schmitt}}]{riff08ifu}
{Riffel}, R.~A., {Storchi-Bergmann}, T., {Winge}, C., {McGregor}, P.~J.,
  {Beck}, T., \& {Schmitt}, H. 2008, \mnras, 385, 1129

\bibitem[{{Rigby} {et~al.}(2011){Rigby}, {Best}, {Brookes}, {Peacock},
  {Dunlop}, {R{\"o}ttgering}, {Wall}, \& {Ker}}]{rigb11radio}
{Rigby}, E.~E., {Best}, P.~N., {Brookes}, M.~H., {Peacock}, J.~A., {Dunlop},
  J.~S., {R{\"o}ttgering}, H.~J.~A., {Wall}, J.~V., \& {Ker}, L. 2011, \mnras,
  416, 1900

\bibitem[{{Rigby} {et~al.}(2008){Rigby}, {Best}, \& {Snellen}}]{rigb08radio}
{Rigby}, E.~E., {Best}, P.~N., \& {Snellen}, I.~A.~G. 2008, \mnras, 385, 310

\bibitem[{{Risaliti} {et~al.}(2002){Risaliti}, {Elvis}, \&
  {Nicastro}}]{risa02agnnh}
{Risaliti}, G., {Elvis}, M., \& {Nicastro}, F. 2002, \apj, 571, 234

\bibitem[{{Risaliti} {et~al.}(1999){Risaliti}, {Maiolino}, \&
  {Salvati}}]{risa99}
{Risaliti}, G., {Maiolino}, R., \& {Salvati}, M. 1999, \apj, 522, 157

\bibitem[{{Roberts}(2007)}]{robe07ulx}
{Roberts}, T.~P. 2007, \apss, 311, 203

\bibitem[{{Roberts} {et~al.}(2003){Roberts}, {Goad}, {Ward}, \&
  {Warwick}}]{robe03ulx}
{Roberts}, T.~P., {Goad}, M.~R., {Ward}, M.~J., \& {Warwick}, R.~S. 2003,
  \mnras, 342, 709

\bibitem[{{Robinson} {et~al.}(2010){Robinson}, {Young}, {Axon}, {Kharb}, \&
  {Smith}}]{robi10recoil}
{Robinson}, A., {Young}, S., {Axon}, D.~J., {Kharb}, P., \& {Smith}, J.~E.
  2010, \apjl, 717, L122

\bibitem[{{Rodighiero} {et~al.}(2010{\natexlab{a}}){Rodighiero}, {Cimatti},
  {Gruppioni}, {Popesso}, {Andreani}, {Altieri}, {Aussel}, {Berta},
  {Bongiovanni}, {Brisbin}, {Cava}, {Cepa}, {Daddi}, {Dominguez-Sanchez},
  {Elbaz}, {Fontana}, {F{\"o}rster Schreiber}, {Franceschini}, {Genzel},
  {Grazian}, {Lutz}, {Magdis}, {Magliocchetti}, {Magnelli}, {Maiolino},
  {Mancini}, {Nordon}, {Perez Garcia}, {Poglitsch}, {Santini},
  {Sanchez-Portal}, {Pozzi}, {Riguccini}, {Saintonge}, {Shao}, {Sturm},
  {Tacconi}, {Valtchanov}, {Wetzstein}, \& {Wieprecht}}]{rodi10sf}
{Rodighiero}, G., et~al.\ 2010{\natexlab{a}},  \aap, 518, L25+

\bibitem[{{Rodighiero} {et~al.}(2010{\natexlab{b}}){Rodighiero}, {Vaccari},
  {Franceschini}, {Tresse}, {Le Fevre}, {Le Brun}, {Mancini}, {Matute},
  {Cimatti}, {Marchetti}, {Ilbert}, {Arnouts}, {Bolzonella}, {Zucca},
  {Bardelli}, {Lonsdale}, {Shupe}, {Surace}, {Rowan-Robinson}, {Garilli},
  {Zamorani}, {Pozzetti}, {Bondi}, {de la Torre}, {Vergani}, {Santini},
  {Grazian}, \& {Fontana}}]{rodi11irlf}
{Rodighiero}, G., et~al.\ 2010{\natexlab{b}}, \aap, 515, A8+

\bibitem[{{Rodr{\'{\i}}guez-Zaur{\'{\i}}n}
  {et~al.}(2011){Rodr{\'{\i}}guez-Zaur{\'{\i}}n}, {Arribas}, {Monreal-Ibero},
  {Colina}, {Alonso-Herrero}, \& {Alfonso-Garz{\'o}n}}]{rodr11ifu}
{Rodr{\'{\i}}guez-Zaur{\'{\i}}n}, J., {Arribas}, S., {Monreal-Ibero}, A.,
  {Colina}, L., {Alonso-Herrero}, A., \& {Alfonso-Garz{\'o}n}, J. 2011, \aap,
  527, A60+

\bibitem[{{Rosario} {et~al.}(2011){Rosario}, {McGurk}, {Max}, {Shields},
  {Smith}, \& {Ammons}}]{rosa10dualagn}
{Rosario}, D.~J., {McGurk}, R.~C., {Max}, C.~E., {Shields}, G.~A., {Smith},
  K.~L., \& {Ammons}, S.~M. 2011, \apj, 739, 44

\bibitem[{{Ross} {et~al.}(2009){Ross}, {Shen}, {Strauss}, {Vanden Berk},
  {Connolly}, {Richards}, {Schneider}, {Weinberg}, {Hall}, {Bahcall}, \&
  {Brunner}}]{ross09qsoclust}
{Ross}, N.~P., et~al.\ 2009, \apj, 697, 1634

\bibitem[{{Rovilos} {et~al.}(2011){Rovilos}, {Fotopoulou}, {Salvato},
  {Burwitz}, {Egami}, {Hasinger}, \& {Szokoly}}]{rovi11xhost}
{Rovilos}, E., {Fotopoulou}, S., {Salvato}, M., {Burwitz}, V., {Egami}, E.,
  {Hasinger}, G., \& {Szokoly}, G. 2011, \aap, 529, A135+

\bibitem[{{Rovilos} \& {Georgantopoulos}(2007)}]{rovi07xhost}
{Rovilos}, E. \& {Georgantopoulos}, I. 2007, \aap, 475, 115

\bibitem[{{Rovilos} {et~al.}(2010){Rovilos}, {Georgantopoulos}, {Akylas}, \&
  {Fotopoulou}}]{rovi10xfaint}
{Rovilos}, E., {Georgantopoulos}, I., {Akylas}, A., \& {Fotopoulou}, S. 2010,
  \aap, 522, A11+

\bibitem[{{Sadler} {et~al.}(2007){Sadler}, {Cannon}, {Mauch}, {Hancock},
  {Wake}, {Ross}, {Croom}, {Drinkwater}, {Edge}, {Eisenstein}, {Hopkins},
  {Johnston}, {Nichol}, {Pimbblet}, {de Propris}, {Roseboom}, {Schneider}, \&
  {Shanks}}]{sadl07radio}
{Sadler}, E.~M., et~al.\ 2007,  \mnras, 381, 211

\bibitem[{{Saintonge} {et~al.}(2008){Saintonge}, {Tran}, \&
  {Holden}}]{sain08clustsf}
{Saintonge}, A., {Tran}, K.-V.~H., \& {Holden}, B.~P. 2008, \apjl, 685, L113

\bibitem[{{Salpeter}(1964)}]{sal64}
{Salpeter}, E.~E. 1964, \apj, 140, 796

\bibitem[{{Salvato} {et~al.}(2011){Salvato}, {Ilbert}, {Hasinger}, {Rau},
  {Civano}, {Zamorani}, {Brusa}, {Elvis}, {Vignali}, {Aussel}, {Comastri},
  {Fiore}, {Le Floc'h}, {Mainieri}, {Bardelli}, {Bolzonella}, {Bongiorno},
  {Capak}, {Caputi}, {Cappelluti}, {Carollo}, {Contini}, {Garilli}, {Iovino},
  {Fotopoulou}, {Fruscione}, {Gilli}, {Halliday}, {Kneib}, {Kakazu},
  {Kartaltepe}, {Koekemoer}, {Kovac}, {Ideue}, {Ikeda}, {Impey}, {Le Fevre},
  {Lamareille}, {Lanzuisi}, {Le Borgne}, {Le Brun}, {Lilly}, {Maier},
  {Manohar}, {Masters}, {McCracken}, {Messias}, {Mignoli}, {Mobasher}, {Nagao},
  {Pello}, {Puccetti}, {Perez-Montero}, {Renzini}, {Sargent}, {Sanders},
  {Scodeggio}, {Scoville}, {Shopbell}, {Silvermann}, {Taniguchi}, {Tasca},
  {Tresse}, {Trump}, \& {Zucca}}]{salv11xcosmos}
{Salvato}, M., et~al.\ 2011, \apj, 742, 61

\bibitem[{{S{\'a}nchez} {et~al.}(2004){S{\'a}nchez}, {Jahnke}, {Wisotzki},
  {McIntosh}, {Bell}, {Barden}, {Beckwith}, {Borch}, {Caldwell},
  {H{\"a}ussler}, {Jogee}, {Meisenheimer}, {Peng}, {Rix}, {Somerville}, \&
  {Wolf}}]{sanc04agnhost}
{S{\'a}nchez}, S.~F., et~al.\ 2004, \apj, 614, 586

\bibitem[{{Sanders} {et~al.}(1988){Sanders}, {Soifer}, {Elias}, {Madore},
  {Matthews}, {Neugebauer}, \& {Scoville}}]{sand88}
{Sanders}, D.~B., {Soifer}, B.~T., {Elias}, J.~H., {Madore}, B.~F., {Matthews},
  K., {Neugebauer}, G., \& {Scoville}, N.~Z. 1988, \apj, 325, 74

\bibitem[{{Sarzi} {et~al.}(2010){Sarzi}, {Shields}, {Schawinski}, {Jeong},
  {Shapiro}, {Bacon}, {Bureau}, {Cappellari}, {Davies}, {de Zeeuw}, {Emsellem},
  {Falc{\'o}n-Barroso}, {Krajnovi{\'c}}, {Kuntschner}, {McDermid}, {Peletier},
  {van den Bosch}, {van de Ven}, \& {Yi}}]{sarz10sauron}
{Sarzi}, M., et~al.\ 2010, \mnras, 402, 2187

\bibitem[{{Satyapal} {et~al.}(2004){Satyapal}, {Sambruna}, \&
  {Dudik}}]{saty04liner}
{Satyapal}, S., {Sambruna}, R.~M., \& {Dudik}, R.~P. 2004, \aap, 414, 825

\bibitem[{{Satyapal} {et~al.}(2008){Satyapal}, {Vega}, {Dudik}, {Abel}, \&
  {Heckman}}]{saty08irs}
{Satyapal}, S., {Vega}, D., {Dudik}, R.~P., {Abel}, N.~P., \& {Heckman}, T.
  2008, \apj, 677, 926

\bibitem[{{Sazonov} {et~al.}(2007){Sazonov}, {Revnivtsev}, {Krivonos},
  {Churazov}, \& {Sunyaev}}]{sazo07agn}
{Sazonov}, S., {Revnivtsev}, M., {Krivonos}, R., {Churazov}, E., \& {Sunyaev},
  R. 2007, \aap, 462, 57

\bibitem[{{Schartmann} {et~al.}(2010){Schartmann}, {Burkert}, {Krause},
  {Camenzind}, {Meisenheimer}, \& {Davies}}]{scha10ngc1068}
{Schartmann}, M., {Burkert}, A., {Krause}, M., {Camenzind}, M., {Meisenheimer},
  K., \& {Davies}, R.~I. 2010, \mnras, 403, 1801

\bibitem[{{Schartmann} {et~al.}(2008){Schartmann}, {Meisenheimer}, {Camenzind},
  {Wolf}, {Tristram}, \& {Henning}}]{scha08torus}
{Schartmann}, M., {Meisenheimer}, K., {Camenzind}, M., {Wolf}, S., {Tristram},
  K.~R.~W., \& {Henning}, T. 2008, \aap, 482, 67

\bibitem[{{Schartmann} {et~al.}(2009){Schartmann}, {Meisenheimer}, {Klahr},
  {Camenzind}, {Wolf}, \& {Henning}}]{scha09torus}
{Schartmann}, M., {Meisenheimer}, K., {Klahr}, H., {Camenzind}, M., {Wolf}, S.,
  \& {Henning}, T. 2009, \mnras, 393, 759

\bibitem[{{Schawinski} {et~al.}(2010){Schawinski}, {Dowlin}, {Thomas}, {Urry},
  \& {Edmondson}}]{scha10merge}
{Schawinski}, K., {Dowlin}, N., {Thomas}, D., {Urry}, C.~M., \& {Edmondson}, E.
  2010, \apjl, 714, L108

\bibitem[{{Schawinski} {et~al.}(2011){Schawinski}, {Treister}, {Urry},
  {Cardamone}, {Simmons}, \& {Yi}}]{scha11xmorph}
{Schawinski}, K., {Treister}, E., {Urry}, C.~M., {Cardamone}, C.~N., {Simmons},
  B., \& {Yi}, S.~K. 2011, \apjl, 727, L31+

\bibitem[{{Schawinski} {et~al.}(2009){Schawinski}, {Virani}, {Simmons}, {Urry},
  {Treister}, {Kaviraj}, \& {Kushkuley}}]{scha09agn}
{Schawinski}, K., {Virani}, S., {Simmons}, B., {Urry}, C.~M., {Treister}, E.,
  {Kaviraj}, S., \& {Kushkuley}, B. 2009, \apjl, 692, L19

\bibitem[{{Scheuer}(1974)}]{sche74radio}
{Scheuer}, P.~A.~G. 1974, \mnras, 166, 513

\bibitem[{{Schmidt}(1968)}]{schm68}
{Schmidt}, M. 1968, \apj, 151, 393

\bibitem[{{Schmidt} \& {Green}(1983)}]{schm83qso}
{Schmidt}, M. \& {Green}, R.~F. 1983, \apj, 269, 352

\bibitem[{{Schmitt} \& {Kinney}(1996)}]{schm96nlr}
{Schmitt}, H.~R. \& {Kinney}, A.~L. 1996, \apj, 463, 498

\bibitem[{{Schnorr M{\"u}ller} {et~al.}(2011){Schnorr M{\"u}ller},
  {Storchi-Bergmann}, {Riffel}, {Ferrari}, {Steiner}, {Axon}, \&
  {Robinson}}]{schn11ifu}
{Schnorr M{\"u}ller}, A., {Storchi-Bergmann}, T., {Riffel}, R.~A., {Ferrari},
  F., {Steiner}, J.~E., {Axon}, D.~J., \& {Robinson}, A. 2011, \mnras, 413, 149

\bibitem[{{Sch{\"o}del} {et~al.}(2007){Sch{\"o}del}, {Eckart}, {Alexander},
  {Merritt}, {Genzel}, {Sternberg}, {Meyer}, {Kul}, {Moultaka}, {Ott}, \&
  {Straubmeier}}]{scho07nucl}
{Sch{\"o}del}, R., et~al.\ 2007, \aap, 469, 125

\bibitem[{{Schramm} {et~al.}(2008){Schramm}, {Wisotzki}, \&
  {Jahnke}}]{schr08qsohost}
{Schramm}, M., {Wisotzki}, L., \& {Jahnke}, K. 2008, \aap, 478, 311

\bibitem[{{Schweitzer} {et~al.}(2008){Schweitzer}, {Groves}, {Netzer}, {Lutz},
  {Sturm}, {Contursi}, {Genzel}, {Tacconi}, {Veilleux}, {Kim}, {Rupke}, \&
  {Baker}}]{schw08qsodust}
{Schweitzer}, M., et~al.\ 2008, \apj, 679, 101

\bibitem[{{Schweitzer} {et~al.}(2006){Schweitzer}, {Lutz}, {Sturm}, {Contursi},
  {Tacconi}, {Lehnert}, {Dasyra}, {Genzel}, {Veilleux}, {Rupke}, {Kim},
  {Baker}, {Netzer}, {Sternberg}, {Mazzarella}, \& {Lord}}]{schw06qsosf}
{Schweitzer}, M., et~al.\ 2006, \apj, 649, 79

\bibitem[{{Sereno} {et~al.}(2011){Sereno}, {Jetzer}, {Sesana}, \&
  {Volonteri}}]{sere11lisa}
{Sereno}, M., {Jetzer}, P., {Sesana}, A., \& {Volonteri}, M. 2011, \mnras, 415,
  2773

\bibitem[{{Serjeant} \& {Hatziminaoglou}(2009)}]{serj09qsosf}
{Serjeant}, S. \& {Hatziminaoglou}, E. 2009, \mnras, 397, 265

\bibitem[{{Sesana} {et~al.}(2007){Sesana}, {Volonteri}, \&
  {Haardt}}]{sesa07bhlisa}
{Sesana}, A., {Volonteri}, M., \& {Haardt}, F. 2007, \mnras, 377, 1711

\bibitem[{{Seymour} {et~al.}(2011){Seymour}, {Symeonidis}, {Page}, {Amblard},
  {Arumugam}, {Aussel}, {Blain}, {Bock}, {Boselli}, {Buat},
  {Castro-Rodr{\'{\i}}guez}, {Cava}, {Chanial}, {Clements}, {Conley},
  {Conversi}, {Cooray}, {Dowell}, {Dwek}, {Eales}, {Elbaz}, {Franceschini},
  {Glenn}, {Solares}, {Griffin}, {Hatziminaoglou}, {Ibar}, {Isaak}, {Ivison},
  {Lagache}, {Levenson}, {Lu}, {Madden}, {Maffei}, {Mainetti}, {Marchetti},
  {Nguyen}, {O'Halloran}, {Oliver}, {Omont}, {Panuzzo}, {Papageorgiou},
  {Pearson}, {P{\'e}rez-Fournon}, {Pohlen}, {Rawlings}, {Rizzo}, {Roseboom},
  {Rowan-Robinson}, {Schulz}, {Scott}, {Shupe}, {Smith}, {Stevens}, {Trichas},
  {Tugwell}, {Vaccari}, {Valtchanov}, {Vigroux}, {Wang}, {Wright}, {Xu}, \&
  {Zemcov}}]{seym11radiosf}
{Seymour}, N., et~al.\ 2011, \mnras, 413, 1777

\bibitem[{{Shakura} \& {Sunyaev}(1973)}]{shak73}
{Shakura}, N.~I. \& {Sunyaev}, R.~A. 1973, \aap, 24, 337

\bibitem[{{Shang} {et~al.}(2010){Shang}, {Bryan}, \& {Haiman}}]{shan10seedbh}
{Shang}, C., {Bryan}, G.~L., \& {Haiman}, Z. 2010, \mnras, 402, 1249

\bibitem[{{Shankar}(2009)}]{shank09}
{Shankar}, F. 2009, New Astronomy Reviews, 53, 57

\bibitem[{{Shankar} {et~al.}(2004){Shankar}, {Salucci}, {Granato}, {De Zotti},
  \& {Danese}}]{shan04agnbh}
{Shankar}, F., {Salucci}, P., {Granato}, G.~L., {De Zotti}, G., \& {Danese}, L.
  2004, \mnras, 354, 1020

\bibitem[{{Shankar} {et~al.}(2009){Shankar}, {Weinberg}, \&
  {Miralda-Escud{\'e}}}]{shan09agnbh}
{Shankar}, F., {Weinberg}, D.~H., \& {Miralda-Escud{\'e}}, J. 2009, \apj, 690,
  20

\bibitem[{{Shao} {et~al.}(2010){Shao}, {Lutz}, {Nordon}, {Maiolino},
  {Alexander}, {Altieri}, {Andreani}, {Aussel}, {Bauer}, {Berta},
  {Bongiovanni}, {Brandt}, {Brusa}, {Cava}, {Cepa}, {Cimatti}, {Daddi},
  {Dominguez-Sanchez}, {Elbaz}, {F{\"o}rster Schreiber}, {Geis}, {Genzel},
  {Grazian}, {Gruppioni}, {Magdis}, {Magnelli}, {Mainieri}, {P{\'e}rez
  Garc{\'{\i}}a}, {Poglitsch}, {Popesso}, {Pozzi}, {Riguccini}, {Rodighiero},
  {Rovilos}, {Saintonge}, {Salvato}, {Sanchez Portal}, {Santini}, {Sturm},
  {Tacconi}, {Valtchanov}, {Wetzstein}, \& {Wieprecht}}]{shao10agnsf}
{Shao}, L., et~al.\ 2010, \aap, 518, L26+

\bibitem[{{Shapiro} \& {Teukolsky}(1983)}]{shap83}
{Shapiro}, S.~L. \& {Teukolsky}, S.~A. 1983, {Black holes, white dwarfs, and
  neutron stars: The physics of compact objects} (New York: Wiley-Interscience)

\bibitem[{{Shen} {et~al.}(2008){Shen}, {Greene}, {Strauss}, {Richards}, \&
  {Schneider}}]{shen08bhmass}
{Shen}, Y., {Greene}, J.~E., {Strauss}, M.~A., {Richards}, G.~T., \&
  {Schneider}, D.~P. 2008, \apj, 680, 169

\bibitem[{{Shen} \& {Kelly}(2010)}]{shen10mbhbias}
{Shen}, Y. \& {Kelly}, B.~C. 2010, \apj, 713, 41

\bibitem[{{Shen} {et~al.}(2011){Shen}, {Liu}, {Greene}, \&
  {Strauss}}]{shen11doubleagn}
{Shen}, Y., {Liu}, X., {Greene}, J.~E., \& {Strauss}, M.~A. 2011, \apj, 735, 48

\bibitem[{{Shen} {et~al.}(2007){Shen}, {Strauss}, {Oguri}, {Hennawi}, {Fan},
  {Richards}, {Hall}, {Gunn}, {Schneider}, {Szalay}, {Thakar}, {Vanden Berk},
  {Anderson}, {Bahcall}, {Connolly}, \& {Knapp}}]{shen07clust}
{Shen}, Y., et~al.\ 2007, \aj, 133, 2222

\bibitem[{{Shi} {et~al.}(2008){Shi}, {Rieke}, {Donley}, {Cooper}, {Willmer}, \&
  {Kirby}}]{shi08xhost}
{Shi}, Y., {Rieke}, G., {Donley}, J., {Cooper}, M., {Willmer}, C., \& {Kirby},
  E. 2008, \apj, 688, 794

\bibitem[{{Shi} {et~al.}(2009){Shi}, {Rieke}, {Ogle}, {Jiang}, \&
  {Diamond-Stanic}}]{shi09qsosf}
{Shi}, Y., {Rieke}, G.~H., {Ogle}, P., {Jiang}, L., \& {Diamond-Stanic}, A.~M.
  2009, \apj, 703, 1107

\bibitem[{{Shields} {et~al.}(2007){Shields}, {Rix}, {Sarzi}, {Barth},
  {Filippenko}, {Ho}, {McIntosh}, {Rudnick}, \& {Sargent}}]{shie07line}
{Shields}, J.~C., et~al.\ 2007, \apj, 654, 125

\bibitem[{{Shlosman} {et~al.}(1990){Shlosman}, {Begelman}, \&
  {Frank}}]{shlo90fuel}
{Shlosman}, I., {Begelman}, M.~C., \& {Frank}, J. 1990, \nat, 345, 679

\bibitem[{{Shlosman} {et~al.}(1989){Shlosman}, {Frank}, \&
  {Begelman}}]{shlo89bars}
{Shlosman}, I., {Frank}, J., \& {Begelman}, M.~C. 1989, \nat, 338, 45

\bibitem[{{Siemiginowska} {et~al.}(2010){Siemiginowska}, {Burke}, {Aldcroft},
  {Worrall}, {Allen}, {Bechtold}, {Clarke}, \& {Cheung}}]{siem10qso}
{Siemiginowska}, A., {Burke}, D.~J., {Aldcroft}, T.~L., {Worrall}, D.~M.,
  {Allen}, S., {Bechtold}, J., {Clarke}, T., \& {Cheung}, C.~C. 2010, \apj,
  722, 102

\bibitem[{{Sijacki} {et~al.}(2007){Sijacki}, {Springel}, {Di Matteo}, \&
  {Hernquist}}]{sija07agnfeed}
{Sijacki}, D., {Springel}, V., {Di Matteo}, T., \& {Hernquist}, L. 2007,
  \mnras, 380, 877

\bibitem[{{Silk} \& {Rees}(1998)}]{silk98}
{Silk}, J. \& {Rees}, M.~J. 1998, \aap, 331, L1

\bibitem[{{Silverman} {et~al.}(2008{\natexlab{a}}){Silverman}, {Green},
  {Barkhouse}, {Kim}, {Kim}, {Wilkes}, {Cameron}, {Hasinger}, {Jannuzi},
  {Smith}, {Smith}, \& {Tananbaum}}]{silv08xlf}
{Silverman}, J.~D., et~al.\ 2008{\natexlab{a}}, \apj,  679, 118

\bibitem[{{Silverman} {et~al.}(2011){Silverman}, {Kampczyk}, {Jahnke},
  {Andrae}, {Lilly}, {Elvis}, {Civano}, {Mainieri}, {Vignali}, {Zamorani},
  {Nair}, {Le Fevre}, {de Ravel}, {Bardelli}, {Bongiorno}, {Bolzonella},
  {Brusa}, {Cappelluti}, {Cappi}, {Caputi}, {Carollo}, {Contini}, {Coppa},
  {Cucciati}, {de la Torre}, {Franzetti}, {Garilli}, {Halliday}, {Hasinger},
  {Iovino}, {Knobel}, {koekemoer}, {Kovac}, {Lamareille}, {Le Borgne}, {Le
  Brun}, {Maier}, {Mignoli}, {Pello}, {Perez Montero}, {Ricciardelli}, {Peng},
  {Scodeggio}, {Tanaka}, {Tasca}, {Tresse}, {Vergani}, {Zucca}, {Comastri},
  {Finoguenov}, {Fu}, {Gilli}, {Hao}, {Ho}, \& {Salvato}}]{silv11agninter_aph}
{Silverman}, J.~D., et~al.\ 2011, \apj\ in press (arXiv:1109.1292)

\bibitem[{{Silverman} {et~al.}(2009{\natexlab{a}}){Silverman}, {Kova{\v c}},
  {Knobel}, {Lilly}, {Bolzonella}, {Lamareille}, {Mainieri}, {Brusa},
  {Cappelluti}, {Peng}, {Hasinger}, {Zamorani}, {Scodeggio}, {Contini},
  {Carollo}, {Jahnke}, {Kneib}, {Le Fevre}, {Bardelli}, {Bongiorno}, {Brunner},
  {Caputi}, {Civano}, {Comastri}, {Coppa}, {Cucciati}, {de la Torre}, {de
  Ravel}, {Elvis}, {Finoguenov}, {Fiore}, {Franzetti}, {Garilli}, {Gilli},
  {Griffiths}, {Iovino}, {Kampczyk}, {Koekemoer}, {Le Borgne}, {Le Brun},
  {Maier}, {Mignoli}, {Pello}, {Perez Montero}, {Ricciardelli}, {Tanaka},
  {Tasca}, {Tresse}, {Vergani}, {Vignali}, {Zucca}, {Bottini}, {Cappi},
  {Cassata}, {Marinoni}, {McCracken}, {Memeo}, {Meneux}, {Oesch}, {Porciani},
  \& {Salvato}}]{silv09agnenv}
{Silverman}, J.~D., et~al.\ 2009{\natexlab{a}}, \apj, 695,  171

\bibitem[{{Silverman} {et~al.}(2009{\natexlab{b}}){Silverman}, {Lamareille},
  {Maier}, {Lilly}, {Mainieri}, {Brusa}, {Cappelluti}, {Hasinger}, {Zamorani},
  {Scodeggio}, {Bolzonella}, {Contini}, {Carollo}, {Jahnke}, {Kneib},
  {LeF{\`e}vre}, {Merloni}, {Bardelli}, {Bongiorno}, {Brunner}, {Caputi},
  {Civano}, {Comastri}, {Coppa}, {Cucciati}, {de la Torre}, {de Ravel},
  {Elvis}, {Finoguenov}, {Fiore}, {Franzetti}, {Garilli}, {Gilli}, {Iovino},
  {Kampczyk}, {Knobel}, {Kova{\v c}}, {LeBorgne}, {LeBrun}, {Mignoli}, {Pello},
  {Peng}, {Montero}, {Ricciardelli}, {Tanaka}, {Tasca}, {Tresse}, {Vergani},
  {Vignali}, {Zucca}, {Bottini}, {Cappi}, {Cassata}, {Fumana}, {Griffiths},
  {Kartaltepe}, {Koekemoer}, {Marinoni}, {McCracken}, {Memeo}, {Meneux},
  {Oesch}, {Porciani}, \& {Salvato}}]{silv09xcosmos_env}
{Silverman}, J.~D., et~al.\ 2009{\natexlab{b}}, \apj, 696, 396

\bibitem[{{Silverman} {et~al.}(2008{\natexlab{b}}){Silverman}, {Mainieri},
  {Lehmer}, {Alexander}, {Bauer}, {Bergeron}, {Brandt}, {Gilli}, {Hasinger},
  {Schneider}, {Tozzi}, {Vignali}, {Koekemoer}, {Miyaji}, {Popesso}, {Rosati},
  \& {Szokoly}}]{silv08host}
{Silverman}, J.~D., et~al.\ 2008{\natexlab{b}}, \apj,  675, 1025

\bibitem[{{Sim{\~o}es Lopes} {et~al.}(2007){Sim{\~o}es Lopes},
  {Storchi-Bergmann}, {de F{\'a}tima Saraiva}, \& {Martini}}]{simo07bhdust}
{Sim{\~o}es Lopes}, R.~D., {Storchi-Bergmann}, T., {de F{\'a}tima Saraiva}, M.,
  \& {Martini}, P. 2007, \apj, 655, 718

\bibitem[{{Simmons} {et~al.}(2011){Simmons}, {Van Duyne}, {Urry}, {Treister},
  {Koekemoer}, {Grogin}, \& {The GOODS Team}}]{simm11edd}
{Simmons}, B.~D., {Van Duyne}, J., {Urry}, C.~M., {Treister}, E., {Koekemoer},
  A.~M., {Grogin}, N.~A., \& {The GOODS Team}. 2011, \apj, 734, 121

\bibitem[{{Simpson}(2005)}]{simp05agn}
{Simpson}, C. 2005, \mnras, 360, 565

\bibitem[{{Smith} {et~al.}(2004){Smith}, {Robinson}, {Alexander}, {Young},
  {Axon}, \& {Corbett}}]{smit04seyfpol}
{Smith}, J.~E., {Robinson}, A., {Alexander}, D.~M., {Young}, S., {Axon}, D.~J.,
  \& {Corbett}, E.~A. 2004, \mnras, 350, 140

\bibitem[{{Smith} {et~al.}(2010){Smith}, {Shields}, {Bonning}, {McMullen},
  {Rosario}, \& {Salviander}}]{smith10binary}
{Smith}, K.~L., {Shields}, G.~A., {Bonning}, E.~W., {McMullen}, C.~C.,
  {Rosario}, D.~J., \& {Salviander}, S. 2010, \apj, 716, 866

\bibitem[{{Smith} {et~al.}(2009){Smith}, {Lucey}, \& {Hudson}}]{smit09galage}
{Smith}, R.~J., {Lucey}, J.~R., \& {Hudson}, M.~J. 2009, \mnras, 400, 1690

\bibitem[{{Smol{\v c}i{\'c}}(2009)}]{smol09radio}
{Smol{\v c}i{\'c}}, V. 2009, \apjl, 699, L43

\bibitem[{{Smol{\v c}i{\'c}} {et~al.}(2011){Smol{\v c}i{\'c}}, {Finoguenov},
  {Zamorani}, {Schinnerer}, {Tanaka}, {Giodini}, \&
  {Scoville}}]{smol11radiohod}
{Smol{\v c}i{\'c}}, V., {Finoguenov}, A., {Zamorani}, G., {Schinnerer}, E.,
  {Tanaka}, M., {Giodini}, S., \& {Scoville}, N. 2011, \mnras, 416, L31

\bibitem[{{Smol{\v c}i{\'c}} \& {Riechers}(2011)}]{smol11radiomol}
{Smol{\v c}i{\'c}}, V. \& {Riechers}, D.~A. 2011, \apj, 730, 64

\bibitem[{{Smol{\v c}i{\'c}} {et~al.}(2009){Smol{\v c}i{\'c}}, {Zamorani},
  {Schinnerer}, {Bardelli}, {Bondi}, {B{\^i}rzan}, {Carilli}, {Ciliegi},
  {Elvis}, {Impey}, {Koekemoer}, {Merloni}, {Paglione}, {Salvato}, {Scodeggio},
  {Scoville}, \& {Trump}}]{smol09radioevol}
{Smol{\v c}i{\'c}}, V., et~al.\ 2009, \apj, 696, 24

\bibitem[{{Sobolewska} {et~al.}(2011){Sobolewska}, {Siemiginowska}, \&
  {Gierli{\'n}ski}}]{sobo11agnstate}
{Sobolewska}, M.~A., {Siemiginowska}, A., \& {Gierli{\'n}ski}, M. 2011, \mnras,
  413, 2259

\bibitem[{{Soltan}(1982)}]{solt82}
{Soltan}, A. 1982, \mnras, 200, 115

\bibitem[{{Somerville} {et~al.}(2008){Somerville}, {Hopkins}, {Cox},
  {Robertson}, \& {Hernquist}}]{some08bhev}
{Somerville}, R.~S., {Hopkins}, P.~F., {Cox}, T.~J., {Robertson}, B.~E., \&
  {Hernquist}, L. 2008, \mnras, 391, 481

\bibitem[{{Spoon} {et~al.}(2004){Spoon}, {Moorwood}, {Lutz}, {Tielens},
  {Siebenmorgen}, \& {Keane}}]{spoo04arp220}
{Spoon}, H.~W.~W., {Moorwood}, A.~F.~M., {Lutz}, D., {Tielens}, A.~G.~G.~M.,
  {Siebenmorgen}, R., \& {Keane}, J.~V. 2004, \aap, 414, 873

\bibitem[{{Springel} {et~al.}(2005{\natexlab{a}}){Springel}, {Di Matteo}, \&
  {Hernquist}}]{spri05merge}
{Springel}, V., {Di Matteo}, T., \& {Hernquist}, L. 2005{\natexlab{a}}, \apjl,
  620, L79

\bibitem[{{Springel} {et~al.}(2005{\natexlab{b}}){Springel}, {Di Matteo}, \&
  {Hernquist}}]{spri05}
---. 2005{\natexlab{b}}, \mnras, 361, 776

\bibitem[{{Stacy} {et~al.}(2010){Stacy}, {Greif}, \& {Bromm}}]{stac10pop3}
{Stacy}, A., {Greif}, T.~H., \& {Bromm}, V. 2010, \mnras, 403, 45

\bibitem[{{Steidel} {et~al.}(1998){Steidel}, {Adelberger}, {Dickinson},
  {Giavalisco}, {Pettini}, \& {Kellogg}}]{stei98ssa}
{Steidel}, C.~C., {Adelberger}, K.~L., {Dickinson}, M., {Giavalisco}, M.,
  {Pettini}, M., \& {Kellogg}, M. 1998, \apj, 492, 428

\bibitem[{{Steidel} {et~al.}(2002){Steidel}, {Hunt}, {Shapley}, {Adelberger},
  {Pettini}, {Dickinson}, \& {Giavalisco}}]{stei02agn}
{Steidel}, C.~C., {Hunt}, M.~P., {Shapley}, A.~E., {Adelberger}, K.~L.,
  {Pettini}, M., {Dickinson}, M., \& {Giavalisco}, M. 2002, \apj, 576, 653

\bibitem[{{Stern} {et~al.}(2005){Stern}, {Eisenhardt}, {Gorjian}, {Kochanek},
  {Caldwell}, {Eisenstein}, {Brodwin}, {Brown}, {Cool}, {Dey}, {Green},
  {Jannuzi}, {Murray}, {Pahre}, \& {Willner}}]{ster05}
{Stern}, D., et~al.\ 2005, \apj, 631, 163

\bibitem[{{Stevens} {et~al.}(2005){Stevens}, {Page}, {Ivison}, {Carrera},
  {Mittaz}, {Smail}, \& {McHardy}}]{stev05submm}
{Stevens}, J.~A., {Page}, M.~J., {Ivison}, R.~J., {Carrera}, F.~J., {Mittaz},
  J.~P.~D., {Smail}, I., \& {McHardy}, I.~M. 2005, \mnras, 360, 610

\bibitem[{{Stobbart} {et~al.}(2006){Stobbart}, {Roberts}, \&
  {Wilms}}]{stob06ulx}
{Stobbart}, A.-M., {Roberts}, T.~P., \& {Wilms}, J. 2006, \mnras, 368, 397

\bibitem[{{Stoklasov{\'a}} {et~al.}(2009){Stoklasov{\'a}}, {Ferruit},
  {Emsellem}, {Jungwiert}, {P{\'e}contal}, \& {S{\'a}nchez}}]{stok09ifu}
{Stoklasov{\'a}}, I., {Ferruit}, P., {Emsellem}, E., {Jungwiert}, B.,
  {P{\'e}contal}, E., \& {S{\'a}nchez}, S.~F. 2009, \aap, 500, 1287

\bibitem[{{Storchi-Bergmann} {et~al.}(2007){Storchi-Bergmann}, {Dors},
  {Riffel}, {Fathi}, {Axon}, {Robinson}, {Marconi}, \&
  {{\"O}stlin}}]{stor07ifu}
{Storchi-Bergmann}, T., {Dors}, Jr., O.~L., {Riffel}, R.~A., {Fathi}, K.,
  {Axon}, D.~J., {Robinson}, A., {Marconi}, A., \& {{\"O}stlin}, G. 2007, \apj,
  670, 959

\bibitem[{{Storchi-Bergmann} {et~al.}(2001){Storchi-Bergmann}, {Gonz{\'a}lez
  Delgado}, {Schmitt}, {Cid Fernandes}, \& {Heckman}}]{stor01seyf}
{Storchi-Bergmann}, T., {Gonz{\'a}lez Delgado}, R.~M., {Schmitt}, H.~R., {Cid
  Fernandes}, R., \& {Heckman}, T. 2001, \apj, 559, 147

\bibitem[{{Storchi-Bergmann} {et~al.}(2010){Storchi-Bergmann}, {Lopes},
  {McGregor}, {Riffel}, {Beck}, \& {Martini}}]{stor10ngc4151}
{Storchi-Bergmann}, T., {Lopes}, R.~D.~S., {McGregor}, P.~J., {Riffel}, R.~A.,
  {Beck}, T., \& {Martini}, P. 2010, \mnras, 402, 819

\bibitem[{{Storchi-Bergmann} {et~al.}(2005){Storchi-Bergmann}, {Nemmen},
  {Spinelli}, {Eracleous}, {Wilson}, {Filippenko}, \& {Livio}}]{stor05ngc1097}
{Storchi-Bergmann}, T., {Nemmen}, R.~S., {Spinelli}, P.~F., {Eracleous}, M.,
  {Wilson}, A.~S., {Filippenko}, A.~V., \& {Livio}, M. 2005, \apjl, 624, L13

\bibitem[{{Strateva} {et~al.}(2001){Strateva}, {Ivezi{\'c}}, {Knapp},
  {Narayanan}, {Strauss}, {Gunn}, {Lupton}, {Schlegel}, {Bahcall}, {Brinkmann},
  {Brunner}, {Budav{\'a}ri}, {Csabai}, {Castander}, {Doi}, {Fukugita}, {Gy{\H
  o}ry}, {Hamabe}, {Hennessy}, {Ichikawa}, {Kunszt}, {Lamb}, {McKay},
  {Okamura}, {Racusin}, {Sekiguchi}, {Schneider}, {Shimasaku}, \&
  {York}}]{stra01galcol}
{Strateva}, I., et~al.\ 2001, \aj, 122, 1861

\bibitem[{{Sturm} {et~al.}(2011){Sturm}, {Gonz{\'a}lez-Alfonso}, {Veilleux},
  {Fischer}, {Graci{\'a}-Carpio}, {Hailey-Dunsheath}, {Contursi}, {Poglitsch},
  {Sternberg}, {Davies}, {Genzel}, {Lutz}, {Tacconi}, {Verma}, {Maiolino}, \&
  {de Jong}}]{stur11qsoco}
{Sturm}, E., et~al.\ 2011, \apjl, 733, L16

\bibitem[{{Sulentic} {et~al.}(2000){Sulentic}, {Zwitter}, {Marziani}, \&
  {Dultzin-Hacyan}}]{sule00eigen}
{Sulentic}, J.~W., {Zwitter}, T., {Marziani}, P., \& {Dultzin-Hacyan}, D. 2000,
  \apjl, 536, L5

\bibitem[{{Sun} {et~al.}(2009){Sun}, {Voit}, {Donahue}, {Jones}, {Forman}, \&
  {Vikhlinin}}]{sun09groups}
{Sun}, M., {Voit}, G.~M., {Donahue}, M., {Jones}, C., {Forman}, W., \&
  {Vikhlinin}, A. 2009, \apj, 693, 1142

\bibitem[{{Symeonidis} {et~al.}(2010){Symeonidis}, {Rosario}, {Georgakakis},
  {Harker}, {Laird}, {Page}, \& {Willmer}}]{syme10spitzer}
{Symeonidis}, M., {Rosario}, D., {Georgakakis}, A., {Harker}, J., {Laird},
  E.~S., {Page}, M.~J., \& {Willmer}, C.~N.~A. 2010, \mnras, 403, 1474

\bibitem[{{Szokoly} {et~al.}(2004){Szokoly}, {Bergeron}, {Hasinger}, {Lehmann},
  {Kewley}, {Mainieri}, {Nonino}, {Rosati}, {Giacconi}, {Gilli}, {Gilmozzi},
  {Norman}, {Romaniello}, {Schreier}, {Tozzi}, {Wang}, {Zheng}, \&
  {Zirm}}]{szok04}
{Szokoly}, G.~P., et~al.\ 2004, \apjs, 155, 271

\bibitem[{{Tacconi} {et~al.}(2010){Tacconi}, {Genzel}, {Neri}, {Cox}, {Cooper},
  {Shapiro}, {Bolatto}, {Bouch{\'e}}, {Bournaud}, {Burkert}, {Combes},
  {Comerford}, {Davis}, {Schreiber}, {Garcia-Burillo}, {Gracia-Carpio}, {Lutz},
  {Naab}, {Omont}, {Shapley}, {Sternberg}, \& {Weiner}}]{tacc10fgas}
{Tacconi}, L.~J., et~al.\ 2010, \nat, 463, 781

\bibitem[{{Tacconi} {et~al.}(2008){Tacconi}, {Genzel}, {Smail}, {Neri},
  {Chapman}, {Ivison}, {Blain}, {Cox}, {Omont}, {Bertoldi}, {Greve},
  {F{\"o}rster Schreiber}, {Genel}, {Lutz}, {Swinbank}, {Shapley}, {Erb},
  {Cimatti}, {Daddi}, \& {Baker}}]{tacc08smg}
{Tacconi}, L.~J., et~al.\ 2008, \apj, 680, 246

\bibitem[{{Tasse} {et~al.}(2008){Tasse}, {Best}, {R{\"o}ttgering}, \& {Le
  Borgne}}]{tass08radiomode}
{Tasse}, C., {Best}, P.~N., {R{\"o}ttgering}, H., \& {Le Borgne}, D. 2008,
  \aap, 490, 893

\bibitem[{{Tasse} {et~al.}(2011){Tasse}, {R{\"o}ttgering}, \&
  {Best}}]{tass11xenv}
{Tasse}, C., {R{\"o}ttgering}, H., \& {Best}, P.~N. 2011, \aap, 525, A127+

\bibitem[{{Teyssier} {et~al.}(2011){Teyssier}, {Moore}, {Martizzi}, {Dubois},
  \& {Mayer}}]{teys11agnfeed}
{Teyssier}, R., {Moore}, B., {Martizzi}, D., {Dubois}, Y., \& {Mayer}, L. 2011,
  \mnras, 414, 195

\bibitem[{{Thompson} {et~al.}(2005){Thompson}, {Quataert}, \&
  {Murray}}]{thom05rad}
{Thompson}, T.~A., {Quataert}, E., \& {Murray}, N. 2005, \apj, 630, 167

\bibitem[{{Tombesi} {et~al.}(2010){Tombesi}, {Cappi}, {Reeves}, {Palumbo},
  {Yaqoob}, {Braito}, \& {Dadina}}]{tomb10outflow}
{Tombesi}, F., {Cappi}, M., {Reeves}, J.~N., {Palumbo}, G.~G.~C., {Yaqoob}, T.,
  {Braito}, V., \& {Dadina}, M. 2010, \aap, 521, A57+

\bibitem[{{Tommasin} {et~al.}(2010){Tommasin}, {Spinoglio}, {Malkan}, \&
  {Fazio}}]{toma10irs}
{Tommasin}, S., {Spinoglio}, L., {Malkan}, M.~A., \& {Fazio}, G. 2010, \apj,
  709, 1257

\bibitem[{{Toomre}(1964)}]{toom64}
{Toomre}, A. 1964, \apj, 139, 1217

\bibitem[{{Tozzi} {et~al.}(2006){Tozzi}, {Gilli}, {Mainieri}, {Norman},
  {Risaliti}, {Rosati}, {Bergeron}, {Borgani}, {Giacconi}, {Hasinger},
  {Nonino}, {Streblyanska}, {Szokoly}, {Wang}, \& {Zheng}}]{tozz06}
{Tozzi}, P., et~al.\ 2006, \aap, 451, 457

\bibitem[{{Trakhtenbrot} {et~al.}(2011){Trakhtenbrot}, {Netzer}, {Lira}, \&
  {Shemmer}}]{trak11bhmass}
{Trakhtenbrot}, B., {Netzer}, H., {Lira}, P., \& {Shemmer}, O. 2011, \apj, 730,
  7

\bibitem[{{Tran} {et~al.}(2001){Tran}, {Lutz}, {Genzel}, {Rigopoulou}, {Spoon},
  {Sturm}, {Gerin}, {Hines}, {Moorwood}, {Sanders}, {Scoville}, {Taniguchi}, \&
  {Ward}}]{tran01ulirg}
{Tran}, Q.~D., et~al.\ 2001, \apj, 552, 527

\bibitem[{{Treister} {et~al.}(2009){Treister}, {Cardamone}, {Schawinski},
  {Urry}, {Gawiser}, {Virani}, {Lira}, {Kartaltepe}, {Damen}, {Taylor}, {Le
  Floc'h}, {Justham}, \& {Koekemoer}}]{trei09obscagn}
{Treister}, E., et~al.\ 2009, \apj, 706, 535

\bibitem[{{Treister} {et~al.}(2010{\natexlab{a}}){Treister}, {Natarajan},
  {Sanders}, {Urry}, {Schawinski}, \& {Kartaltepe}}]{trei10bhmerge}
{Treister}, E., {Natarajan}, P., {Sanders}, D.~B., {Urry}, C.~M., {Schawinski},
  K., \& {Kartaltepe}, J. 2010{\natexlab{a}}, Science, 328, 600

\bibitem[{{Treister} \& {Urry}(2006)}]{trei06evol}
{Treister}, E. \& {Urry}, C.~M. 2006, \apjl, 652, L79

\bibitem[{{Treister} {et~al.}(2010{\natexlab{b}}){Treister}, {Urry},
  {Schawinski}, {Cardamone}, \& {Sanders}}]{trei10obscagn}
{Treister}, E., {Urry}, C.~M., {Schawinski}, K., {Cardamone}, C.~N., \&
  {Sanders}, D.~B. 2010{\natexlab{b}}, \apjl, 722, L238

\bibitem[{{Tremaine} {et~al.}(2002){Tremaine}, {Gebhardt}, {Bender}, {Bower},
  {Dressler}, {Faber}, {Filippenko}, {Green}, {Grillmair}, {Ho}, {Kormendy},
  {Lauer}, {Magorrian}, {Pinkney}, \& {Richstone}}]{trem02}
{Tremaine}, S., et~al.\ 2002, \apj, 574, 740

\bibitem[{{Tremonti} {et~al.}(2007){Tremonti}, {Moustakas}, \&
  {Diamond-Stanic}}]{trem07psb}
{Tremonti}, C.~A., {Moustakas}, J., \& {Diamond-Stanic}, A.~M. 2007, \apjl,
  663, L77

\bibitem[{{Trenti} \& {Stiavelli}(2009)}]{tren09pop3}
{Trenti}, M. \& {Stiavelli}, M. 2009, \apj, 694, 879

\bibitem[{{Trichas} {et~al.}(2009){Trichas}, {Georgakakis}, {Rowan-Robinson},
  {Nandra}, {Clements}, \& {Vaccari}}]{tric09agnsf}
{Trichas}, M., {Georgakakis}, A., {Rowan-Robinson}, M., {Nandra}, K.,
  {Clements}, D., \& {Vaccari}, M. 2009, \mnras, 399, 663

\bibitem[{{Trippe} {et~al.}(2007){Trippe}, {Paumard}, {Ott}, {Gillessen},
  {Eisenhauer}, {Martins}, \& {Genzel}}]{trip07sgra}
{Trippe}, S., {Paumard}, T., {Ott}, T., {Gillessen}, S., {Eisenhauer}, F.,
  {Martins}, F., \& {Genzel}, R. 2007, \mnras, 375, 764

\bibitem[{{Tristram} {et~al.}(2007){Tristram}, {Meisenheimer}, {Jaffe},
  {Schartmann}, {Rix}, {Leinert}, {Morel}, {Wittkowski}, {R{\"o}ttgering},
  {Perrin}, {Lopez}, {Raban}, {Cotton}, {Graser}, {Paresce}, \&
  {Henning}}]{tris07torus}
{Tristram}, K.~R.~W., et~al.\ 2007, \aap, 474, 837

\bibitem[{{Tristram} {et~al.}(2009){Tristram}, {Raban}, {Meisenheimer},
  {Jaffe}, {R{\"o}ttgering}, {Burtscher}, {Cotton}, {Graser}, {Henning},
  {Leinert}, {Lopez}, {Morel}, {Perrin}, \& {Wittkowski}}]{tris09seyf}
{Tristram}, K.~R.~W., et~al.\ 2009, \aap, 502, 67

\bibitem[{{Trouille} {et~al.}(2011){Trouille}, {Barger}, \&
  {Tremonti}}]{trou11optx}
{Trouille}, L., {Barger}, A.~J., \& {Tremonti}, C. 2011, \apj, 742, 46

\bibitem[{{Tueller} {et~al.}(2010){Tueller}, {Baumgartner}, {Markwardt},
  {Skinner}, {Mushotzky}, {Ajello}, {Barthelmy}, {Beardmore}, {Brandt},
  {Burrows}, {Chincarini}, {Campana}, {Cummings}, {Cusumano}, {Evans},
  {Fenimore}, {Gehrels}, {Godet}, {Grupe}, {Holland}, {Kennea}, {Krimm},
  {Koss}, {Moretti}, {Mukai}, {Osborne}, {Okajima}, {Pagani}, {Page}, {Palmer},
  {Parsons}, {Schneider}, {Sakamoto}, {Sambruna}, {Sato}, {Stamatikos},
  {Stroh}, {Ukwata}, \& {Winter}}]{tuel10bat}
{Tueller}, J., et~al.\ 2010, \apjs, 186, 378

\bibitem[{{Tueller} {et~al.}(2008){Tueller}, {Mushotzky}, {Barthelmy},
  {Cannizzo}, {Gehrels}, {Markwardt}, {Skinner}, \& {Winter}}]{tuel08bat}
{Tueller}, J., {Mushotzky}, R.~F., {Barthelmy}, S., {Cannizzo}, J.~K.,
  {Gehrels}, N., {Markwardt}, C.~B., {Skinner}, G.~K., \& {Winter}, L.~M. 2008,
  \apj, 681, 113

\bibitem[{{Turk} {et~al.}(2009){Turk}, {Abel}, \& {O'Shea}}]{turk09pop3}
{Turk}, M.~J., {Abel}, T., \& {O'Shea}, B. 2009, Science, 325, 601

\bibitem[{{Turnbull} {et~al.}(1999){Turnbull}, {Bridges}, \&
  {Carter}}]{turn99shell}
{Turnbull}, A.~J., {Bridges}, T.~J., \& {Carter}, D. 1999, \mnras, 307, 967

\bibitem[{{Ueda} {et~al.}(2003){Ueda}, {Akiyama}, {Ohta}, \& {Miyaji}}]{ueda03}
{Ueda}, Y., {Akiyama}, M., {Ohta}, K., \& {Miyaji}, T. 2003, \apj, 598, 886

\bibitem[{{Ulrich} {et~al.}(1997){Ulrich}, {Maraschi}, \& {Urry}}]{ulri97}
{Ulrich}, M.-H., {Maraschi}, L., \& {Urry}, C.~M. 1997, \araa, 35, 445

\bibitem[{{Ulvestad} \& {Ho}(2002)}]{ulve02agn}
{Ulvestad}, J.~S. \& {Ho}, L.~C. 2002, \apj, 581, 925

\bibitem[{{Urrutia} {et~al.}(2008){Urrutia}, {Lacy}, \&
  {Becker}}]{urru08qsohost}
{Urrutia}, T., {Lacy}, M., \& {Becker}, R.~H. 2008, \apj, 674, 80

\bibitem[{{Urry} \& {Padovani}(1995)}]{urry95}
{Urry}, C.~M. \& {Padovani}, P. 1995, \pasp, 107, 803

\bibitem[{{Uttley} {et~al.}(2005){Uttley}, {McHardy}, \&
  {Vaughan}}]{uttl05xvar}
{Uttley}, P., {McHardy}, I.~M., \& {Vaughan}, S. 2005, \mnras, 359, 345

\bibitem[{{Valiante} {et~al.}(2007){Valiante}, {Lutz}, {Sturm}, {Genzel},
  {Tacconi}, {Lehnert}, \& {Baker}}]{vali07smg}
{Valiante}, E., {Lutz}, D., {Sturm}, E., {Genzel}, R., {Tacconi}, L.~J.,
  {Lehnert}, M.~D., \& {Baker}, A.~J. 2007, \apj, 660, 1060

\bibitem[{{van Wassenhove} {et~al.}(2010){van Wassenhove}, {Volonteri},
  {Walker}, \& {Gair}}]{vanw10bhrelic}
{van Wassenhove}, S., {Volonteri}, M., {Walker}, M.~G., \& {Gair}, J.~R. 2010,
  \mnras, 408, 1139

\bibitem[{{Vasudevan} \& {Fabian}(2007)}]{vasu07bolc}
{Vasudevan}, R.~V. \& {Fabian}, A.~C. 2007, \mnras, 381, 1235

\bibitem[{{Vasudevan} \& {Fabian}(2009)}]{vasu09sed}
---. 2009, \mnras, 392, 1124

\bibitem[{{Vaughan} {et~al.}(2003){Vaughan}, {Edelson}, {Warwick}, \&
  {Uttley}}]{vaug03xvar}
{Vaughan}, S., {Edelson}, R., {Warwick}, R.~S., \& {Uttley}, P. 2003, \mnras,
  345, 1271

\bibitem[{{Veilleux}(1991)}]{veil91seyfline}
{Veilleux}, S. 1991, \apjs, 75, 383

\bibitem[{{Veilleux} {et~al.}(2005){Veilleux}, {Cecil}, \&
  {Bland-Hawthorn}}]{veil05}
{Veilleux}, S., {Cecil}, G., \& {Bland-Hawthorn}, J. 2005, \araa, 43, 769

\bibitem[{{Veilleux} {et~al.}(2009){Veilleux}, {Kim}, {Rupke}, {Peng},
  {Tacconi}, {Genzel}, {Lutz}, {Sturm}, {Contursi}, {Schweitzer}, {Dasyra},
  {Ho}, {Sanders}, \& {Burkert}}]{veil09qsomorph}
{Veilleux}, S., et~al.\ 2009,  \apj, 701, 587

\bibitem[{{Veilleux} {et~al.}(1999){Veilleux}, {Kim}, \&
  {Sanders}}]{veil99ulirg}
{Veilleux}, S., {Kim}, D.-C., \& {Sanders}, D.~B. 1999, \apj, 522, 113

\bibitem[{{Veilleux} \& {Osterbrock}(1987)}]{veil87line}
{Veilleux}, S. \& {Osterbrock}, D.~E. 1987, \apjs, 63, 295

\bibitem[{{Vestergaard}(2002)}]{vest02bhmass}
{Vestergaard}, M. 2002, \apj, 571, 733

\bibitem[{{Vestergaard} \& {Peterson}(2006)}]{vest06}
{Vestergaard}, M. \& {Peterson}, B.~M. 2006, \apj, 641, 689

\bibitem[{{Vignali} {et~al.}(2010){Vignali}, {Alexander}, {Gilli}, \&
  {Pozzi}}]{vign10qso2}
{Vignali}, C., {Alexander}, D.~M., {Gilli}, R., \& {Pozzi}, F. 2010, \mnras,
  404, 48

\bibitem[{{Vignali} {et~al.}(2009){Vignali}, {Pozzi}, {Fritz}, {Comastri},
  {Gruppioni}, {Bellocchi}, {Fiore}, {Brusa}, {Maiolino}, {Mignoli}, {La
  Franca}, {Pozzetti}, {Zamorani}, \& {Merloni}}]{vign09qso2}
{Vignali}, C., et~al.\ 2009, \mnras,  395, 2189

\bibitem[{{Vollmer} {et~al.}(2008){Vollmer}, {Beckert}, \&
  {Davies}}]{voll08torus}
{Vollmer}, B., {Beckert}, T., \& {Davies}, R.~I. 2008, \aap, 491, 441

\bibitem[{{Volonteri}(2007)}]{volo07recoil}
{Volonteri}, M. 2007, \apjl, 663, L5

\bibitem[{{Volonteri}(2010)}]{volo10bhform}
---. 2010, \aapr, 18, 279

\bibitem[{{Volonteri} \& {Begelman}(2010)}]{volo10quasistar}
{Volonteri}, M. \& {Begelman}, M.~C. 2010, \mnras, 409, 1022

\bibitem[{{Volonteri} {et~al.}(2008){Volonteri}, {Lodato}, \&
  {Natarajan}}]{volo08bhseed}
{Volonteri}, M., {Lodato}, G., \& {Natarajan}, P. 2008, \mnras, 383, 1079

\bibitem[{{Volonteri} \& {Natarajan}(2009)}]{volo09lowmass}
{Volonteri}, M. \& {Natarajan}, P. 2009, \mnras, 400, 1911

\bibitem[{{Volonteri} {et~al.}(2011){Volonteri}, {Natarajan}, \&
  {G{\"u}ltekin}}]{volo11mbhmhalo}
{Volonteri}, M., {Natarajan}, P., \& {G{\"u}ltekin}, K. 2011, \apj, 737, 50

\bibitem[{{Volonteri} \& {Rees}(2005)}]{volo05highzbh}
{Volonteri}, M. \& {Rees}, M.~J. 2005, \apj, 633, 624

\bibitem[{{Volonteri} {et~al.}(2006){Volonteri}, {Salvaterra}, \&
  {Haardt}}]{volo06accr}
{Volonteri}, M., {Salvaterra}, R., \& {Haardt}, F. 2006, \mnras, 373, 121

\bibitem[{{Wada} \& {Norman}(2002)}]{wada02agnsf}
{Wada}, K. \& {Norman}, C.~A. 2002, \apjl, 566, L21

\bibitem[{{Wada} {et~al.}(2009){Wada}, {Papadopoulos}, \&
  {Spaans}}]{wada09agnmol}
{Wada}, K., {Papadopoulos}, P.~P., \& {Spaans}, M. 2009, \apj, 702, 63

\bibitem[{{Wake} {et~al.}(2008){Wake}, {Croom}, {Sadler}, \&
  {Johnston}}]{wake08radio}
{Wake}, D.~A., {Croom}, S.~M., {Sadler}, E.~M., \& {Johnston}, H.~M. 2008,
  \mnras, 391, 1674

\bibitem[{{Walter} {et~al.}(2004){Walter}, {Carilli}, {Bertoldi}, {Menten},
  {Cox}, {Lo}, {Fan}, \& {Strauss}}]{walt04qsoco}
{Walter}, F., {Carilli}, C., {Bertoldi}, F., {Menten}, K., {Cox}, P., {Lo},
  K.~Y., {Fan}, X., \& {Strauss}, M.~A. 2004, \apjl, 615, L17

\bibitem[{{Wardlow} {et~al.}(2011){Wardlow}, {Smail}, {Coppin}, {Alexander},
  {Brandt}, {Danielson}, {Luo}, {Swinbank}, {Walter}, {Wei{\ss}}, {Xue},
  {Zibetti}, {Bertoldi}, {Biggs}, {Chapman}, {Dannerbauer}, {Dunlop},
  {Gawiser}, {Ivison}, {Knudsen}, {Kov{\'a}cs}, {Lacey}, {Menten}, {Padilla},
  {Rix}, \& {van der Werf}}]{ward11less}
{Wardlow}, J.~L., et~al.\ 2011, \mnras, 415, 1479

\bibitem[{{Waskett} {et~al.}(2005){Waskett}, {Eales}, {Gear}, {McCracken},
  {Lilly}, \& {Brodwin}}]{wass05xenv}
{Waskett}, T.~J., {Eales}, S.~A., {Gear}, W.~K., {McCracken}, H.~J., {Lilly},
  S., \& {Brodwin}, M. 2005, \mnras, 363, 801

\bibitem[{{Watson} {et~al.}(2009){Watson}, {Schr{\"o}der}, {Fyfe}, {Page},
  {Lamer}, {Mateos}, {Pye}, {Sakano}, {Rosen}, {Ballet}, {Barcons}, {Barret},
  {Boller}, {Brunner}, {Brusa}, {Caccianiga}, {Carrera}, {Ceballos}, {Della
  Ceca}, {Denby}, {Denkinson}, {Dupuy}, {Farrell}, {Fraschetti}, {Freyberg},
  {Guillout}, {Hambaryan}, {Maccacaro}, {Mathiesen}, {McMahon}, {Michel},
  {Motch}, {Osborne}, {Page}, {Pakull}, {Pietsch}, {Saxton}, {Schwope},
  {Severgnini}, {Simpson}, {Sironi}, {Stewart}, {Stewart}, {Stobbart}, {Tedds},
  {Warwick}, {Webb}, {West}, {Worrall}, \& {Yuan}}]{wats09twoxmm}
{Watson}, M.~G., et~al.\ 2009,  \aap, 493, 339

\bibitem[{{Wild} {et~al.}(2010){Wild}, {Heckman}, \& {Charlot}}]{wild10agnsf}
{Wild}, V., {Heckman}, T., \& {Charlot}, S. 2010, \mnras, 405, 933

\bibitem[{{Wild} {et~al.}(2007){Wild}, {Kauffmann}, {Heckman}, {Charlot},
  {Lemson}, {Brinchmann}, {Reichard}, \& {Pasquali}}]{wild07agnsf}
{Wild}, V., {Kauffmann}, G., {Heckman}, T., {Charlot}, S., {Lemson}, G.,
  {Brinchmann}, J., {Reichard}, T., \& {Pasquali}, A. 2007, \mnras, 381, 543

\bibitem[{{Willott} {et~al.}(2010){Willott}, {Delorme}, {Reyl{\'e}}, {Albert},
  {Bergeron}, {Crampton}, {Delfosse}, {Forveille}, {Hutchings}, {McLure},
  {Omont}, \& {Schade}}]{will10z6qso}
{Willott}, C.~J., et~al.\ 2010, \aj, 139, 906

\bibitem[{{Willott} {et~al.}(2001){Willott}, {Rawlings}, \&
  {Blundell}}]{will01radiohost}
{Willott}, C.~J., {Rawlings}, S., \& {Blundell}, K.~M. 2001, \mnras, 324, 1

\bibitem[{{Wilson} {et~al.}(1988){Wilson}, {Ward}, \& {Haniff}}]{wils88seyf}
{Wilson}, A.~S., {Ward}, M.~J., \& {Haniff}, C.~A. 1988, \apj, 334, 121

\bibitem[{{Winter} {et~al.}(2010){Winter}, {Lewis}, {Koss}, {Veilleux},
  {Keeney}, \& {Mushotzky}}]{wint10batagn}
{Winter}, L.~M., {Lewis}, K.~T., {Koss}, M., {Veilleux}, S., {Keeney}, B., \&
  {Mushotzky}, R.~F. 2010, \apj, 710, 503

\bibitem[{{Winter} {et~al.}(2009){Winter}, {Mushotzky}, {Reynolds}, \&
  {Tueller}}]{wint09batagn}
{Winter}, L.~M., {Mushotzky}, R.~F., {Reynolds}, C.~S., \& {Tueller}, J. 2009,
  \apj, 690, 1322

\bibitem[{{Wise} {et~al.}(2008){Wise}, {Turk}, \& {Abel}}]{wise08protogal}
{Wise}, J.~H., {Turk}, M.~J., \& {Abel}, T. 2008, \apj, 682, 745

\bibitem[{{Woo} {et~al.}(2008){Woo}, {Treu}, {Malkan}, \&
  {Blandford}}]{woo08qsombh}
{Woo}, J., {Treu}, T., {Malkan}, M.~A., \& {Blandford}, R.~D. 2008, \apj, 681,
  925

\bibitem[{{Woo} {et~al.}(2010){Woo}, {Treu}, {Barth}, {Wright}, {Walsh},
  {Bentz}, {Martini}, {Bennert}, {Canalizo}, {Filippenko}, {Gates}, {Greene},
  {Li}, {Malkan}, {Stern}, \& {Minezaki}}]{woo10revmap}
{Woo}, J.-H., et~al.\ 2010, \apj, 716, 269

\bibitem[{{Xue} {et~al.}(2010){Xue}, {Brandt}, {Luo}, {Rafferty}, {Alexander},
  {Bauer}, {Lehmer}, {Schneider}, \& {Silverman}}]{xue10xhost}
{Xue}, Y.~Q., et~al.\ 2010, \apj, 720, 368

\bibitem[{{Xue} {et~al.}(2011){Xue}, {Luo}, {Brandt}, {Bauer}, {Lehmer},
  {Broos}, {Schneider}, {Alexander}, {Brusa}, {Comastri}, {Fabian}, {Gilli},
  {Hasinger}, {Hornschemeier}, {Koekemoer}, {Liu}, {Mainieri}, {Paolillo},
  {Rafferty}, {Rosati}, {Shemmer}, {Silverman}, {Smail}, {Tozzi}, \&
  {Vignali}}]{xue11cdfs}
{Xue}, Y.~Q., et~al.\ 2011, \apjs, 195, 10

\bibitem[{{Yamada} {et~al.}(2009){Yamada}, {Itoh}, {Makishima}, \&
  {Nakazawa}}]{yama09ngc4258}
{Yamada}, S., {Itoh}, T., {Makishima}, K., \& {Nakazawa}, K. 2009, \pasj, 61,
  309

\bibitem[{{Yan} {et~al.}(2011){Yan}, {Ho}, {Newman}, {Coil}, {Willmer},
  {Laird}, {Georgakakis}, {Aird}, {Barmby}, {Bundy}, {Cooper}, {Davis},
  {Faber}, {Fang}, {Griffith}, {Koekemoer}, {Koo}, {Nandra}, {Park},
  {Sarajedini}, {Weiner}, \& {Willner}}]{yan11aegis}
{Yan}, R., et~al.\ 2011, \apj, 728, 38

\bibitem[{{York} {et~al.}(2000){York}, {Adelman}, {Anderson}, {Anderson},
  {Annis}, {Bahcall}, {Bakken}, {Barkhouser}, {Bastian}, {Berman}, {Boroski},
  {Bracker}, {Briegel}, {Briggs}, {Brinkmann}, {Brunner}, {Burles}, {Carey},
  {Carr}, {Castander}, {Chen}, {Colestock}, {Connolly}, {Crocker}, {Csabai},
  {Czarapata}, {Davis}, {Doi}, {Dombeck}, {Eisenstein}, {Ellman}, {Elms},
  {Evans}, {Fan}, {Federwitz}, {Fiscelli}, {Friedman}, {Frieman}, {Fukugita},
  {Gillespie}, {Gunn}, {Gurbani}, {de Haas}, {Haldeman}, {Harris}, {Hayes},
  {Heckman}, {Hennessy}, {Hindsley}, {Holm}, {Holmgren}, {Huang}, {Hull},
  {Husby}, {Ichikawa}, {Ichikawa}, {Ivezi{\'c}}, {Kent}, {Kim}, {Kinney},
  {Klaene}, {Kleinman}, {Kleinman}, {Knapp}, {Korienek}, {Kron}, {Kunszt},
  {Lamb}, {Lee}, {Leger}, {Limmongkol}, {Lindenmeyer}, {Long}, {Loomis},
  {Loveday}, {Lucinio}, {Lupton}, {MacKinnon}, {Mannery}, {Mantsch}, {Margon},
  {McGehee}, {McKay}, {Meiksin}, {Merelli}, {Monet}, {Munn}, {Narayanan},
  {Nash}, {Neilsen}, {Neswold}, {Newberg}, {Nichol}, {Nicinski}, {Nonino},
  {Okada}, {Okamura}, {Ostriker}, {Owen}, {Pauls}, {Peoples}, {Peterson},
  {Petravick}, {Pier}, {Pope}, {Pordes}, {Prosapio}, {Rechenmacher}, {Quinn},
  {Richards}, {Richmond}, {Rivetta}, {Rockosi}, {Ruthmansdorfer}, {Sandford},
  {Schlegel}, {Schneider}, {Sekiguchi}, {Sergey}, {Shimasaku}, {Siegmund},
  {Smee}, {Smith}, {Snedden}, {Stone}, {Stoughton}, {Strauss}, {Stubbs},
  {SubbaRao}, {Szalay}, {Szapudi}, {Szokoly}, {Thakar}, {Tremonti}, {Tucker},
  {Uomoto}, {Vanden Berk}, {Vogeley}, {Waddell}, {Wang}, {Watanabe},
  {Weinberg}, {Yanny}, \& {Yasuda}}]{york00sdss}
{York}, D.~G., et~al.\ 2000, \aj, 120, 1579

\bibitem[{{Young} {et~al.}(2002){Young}, {Wilson}, {Terashima}, {Arnaud}, \&
  {Smith}}]{youn02cyga}
{Young}, A.~J., {Wilson}, A.~S., {Terashima}, Y., {Arnaud}, K.~A., \& {Smith},
  D.~A. 2002, \apj, 564, 176

\bibitem[{{Young}(2000)}]{youn00scat}
{Young}, S. 2000, \mnras, 312, 567

\bibitem[{{Young} {et~al.}(2007){Young}, {Axon}, {Robinson}, {Hough}, \&
  {Smith}}]{youn07qsowind}
{Young}, S., {Axon}, D.~J., {Robinson}, A., {Hough}, J.~H., \& {Smith}, J.~E.
  2007, \nat, 450, 74

\bibitem[{{Yuan} {et~al.}(2003){Yuan}, {Quataert}, \& {Narayan}}]{yuan03sgra}
{Yuan}, F., {Quataert}, E., \& {Narayan}, R. 2003, \apj, 598, 301

\bibitem[{{Yuan} {et~al.}(2010){Yuan}, {Kewley}, \& {Sanders}}]{yuan10lirg}
{Yuan}, T.-T., {Kewley}, L.~J., \& {Sanders}, D.~B. 2010, \apj, 709, 884

\bibitem[{{Zakamska} {et~al.}(2003){Zakamska}, {Strauss}, {Krolik}, {Collinge},
  {Hall}, {Hao}, {Heckman}, {Ivezi{\'c}}, {Richards}, {Schlegel}, {Schneider},
  {Strateva}, {Vanden Berk}, {Anderson}, \& {Brinkmann}}]{zaka03}
{Zakamska}, N.~L., et~al.\ 2003, \aj, 126,  2125

\bibitem[{{Zhang} {et~al.}(2009){Zhang}, {Soria}, {Zhang}, {Swartz}, \&
  {Liu}}]{zhang09census}
{Zhang}, W.~M., {Soria}, R., {Zhang}, S.~N., {Swartz}, D.~A., \& {Liu}, J.~F.
  2009, \apj, 699, 281

\bibitem[{{Zheng} {et~al.}(2004){Zheng}, {Mikles}, {Mainieri}, {Hasinger},
  {Rosati}, {Wolf}, {Norman}, {Szokoly}, {Gilli}, {Tozzi}, {Wang}, {Zirm}, \&
  {Giacconi}}]{zhen04}
{Zheng}, W., et~al.\ 2004, \apjs, 155, 73


\end{thebibliography}










\end{document}